\documentstyle[suthesis,12pt,a4,epsfig,wrapfig]{report}

\begin{document}

\newcommand{\lesssim}{{\raisebox{-.8ex}{$<$}\atop\raisebox{.4ex}{$\sim$}}}
\newcommand{\gtrsim}{{\raisebox{-.8ex}{$>$}\atop\raisebox{.4ex}{$\sim$}}}

\def\ra{\rightarrow}
\def\mco{\multicolumn}
\def\h{\hbar}
\def\th{\theta}
\def\b{\bar}
\def\U{\Upsilon}

\def\mb{\bar{\mu}}
\def\da{\downarrow}
\def\up{\uparrow}
\def\be{\begin{equation}}
\def\ee{\end{equation}}
\def\bea{\begin{eqnarray}}
\def\eea{\end{eqnarray}}
\def\benm{\begin{enumerate}}
\def\eenm{\end{enumerate}}
\def\eps{\varepsilon}
\def\g{\gamma}
\def\d{\delta}
\def\a{\alpha}
\def\e{\varepsilon}
\def\l{\lambda}
\def\lp{\lambda^\prime}
\def\o{\omega}
\def\ho{\hbar\omega}
\def\s{\sigma}
\def\La{\Lambda}
\def\nn{\nonumber\\}
\def\gt{\widetilde{G}}
\def\G{\Gamma}
\def\w{\langle W|}
\def\tA{\tilde A}
\def\tB{\tilde B}
\def\tr{\widetilde{\rho}}
\def\vr{\varrho}
\def\v{|V\rangle}
\def\wt{\langle {\tilde W}|}
\def\vt{|{\tilde V}\rangle}
\def\r#1{(\ref{#1})}
\def\la#1{\label{#1}}
\def\sm{{\bar Q}}
\def\cq{{\cal Q}}
\def\t#1{\langle\tau_{#1}\rangle}
\def\2t#1#2{\langle\tau_{#1}\tau_{#2}\rangle}
\def\up{\uparrow}
\def\1l{\lambda^{(1)}}
\def\lp{\Lambda_p}
\def\lh{\Lambda_h}
\def\ph{\lp,\lh}


\thispagestyle{empty}

\begin{center}
{{\Large\bf YEREVAN STATE UNIVERSITY}\\ 
MINISTRY OF EDUCATION AND SCIENCE, REPUBLIC OF ARMENIA}
\end{center}

\vskip 1.5cm


\vskip 2cm

\begin{center}
{\Large \bf {Badalian Samvel Michael}}
\end{center}
\vskip 1.5cm

\begin{center}
{\LARGE \bf Carrier Interaction in Quantum Nanosystems}
\end{center}
\vskip 1 cm

\noindent
{Speciality: {\bf A.04.10 -- Semiconductor and dielectric physics}}
\vskip 2 cm

\begin{center}
{\large A dissertation submitted to the $049$ Specialized Council of\\
Yerevan State University for the scientific degree of \\ 
\vskip 0.5 cm
{\bf Doctor of Physical and Mathematical Sciences}}
\end{center}

\vfill

\begin{center}
{\Large {\bf Yerevan -- 1997}}
\end{center}
\newpage
\noindent
The work is done at Yerevan State University

\vskip 1cm

\begin{tabular}{lp{3.2in}}
Official opponents&\\ 
&Dr. phys. math. sc. A A Kirakosyan\\
\\
&Dr. phys. math. sc. H R Minasyan\\
\\
&Dr. phys. math. sc. A Ya Shik\\
\\
\\
Leading institution& Institute for Radiophysics and Electronics,\\
& the National Academy of Sciences of Armenia\\
\\
\end{tabular}

\vskip 2cm

The public defense will take place on December 27, 1997 at 13:00 at the 
session of YSU Specialized Council $049$ (1 A Manoogian St., 
Yerevan, Armenia, 375049).

\vskip 2.5cm

The dissertation is available at YSU library.
\vskip 1.5cm 
The abstract of the dissertation is sent out on October 27, 1997.
\vfill

Secretary of the Specialized Council $049$,

cand. phys. math. sc. \hfill V P Qalantaryan

\newpage
\null\vfill
\begin{center}
{\Large\bf \copyright Copyright 1997\\ by \\ Badalian Samvel Michael}
\end{center}
\vfill
\newpage
%

\tableofcontents
\listoffigures
\listoftables
\newpage

\chapter{General}
\la{general}
\section{Introduction}
For decades, the microelectronics is the key-technology whose device
structures are widely adopted for industry production being in great 
market demand. Giant efforts and resources are directed to the growth
of the microelectronics
in such industry developed countries as Japan, Germany and the USA.
Alone in Germany, expenditures in the field of the microelectronics 
account for more than a third of the annual turnover ({\it circa}
600 milliard DM) with 3 million workers involved in this sphere
\cite{spiegel}.
Nowadays, development of submicron electronic structures is being
pursued in many countries because these structures give promise of new 
material systems with enhanced optical, transport and thermalization
properties that could make an impact in a variety of technologies,
including semiconductor lasers and modulators, detectors, and optical
filters.

Recent rapid advances in semiconductor technology of ultrafine 
lithography and modern etching technique have enable the fabrication
of a large diversity of ultra-narrow synthetic structures with perfect 
atomic interfaces separating different materials including silicon, III-V 
and II-IV semiconductors and others (\cite{esaki1}, for a review
see \cite{esaki2,capasso1,capasso2}). 
In these material systems, both composition and doping can be controlled
on a scale of the order of the de Broglie wavelength of carriers.
Such a class of artificial semiconductor nanostructures (contacts 
of metal-insulator-semiconductor types, heterojunctions, 
superlattices, {\it etc.})
with dimensions in the range of $1-100$ nm has opened up a new 
dimension in solid state and semiconductor physics in the regime when
the quantum size effect appears.
This new freedom in engineering the electronic states and their 
properties offer a feasibility to study physics laws in real low 
dimensional systems and an exciting potential for electronic
applications as well for investigations of new tantalizing phenomena 
principal for fundamental physics.

During the last decade, the semiconductor nanoscale systems with
carrier confinement in one (quantum wells [QW]), two (quantum 
wires [QWr]), and all three dimensions (quantum dots [QD]) have 
been studied intensively in theory and experiment 
\cite{afs,esipov,butcher}
to characterize, understand conceptually, and exploit quantum effects
in technologically important semiconductor device structures. 
Already, a set of novel results such as the weak and Anderson localization 
\cite{schmid1}, the effect of very high electron
mobility in modulation doped heterojunctions  \cite{kalfa,karpus1}
has been discovered in 2D nanosystems.
Quantum effects in 2D nanosystems have been exploited to provide
enhanced semiconductor lasers, a new class of infrared detectors, 
and resonant tunneling diodes.
The high electron mobility transistors on GaAs heterostructures are
used in a source-drain channel to construct high frequency amplifiers
with the limiting frequency above than $100$ GHz \cite{weisbuch}.

However, investigations of nanosystems with a two dimensional electron
gas (2DEG) is frequently connected with the use of high magnetic fields.
For example, high mobility samples with the 2DEG of nanoscale 
dimensions exposed to the normal high magnetic field exhibit
quantized resistance (the integer quantum Hall effect [QHE], 
\cite{klitzing}, for a review see \cite{hajdu}) and exotic charge 
correlation (the fractional QHE, \cite{tsui1,tsui2}, for a review see 
\cite{chakrobarty}). 
Understanding the integer and the fractional QHE has intrigued and 
challenged researchers for more than 10 years now, both for the wide 
range of new phenomena discovered \cite{landwehr88,landwehr92,landwehr96,prange,komhir} and for 
the potential for high precision metrology based on the QHE 
\cite{klitzing1,braun}.

Another key feature of the 2DEG in the QHE geometry is the
existence of quantum edge states localized in the vicinity of 
the sample boundary.
The quantization of the Hall resistance into the $h/e^2$ portions can be 
explained easily under assumption that the edge states are not
influenced by scattering and the transport is adiabatic 
(see reviews on the edge states and QHE \cite{prange,buttiker,haug}
and references cited therein).

QWrs and QDs give promise to provide even greater enhancement
in optical and relaxation properties of semiconductor device structures
because the wave functions are further compressed by lateral confinement 
and the quantum effects can be used to concentrate the optical oscillator
strengths at the active transitions \cite{bryant,wegscheider,sakaki95}. 
In this limit these structures behave as manmade artificial {\it quasi}atoms 
\cite{leo}. Furthermore, these structures under different new environments 
exhibit discrete charging effects and give promise of devices operating in
the limit of single electron transport. 

Carrier scattering is a fundamental process that always exist in crystals.
Characterizing and understanding carrier scattering processes in the
quantum nanosystems are critical both for unraveling the basic phenomena
of quantum effects in nanophysics and for controlling carrier dynamics 
in the nanoscale device structures, such as relaxation rates for carrier 
thermalization, equilibration and Auger scattering rates for carrier 
distribution, and diffusion rates for carrier transport. Carrier scattering
is strategic for identifying the lateral confinement effects and for 
developing enhanced, useful device structures. For example,
blueshifts in QWr photoluminescence are typically attributed to lateral
confinement. Recently, blueshifts observed in QWr 
magneto-photoluminescence
have been attributed to suppression of carrier transport and incomplete
relaxation in QWrs \cite{lehr} rather than to lateral confinement induced
level shifts. 
Small deep-dry-etched QDs often exhibit poor luminescence efficiency.
It has been suggested that this poor luminescence has an intrinsic origin, 
resulting from a suppression of phonon scattering in small QDs that leads
to a "phonon bottleneck" for electron relaxation \cite{bockelmann,%
benisty}. If this intrinsic mechanism for suppressing luminescence 
dominates extrinsic effects due to processing, then the possibilities for 
using the effectively 0D nanosystems (QDs, QWrs and 2D nanostructures 
in normal quantizing magnetic fields) in any application that relies on 
carrier relaxation, such as injection lasers, would be severely limited. 
Thus it is critical that the character of carrier scattering, relaxation and 
transport in effectively 0D nanostructures be established thoroughly to 
separate transport from confinement effects and better control and relieve
the suppression of transport.
Multiphonon processes \cite{sheregii} provide additional evidence 
for this phonon bottleneck in these 0D systems \cite{torres}.
It has been suggested that multiphonon processes \cite{inoshita} and 
Auger scattering \cite{bockel} can break this "bottleneck".
 in deep-etched QDs. 
Thus a careful analysis of all of these scattering mechanisms for low 
dimensional nanostructures is imperative.
Clearly, recent theoretical progress, and the many experiments in this hot 
topic will lead to a number of novel questions. 

In this work I present theoretical investigations of carrier scattering
in quantum nanosystems carried out during the last ten years by 
coworkers and me.
Carrier interaction with phonons, photons, impurities, and electrons 
in semiconductor nanoscale systems with carrier confinement in one 
and two dimensions in zero and quantizing magnetic fields has been 
addressed. Most importantly, our calculations allow to better understand
phonon signature in optical, thermalization, and transport experiments
that can be used to identify and characterize the basic phenomena of 
quantum confinement in these quantum nanostructures.

\section{Scientific novelty and practical value of the dissertation}
\la{novelty}
We have investigated peculiarities of the polaron (elementary excitations 
in the electron+phonon system) spectrum near
the longitudinal optical phonon emission threshold. In spite of weak 
electron-phonon coupling, we have obtained that in the 2DEG in the 
QHE geometry, new complex quasiparticles, electron-phonon bound 
states, appear in magneto-polaron spectrum. They constitute an infinite 
set which is coagulated to the threshold both above and below 
it  \cite{badalo2a,badalbs}.
In contrast to the virtual phonons taking part in the formation 
of the usual Fr\"olich polaron, phonons in the bound states are almost 
real. The characteristic scale of the binding energies is essentially 
greater than the corresponding scale in massive samples.

The existence of the electron-phonon bound states results to the fine 
structure of the cyclotron-phonon resonance \cite{badalcpr}.
According to the perturbation theory, photon absorption was to be 
expected at the phonon emission threshold. In reality, the perturbation 
theory becomes inapplicable in the immediate vicinity of the threshold. 
The true spectrum is obtained from the solution of an integral equation 
for the electron-phonon scattering amplitude.
Absorption entirely governed by the bound states with the total angular 
momentum $\pm1$. The absorption spectrum consists of two groups of 
peaks which are approximately of the same amplitude and are located 
approximately asymmetrically relative to the "threshold".

Our calculations explain the dynamics of carrier relaxation in the 
1D and 2D quantum nanosystems both in zero and quantizing magnetic 
fields. Understanding and controlling this dynamics is critical because 
rapid carrier relaxation is crucial for many of the technological 
applications proposed for semiconductor nanoscale quantum devices.
We have investigated thoroughly the main relaxation characteristics
of 1D and 2D electron systems due to acoustic phonon (deformation DA
and piezoelectric PA) and polar optical PO phonon scattering. 

Special attention has been paid to the presence of various interfaces
separating different materials in 2D nanostructures. These interfaces
influence the acoustic phonon normal modes and can affect essentially
electron-phonon interaction. We have proposed a new method for 
calculating the probability of electron scattering from the deformation
 potential of acoustic phonons \cite{badalinteff}. 
Such a probability summed over all phonon modes 
of the layered elastic medium can be expressed in terms of the elasticity
theory Green function which contains all information about structure
geometry.

Exploiting this method, the energy and momentum relaxation times of 
a test electron as well as the relaxation rate of electron temperature for 
the whole Fermi 2DEG located in the vicinity of an interface between 
elastic semi-spaces have been cablculated \cite{badalphd,badalsemispc}.
Analysis of limiting cases for an interface between solid and liquid 
semi-spaces, for a free and rigid surfaces has shown that there are situations 
when the phonon reflection from various interfaces alters the energy 
(or electron temperature) dependence of the relaxation times and 
leads to a strong reduction of the relaxation rates.

To illustrate the interface effect in quantizing magnetic fields, 
electron relaxation between discrete Landau levels in 2DEG
located near free crystal surface has been studied \cite{badaldon,badalfree}. 
The interface effect
is obtained to be highest in the magnetic field since the 2DEG interacts
with almost monochromatic cyclotron phonons in this case. The electron
transition probability has an oscillating behavior of the 
magnetic field and of the distance from the 2DEG to the interface.
Scattering from the deformation potential of acoustic phonons has been 
only considered since in quantizing magnetic fields, scattering from 
piezoelectric potential is strongly suppressed \cite{badalrspt}. 

Optical phonons determine the character of electron relaxation at
high temperatures and in various experiments with an optical excitation
or Auger processes. 
In the 2DEG subjected to strong magnetic fields 
with rather thin electron layers and subjected to rather strong magnetic 
fields, a large separation between Landau levels cannot be covered 
by an acoustical LA phonon \cite{badalfree} so the multiphonon 2LA 
\cite{falko} or an optical phonon assisted \cite{badalpoda} processes 
become more efficient.
Longitudinal optical phonon assisted inter Landau level transitions 
via one-phonon emission mechanism requires a precise resonance. 
Away from the resonance, efficiency of this process falls
steeply. Non-resonant optical phonon emission in the effectively 0D 
systems should be accompanied by acoustic phonon emission via 
the two-phonon emission mechanism. We have calculated polar 
optical PO phonon assisted electron relaxation as a function of the 
inter Landau level spacing in the 2DEG in the QHE geometry
 \cite{badalpoda}. The 
interface optical SO phonon relaxation has been found to be at least by
an order weaker than relaxation via polar optical PO phonon emission. 
To obtain a finite relaxation rate associated with one-phonon emission, 
the allowance for the Landau level broadening and for the PO phonon 
dispersion has been made.
Immediately below the phonon energy, $\hbar\omega_{PO}$,
the PO phonon dispersion contribution gives rise to a sharp peak with 
the peak value approximately $0.17$ fs$^{-1}$. The Landau level
broadening contribution has a rather broad peak with relatively lower 
peak value. Below $\h\omega_{PO}$ within an energy range of the order 
of $\hbar\sqrt{\omega_{B}/\tau}$, the one-phonon relaxation rate 
exceeds $1$ ps$^{-1}$ ($\tau$ is the relaxation time deduced from the 
mobility. In GaAs/AlGaAs with mobility 
$\mu=25$ V$^{-1}$ s$^{-1}$ m$^2$ this range makes up $0.7$ meV).

Two-phonon emission is a controlling relaxation mechanism above 
$\h\o_{PO}$ \cite{badalpoda}. For energies $\Delta l\h\o_{B}$ 
immediately above $\h\o_{PO}$, PO+DA phonon relaxation has a 
sharp onset. The relaxation rate increases as a fifth power in the 
magnetic field achieving to the peak value exceeding $1$
ps $^{-1}$ at energy separations of the order of $\h s/a_{B}$ 
($s$ is the sound velocity, $a_B$ is the magnetic length. 
In GaAs at $B=7$ T we have $\hbar s/a_{B}\approx 0.4$ meV). 
At higher magnetic fields in
the energy range $sa^{-1}_{B}\lesssim\Delta l\omega_{B}-\o_{PO}
\lesssim sd^{-1}$ (in GaAs with $d=3$ nm we have $\h s/d \approx 1.2$
meV), the two-phonon peak decreases linearly in the magnetic field. 
Above 
$\h\o_{PO}$ within the wide energy range (in GaAs this range makes
up approximately $5$ meV), the magnetic field dependence of
the relaxation rate is rather weak and the subnanosecond relaxation 
between Landau levels can be achieved via the two-phonon emission 
mechanism.
Our analysis has demonstrated that in some experimental situations the
PO+DA-phonon emission mechanism is more efficient than relaxation
in two consecutive emission acts: PO phonon emission (even under the 
sharp resonance) with subsequent emission of either LA- or 2LA-phonons.

As distinct from the conventional relaxation experiments where
the relaxation rates are measured directly, in the ballistic phonon
emission experiments, the intensity and the angular distribution of 
the phonon signal are detected on the sample reverse face which
provide a worthy source of information on electronic properties.
Such experiments are especially valuable in systems of reduced 
dimensionality since carriers are confined to small nanoscale regions 
while other quasiparticles such as phonons can be not localized and
more reachable for study.

In the dissertation I present calculations of ballistic acoustic phonon 
emission at electron transitions between fully discrete Landau levels
in a 2DEG with account of the phonon reflection from a GaAs/AlGaAs
type interface \cite{badalbal,badalsurf}. 
In accordance with the experimental results, we have 
obtained that the angular distribution of emitted phonons has a distinctly
expressed peak for small angles. Account for the interface effect affects 
essentially the intensity and the composition of the detected phonon field.
Upon their {\it reflection} and {\it conversion} at the crystal surface, 
longitudinal acoustic LA phonons propagate backwards in the forms of 
LA and transverse acoustic TA phonons. The reflected LA phonons 
interfere with the initial LA phonons emitted by the 2DEG in the same 
direction. Therefore, we have obtained that under the deformation 
electron-phonon interaction,
the detector records on the sample reverse face both the {\it interference}
field of the LA phonons and, which is most intriguing for experiment, 
the {\it conversion} field of the TA phonons \cite{badalbal}.
Our calculations of emission spectrum for surface acoustic phonons 
show that an exponential suppression of the emission of surface acoustic
phonons occurs in a wide range of the magnetic field variation 
 \cite{badalsurf}.
So, the cooling of the heated 2DEG is only at the expense of bulk
LA and TA phonons.

Investigation of edge state scattering is one of key stages of the 
dissertation \cite{bad,maslov,badaledbal1,badaledbal}.
The single particle energy spectrum in a 2DEG exposed to a 
homogeneous magnetic field normal to the electron plane is 
separated into the edge states and the bulk Landau states 
The edge states correspond to the classical skipping orbits and 
are confined near sample edges.
Edge states exist also in QWrs. If the wire width $L$ is much 
greater than the magnetic length $a_B$,  $L\gg a_B$, then the 
edge states both in 2DEGs and QWrs can be treated in the 
same way as a 1D electron system.
The edge states play an important role both in conventional transport
measurement experiments and in ballistic phonon emission and 
absorption experiments. In the latter case, equally with the bulk 
Landau states, the quantum edge states also give a contribution to
emission and absorption of ballistic phonons.
We have calculated ballistic acoustic (both for deformation DA and 
piezoelectric PA interactions) \cite{badaledbal} and 
polar optical phonon emission by the quantum edge states. An analytic 
expression for the ballistic acoustic energy flux emitted by the quantum 
edge states has been derived. Detailed analysis of the phonon emission 
intensity distribution has been made in low and high temperature regimes
and for different positions of the Fermi level.
At the same time as phonon emission by the bulk Landau states is 
concentrated within a narrow cone around the magnetic field, at the inter
edge state transitions and at low temperatures, the emitted phonon field is
predominantly concentrated within a narrow cone around the direction of 
edge state propagation, while at high temperatures -- around the magnetic
field normal to the electron plane.
At low temperatures the emission intensity decreases exponentially with 
decreasing filling of Fermi level.
In contrast to the case of bulk Landau states where piezoelectric
interaction is always suppressed with respect to the deformation 
interaction, in the edge state case, the relative contributions of
piezoelectric and deformation interactions depend on the magnetic
field and temperature.

We have studied an optical phonon assisted edge state relaxation 
for a test electron in QWrs with a rectangular cross section exposed 
to the normal magnetic field. 
The intrasubband scattering rates as a function of the initial
electron energy for different values of the magnetic field has been
calculated. By considering 
different limiting cases of the ratio of the cyclotron frequency 
to the strength of the lateral confinement, results for edge 
state relaxation both in 2DEGs and QWrs as well as for 
the magnetic field free case can be obtained.

We have calculated the inter edge state scattering length for an 
arbitrary confining potential \cite{bad,maslov}. Phonon (deformation 
acoustic DA and piezoelectric
PA interactions) and impurity scatterings are discussed and analytical 
expressions for scattering lengths are derived. As follows from energy 
and momentum 
conservation, only phonons with frequencies above some threshold can 
participate in the transitions between edge states. As a result, phonon 
scattering is exponentially suppressed at low temperatures. According 
to our evaluations, the observed temperature dependence of the scattering
length cannot be attributed to phonon scattering.

Auger scattering in semiconductors is well known from investigations of
non-radiative recombination. Free electrons and holes are a
prerequisite for this process: the energy obtained in the recombination of
an electron-hole pair is taken to excite another electron. The latter
electron may loose its excess energy by electron-lattice relaxation; thus
the recombination energy is converted into heat. 
In quantum nanosystems, Auger processes become possible between
different subbands.
The reduction of Auger scattering rate in effectively 0D nanosystems 
due to the discreteness of the electronic states has been
used as an argument to propose quantum dots lasers.

In recent magneto-luminescence experiments by Potemski {\it et al.} 
\cite{potemski1,potemski2} on one-side modulation doped GaAs/AlGaAs
quantum wells, an up-conversion has been observed and interpreted as 
being due to an Auger process. 
The luminescence spectrum under interband excitation at low 
temperatures and for low excitation powers shows two peaks:
besides the luminescence due to recombination of an electron from
lowest Landau level with a hole in a valence band, a second peak is
observed above the exciting laser energy and is related to
recombination of an Auger up-converted electron with a hole. 
We have developed a theory of Auger up-conversion in quantum 
wells in quantizing magnetic fields to explain these experimental
results \cite{badalauger,badalrspt}.
We have calculated the characteristic times of electron-electron 
scattering processes between Landau levels of the lowest electric
subband and of electron-acoustic phonon scattering between Landau 
levels of the two lowest electric subbands as well as the lifetime of
a test hole both with respect to the Auger process and phonon 
emission. By analyzing rate equations for these processes as well as
for the pumping by interband excitation and the recombination of 
electrons with photo-induced holes, we have found the Auger process
time. As well the magnetic field and the excitation power 
dependencies of the two luminescence peaks have been obtained
which are consistent with the experimental findings. Thus, an 
understanding of the Auger up-conversion observed in the 
magneto-luminescence in quantum wells is provided.

Recently a research group from the Toshiba Cambridge Research 
Center Ltd. and Cavendish Laboratory have proposed a new technique
to produce non-homogeneous magnetic fields 
\cite{foden1,mark1,mark2,mark3,foden2}.
A remotely doped GaAs/Al$_x$Ga$_{1-x}$As heterojunction is grown 
over wafer previously patterned with series of facet. 
The use of {\it in situ} cleaning technique enables to regrow uniform
high quality 2DEGs which is no longer planar but follows the contour of 
the original wafer.
Application of a homogeneous magnetic field to this structure results
a spatially varying field component normal to the 2DEG. 
Thus, this technique offers possibilities to investigate the effect of varying 
{\it the topography} of an electron gas in addition to varying the 
{\it dimensionality}. This new technology will open up a new dimension 
in nanophysics. 

We have investigated theoretically the magneto-transport of the 
non-planar 2DEG  \cite{badaltrs1,badaltrs2,badaltrs3,badaltrs4,ibrahim}.
As an example the electric field distribution of the 2DEG with a magnetic
field barrier is calculated. The system satisfies the Poisson equation in
which linear charges develop at the magnetic/non-magnetic field interface.
The magneto-resistance across the facet as well as in the planar regions 
of the 2DEG have been calculated which explain the main features of the 
magnetic field dependencies observed experimentally 
by M L Leadbeater {\it et al} \cite{mark1,mark2}.

\section{Main goals and tasks of the dissertation}
\begin{enumerate}
\item{Investigation of threshold peculiarities of the polaron spectrum in
the 2DEG in a quantizing magnetic field normal to the electron sheet. 
De\-monst\-ration of the existence of the spectrum new branches which 
describe the bound states of an electron and an optical phonon.}
\item{Study of photon absorption on these electron-phonon bound states.
Establishment of the fine structure of the cyclotron-phonon resonance 
in quantum wells and heterostructures. Consideration both bulk and 
surface optical phonons in formation of the bound states.}

\item{Development of a new method for calculating the probability
of electron scattering from the deformation potential of acoustic 
phonons. Use of this method to treat the phonon reflection from
a crystal surface and interfaces separating different materials.}

\item{Exploiting the proposed method to calculate the electron 
energy and momentum relaxation times for a test electron as
well as the relaxation rate of electron temperature for the whole 
Fermi 2DEG taking into account the phonon reflection from 
the interface between semi-infinite elastic media.}

\item{
Calculation of the transition probability between fully quantized
Landau levels and taking into account the crystal free surface 
effect on interaction of the 2DEG with acoustic phonons.}

\item{
Study of ballistic acoustic phonon emission in quantizing magnetic 
fields normal to the 2DEG plane when reflection of these 
phonons from a GaAs/AlGaAs type interface is taken into account.}

\item{Investigation of longitudinal polar PO optical phonon assisted 
electron relaxation in the 2DEG in the QHE geometry.
Consideration of inter Landau level relaxation via one phonon (for bulk 
PO and interface SO phonons) and two PO+DA phonon (for deformation
acoustic DA and PO phonons) emission processes.}

\item{
Calculation of the scattering length for inter edge states transitions
in quantizing magnetic fields due to acoustic phonons (DA and PA 
interactions) and short- and long-range impurities, 
assuming that the shape of the confining potential is arbitrary.}

\item{
Study of emission of ballistic acoustic phonons (due to deformation and
piezoelectric interactions) by quantum edge states in quantizing magnetic
fields. Detailed analysis of the emission intensity and the angular 
distribution in low and high temperature regimes. Consideration of 
different positions of the Fermi level.}

\item{Calculation of the optical phonon assisted edge state relaxation 
rates in quantum wires exposed to quantizing magnetic fields. 
Derivation of the PO phonon emission rate dependence on the electron
initial energy and magnetic field.}

\item{Construction of a theory of the Auger up-conversion observed
in recent magneto-luminescence experiment.
Calculation of the relaxation rates of the carrier-carrier (between 
discrete Landau levels of the lowest electric subband) and the
carrier-phonon (between discrete Landau levels of the two lowest 
electric subbands) scattering processes in quantum wells exposed 
to the normal quantizing magnetic field.}

\item{
Investigation of the magneto-transport of a non-planar 2DEG. Calculation 
of the electric field distribution in the presence of a magnetic tunnel 
barrier of $\mu$m width. Derivation of the magnetic field dependence
of the magneto-resistance both across the facet and in the planar regions
of the 2DEG.}
\end{enumerate}

\section{Basic theses of the dissertation submitted to the defense}
\begin{enumerate}

\item{Despite to the weak electron-phonon coupling, an infinite set
of bound states of an electron with an optical phonon exists both 
above and below the threshold of optical phonon emission in the 
magneto-polaron spectrum in the 2DEG.}

\item{The spectrum of the cyclotron-phonon resonance has a fine
structure which is governed entirely by the electron-phonon bound 
states. Therefore,
electromagnetic absorption is concentrated not at the threshold but 
below and above it at the separations of the binding energies. 
The absorption spectrum consists of two groups of peaks 
which constitute an "asymmetric doublet".}

\item{It is easy to account for the phonon reflection from various
interfaces separating different materials using the proposed method 
for calculating the probability of electron scattering from the 
deformation potential of acoustic phonons.}

\item{There are situations when the presence of an interface, near which
the Fermi 2DEG is located, leads to a strong reduction of the energy and
the momentum relaxation and alters the dependence of the relaxation rates
on the electron energy (or on electron temperature).}

\item{The interface effect in the 2DEG is strongest in quantizing magnetic
fields normal to the plane of electrons and for the free crystal surface.
Surface effect becomes weaker with a distance $z_0$ between 
the 2DEG and the crystal surface for two reasons: because of the 
spread of momenta and the spread of frequencies of emitted phonons. 
Even in high-quality heterostructures, the Landau level broadening is 
essential.}

\item{The transition probability between two Landau levels via deformation 
acoustic phonons is an oscillating function both of the magnetic field $B$ and
of the distance $z_0$.
The magnetic field oscillations are on top of a smooth background of the 
transition probability which has a subnanosecond peak for intermediate 
values of $B$ (in GaAs, for fields of the order of $1$ T) and decreases 
as a fourth power of $B$ for large $B$. For vanishing fields, selection 
rules force the transition probability to fall to zero.}

\item{In quantizing magnetic fields, inter Landau level electron transitions
in the 2DEG via piezoelectric acoustic phonon interaction suppressed with
respect to the deformation interaction mechanism.}

\item{Acoustic energy flux of ballistic cyclotron phonons emitted from the 
2DEG in a normal quantizing magnetic field is concentrated in a narrow cone
around the magnetic field. The interface affects essentially both the intensity
and composition of the emitted phonon field so on the sample reverse face,
the detector records an {\it interference} field of longitudinal LA phonons 
and a {\it conversion} field of transverse TA phonons.}

\item{Exponential suppression of emission of cyclotron surface acoustic SA 
phonons occurs in a wide range of the magnetic field variation, 
$\h\o_B\gg 2mc_R^2$ ($\o_B$ is the cyclotron frequency, $m_c$ is the 
electron mass and $c_R$ is the velocity of surface waves). So cooling of the 
heated 2DEG is only at the expense of emission of bulk LA and TA phonons.}

\item{An allowance for the polar optical PO phonon dispersion and the 
Landau level broadening yield a finite relaxation rate associated with 
one-phonon emission.
Immediately below the PO phonon energy, $\h\omega_{LO}$, the PO 
phonon dispersion contribution give rise to a very sharp peak with peak
value approximately $0.17$ fs$^{-1}$. The Landau level broadening 
contribution has a rather broad peak with relatively lower peak value. 
Within an energy range of the order of $\h\sqrt{\o_{B}/\tau}$ ($\tau$ 
is a relaxation time deduced from the mobility.), the one-phonon 
relaxation rate exceeds $1$ ps$^{-1}$.
Relaxation via surface SO phonon emission mechanism at least by an 
order is weaker than via bulk phonons.}

\item{The two-phonon emission is a controlling relaxation mechanism 
above $\h\o _{PO}$. For energies $\Delta l\h\o _B$ immediately above 
$\h\o _{PO}$, PO+DA phonon relaxation rate increases
as a fifth power in the magnetic field $B$. At energy separations of
the order of $sa^{-1}_{B}$ ($s$ is the sound velocity, $a_{B}$ is the 
magnetic length), PO+DA
phonon emission provides a mechanism of subpicosecond relaxation.
At higher $B$, the two-phonon relaxation peak decreases 
linearly in $B$ and within a wide energy range of the order
of $\hbar\omega_{B}$, subnanosecond relaxation can be achieved.}

\item{Inter edge state relaxation due to acoustic phonon scattering is strongly 
suppressed at low temperatures in comparison with short- and long-range 
impurity scattering since only phonons with frequencies above some 
threshold can cause transitions.}

\item{At low temperatures emission of ballistic acoustic phonons (due to 
deformation and piezoelectric interactions) at inter edge state transitions
is predominantly concentrated within a narrow cone around the direction of 
edge state propagation while at high temperatures -- around the magnetic 
field normal to the electron plane. The emission intensity decreases with 
decreasing filling of the Fermi level. This diminution is exponential
at low temperatures.}

\item{In contrast to the bulk Landau state, relative contributions of 
piezoelectric and deformation interactions depend on the magnetic field, 
electron temperature, and on the shape of the confining potential. 
At low temperatures and for the non-smooth confining potential,
DA interaction suppressed with respect to PA while for the smooth
potential as well as for both cases at high temperatures, 
DA and PA interactions give roughly the same contribution to 
the emitted phonon field.}

\item{Polar optical phonon emission provides a picosecond edge state
relaxation for a test electron in quantum wires. The scattering rate 
as functions of the electron initial energy has a peak for magnetic fields 
not far from the resonance field while for lower fields, the scattering rate
shows a monotonous increase in energy.}

\item{Our theory of the Auger up-conversion in a 2DEG in a 
normal quantizing magnetic field provides an understanding of the 
up-conversion observed in magneto-luminescence of one-side modulation
doped GaAs/AlGaAs quantum wells \cite{potemski1}, 
in particular, its dependence on the excitation power and the magnetic field.}

\item{According to our theory of the magneto-transport in a non-planar
2DEG, the system satisfies the Poisson equation in which a line charges 
develop at the magnetic/non-magnetic field interface.
The magneto-resistance calculated across the facet and in planar regions 
of the 2DEG explains the main features of the magnetic field dependencies 
observed in the experiment \cite{mark1,mark2}.}
\end{enumerate}

\section{Approbation of the work}
The results of the investigation presented in the dissertation have been 
reported and discussed at the following conferences, schools, and seminars:

\begin{tabular}{lp{3.8in}}

1996&Seminar, Institute for Theoretical Physics, University of Regensburg, 
Germany\\[4pt] 
1996 &12th International Conference "on the Application of High Magnetic 
Fields", W\"urzburg, Germany\\[4pt] 
1996 &9th International Conference on "Superlattices, Microstructures, 
and Microdevices, July, 14-19 1996, Li\'ege, Belgium.\\[4pt] 
1996 & Research Workshop on Condensed Matter Physics, ICTP, Italy\\[4pt]
1996 & Seminar, Department of Condensed Matter Theory, University of
 Antwerp, Belgium\\[4pt]
1995& Research Workshop on Condensed Matter Physics, ICTP, Italy\\[4pt]
1993& Winter school, University of Wroclaw, Poland\\[4pt]
1992& NATO ASI, Ultrashort Processes in, Lucca, Italy\\[4pt]
1992 & Seminar, Institute for Theoretical Physics, University of
Regensburg, Germany\\[4pt]
1993& 13th General Conference of Condensed Matter Division, 
European Physical Society, Regensburg, Germany\\[4pt]
1989& 14th All-Union Conference on "Theory of Semiconductors", Donetsk,
The Ukraine\\[4pt]
1988&Seminar, Institute of Physical Investigations, Ashtarak, Armenia\\[4pt]
1988&Seminar, Landau Institute for Theoretical Physics, Institute of 
Solid State Physics, and Institute of Microelectronics Technology and 
High Purity Materials, Chernogolovka, Moscow district, Russia\\[4pt]
1987& All-Union Conference on "Non-classical crystals", Se\-van, Armenia\\[4pt]
\end{tabular}

\section{Structure and volume of the dissertation}
The dissertation is presented on $227$ pages including $41$ figures 
and $5$ tables. It consists of $7$ Chapters, the Summary, $3$ 
Appendixes, and the Bibliography of $327$ names.

\chapter{Electron-optical phonon bound states}

\section{Introduction}
\label{2intrdctn} 

Over a long period of time polarons, elementary excitations in the 
electron+phonon system, have been extensively studied both
theoretically and experimentally in polar semiconductors and ionic 
crystals \cite{devreese1,devreese2,mitra}. Recent rapid developments
in the semiconductor nanotechnology have made it possible to create 
nanosystems of reduced dimensionality and have caused growing interest 
in polaron effects in such nanosystems, particularly in the 2DEG exposed 
to a normal quantizing external magnetic field. In such systems, due to 
the combined effect of the spatial confinement and Landau quantization, 
the electron energy becomes quantized into a series of completely discrete
Landau levels with an infinite degeneracy. Interaction of such electrons 
with polar optical phonons leads to a formation of 2D 
magneto-polarons. Properties of the 2D magneto-polarons strongly depend
on energy ranges considered in the exact spectrum and on a density of states
on a Landau level. As far as in the 2DEG the density of states on the 
bottom of Landau band is stronger than in massive samples, it
changes from square-root singularity to the delta-function, then the polaron
binding energies and cyclotron resonance peaks should be greater and clearer
for the 2DEG than in bulk samples. One can expect also qualitatively new
effects in the 2D magneto-polaron spectrum.

Studies of the magneto-polaron problem in massive samples (see reviews 
\cite{levrpp73,levufn73}) have shown that despite to the weak electron-optical
phonon coupling, $\alpha$, branches of the magneto-polaron spectrum can
be classified into three groups according to the single-phonon states in the
magneto-polaron, {\it i.e.} according to the effective number, $N$, of
phonons contributing to the magneto-polaron. The first group includes states
with $N\sim\alpha$. The exact spectrum of these branches differs from
priming one only by a simple shift of the bottom of Landau band and by a
polaron renormalization of the electron effective mass. Both these effects
are of the order of $\alpha$. The second and third groups are hybrid and
bound states of electrons and phonons. For these states we have $%
N\approx1/2$ and $N\approx1$, respectively. Peculiarities of the polaron
spectrum near the threshold of an optical phonon emission are responsible
for existence of these two groups. These threshold non-analyticities of the
spectrum are not connected with particular models but have an universal
character \cite{landaukm}. For instance, the non-analyticity of the phonon
spectrum near the disintegration threshold of an optical phonon into two 
acoustic phonons leads to the formation of the hybrid and bound states of
the optical and acoustic phonons \cite{ruvalds,badalo2a}. Notice that for
the formation of the hybrid states, a resonance situation is necessary.

The problem of calculation of the electron and optical phonon hybrid state
spectrum (the spectrum of the resonance magneto-polaron) has been arisen
in connection with experiments of the inter-band magneto-absorption and
of the cyclotron resonance \cite{larsen66,summharper67}, in which a 
splitting of the absorption peak near the resonance $\omega_B=\omega_{0}$
has been observed ($\omega_B$ and $\omega_0$ are the cyclotron and the
longitudinal optical LO phonon frequencies, respectively). The first 
theoretical studies of the resonance magneto-polaron spectrum have been
based on the single-phonon model, in which only no-phonon and 
one-phonon states have been taking into account. It has been shown 
\cite{larsenjon66,harper67,korov67} that if no
interaction is taken into account, two terms of the electron+phonon system
(the first is the electron in the Landau level $l+1$, the second is the
electron in the level $l$ and one LO phonon) are intersected at the
resonance. Switching on the electron-phonon interaction takes off the 
degeneracy and leads to the splitting of the absorption peak which is of 
the order of $\alpha^{3/2}\hbar\omega_{0}$. The above peak, as 
distinct from the peak below, is smeared by the electron-phonon 
interaction. At present the resonant magneto-polaron effect is well
studied in massive samples \cite{harper72,stradling73} and is an unique 
tool to study the electron-phonon interaction.

In the last years in view of its importance, the resonant magneto-polaron
in the 2D nanosystems has attracted considerable attention. Theoretical
investigations of the problem show \cite{korov80,sarma80,zawad85,peet85}
that electron-phonon interaction in the 2D systems is stronger than in
the massive samples.
There are two reasons to account for such situation. The first is, {\it ut
sup}, the singularity of the density of states on a Landau level is stronger
for 2D electrons than that of in the bulk samples. The second is that the
presence of interfaces separating different materials in layered
nanostructures violates the translation invariance in the growing direction.
Hence, in acts of electron-phonon interaction, phonon modes with any 
value of the momentum component in that dimension can take part. Such 
reinforcement of electron-phonon interaction leads to the 2D 
magneto-phonon resonance with the characteristic binding energy scale $%
\alpha^{1/2}\hbar\omega_{0}$ which is for $\alpha\ll1$ essentially greater
than the analogous scale $\alpha^{3/2}\hbar\omega_{0}$ in the massive
samples. Moreover, the branches of the spectrum are rigorously stationary
both above and below the phonon emission threshold so that two infinitely
narrow peaks correspond them in the absorption spectrum. 

Recently the 2D resonance magneto-polaron effect has been observed
in the cyclotron resonance experiments in electron inversion layers on 
InSb \cite{horst83}, accumulation layers on HgCdTe \cite{nicholas86},
and in II-IV CdTe/CdMgTe quantum wells \cite{landwehr}, as well as
in GaAs/AlGaAs heterostructures and quantum wells 
\cite{horst85,sigg85,singleton,mccombe1}.
Analysis of experimental results \cite{horstmerkt86} shows that the 2D
resonance magneto-polaron effect is approximately by $3-6$ times stronger
than that of for massive samples which is in good agreement with the
theoretical calculations \cite{korov80,zawad85}.

The characteristic feature of the resonance magneto-polaron is that the
number of hybrid states in the exact spectrum do not exceed the number of
states in the priming spectrum of the electron+phonon system. However, 
further theoretical developments in the problem of the magneto-polaron 
have shown that the exact spectrum near the LO phonon emission 
threshold can be considerably richer \cite{lev70}. When the magnetic field
decreases or increases, the resonance condition ceases to be fulfilled and, 
one would think that the hybrid states should be transformed into the usual
magneto-polaron states. As a matter of fact it does not take place. In the
exact spectrum of the 3D magneto-polaron in the energy range of the order
of $W_{3D}=\alpha^2\hbar\omega_{0}$ below the threshold $\e_n=E_n+%
\h\o_{0}$ there appear new branches of the electron+phonon system ($E_n$
is the electron priming energy, $n$ is the Landau index.).
These terms are impossible to obtain in the framework of the single-phonon
model which is equivalent to account for only the simplest diagram of the
electron mass operator expanded in a power series in the electron-phonon 
interaction coupling constant. Moreover, the analysis of the perturbation 
theory shows \cite{levi71} that despite to the small coupling constant, the 
sum of the perturbation theory series diverges in the energy range 
$|\e-\e_n|\lesssim W_{3D}$ and it nowise is connected with the resonance 
$\omega_B=\omega_{0}$. To find the exact spectrum in this energy range,
it is necessary to sum infinite number of diagrams for the mass operator, or
that is the same, to solve an integral equation for the electron-phonon vertex
\cite{pit}. When this procedure has been explicitly completed by 
Y B Levinson
\cite{lev70}, it has become clear that in the exact spectrum of the 
electron+phonon system there exists an infinite sequence of the electron and 
phonon bound states with the LO phonon emission threshold as a limit below.
The maximum distance between two neighboring terms is of the order of 
$W_{3D}$ for $\omega_B\sim\omega_{0}$ even if it depends on the 
magnetic field. The bound states ought to realize as a local lattice 
perturbation arisen at the center of the cyclotron orbit.

It was natural to expect that the bound states will become apparent in the
cyclotron phonon resonance spectrum. The cyclotron phonon resonance 
has been predicted theoretically by F G Bass and Y B Levinson \cite{lev65}
(see also reviews \cite{lev78,ivanov78}). It is observed in semiconductors 
as a result of optical transitions of an electron from one Landau level to 
another accompanied by the emission or absorption of an optical LO phonon.
Actuality of the cyclotron phonon resonance has been verified experimentally
in the electromagnetic absorption \cite{mccombe}. The cyclotron phonon
resonance phenomenon would be combination of two processes, of the 
cyclotron resonance observed at the frequency of electromagnetic radiation 
$\nu$ divisible by $\omega_B$ and of the Gurevich-Firsov magneto-phonon
resonance \cite{gurevich61,gurevich62}, coming at the magnetic fields for 
which LO phonon frequency $\omega_0$ is divisible to $\omega_B$. 
According to the simplest ideas based on perturbation theory 
\cite{lev65,enck69}, the cyclotron phonon resonance absorption peak
corresponding to the LO phonon emission should be observed for photons
of energies $\hbar\nu=E_n-E_{n^{\prime}}+\h\o_{LO}$. A more rigorous
analysis \cite{bakanas73} has shown that the cyclotron phonon resonance 
peak should have a fine structure due to appearance of the bound states of
an electron and a newly created LO phonon. In bulk samples, the binding 
energies $W$ and oscillatory strengths $f$ of these states are very small: in
magnetic fields such that $\omega_B\sim\omega_{0}$ we have $W\sim 
W_{3D}$ and $f\sim\alpha^2$. In the case of $n$-type InSb this gives 
$W\approx0.01$  meV in a field $B = 3.5$ T. This makes hard for the
experimental resolution of the bound states in the cyclotron phonon 
resonance peaks. The revealing of the bound states in massive samples 
is conjugated by additional difficulties for following reasons. In the 
simplest case, {\it e.g.}, the bound states could be detected in the light 
absorption at the photon energy $\hbar\nu=\hbar\omega_{0}-W$ 
(the electron transitions absent; $n=n^{\prime}=0$) \cite{lev72ii}. 
However, in this case bound states lie below the threshold and their 
observation is hindered by the strong lattice reflection in the energy range
between the longitudinal and transverse optical phonon energies. The 
absorption spectrum seemed more convenient to study when $n\neq0$ at 
the photon energy $\hbar\nu=\h\omega_B+\h\omega_{0}-W$. This energy 
range of the spectrum, however, lays upon continuum of the 
electron+phonon states so that there cannot exist rigorously stationary 
bound states.

Taking into account the situation described above for the magneto-polaron
problem, particularly, the transformations in the resonant magneto-polaron
effect at the dimensional reduction from 3D to 2D, one can expect that the
environmental factors for the electron-phonon bound state creation and their
experimental observation in the 2DEG should be more favorably inclined.

Subject of this chapter is the bound states of the electron and LO phonon in
the 2D nanostructures exposed to the quantizing magnetic field normal to the
electron sheet. It will be shown that in the vicinity of the LO phonon emission
"threshold" $\e_n$, the spectrum of the magneto-polaron is determined by
an infinite sequence of the bound states coagulated to the "threshold" 
having it as a limit both above and below from it \cite{badalbs}.
(In the 2DEG there is no continuum in the spectrum of the electron+phonon
system, and therefore in this case, the "threshold" has no strict sense. To 
stress this fact, the inverted commas are used.) These bound states are always
rigorously stationary and are manifested in the cyclotron phonon resonance
spectrum as two groups of narrow peaks located above and below the
"threshold" on the distance of the order of $W_{2D}= \alpha\hbar\h\omega_{0}$
\cite{badalcpr}.

Existence of the bound states is determined by the competition of the
kinetic energy of electrons and optical phonons, and their interaction
energy. Therefore, the character of coupling and the energetic scale of the
spectrum strongly depend on the localization extend of electrons and 
phonons. In 2D nanostructures besides the electron localization effect, 
there are certain special features associated with the optical phonon 
localization. Two situations are discussed below: electrons in an
isolated quantum well located deep inside a sample and electrons at a
heterojunction between two semiconductors. Low-momentum LO 
phonons do not usually penetrate from one material to another and, 
therefore, electrons in a quantum well or at a heterojunction interact solely
with LO phonons in the semiconductor where they are located. Further
these bulk-like polar LO phonons we denote as PO phonons.
\begin{figure}[htb]
\epsfxsize=13cm
\epsfysize=13cm
\mbox{\hskip 0cm}\epsfbox{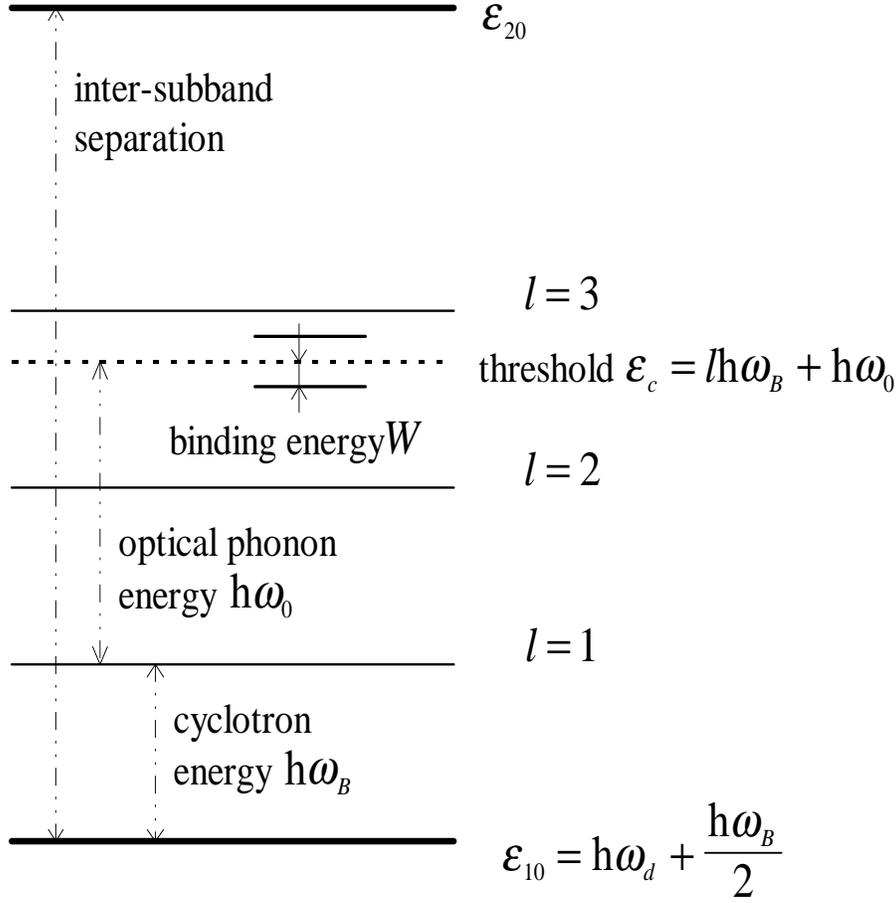}
\caption{Energy levels $\e_{nl}=\e_n+(l+1/2)\h\o_B$ in the 2DEG 
exposed to the quantizing magnetic field normal to the electron sheet.
$\e_n$ is the energy of subband $n$.}
\label{fg1}
\end{figure}
In addition to the bulk-like PO phonons, there exist also surface SO phonons
with an electric field concentrated on both sides near an interface
separating different materials. Recently polaron effect with such SO
phonons in quantum wells and heterojunctions have been considered 
theoretically in several papers 
\cite{lassnig,wendler1,wendler2,deg87,deg88,shi87,badalcpr,moriando,wei,hai}.
It is obtained that the SO phonon contribution to the energy and effective
mass
is essential. In the recent magneto-phonon resonance experiment 
\cite{nicholas85}, the observed states have been interpreted as resonance 2D
magneto-polarons with SO phonons \cite{sarma86}. Therefore, one can 
expect
that SO phonons will also play an important role in the formation of the
electron-phonon bound states. In the case of a heterojunction the SO 
phonons do not exhibit dispersion and, therefore, they behave in the same
way as LO phonons. On the other hand, the SO phonons in a quantum 
well have a strong linear dispersion which smears out the threshold
singularity. Such SO phonons are ignored in this work.

The energy levels of the considered system are shown on Fig.~ \ref{fg1}. 
The energy is reckoned from the lower electron level, whose height above 
the bottom is $\h\o_d+\hbar\omega_B/2$, $\h\o_d$ is the energy scale in 
the growing direction. The dashed line shows the "threshold" 
$\e_n=E_n+\hbar\omega_{0}$ referred to the Landau level $n$, near which 
the bound states of interest to us are located.

Following assumptions have been done in calculations of the bound state
spectrum:

\begin{enumerate}
\item[i)]  {only the polar electron-phonon interaction is considered and the 
priming electron-phonon coupling constant is assumed to be small,
$\alpha\ll1$,}

\item[ii)]  {optical phonons have no dispersion, $\omega({\bf q})=
\omega_{0}$},

\item[iii)]  {crystal temperature is low enough, $ T\ll\hbar\omega_{0}$
so that the phonon absorption processes, proportional to 
$\exp(-\hbar\omega_{0}/T)$, can be neglected (The units are used such 
that the Boltzmann constant $k_B=1$.),}

\item[iv)]  {concentration of electrons is small so that their influence on
phonons as well as interaction between them can be neglected,}

\item[v)]  {interaction of electrons with impurities and acoustic phonons is
absent,}

\item[vi)]  {electrons are confined in a potential well with infinitely high walls,
the well is also narrow enough so that spatial quantization energy scale 
is the largest parameter of the problem, {\it i.e.} the intersubband separation
$\Delta \e_d\sim\h\o_d\gg\h\omega_B,\h\omega_{0}$ .}
\end{enumerate}

The present chapter is organized as follows. In the next section the
threshold behavior of the magneto-polaron spectrum is analyzed and the
integral equation for the electron-phonon scattering amplitude is derived
from which we obtain the bound state spectrum. In Sec.~ \ref{cpr} the
absorption coefficient on these bound state is studied. The allowance for
the presence of bulk and surface optical phonons is made. The actual
calculations of the bound state binding energies and oscillator strengths
are carried out in Sec.~\ref{beos}. Our conclusions and numerical estimates
are presented in Sec.~ \ref{summ}.

\section{The 2D magneto-polaron spectrum}
\label{2dmagnetpolrn}

\subsection{"Threshold" behavior of the polaron spectrum}

The spectrum of two-particle elementary excitations in the electron-phonon
system is determined by the poles of scattering amplitude $\Sigma$ in the
total energy parameter $\varepsilon$. Because the number of phonons is not
conserved, the spectrum obtained from the poles of $\Sigma$ and of the
single-particle Green function are the same. The calculation of $\Sigma$ by
means of the perturbation theory cannot be acceptable since in the energy
range of interest to us, $\varepsilon\approx\varepsilon_n$, the series of
the perturbation theory stops to converge. In terms of the diagram
\begin{figure}[htb]
\epsfxsize=13cm
\epsfysize=9cm
\mbox{\hskip 0cm}\epsfbox{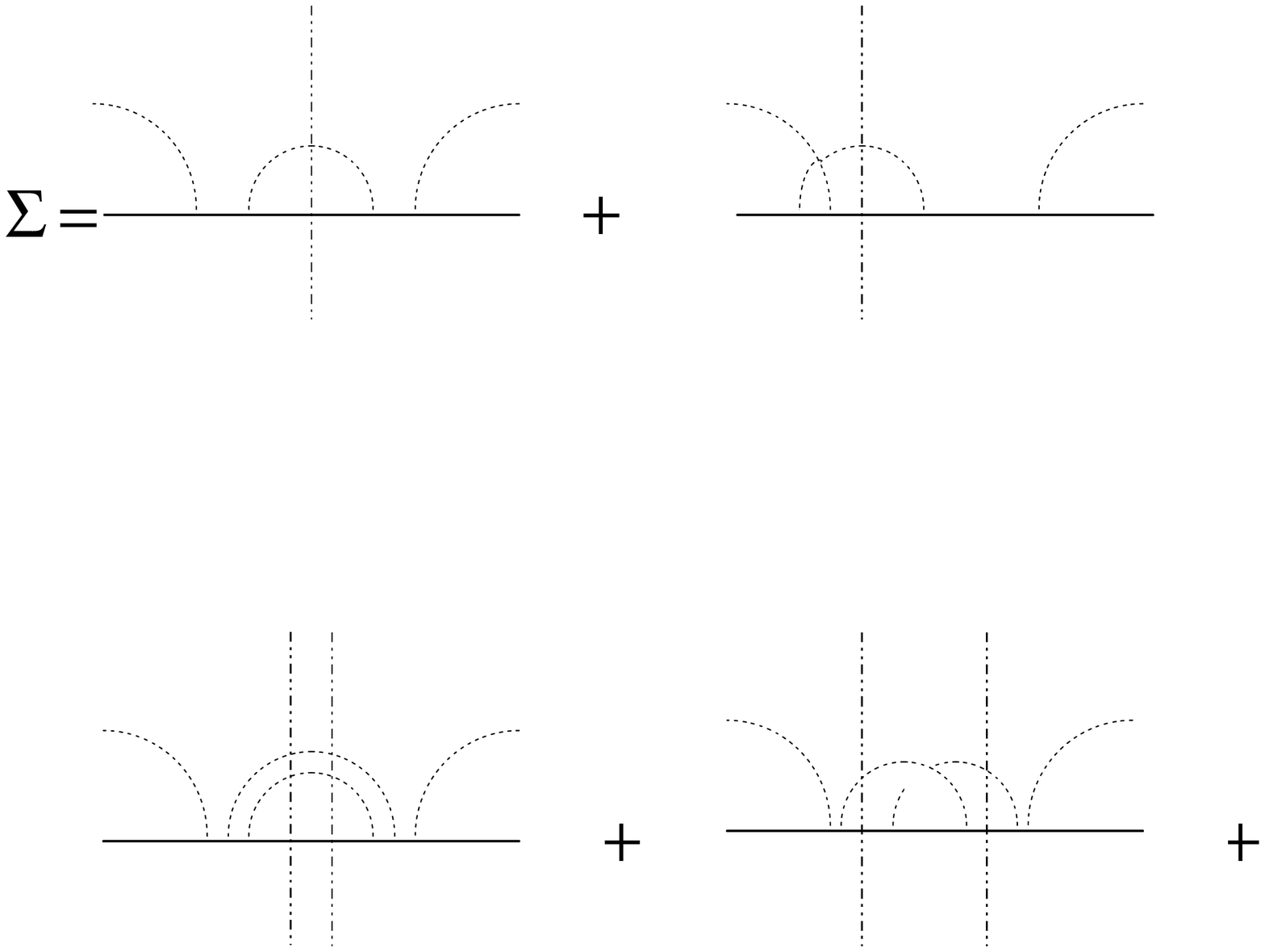}
\caption{Simplest diagrams for the electron-phonon scattering amplitude $%
\Sigma$ with dangerous intersections in one electron and one phonon lines.}
\label{fg2}
\end{figure}
technique, the diagrams with one phonon and one electron intersections are
responsible for this divergence. Such diagram correspond to the processes in
which phonons are almost real \cite{lifpit}. Some diagrams for the amplitude 
$\Sigma$ with such dangerous intersections are shown in Fig.~ \ref{fg2}. It
is seen that the number of the dangerous intersections increases in the
diagram order. Direct calculations of these diagrams show that the
perturbation theory is inapplicable in the energy range $|\varepsilon-%
\varepsilon_n| \sim\alpha\h\omega_{0}$ where both diagrams with and without
intersecting phonon lines (Fig.~ \ref{fg2}) become essential. Contributions
of all diagrams with dangerous intersections tend to infinity when $%
\varepsilon\to\varepsilon_n$ and the growing velocity is proportional to the
number of dangerous intersections in the diagram.
\begin{figure}[htb]
\epsfxsize=14cm
\epsfysize=11cm
\mbox{\hskip 0cm}\epsfbox{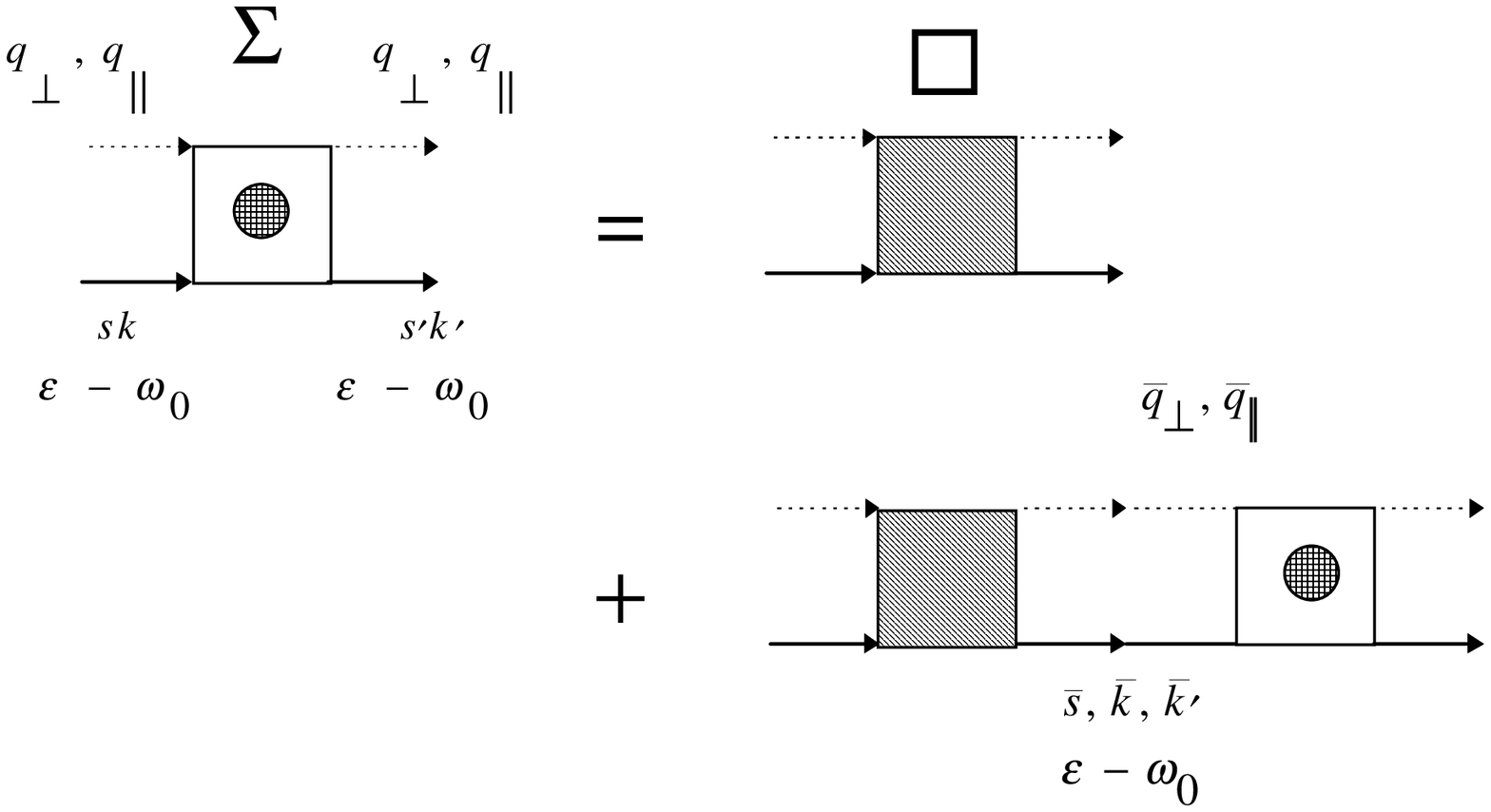}
\caption{Integral equation for the scattering amplitude $\Sigma$ of an
electron by a phonon.}
\label{fg3}
\end{figure}
Thus, to find the exact
spectrum of the system in the region $|\varepsilon-\varepsilon_n|\sim\alpha%
\h\omega_{0}$, the whole series of the perturbation theory should be summed.
This leads to an integral equation for the electron-phonon scattering
amplitude $\Sigma$ drawn on Fig.~ \ref{fg3}. 

\subsection{Integral equation for the electron-phonon scattering amplitude
and bound states}

This equation is written in the gauge invariant diagram technique \cite
{levins71}, therefore only Landau index $s$ is used for the electrons (solid
lines) while the gauge non-invariant quantum number $m$ of the spatial
quantization is absent. The exact electron Green function, $G$ is diagonal
in $s$ \cite{levins71,holstein64,dworin65} but $G$ is not diagonal in $m$
because lack of the translation invariance in the growth $z$-direction. The
assumption $\Delta E_d\gg\h\omega_B,\h\omega_{0}$ permits, however, to neglect
the off-diagonal elements of $G_{mm^{\prime}}$ which simplifies essentially
the integral equation for $\Sigma$. As far as the bound states are sought in
the energy range near the "threshold" then the energetic parameter of the
Green function is $\varepsilon_n-\h\omega_{0}$ in the dangerous intersection
in the second term of the right-hand side of the equation in Fig.~ \ref{fg3}%
. It lies far from the "threshold" where the perturbation theory is valid.
Therefore, the function $G$ can be taken without allowance for the
interaction with the phonons, and only the term ${\bar m}=1$ is to be
retained in the summation over ${\bar m}$. 
All the foregoing allows us to put everywhere $m=1$ (and omit this index
hereafter) so that 
\begin{equation}
G_s(\varepsilon)=(\varepsilon-s\hbar\omega_B+i0)^{-1}.  \label{greenel}
\end{equation}
By force of the assumption of the electron concentration smallness, all
electron Green functions are retarded. For this reason as well as because
system temperature is low, the renormalization of the phonon Green functions
can be also neglected. Then the following expression corresponds to the
phonon propagator $D$ (the dashed lines) 
\begin{equation}
D(\omega,{\bf q})=\hbar^{-1}B({\bf q})\left[(\omega-\omega_{0}+i0)^{-1}-(%
\omega+\omega_{0}- i0)^{-1}\right].  
\label{greenph}
\end{equation}
where $B({\bf q})$ is the modulus squared of the electron-phonon matrix
element. The longitudinal, $q_{\parallel}$ and the transverse, $q_{\perp}$
phonon momenta are shown separately in Fig.~ \ref{fg3}. It is convenient
previously to carry out integration over the phonon energetic parameter $%
\omega$, using the formula 
\begin{equation}
\int {\frac{d\omega}{2\pi}}iD(\omega,{\bf q})F(\omega)=B({\bf q}%
)F(\omega_{0}-i0)  \label{phint}
\end{equation}
where $F(\omega)$ is an analytical function for Im $\omega<0$. Now it is
clear that all $\omega$ ought to replace to $\omega_{0}$ in all $G$ so that
only the integration over the phonon momenta $q_{\parallel}$ and $q_{\perp}$
as well the factor $B({\bf q})$ correspond to all internal phonon lines. For
the more important polar electron-phonon interaction we obtain 
\begin{eqnarray}
B({\bf q})=B_{0}\,\Phi({\bf q_\perp},q_\parallel),\; B_{0}=\pi\alpha
v_0^2,\; v_0={\frac{\hbar p_{0}}{m_c}},\; \hbar p_{0}=\sqrt{%
2m_c\hbar\omega_{0}},  \label{matrix}
\end{eqnarray}
where $m_c$ is the electron effective mass. The factor $B_{0}$ and the form
factor $\Phi({\bf q_\perp},q_\parallel)$ depend on the nature of the
PO phonon localization. For the bulk polar optical PO phonon $\alpha$ is the
usual Fr\"olich coupling constant \cite{frolich}
\begin{equation}
\alpha_{PO}={\frac{e^2}{{\bar \kappa}_{PO}v_{PO}}} ,\,\: {\bar
\kappa^{-1}_{PO}}=\kappa^{-1}_{\infty}-\kappa^{-1}_{0},\, \kappa_{\infty}={%
\frac{\omega^2_{TO}}{\omega^2_{PO}}}\kappa_{0},  \label{fcoupling}
\end{equation}
where $\omega_{PO}$ and $\omega_{TO}$ are frequencies of the longitudinal
and transverse optical phonons, $\kappa_{\infty}$ and $\kappa_{0}$ are the
high and low frequency dielectric permittivities. In this case the form
factor $\Phi({\bf q_\perp},q_\parallel)$ is given by 
\begin{equation}
\Phi({\bf q_\perp},q_\parallel)={\frac{p_{PO}}{q^2_\perp+q^2_\parallel}}
\left|\int_0^d dz |\psi(z)|^2\varphi_{q_\parallel}(z)\right|^2  \label{ff}
\end{equation}
where $\psi$ and $\varphi$ are the wave functions of the electron and
PO phonon for motion along $z$-direction.

For the surface SO phonons, the electron-phonon interaction coupling constant $\alpha_{SO}$ in a
single heterostructure is defined on the analogy of the Fr\"olich constant 
\begin{equation}
\alpha_{SO}={\frac{e^2}{\bar{\kappa}_{SO}v_{SO}}}   \label{scoupling}
\end{equation}
where $\bar{\kappa}_{SO}$ has the form \cite{lassnig,badalcpr,moriando} 
\begin{equation}
\bar{\kappa}_{SO}={\frac{\omega^2_{SO}}{2}}\sum_{\nu=1}^2 \kappa_{\infty\nu}{%
\frac{\omega^2_{PO \nu}-\omega^2_{TO \nu} }{(\omega^2_{SO}-\omega^2_{TO
\nu})^2}}  \label{dielconst}
\end{equation}
Here index $\nu=1,2$ denotes the different media in contact at the
heteroface. The frequency of the interface SO phonon is the root of the
following equation \cite{klein,klein1} 
\begin{equation}
\sum_{\nu=1}^2 \epsilon_{\nu}(\omega)=0,\,
\epsilon_{\nu}(\omega)=\kappa_{\infty\nu}{\frac{\omega^2-\omega^2_{LO \nu}}{%
\omega^2-\omega^2_{TO \nu}}}  \label{dieleq}
\end{equation}
where $\epsilon_{\nu}$ is the dielectric function of the medium $\nu$. In
this case, the form factor $\Phi({\bf q_\perp})$ is given by 
\begin{equation}
\Phi({\bf q_\perp})={\frac{p_{SO}}{2q_\perp}} \left|\int_0^d dz
|\psi(z)|^2\exp{(-q_\perp z)}\right|^2  \label{sff}
\end{equation}
In formulas (\ref{fcoupling})-(\ref{sff}), $v_{PO}, \: v_{SO}$ and $p_{PO},
\: p_{SO}$ are defined in the same way as $v_0$ and $p_0$ in the formula (%
\ref{matrix}). Below  $\alpha$ and $\omega_0$ will denote either $%
\alpha_{PO}$ and $\omega_{PO}$ or $\alpha_{SO}$ and $\omega_{SO}$.

\begin{figure}[htb]
\epsfxsize=14cm
\epsfysize=5cm
\mbox{\hskip 0cm}\epsfbox{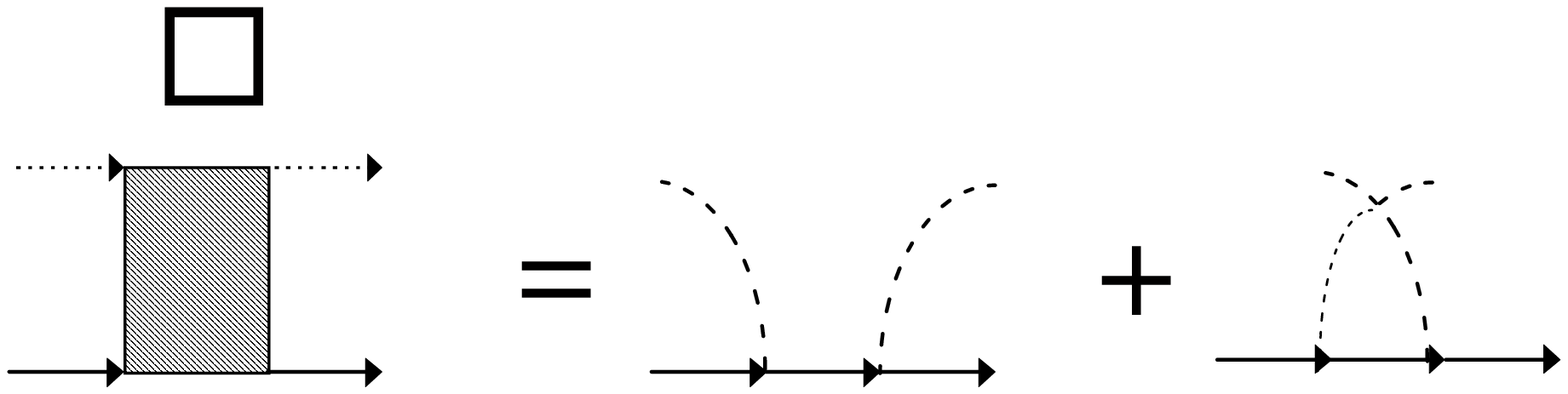}
\caption{Irreducible electron-phonon scattering amplitude $\Box$.}
\label{fg4}
\end{figure}

In Fig.~ \ref{fg3} the shaded box shows the amplitude $\Box$ which has no
dangerous intersections with one electron and one phonon lines, {\it i.e} $%
\Box$ is regular at $\varepsilon\to\varepsilon_n$, and therefore, can be
calculated using the expansion in coupling constant $\alpha$. The
irreducible amplitude $\Box$ can have an intersection in one electron line,
therefore in the lowest order in $\alpha$, the equation shown in 
Fig.~\ref{fg4} takes place. 
The point corresponds to a gauge invariant part of the electron-phonon interaction vertex: 
\begin{eqnarray}
&&\Lambda_{ss^{\prime}}({\bf q_\perp})=e^{-i\phi(s-s^{\prime})}Q_{ss^{%
\prime}}(t), \:\, t={\frac{1}{2}}a^2_Bq^2_\perp,\: a_B=\sqrt{{\frac{\hbar c}{%
eB}}}, \\
&&Q_{ss^{\prime}}(t)=\sqrt{{\frac{s!}{s^{\prime}!}}} t^{\frac{s-s^{\prime}}{2%
}}e^{-{\frac{t }{2}}}L_s^{s-s^{\prime}}(t)
=(-1)^{s-s^{\prime}}Q_{s^{\prime}s}(t)  \label{ffmagnetic}
\end{eqnarray}
where $L_s^{s-s^{\prime}}(t)$ are the associated Laguerre polynomials and $%
\phi$ is the polar angle of the vector ${\bf q_\perp}$.

Since we are interested in the spectrum region near the "threshold" referred
to the Landau level $n$ then it is possible to avoid of the summation over
the Landau quantum number ${\bar s}$ in the dangerous intersection in the
second term of the right-hand side of the equation for $\Sigma$. In this
energy region $\varepsilon\approx\varepsilon_n$, only the non-analytical
contributions from $G$ are essential in the sum over quantum numbers of the
internal electron lines. Thus in the equation for $\Sigma$ only the term ${%
\bar s}=n$ should be retained in the sum over ${\bar s}$. For the same
reason, in the equation for $\Sigma$ in all electron external lines, we
ought to put $s=s^{\prime}=n$. Notice that in the gauge invariant diagram
technique it is possible to choose for $\Sigma$ and $\Box$ "four-tails" such
a definition that these quantities will depend only on the difference $%
\phi-\phi^{\prime}$ of the polar angles of the vectors ${\bf q_\perp}$ and $%
{\bf q^{\prime}_\perp}$. Therefore, expanding $\Sigma$ and $\Box$ into the
Fourier series, the angle variables $l$ in the equation for $\Sigma$ can be
separated. Then the Fourier components 
\begin{equation}
\Sigma_{n}^l(\varepsilon;t,t^{\prime},q_\parallel,
q_\parallel^{\prime})=\int^{2\pi}_0\! {\frac{d\phi}{2\pi}},
e^{-il(\phi-\phi^{\prime})}\Sigma_{n}^l(\varepsilon;{\bf q_\perp},{\bf %
q_\perp^{\prime}}; q_\parallel,q_\parallel^{\prime}) \exp\{{\frac{ia^2_B}{2}}%
[{\bf q_\perp}{\bf q_\perp^{\prime}}]\}  \label{fourier}
\end{equation}
with $l=0,\pm1,\pm2,\dots$ will satisfy the independent equations 
\begin{equation}
\Sigma_{n}^l(\varepsilon;t,t^{\prime}; q_\parallel,
q_\parallel^{\prime})=\Box_n^l(t,t^{\prime})+ \tilde{M}_n(\varepsilon) \sum_{%
\bar{q}_\parallel} \int^{\infty}_0\!{d\bar{t}}\, \Box_n^l(t\bar{t})\Phi(t,%
\bar{q}_\parallel)\Sigma_{n}^l(\varepsilon;\bar{t},t^{\prime}; \bar{q}%
_\parallel,q_\parallel)  \label{inteq}
\end{equation}
where $\Box_n^l$ are the Fourier components the of the kernel $\Box_n$
defined by analogy with $\Sigma_n^l$. The appearance of the phase factor $%
\exp\{{\frac{ia^2_B}{2}}[{\bf q_\perp}{\bf q_\perp^{\prime}}]\}$ in Eq.~ (%
\ref{fourier}) is connected with use of the gauge invariant diagram
technique and is the result of the integration over gauge non-invariant
quantum number $k$ in usual diagram technique. The square brackets in this
factor mean the longitudinal (along the magnetic field {\bf B}) component of
the vector product. The quantity $\tilde{M}(\varepsilon)$ in Eq.~ (\ref
{inteq}) is proportional to the priming electron-phonon interaction coupling constant $\alpha$ and
the density of states on the Landau level 
\begin{equation}
\tilde{M}_n(\varepsilon)=-{\frac{\alpha\omega_B\omega_{0}}{\varepsilon-
\varepsilon_n+i0}},  \label{effcc}
\end{equation}
consequently, it determines the effective electron-phonon interaction in the region near the
"threshold".

It is seen that the integral equation (\ref{inteq}) is in two variables, $t$
and $q_\parallel$. Moreover, the electron-phonon interaction vertex part $\Phi$, arising from the
integration over $z$ coordinate of the electron-phonon interaction nodule and depending only on the
phonon momenta, enters in the equation in non-explicit form. (Recall that
based on the assumption $\Delta E_d\gg\omega_B,\omega_0$, we put $m=1$ in
all lines.) This vertex part with an external phonon line, however, enter as
a factor into the amplitudes $\Sigma$ and $\Box$ (the summation over $m$ is
absent in one electron line of the vertex in order to catch on the vertex
with internal lines). Therefore, the integral equation (\ref{inteq}) for the
amplitude $\Sigma_n^l$ can be simplified by symmetrizing it and averaging
over $q_\parallel$. We introduce for this purpose the averaged form factors 
\begin{equation}
{\hat \Phi}(t)=\sum_{q_\parallel}\Phi({\bf q_\perp},q_\parallel).
\label{hff}
\end{equation}
and new amplitudes 
\begin{eqnarray}
&&\hat{\Sigma}_n^l(\varepsilon;t,t^{\prime})=[\hat{\Phi}(t)\hat{\Phi}%
(t^{\prime})]^{-1/2} \sum_{q_\parallel ,
q_\parallel^{\prime}}\Phi(t,q_\parallel)
\Phi(t^{\prime},q_\parallel^{\prime}) \Sigma^l_{n}(\varepsilon;t,t^{\prime};
q_\parallel , q_\parallel^{\prime}), \\
&&\hat{\Box}_n^l(t,t^{\prime})=[\hat{\Phi}(t)\hat{\Phi}(t^{\prime})]^{1/2}%
\Box_n^l(t,t^{\prime})  \label{newffampl}
\end{eqnarray}
Transforming to the dimensionless quantities 
\begin{equation}
R_n^l(\varepsilon;t,t^{\prime})=-\omega_{0}\hat{\Sigma}_n^l(%
\varepsilon;t,t^{\prime}),\, K_n^l(t,t^{\prime})=-\omega_{0}\hat{\Box}%
_l^l(t,t^{\prime}),\, \lambda_l(\varepsilon)=-\omega_{0}^{-1} \tilde{M}%
_l(\varepsilon)  \label{eq15}
\end{equation}
we obtain the final integral equation 
\begin{equation}
R_l^l(\varepsilon;t,t^{\prime})=K_l^l(t,t^{\prime})+\lambda_l(\varepsilon)
\int_{0}^{\infty} d\bar{t} K_l^l(t,\bar{t})R_l^l(\varepsilon;\bar{t}%
,t^{\prime}).  \label{eq16}
\end{equation}
Thus, $R$ is the Fredholm resolvent of the kernel $K$, which we now write
out in explicit form 
\begin{eqnarray}
K_{n}^{l}(t,t^{\prime})&=&\left[\hat{\Phi}(t) \hat{\Phi}(t^{\prime})%
\right]^{1/2} \sum_{s=0}^{\infty}Q_{ns}(t)Q_{ns}(t^{\prime})  \nonumber \\
&\times& \left[{\frac{\sigma}{\sigma+s-n}}J_{l+s-2n}(2\sqrt{tt^{\prime}})-{%
\frac{\sigma}{\sigma-s+n}}\delta_{ls}\right].  \label{kernel}
\end{eqnarray}
Here the $J_{l}$ is the Bessel function, $\sigma=\omega_{LO}/\omega_B$, and
the two terms in the square brackets are connected with the two terms in
Fig.~ \ref{fg4}.

Now it is clear that the scattering amplitude $\Sigma$ has a pole in $%
\varepsilon$ if $\lambda_l$ coincides with an eigenvalue of the kernel $K_l^l
$. In other words, the equation for the energies of the bound states of the
electron and optical phonon is 
\begin{equation}
\lambda_l(\varepsilon)=\lambda_{n,r}^l,\,r=1,2,3,\dots  \label{bas}
\end{equation}
where the subscript $r$ numbers different eigenvalues $\lambda_{n,r}^l$ of
the kernel $K_n^l$. It follows from Eq.~ \ref{bas} that the energies of the
bound states are 
\begin{equation}
\varepsilon^l_{n,r}=\varepsilon_n-{\frac{\alpha\omega_B}{\lambda_{n,r}^l}}%
\equiv \varepsilon_n-W^l_{n,r}  \label{spec}
\end{equation}
where $W^l_{n,r}$ are binding energies of the bound state referred to the
Landau level $n$. It is seen that for any $n=1,2,3,\dots$, all bound states
are rigorously stationary. The bound states are identified by the total
angular momentum $l$ of the rotation around ${\bf B}$. It takes both
non-negative and negative integer values. For the wealth of the spectrum for
a given $n$ and $l$, participation in the formation of bound states of
phonons with any momentum is liable. In the case of a 3D samples, the bound
states appeared only below the threshold, for in this case a continuum of
two-particle electron+phonon states is located above the threshold. There is
no such continuum for the 2D electron systems so that bound states are
present both below the "threshold" ($W> 0$) and above it ($W< 0$). This
means that the energies of the bound states are determined, according to
Eq.~ \ref{spec}, by eigenvalues $\lambda_{n,r}^l$ of both signs.

\section{Cyclotron-phonon resonance}

\label{cpr} In this section, the effect of the electron and optical phonon
bound states on the spectrum of the cyclotron phonon resonance in the 2DEG is studied. The cyclotron phonon resonance is
observed in semiconductors as a result of optical transitions of an electron
from one Landau level to another accompanied by the emission and absorption
of an optic phonon. According to the simplest ideas based on the
perturbation theory \cite{lev65,enck69}, the absorption peaks corresponding
to the emission of the optical phonons should observed for photons of
frequencies $\nu_n=n\omega_{B}+\omega_0$. The more rigorous analysis \cite
{bakanas73} shows, however, the cyclotron phonon resonance peaks should have fine structure due to
appearance of the bound states of an electron and a newly created optical
phonon. Thus, the bound states in the 2DEG should be developed in the
absorption spectrum at energies 
\begin{equation}
\hbar\nu=\hbar\nu_n-W^l_{nr},\;\nu_n=n\omega_{B}+\omega_0.  \label{cprfrq}
\end{equation}
On absorbing a photon of this frequency, the electron is transferred from
the initial $s=0$ Landau level to the $s=n$ level and simultaneously creates
the optical phonon which is bound to the electron in the final state.

In calculating the absorption coefficient it is assumed that the temperature
and density of the carrier are low enough so that $ T\ll W$ and the Fermi
energy $\varepsilon_F\ll W$. Therefore before the light absorption, all
electrons are in the level $s=0$. The capacitance of one Landau level
allowing for two spin orientations in a magnetic field $B=5$ T amounts $%
N_0=2.4\cdot10^{15}$ m$^{-2}$. We assume also that electron transitions
occur between the Landau levels corresponding to the lower transverse
quantization level, $k=1$, and the higher transverse quantization levels, $%
k>1$ are located so far away that they can be ignored. The thickness of the
2DEG sheet, for which spatial quantization effects become apparent, is so
small that the spatial dispersion of the light. can be neglected.

\subsection{Absorption coefficient and bound states}

\begin{figure}[htb]
\epsfxsize=13cm
\epsfysize=7cm
\mbox{\hskip 0cm}\epsfbox{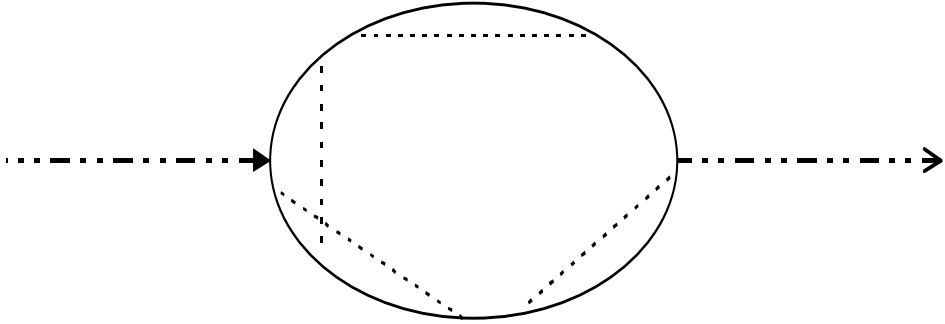}
\caption{Photon polarization operator $\Pi^\pm(\nu)$.}
\label{fg5}
\end{figure}

\begin{figure}[htb]
\epsfxsize=14cm
\epsfysize=5cm
\mbox{\hskip 0cm}\epsfbox{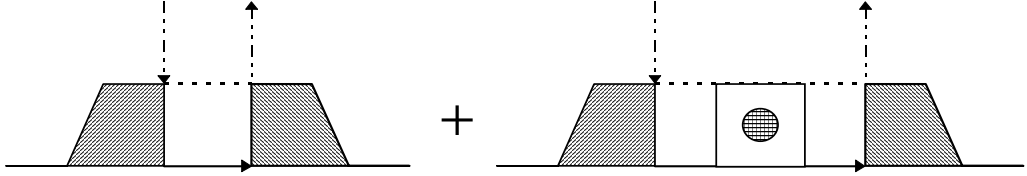}
\caption{Sum of diagrams for the photon polarization operator 
$-iP^\pm(\nu)$.}
\label{fg6}
\end{figure}

The absorption coefficient can be defined from Maxwell equations as a light
energy fraction $w$ at the frequency $\nu$ absorbed in the 2DEG in the ${\bf %
z}$-direction. It is convenient to express the electron current density in
Maxwell equations in terms of the photon polarization operator $\Pi$. Then
using the calculation methods in \cite{korov79}, the fraction $w$ sought for
can be represented in the form 
\begin{eqnarray}
&&w^\pm(\nu)=-{\frac{4\pi c}{\nu\sqrt{\kappa(\nu)}}}\mbox{Im}\Pi^\pm(\nu),
\label{fraction} \\
&&\Pi^\pm(\nu)=\int dz_1dz_2
\Pi^{\pm}(\nu;z_1,z_2)|\psi(z_1)|^2|\psi(z_2)|^2.  \label{polop}
\end{eqnarray}
Here $\Pi^\pm (\nu;z_1,z_2)$ is the photon polarization operator in the
coordinate representation. It is assumed that the incident light is along $%
{\bf z}$-direction and has circular polarization. The plus and minus signs
pertain here to right-hand and left-hand polarization of the light, and $%
\kappa(\nu)$ is the dielectric permeability of the well walls at the
frequency $\nu$. In the diagram technique, the sum of two diagrams in Fig.~ 
\ref{fg5} corresponds to the polarization operator $\Pi^\pm(\nu)$. 
The diagrams consist of electron loops with all possible phonon lines and,
it is convenient, to consider that all the electron and phonon lines are
"bold", {\it i.e.} the exact electron and phonon Green functions correspond
them. Wave lines on this figure are referred to the photon lines. Each
electron Green function, in general, is the sum of the retarded and advanced
Green functions, $G=G^R+G^A$. However, $G^A$ is finite only for a small
range of electronic states, the volume of which is proportional to the
electron concentration. If put $G= G^R$ in all lines and to carry out
integration over the electron energetic parameter $\varepsilon$, then one
can be convinced that $\Pi=0$ as far as the expression to be integrated is
analytical in upper half plane Im$\varepsilon>0$. In order to obtain an
expression for $\Pi$ in the lowest order in the electron concentration it is
necessary in each diagram to replace $G$ by $G^A$ in one electron line and
to replace $G$ by $G^R$ in all remaining lines. The line $G^A$ picked up in
this way is referred to the initial electron state in the lowest energy
level $E_0$ and so far as the perturbation theory is valid in this energy
range then the free Green function for $G^A$ can be taken, 
\begin{equation}
G^A_s(\varepsilon)={\frac{2\pi a_B^2N}{\varepsilon-s\hbar\omega_B-i0}}%
\delta_{s0}.  \label{greenadv}
\end{equation}
Here $N$ is the carrier density for $1\mbox{cm}^2$ area of the 2DEG plane.
Now taking the integral over parameters of the function Eq.~ (\ref{greenadv}%
), $\Pi$ can be represented in the form 
\begin{equation}
\Pi^\pm(\nu)=N P^\pm(\nu),  \label{npolop}
\end{equation}
where the quantity $-iP^\pm(\nu)$ is determined by the sum of diagrams shown
in Fig.~ \ref{fg6}. 
In this figure shaded parts are represented by the compact diagrams without
dangerous intersections in one electron and one phonon lines referred to the
Landau level $n$. As far as we are interested in the frequency range $%
\nu\approx\nu_n$, it is clear that the main contribution to $-iP^\pm(\nu)$
comes from diagrams with dangerous intersections. Therefore in the sum in
Fig.~ \ref{fg6}, the diagrams without dangerous intersections giving a
smoothly varying contribution in $-iP^\pm(\nu)$ and coinciding with the free
electron contribution in the absorption are not depicted.

Now making use of the gradient invariant diagram technique, the
contributions of diagrams in Fig.~ \ref{fg6} can be calculated explicitly.
In terms of the scattering amplitude $\Sigma^l$ with $l=\pm 1$, it gives 
\begin{eqnarray}
&&\Pi^\pm(\nu)={\frac{N}{2}}{\frac{e^2}{mc^2}}{\frac{\omega_B\omega_{LO} }{%
[\omega_0+(n\mp1)\omega_{B}]^2}}\left\{\lambda(\nu)\bar{\Phi}_n+
\lambda(\nu)^2\bar{R}_n^\pm(\nu)\right\},  \label{polarop} \\
&&\bar{R}_n^\pm(\nu)= {\frac{1}{n!}}\int_0^\infty dt
dt^{\prime}e^{-(t+t^{\prime})/2}(tt^{\prime})^{(n+1)/2} [\hat{\Phi}(t)\hat{%
\Phi}(t^{\prime})]^{1/2}R^{\pm 1}_{n}(\nu;t,t^{\prime}).  \label{resolvent}
\\
&&\bar{\Phi}_n={\frac{1}{n!}}\int_0^\infty dt e^{-t}t^{n+1}e^{-t}\hat{\Phi}%
(t).  \label{bff}
\end{eqnarray}
Notice that two terms in the figure brackets in Eq.~ (\ref{polarop}) are
connected with the two terms of Fig.~ \ref{fg6}. Expanding the resolvent $R$
as a sum over the poles, we bring $\bar{R}_n^\pm(\nu)$ to the form 
\begin{eqnarray}
&&\bar{R}_n^\pm(\nu)=-\sum_{r=0}^{\infty}{\frac{|d_{nr}^{\pm}|^2}{%
\lambda(\nu)-\lambda_{nr}^{\pm1}+i0}}, \\
&&d_{nr}^{\pm}={\frac{1}{\sqrt{n!}}}\int_0^\infty dt e^{-t/2}t^{(n+1)/2} 
\hat{\Phi}(t)^{1/2}\chi^{\pm 1}_{nr}(t),  \label{poleresolvent}
\end{eqnarray}
where $\chi^{\pm 1}_{nr}(t)$ are the eigenfunctions of the kernel $K^{\pm
1}_{n}(t,t^{\prime})$. Now for the absorbed fraction of the radiation energy 
$w$, we obtain ultimately 
\begin{equation}
w^\pm(\nu)=2\pi N{\frac{e^2}{m c^2}}{\frac{c}{\sqrt{\kappa(\nu)}}} {\frac{%
\omega_B\omega_{LO}}{[\omega_0+(n\mp1)\omega_{B}]^2}} {\frac{1}{\nu_n}} %
\mbox{Im}{\cal F}_n^\pm(\nu),  \label{absrpt0}
\end{equation}
where 
\begin{equation}
{\cal F}_n^\pm(\nu)= -\lambda(\nu)\sum_{r=0}^{\infty} {\frac{%
\lambda_{nr}^{\pm1}|d_{nr}^{\pm}|^2 }{\lambda(\nu)-\lambda_{nr}^{\pm1}+i0}}.
\label{F}
\end{equation}
Here we use the completeness of eigenfunctions $\chi^{\pm 1}_{nr}(t)$ and
unify both terms in the figure brackets in Eq.~ (\ref{polarop}). From Eq.~ (%
\ref{F}) we see that for $\nu<\nu_n$, only terms of the sum contribute to $%
\mbox{Im}{\cal F}$ for which $\lambda_{nr}^{\pm 1}>0$, and on the contrary,
for $\nu>\nu_n$, only the terms, for which $\lambda_{nr}^{\pm1} <0$. It is
easy to be convinced that $\mbox{Im}{\cal F}=0$ for $\nu=\nu_n$. Taking the
imaginary part of ${\cal {F}}$ and using up the delta-function in $\nu$
variable, we obtain the following final formula for the absorption
coefficient 
\begin{equation}
w^\pm(\nu)=2\pi N{\frac{e^2}{m c^2}}{\frac{c}{\sqrt{\kappa(\nu_n)}}}
\sum_{r=0}^{\infty} f_{nr}^\pm\pi\delta(\nu-\nu_n+W_{nr}^{\pm 1}).
\label{absrpt}
\end{equation}
The oscillator strength for the transition to the bound state $nr$ with $l=
\pm 1$ is here 
\begin{equation}
f_{nr}^\pm=\alpha{\frac{\omega_{0}}{\omega_0+n\omega_B}} {\frac{%
\omega_B\omega_{LO}}{[\omega_0+(n\mp1)\omega_{B}]^2}} |d_{nr}^{\pm}|^2.
\label{oscillstr}
\end{equation}
Recall that in Eq.~ (\ref{absrpt}), $W_{nr}^{\pm 1}$ denote the binding
energies of the bound states with $l=\pm 1$, located near the "threshold" $%
\nu=\nu_n$. So far as for the initial state $l=0$ then it is seen that the
same selection rules as for free electrons are remained for the total
momentum $l$. At the absorption of right-hand polarized light, the bound
states with $l=+1$ are manifested while for left-hand polarized light, the
bound states with $l=-1$ are actual. Namely on this ground, one can consider 
$l<0$ as a generalization of the Landau quantum number.

Using the identity 
\begin{equation}
\sum_{r=0}^{\infty}|d_{nr}^{\pm}|^2={\bar \Phi}_n,  \label{equiv}
\end{equation}
one can find for the total oscillatory strength 
\begin{equation}
f_n^\pm\equiv\sum_{r=0}^{\infty}f_{nr}^{\pm}=\alpha{\frac{\omega_{0}}{%
\omega_0 +n\omega_B}}{\frac{\omega_B\omega_{LO}}{[\omega_0+(n\mp1)%
\omega_{B}]^2}} {\bar \Phi}_n.  \label{totstrg}
\end{equation}
It is seen that at $n=1$ in the strong magnetic fields, $\omega_B\gg\omega_0$%
, there is an asymmetry in the absorption with respect to the left- and
right-hand polarizations.

Notice that the first term in the figure brackets in Eq.~ \ref{polarop}
corresponds to a perturbation-theory calculation of the absorption and
yields a delta-like absorption peak at the "threshold" for $\nu=\nu_n$.
Actually, there is no such peak in absorption. It is canceled out when the
second term is taken into account so that absorption is concentrated on the
bound states below and above the "threshold" $\nu=\nu_n$. In the 3D case,
this corresponds to the transformation of the threshold singularity $%
(\nu-\nu_n)^{-1/2}$, which becomes infinite at the threshold, into the
singularity $(\nu-\nu_n)^{1/2}$, which goes to zero \cite{lev72ii}.

\section{Binding energies and oscillator strengths}

\label{beos} The investigation of the kernel (\ref{kernel}) cannot be done
in general case. Therefore in this section, the actual calculation of the
binding energies and of the oscillator strengths are carried out by
considering separately the limiting cases of strong and weak magnetic
fields, {\it i.e.} $\omega_B\gg\omega_0$ and $\omega_B\ll\omega_0$.

\subsection{The averaged form factors}

We find first the averaged form factors ${\bar \Phi}$ and ${\hat \Phi}(t)$
for the bulk PO phonons inside quantum wells and in heterostructures as well
as for the surface SO phonons created in single heterostructures near
interfaces separating different materials. These form factors, just as the
electron-phonon interaction coupling constants $\alpha$, depend on the interaction mechanism and on
the nature of the phonon localization. According to the experimental
situation, we consider that in the quantum wells, electrons are confined in
the infinitely high quantum well (par example, in the $AlAs/GaAs/AlAs$ type
quantum wells, the height of the well walls is circa $1500$ meV at the same
time as the energy of the lowest level of the spatial quantization is around 
$50$ meV). Hence the electron wave functions and energies for an electrons
moving along ${\bf z}\parallel{\bf B}$-direction are 
\begin{equation}
\psi_k(z)=\sqrt{\frac{2}{d}}\sin{\frac{\pi k}{d}}z,\: k=1,2,3,\dots, \;
\varepsilon_k=k^2\hbar\omega_d,\:\hbar\omega_d={\frac{\pi^2\hbar^2}{2m_cd^2}}
\label{wf}
\end{equation}
where $d$ is the width of the well. The bulk-like PO phonons in the quantum
wells are confined between well facings \cite{klein,klein1} and, therefore, their
wave functions for the moving along ${\bf z}$ are 
\begin{eqnarray}
\varphi_{q_\parallel}(z)=\sqrt{\frac{2}{d}}\sin q_\parallel z, \:
q_\parallel={\frac{\pi}{d}}j, \: j=1,2,3,\dots.  \label{pwf}
\end{eqnarray}
In the case of the heterostructure, we use the variation electron wave
function for the lowest level of the spatial quantization suggested in \cite
{fanghow,sternhow}, 
\begin{equation}
\psi(z)= \left ({\frac{b^3}{2}}\right)^{1/2}z\exp\left(-{\frac{bz}{2}}\right)
\label{varywf}
\end{equation}
where the parameter $b$ is determined from the minimum of the total electron
energy. The average distance $\bar{z}$ of electrons from the heteroface is $%
\bar{z}=3/b$.

In this case, the bulk PO phonon motion along ${\bf z}$ is confined from one
side by the heteroface (PO phonons do not penetrate from one material of the
heterostructures to another.). Therefore, wave functions are given by 
\begin{equation}
\varphi_{q_\parallel}(z)=\sqrt{\frac{2}{L}}\sin q_\parallel z, \:
q_\parallel={\frac{\pi}{L}}j, \: j =1,2,3,\dots,  \label{heterowf}
\end{equation}
where $L$ is the normalizing length in that semiconductor where the 2DEG is
located.

Now using the condition that the well be narrow, which is equivalent to $%
d\ll a_B, p_0$, we find the averaged form factors (\ref{hff}) and (\ref{bff}%
):

for the PO phonons in the quantum wells 
\begin{eqnarray}
&&\hat{\Phi}(t)= \left({\frac{2}{\pi}}\right)^3\sum_{n=1}^\infty{\frac{1}{n^2%
}} \left|\int_0^\pi d\zeta \sin^2\zeta\sin n\zeta\right|^2\delta^{1/2}=
\left({\frac{\pi}{12}}+{\frac{5}{8\pi}}\right)\delta^{1/2}\approx
0.46\delta^{1/2},  \label{hffqw} \\
&&\bar{\Phi}_n=(1+n)\hat{\Phi}(t), \: \delta={\frac{\omega_{PO}}{\omega_d}}%
=(p_{PO}d/\pi)^2,  \label{bffqw}
\end{eqnarray}

for the PO phonons in the heterostructures 
\begin{equation}
\hat{\Phi}(t)={\frac{11\pi}{16}}\delta^{1/2}\approx 2,16\delta^{1/2}, \; 
\bar{\Phi}_n=(1+n)\hat{\Phi}(t), \: \delta=(p_{PO}\bar{z}/\pi)^2.
\label{ffhs}
\end{equation}
For the surface SO phonons created at the single heterofaces one can obtain 
\begin{equation}
\hat{\Phi}(t)={\frac{1}{2}}\left({\frac{\sigma}{t}}\right)^{1/2},\; \bar{\Phi%
}_n={\frac{(2n+1)!!}{2^{n+2}n!}}\sqrt{\pi\sigma}, \: \sigma={\frac{%
\omega_{SO}}{\omega_B}}.  \label{ffso}
\end{equation}
Notice that form factors $\hat{\Phi}(t)$ and $\bar{\Phi}_n$ which enter in
the kernel (\ref{kernel}) and in the polarization operator (\ref{polarop}),
respectively, do not depend on the width of the 2DEG in the case of the
SO phonons (so long as $d\ll a_B,p^{-1}_{SO})$ while in the case of the bulk
PO phonons, the width $d$ enters in the form factors only as a factor. For
the latter reason, the binding energies and oscillator strengths of the
PO phonon assisted bound states in the quantum wells and heterostructures
differ only by numerical factors appeared in the form factors $\hat{\Phi}(t)$
in Esq.~ (\ref{hffqw}) and (\ref{ffhs}). For brevity, therefore, below we
adduce only results for PO phonons in quantum wells, bearing in mind that
the binding energies and oscillator strengths for PO phonons in
heterostructures can be obtained from $W$ and $f$ in the quantum wells by
multiplying them with a numerical factor $4.7$.

\subsection{Strong magnetic fields}

In the following we will consider the bound states for $l=\pm1$, since only
these branches of the bound states contribute to the light absorption.

In the strong fields we have $\sigma\equiv\omega_0/\omega_B\ll1$, and
therefore, only the term $s=0$ is to be retained in the sum for the kernel (%
\ref{kernel}), so that 
\begin{equation}
K_n^{\pm 1}(t,t^{\prime})=\pm[\hat{\Phi}(t)\hat{\Phi}(t^{\prime})]^{1/2}
Q_{nn}(t)Q_{nn}(t^{\prime})\left[J_{\pm1-n}(2\sqrt{tt^{\prime}}%
)-\delta_{\pm1n}\right].  \label{kernelsf}
\end{equation}
For the PO phonons at $n=0$, one is managed to find the solution of this
kernel using the invariance of the Laguerre polynomials with respect to the
Bessel transformation. The kernel (\ref{kernelsf}) differs only by a factor
from the kernel investigated in Ref.~ \cite{lev72i}. Therefore, borrowing
the eigenvalues and eigenfunctions from this reference, we get the binding
energies and oscillator strengths in this case 
\begin{eqnarray}
&&W^{\pm}_r=\pm(-1)^r{\left({\frac{\pi}{12}}+{\frac{5}{8\pi}}\right)}
\rho^{2(r+1)}\delta^{1/2}\sigma^{-1} \alpha\omega_{LO},  \label{besf} \\
&&f_r^\pm=5(r+1)\left({\frac{\pi}{12}}+{\frac{5}{8\pi}}\right)
\rho^{4(r+1)}\delta^{1/2} \alpha,  \label{ossf}
\end{eqnarray}
where $\rho=(\sqrt{5}-1)/2=0.618\dots$ coincides with the golden ratio, $%
r=1,2,3,\dots$. We see from here that the spectra and oscillator strengths
of the branches $l=1$ and $l=-1$ are obtained from each other by mirror
reflection about the threshold. This is true also for the case of the
SO phonons at $n=0$. Comparison of the formulas (\ref{besf}) and (\ref{ossf}%
) with corresponding formulas from \cite{lev72i,lev72ii} shows that the
binding energies and oscillator strengths fall slowly in the series than in
the three dimensional case. As before, however, the first line contains the
main part of the total absorption and is mostly removed from the
"threshold". For the case of the SO phonons as well as for the PO phonons at 
$n\neq0$, one is not succeeded to find the exact solution of the kernel $%
K_n^{\pm 1}$. Nonetheless, one can see from Eq.~ (\ref{kernelsf}) that in
this case, all the physical parameters, particularly $d$ and $B$, enter in
the kernel $K_n^{\pm 1}$ only in the form of multipliers (via the from
factors). This means that the eigenvalues $\lambda^{\pm1}_{nr}$ are
proportional to these factors and the eigenfunctions $\chi^{\pm1}_{nr}$ do
not depend on them. As a result, the dependencies of the binding energies
and the oscillator strengths on $d$ and $B$ are easily determined. Applying
the methods related with the Sylvester determinants \cite{gelfand}, one can
show that eigenvalues of the kernel \ref{kernelsf} of either sign exist, and
their number in both signs is infinite. The increasing velocity of the
eigenvalues is determined by the analytical features of the kernel \cite
{gelfand2}; as far as the later are the same both for the case of the
PO phonons at $n=0$ and for the case of the PO phonons at $n\neq0$ as well
as for the case of the SO phonons, then one can anticipate that the
eigenfunctions of the kernel (\ref{kernelsf}) will increase exponentially.
This means that the formulas (\ref{besf}) and (\ref{ossf}), obtained in the
particular case, correctly illustrate the dependencies of the binding
energies and the oscillator strengths on the series number $r$. Formulas for
the binding energies and the oscillator strengths of the most removed bound
states are collected in the Table~ \ref{tb1}. 
\begin{table}[h]
\caption{Binding energies and oscillator strengths in the strong magnetic
fields. Recall that $W_{2D}=\alpha\hbar\omega_{0}$ denotes the
characteristic binding energy scale for the 2DEG. The bound state index $%
r=2,3,4,\dots$.}
\label{tb1}
\begin{center}
\begin{tabular}{|c|c|c|}
\hline\hline
{Phonon} & {Binding energies} & {Oscillator strengths} \\ 
mode &  &  \\ \hline
&  &  \\ 
& {$W^{\pm}_{0r}=\mp(-1)^r \rho^{2r}\delta^{\frac{1}{2}%
}\sigma^{-1}W_{2D}^{PO}$} & {$f_{0r}^\pm= r\rho^{4r}\delta^{\frac{1}{2}%
}\alpha_{PO}$} \\ 
&  &  \\ 
PO & {$|W^{\pm}_{nr}|\sim\delta^{\frac{1}{2}}\sigma^{-1}W_{2D}^{PO},\, n\geq1
$} & {$f_{1r}^{+}\sim\delta^{\frac{1}{2}}\sigma^{-1}\alpha_{PO},\,
f_{1r}^{-}\sim\delta^{\frac{1}{2}}\sigma\alpha_{PO} $} \\ 
&  &  \\ 
&  & {$f_{nr}^{\pm}\sim\delta\sigma\alpha_{PO}, \, n\geq2 $} \\ 
&  &  \\ \hline
&  &  \\ 
&  & {$f_{0r}^\pm\sim{\sigma}^{\frac{1}{2}}\alpha_{SO}$} \\ 
&  &  \\ 
SO & {$|W^{\pm}_{nr}|\sim\sigma^{-{\frac{1}{2}}}W_{2D}^{SO}$} & {$%
f_{1r}^{+}\sim{\sigma}^{-{\frac{1}{2}}}\alpha_{SO},\, f_{1r}^{-}\sim{\sigma}%
^{\frac{3}{2}}\alpha_{SO}$} \\ 
&  &  \\ 
&  & {$f_{nr}^{\pm}\sim{\sigma}^{\frac{3}{2}}\alpha_{SO}, \, n\geq2 $} \\ 
&  &  \\ \hline\hline
\end{tabular}
\end{center}
\end{table}
Thus, in the strong magnetic fields for any $l=\pm1$ both for PO  and
SO phonons, there exist two sequences of bound states concentrated to the
"threshold" both from above and below it.

\subsection{Weak magnetic fields}

In weak fields ($\sigma\equiv\omega_0/\omega_B\gg 1$), the kernel (\ref
{kernel}) can be approximated by a sequence of degenerate kernels \cite
{lev72i,lev72ii}, {\it i.e.} terms of the sum in (\ref{kernel}) can be
expanded in powers of $\sigma^{-1}$ so that this (\ref{kernel}) can be
represented by 
\begin{equation}
K^\pm_n(t,t^{\prime})=\sum_{i,j=0}^{\infty}\left(K^\pm_n\right)_{ij}
\left(g_n(t)\right)_{i}\left(g_n(t^{\prime})\right)_{j}  \label{kerneldeg}
\end{equation}
where 
\begin{equation}
\left(g_n(t)\right)_{i}=t^{\frac{1+n+2i}{2}}e^{-{\frac{t}{2}}}\sqrt{\hat{\Phi%
}(t)}.  \label{coeff}
\end{equation}
The coefficients $\left(K^\pm_n\right)_{ij}$ are constructed in such a way
that if we retain a finite number of terms in the expansion in terms of $%
\sigma^{-1}$, an infinite matrix reduces to a finite one while the integral
equation--to a finite system of linear equations. Solving the obtained
finite system of linear equations one can find the eigenvalues $%
\lambda^{\pm1}_{nr}$ and the integral (\ref{resolvent}) from the resolvent $%
R_n^{\pm1}$. For example, in the principal order retaining the terms of the
order of $\sigma^{-1}$ and dropping the terms of the higher order of $%
\sigma^{-2}, \sigma^{-3}, \dots$ we find for:

$n=0$ 
\begin{equation}
\left|\left|K^{+}_0\right|\right|=
\begin{array}{||cc||}
\sigma^{-1} & 0 \\ 
0 & \sigma^{-1}
\end{array}
\, , \; \left|\left|K^{-}_0\right|\right|=
\begin{array}{||c||}
-\sigma^{-1}
\end{array}
\, ,  \label{coeff0}
\end{equation}

for $n=1$ 
\begin{eqnarray}
\left|\left|K^{+}_1\right|\right|=
\begin{array}{||cc||}
-3\sigma^{-1} & \sigma^{-1} \\ 
\sigma^{-1} & -{\frac{1}{2}}\sigma^{-1}
\end{array}
, , \; \left|\left|K^{-}_1\right|\right|=
\begin{array}{||c||}
-\sigma^{-1}
\end{array}
\, .  \label{coeff1}
\end{eqnarray}
In this lowest order in $B$ we get three bound states above the "threshold":
two with $l=+1$ and one with $l =-1$. Expressions obtained for the binding
energies and oscillator strengths at $n=0$ and $n=1$ are collected in Table~ 
\ref{tb2} and \ref{tb3}, respectively for the bound states with assistance
of the PO phonons in the quantum wells and heterostructures and of the
SO phonons created at the single heterofaces. 
\begin{table}[h]
\caption{Binding energies and oscillator strengths in the weak magnetic
fields for the PO phonon assisted bound states in quantum wells.}
\label{tb2}
\begin{center}
\begin{tabular}{|c|c|c|}
\hline\hline
{n} & {Binding energies} & {Oscillator strengths} \\ 
&  &  \\ \hline
&  &  \\ 
& {$W^{+}_{01,3} =(2\pm\sqrt{3})\left({\frac{\pi}{12}}+{\frac{5}{8\pi}}%
\right) \delta^{\frac{1}{2}}\sigma^{-2}W_{2D}^{PO}$} & {$f_{01,3}^{+}={\frac{%
\sqrt{3}\pm1}{2\sqrt{3}}} \left({\frac{\pi}{12}}+{\frac{5}{8\pi}}\right)
\delta^{\frac{1}{2}}\sigma^{-2}\alpha_{PO}$} \\ 
n=0 &  &  \\ 
& {$W^{-}_{02}= -\left({\frac{\pi}{12}}+{\frac{5}{8\pi}}\right) \delta^{%
\frac{1}{2}}\sigma^{-2}W_{2D}^{PO}$} & {$f_{02}^{-}=\left({\frac{\pi}{12}}+{%
\frac{5}{8\pi}}\right) \delta^{\frac{1}{2}}\sigma^{-2}\alpha_{PO}$} \\ 
&  &  \\ \hline
&  &  \\ 
& {$W^{+}_{11,3}=-(3\pm\sqrt{3})\left({\frac{\pi}{12}}+{\frac{5}{8\pi}}%
\right) \delta^{\frac{1}{2}}\sigma^{-2}W_{2D}^{PO}$} & {$f_{11,3}^{+}=\left({%
\frac{\pi}{12}}+{\frac{5}{8\pi}}\right) \delta^{\frac{1}{2}%
}\sigma^{-2}\alpha_{PO}$} \\ 
n=1 &  &  \\ 
& {$W^{-}_{12}= -2\left({\frac{\pi}{12}}+{\frac{5}{8\pi}}\right) \delta^{%
\frac{1}{2}}\sigma^{-2}W_{2D}^{PO}$} & {$f_{12}^{-}=2\left({\frac{\pi}{12}}+{%
\frac{5}{8\pi}}\right) \delta^{\frac{1}{2}}\sigma^{-2}\alpha_{PO}$} \\ 
&  &  \\ \hline\hline
\end{tabular}
\end{center}
\end{table}
\begin{table}[h]
\caption{Binding energies and oscillator strengths in the weak magnetic
fields for the SO phonon assisted bound states at the heterofaces.}
\label{tb3}
\begin{center}
\begin{tabular}{|c|c|c|}
\hline\hline
{n} & {Binding energies} & {Oscillator strengths} \\ 
&  &  \\ \hline
&  &  \\ 
& {$W^{+}_{01,3}=-{\frac{23\pm\sqrt{337}}{64}}\sqrt{\pi}\, \sigma^{-{\frac{3%
}{2}}}W_{2D}^{SO}$} & {$f_{01,3}^{+}={\frac{\sqrt{337}\pm11}{8\sqrt{337}}}%
\sqrt{\pi}\, \sigma^{-{\frac{3}{2}}}\alpha_{SO}$} \\ 
n=0 &  &  \\ 
& {$W^{-}_{02}= -{\frac{\sqrt{\pi}}{4}}\,\sigma^{-{\frac{3}{2}}}W_{2D}^{SO}$}
& {$f_{02}^{-}={\frac{\sqrt{\pi} }{4}}\,\sigma^{-{\frac{3}{2}}}\alpha^{SO}$}
\\ 
&  &  \\ \hline
&  &  \\ 
& {$W^{+}_{11,3}=-{\frac{24}{4(19\pm\sqrt{41})}}\sqrt{\pi}\, \sigma^{-{\frac{%
3}{2}}}W_{2D}^{SO}$} & {$f_{11,3}^{+}={\frac{3(\sqrt{41}\pm1)}{16\sqrt{41}}}%
\sqrt{\pi}\, \sigma^{-{\frac{3}{2}}}\alpha_{SO}$} \\ 
n=1 &  &  \\ 
& {$W^{-}_{12}= -{\frac{3\sqrt{\pi}}{8}}\,\sigma^{-{\frac{3}{2}}}W_{2D}^{SO}$%
} & {$f_{12}^{-}={\frac{3\sqrt{\pi} }{8}}\,\sigma^{-{\frac{3}{2}}}\alpha^{SO}
$} \\ 
&  &  \\ \hline\hline
\end{tabular}
\end{center}
\end{table}
It is seen from Tables~ \ref{tb2} and \ref{tb3} that the state with $l=-1$
is located between the two states with $l= + 1$ and the oscillator strength
of the state with $l=-1$ is the sum of the oscillator strengths of the two
states with $l=+1$.
If retain also terms of the orders of $\sigma^{-2}$, the order of the
matrixes $\left(K^\pm_n\right)_{ij}$ will increase by one, and new bound
states will appear with the binding energies and oscillator strengths
proportional to $\sigma^{-2}$. Omitting highly cumbersome computations, we
adduce only final formulas for the binding energies and the oscillator
strengths at $n=0$ of higher order in $\sigma^{-1}$:

for the PO phonon assisted bound states 
\begin{eqnarray}
&& W^{\pm}_{0r} =(-1)^r\left({\frac{\pi}{12}}+{\frac{5}{8\pi}}\right)r!
\delta^{\frac{1}{2}}\sigma^{-r-1}W_{2D}^{PO}, \\
&& f_{0r}^{\pm}=\left({\frac{\pi}{12}}+{\frac{5}{8\pi}}\right) \left\{\!
\begin{array}{c}
r+1 \\ 
r
\end{array}
\!\right\} (r!)^2\delta^{\frac{1}{2}}\sigma^{-2r}\alpha_{PO},  \label{pobeos}
\end{eqnarray}

for the SO phonon assisted bound states 
\begin{eqnarray}
&&W^{\pm}_{0r}=(-1)^r\sqrt{\pi}{\frac{(2r-1)!!}{2^{r+1}}} \left\{\!
\begin{array}{c}
{\frac{2r+1}{2(r+1)}} \\ 
1
\end{array}
\!\right\} \sigma^{-r-{\frac{1}{2}}}W_{2D}^{SO}, \\
&& f_{0r}^{\pm}=\sqrt{\pi}{\frac{\left[(2r-1)!!\right]^3}{2 ^{3r-1}r!}}
\left\{\!
\begin{array}{c}
\left[{\frac{2r+1}{2(r+1)}}\right]^3 \\ 
r
\end{array}
\!\right\}\sigma^{-2r+{\frac{1}{2}}}\alpha_{SO}.  \label{sobsos}
\end{eqnarray}
In these formulas $r=2,3,4,\dots$. The upper and lower values in the figure
brackets pertain to the right- and left-hand polarization of the light.
Recall that $W_{2D}=\alpha\hbar\omega_{0}$ denotes the characteristic
binding energy scale for the 2DEG. Now it is seen that upon increasing the
bound state index $r$, the binding energies and oscillator strengths fall
sharply both for PO  and SO phonons. Furthermore, in weak fields, this
decrease of $W$ and $f$ is faster than that of in the strong fields. In the
weak fields, the following relation takes place 
\begin{equation}
f^\pm_{nr}\sim{\frac{1}{\hbar\omega_0}} W^\pm_{nr}.  \label{sfoscl}
\end{equation}
In the strong fields, this relation holds at $n=0$ only for the right-hand
polarization. The corresponding expression for the left-hand polarization at 
$n=1$ as well as for both polarizations at $n\geq2$ are 
\begin{equation}
f^\pm_{nr}\sim{\frac{1}{\hbar\omega_0}}\sigma^2 W^\pm_{nr}
\end{equation}
and at $n=0$ 
\begin{equation}
f^\pm_{nr}\sim{\frac{1}{\hbar\omega_0}}\sigma W^\pm_{nr},
\end{equation}
{\it i.e.} in these cases the oscillator strengths are essentially weaker.

\begin{figure}[htb]
\epsfxsize=12cm
\epsfysize=10cm
\mbox{\hskip 1cm}\epsfbox{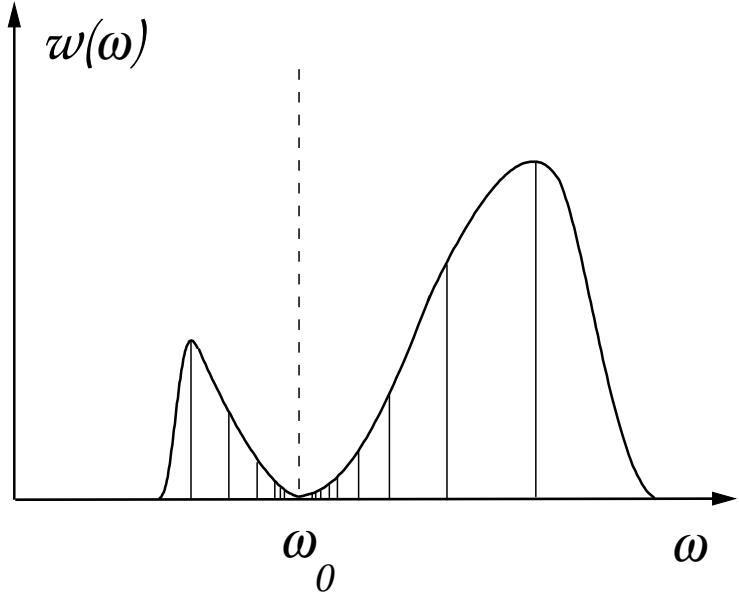}
\caption{Cyclotron phonon resonance spectrum.
The dashed line shows the absorption spectrum when no bound states are 
taken into account. The bold line is the skirt.}
\label{fg7}
\end{figure}

From the results of the calculations for $W$ and $f$, thus, it becomes clear
that the absorption spectrum in both the right- and left-hand polarization
should constitute an "asymmetric doublet" (Fig.~ \ref{fg7}). In the strong
fields this asymmetry is with respect to a numerical parameter of type $\rho$
and in the weak fields -- with respect to the large parameter $\sigma$.

\section{Discussion of results}

\label{summ}

\subsection{Numerical estimates}

Both from the fundamental and technological point of view, the most
important nanostructures are based on GaAs/Al$_x$Ga$_{1-x}$As heterojunction.
Therefore, for numerical estimates, we choose a GaAs/Al$_x$Ga$_{1-x}$As 
single-heterostructure or quantum well as our calculation model. The values
of physical and material \cite{adachi} parameters for these structures are
given in Table~ \ref{tb4} in Appendix~ \ref{appa}. The energies and coupling
constants for the SO phonons arising at the GaAs/Al$_x$Ga$_{1-x}$As 
heteroface for $x=0.3$ and $x=1$ are obtained from formulas 
(\ref{scoupling})-(\ref{dieleq}). 
There exist two, GaAs and AlGaAs like SO phonon modes for a given 
$x$ with energies below the GaAs like and AlGaAs like bulk PO phonon
energies, respectively. The coupling constant $\alpha_{SO}$ for the electron
interaction with GaAs like SO phonons is much greater than that of for the
AlGaAs like SO modes. Therefore we ignore the electron coupling to such
AlGaAs like SO phonons. Notice that the SO phonon energies and coupling
constants for each mode are changed weakly in $x$.

Let discuss in more detailed the bulk PO phonon localization effect across
the GaAs/Al$_x$Ga$_{1-x}$As heteroface. Using the experimental results 
\cite{leviklein}, we find that the PO phonon dispersion law in GaAs can be
described by 
\begin{equation}
\omega_{LO}(q)=\omega_{LO}\left(1-{\frac{q^2}{q_0^2}}\right),\;\,\:
\omega_{LO}=295\;\mbox{cm}^{-1},\; q_0=18.5\;\mbox{nm}^{-1}.
\label{dispers}
\end{equation}
The phonon momenta important in the formation of bound states obey to the
restrictions, $q_\perp\lesssim a_B^{-1}$ and $q_\parallel\lesssim\pi/d$.
Even in strongest fields ($B=20$ T) and for narrow wells ($d=2.5$ nm) we
have $q\lesssim 1$ nm$^{-1}$ so that $|\o_{PO}(q)-\o_{PO}|
\lesssim1$ cm$^{-1}$. However, the GaAs like PO mode in 
Ga$_{0.7}$Al$_{0.3}$As has the frequency $284$ cm$^{-1}$.
Therefore, the PO vibrations in GaAs are localized in a well with walls of 
height $\delta\omega_{PO}=11$ cm$^{-1}$ which is considerably
greater than the dispersion $|\o_{PO}(q)-\o_{PO}|$. The depth of 
penetration of the PO vibrations of GaAs into Ga$_{0.7}$Al$_{0.3}$As 
can be estimated as $q^{-1}\sqrt{\o_{PO}/\delta\o_{PO}}\approx 
0 .3$ nm. This means that the bulk PO phonons are strongly 
localized in quantum wells. This result is in full agreement with the 
experimental Raman scattering data. \cite{jusser,sood,colvard}.

The assumption of the narrow well, $\Delta E_d\equiv
E_2-E_1\gg\hbar\omega_B, \hbar\omega_{0}$, is well justified for the well
width $d$ less than $20$ nm. Par example, at $d=10$ nm, the transverse
quantization separation between the first and second highest levels is 
$\Delta E_d=170$ meV which is notably greater than the phonon energies, 
$\hbar\omega_{PO}=36.62$ meV, $\hbar\omega_{SO}= 34.57$ meV at 
$x=0.3$, and $\hbar\omega_{SO}=34.82$ at $x=1$. In the 
GaAs/Al$_x$Ga$_{1-x}$As quantum wells and heterostructures, the resonance
fields are $B_{PO}=20.87$ T, $B_{SO}=19.70$ T at $x=0.3$, and $B_{SO}
=19.85$ T at $x=1$ so that we usually obtain the situation 
$\omega_B\lesssim\omega_0$ and, therefore, in estimates we use formulas 
obtained in the case of the weak magnetic fields. Thus, for 
GaAs/Al$_x$Ga$_{1-x}$As quantum wells
and heterostructures using formulas in Tables~ \ref{tb2} and \ref{tb3}, as
well the numerical values from Table~ \ref{tb4}, we obtain ultimately the $W$
and $f$. The results for the most removed states bound to the PO  and $GaAs$
like SO phonons ($x=0.3$) in the case of $n=1$ are collected in Table~ \ref
{tb5}. One can see that for $d\sim10$ nm and $B\sim B_0$ the binding energies
are $W\sim 1$ meV.

\begin{table}[tbp]
\caption{Binding energies and oscillator strengths in the weak magnetic
fields for the PO phonon assisted bound states in quantum wells. In these
formulas $d$, $B$ and $W$ should be taken in nm, T, and  meV, respectively.}
\label{tb5}
\begin{center}
\begin{tabular}{|c|c|c|}
\hline\hline
{Phonon} & {Binding energies} & {Oscillator strengths} \\ 
mode &  &  \\ \hline
&  &  \\ 
& {$W^{+}_{11,3}= \left\{\!
\begin{array}{c}
4.4 \\ 
1.2
\end{array}
\!\right\} \left({d/10}\right)\left({B/20.87}\right)^2$} & {$%
f_{11,3}^{+}=2.6\cdot 10^{-2}\left({d/10}\right) \left({B/20.87}\right)^2$}
\\ 
PO &  &  \\ 
& {$W^{-}_{12}=1.9\left({d/10}\right)\left({B/20.87}\right)^2$} & {$%
f_{12}^{-}=5.2\cdot 10^{-2}\left({d/10}\right) \left({B/20.87}\right)^2$} \\ 
&  &  \\ \hline
&  &  \\ 
& {$W^{+}_{11,3}=-\left\{\!
\begin{array}{c}
0.6 \\ 
0.3
\end{array}
\!\right\}\left(B/19.7 \right)^{\frac{3}{2}}$} & {$f_{11,3}^{+}=\left\{\!
\begin{array}{c}
0.8 \\ 
0.6
\end{array}
\!\right\} \cdot 10^{-2}\left({B/19.7}\right)^2 $} \\ 
SO &  &  \\ 
& {$W^{-}_{12}=0.5\left({B/19.7} \right)^{\frac{3}{2}} $} & {$%
f_{12}^{-}=1.4\cdot 10^{-2}\left({B/19.7}\right)^{\frac{3}{2}}$} \\ 
&  &  \\ \hline\hline
\end{tabular}
\end{center}
\end{table}

For actual estimates of the absorbed energy fraction, $w$, we have to allow
for the Landau level broadening, {\it i.e.}, for the smearing of the
delta-function in Eq.~ (\ref{absrpt}). In rough estimates we can replace the
delta-function with a Lorentzian characterized by the total width $\hbar/\tau
$, where $\tau$ is the relaxation time deduced from the mobility. At the
absorption maximum we have then $\pi\delta(\omega)\rightarrow\tau$, so that 
\begin{equation}
w=4\pi N{\frac{e^2}{m_c c^2}}{\frac{c}{\sqrt{\kappa(\nu_n)}}} f\tau.
\label{absrpt1}
\end{equation}
In quantum wells of good (but not exceptional) quality we have $\mu=10^5$
V s m$^{-1}$ which corresponds to $\tau = 4$ ps and a line width $%
\hbar/\tau=0.15$ meV. Therefore, in a field of $B = 4$ T and for $d = 10$ nm
the bound states from Table~ \ref{tb5} can hardly be resolved. However, if
the field is increased up to $10$ T and the well width up to $15$ nm, we can
increase the binding energies and oscillator strengths by an order of
magnitude, {\it i.e.} we can obtain $W\approx1$ meV and $f\approx10^{-2}$. We
will estimate the absorption under these conditions on the basis of Eq.~ (%
\ref{absrpt1}). Assuming that $N = 4\cdot10^{15}$ m$^{-2}$, we find that 
$w = 5\cdot10^{-3}$. This means that a ten-layer superlattice can give rise
to the fully perceptible absorption amounting to few percent. Recalculation
to the bulk absorption coefficient $\gamma=w/d$ yields 
$\gamma\approx10^7$ m$^{-1}$ (at a bulk density $N/d= 5\cdot 10^{23}$
m$^{-3}$).

The above estimates for the phonon dispersion $|\omega_{PO}(q)- \omega_{PO}|$
and for the binding energies $W$ shows that $|\omega_{PO}(q)-
\omega_{PO}|\ll W$ at the phonon momenta important in the bound state
formation. This means that the neglect of the phonon dispersion is justified
in this theory.

\subsection{Conclusion}
\label{2canl} 

From the results of the calculations it follows that the
branches of the spectrum of two-particle elementary excitations in the 2DEG
exposed to the quantizing magnetic field normal to the electron sheet plane
describe the bound states of the electron and the optic phonon and there are
infinite number of them. The bound states are characterized by the total
angular momentum $l=0,\pm1,\pm2,\dots$ of rotating around ${\bf B}$. The
states with a given $l$ are additionally numbered by indexes $n$ and $r$.
Index $n$ indicates the number of the "threshold" near which an infinite
sequence of the bound states numbered by the index $r$ is located. Upon
increasing $r$, causing the binding energy $W^{l}_{nr}$ to fall, there is a
simultaneous fall also in the oscillator strength $f^{l}_{nr}$, approaching
to zero at $r\to\infty$. The absorption is governed by the bound states with 
$l=\pm1$. The binding energies, generally speaking, can be of any sign.
Therefore the absorption concentrated not at the "threshold" but below and
above it at the separation of $W^{\pm1}_{nr}$. The characteristic scale of
the binding energies $W_{2D}=\alpha\omega_0$ is essentially greater than
corresponding scale in the massive samples, $W_{2D}=\alpha^2\omega_0$.

The bound states and the cyclotron phonon resonance are determined by 
the density of states of a system of two particles: an electron at the level $n$
and an optic phonon.
Since the phonon dispersion is ignored, it follows that both particles have
an infinite mass so that this system does not have a continuous spectrum.
All the electron and phonon states are bound and rigorously stationary. This
is the fundamental difference between the cyclotron phonon resonance in a 
2DEG and that in a 3D eectron gas. For this reason the absorption 
coefficient for the 2D cyclotron phonon resonance cannot be
calculated from perturbation theory.

In strong fields, the binding energies and the oscillator strengths of the
states above the threshold are of the same order of magnitude as for the
states below the threshold. Therefore, the absorption spectrum should
consist of two groups of peaks, which are of approximately the same
amplitude and are located approximately asymmetrically relative to the
"threshold". The separation between these two groups of peaks is of the 
same order of magnitude for the right- and left-hand polarizations. The
absorption depends on the number $n$ of the final Landau level. The 
maximal absorption is for the right-hand polarization for the bound states 
arising near the $n=1$ level. The absorption for left-hand
polarization at $n=1$ and for both left- and right-hand polarizations at $%
n\geq2$ are $(B/B_0)^2$ times less. At $n=0$, the
absorption is suppressed for both polarizations in $(B/B_0)$ times
In weak fields the bound states with the maximum binding.
energies and oscillator strengths lie above the thresholds, because the
peaks below the threshold (at $\nu<\nu_n$) should be weaker and closer to
the "threshold". 
However, in contrast to the case of strong fields in the weak fields, the
absorption is of the same order of magnitude irrespective of the
polarization and of the index $n$. 
In the 2DEG, all bound states in the case of $n\geq1$ and part of states at $%
n=0$ lie above the PO phonon energy, hence their observation should not be
hindered by strong lattice reflection. 

From comparison the binding energies for PO  and SO phonons in 
Tables~ \ref{tb2} and \ref{tb3} it is clear that 
\begin{equation}
{\frac{W_{LO}}{W_{SO}}}\sim\delta\sigma^{-{\frac{1}{2}}} \left({\frac{%
\omega_B}{\omega_d}}\right)^{\frac{1}{2}}\ll 1,  \label{compbe}
\end{equation}
In this estimate the difference between the numerical values of the coupling
constants $\alpha$, the phonon frequencies $\omega_0$ for the PO  and
SO phonons are ignored. The relationship (\ref{compbe}) applies in strong
and weak fields. A similar relationship describes the oscillator strengths.
Thus, the states bound to SO phonons should be easier to observe
experimentally.

\chapter{Electron-phonon relaxation. Interface ef\-fect}
\section{Introduction}
\label{3intrdctn}

Scattering of electrons in massive samples by lattice vibrations in metals
and semiconductors has been intensively investigated in the last several
decades and at present it is well studied. Comprehensive reviews for a
variety of scattering phenomena are given in \cite{abricosov} and \cite
{gantlev}. In the last years semiconductor nanoscale systems with carrier
confinement in one, two, and all three dimensions are attracting increasing
interest (see reviews \cite{afs,esipov,butcher,roessler}). The wave 
functions compressed by the lateral confinement
may lead to a substantial enhancement in the optical properties of device
nanostructures because the quantum effects can be used to concentrate the
optical oscillator strengths at the active transitions. Furthermore, these
structures under different new environments exhibit discrete charging effects
and give promise of devices operating in the limit of single electron transport.
In GaAs/AlGaAs type heterostructures, Coulomb scattering
of electrons on the ionized impurities can be suppressed by concentrating
the impurities in layers with wide bands while charge carriers $-$ in narrow
bands. Such a spatial separation of the impurities and charge carriers is
attained by use of a modulated doping technique \cite{stroemer78}. This
leads to the strong enhancement of the carriers mobility. Even more higher
mobilities can be achieved separating the carrier and impurities with an
additional non-doped spacer \cite{stroemer81}. A comparison of experimental
results with theories of electron scattering in modulated doped
heterostructures with spacers shows \cite{hess81} that in temperatures some
below $77$K, the acoustic phonon scattering dominates while at helium
temperatures $-$ Coulomb scattering. In the higher temperature range, the
electron scattering goes mainly due to the electron-optic phonon interaction 
\cite{arora}. In the intermediate temperature range $4.22-77$ K, the 
piezoelectric scattering of acoustic phonons is suppressed \cite{basu} hence
the scattering of the 2DEG on the deformation potential of acoustic phonons
becomes more important. Investigations of the carrier scattering in the 2DEG
have been carried out by many authors 
\cite{fang70,ridley82,sakaki84,price84,kawaguchi}. The electron
scattering has been considered in inversion layers \cite{fang70} and
heterostructures \cite{ridley82,sakaki84}. Various relaxation times verified 
experimentally have been calculated \cite{price84,kawaguchi}. However, 
the effect of the phonon reflection from various interfaces on the scattering 
of the 2DEG from acoustic phonons has not been considered. Interfaces 
always exist in real structures such as inversion layers, heterostructures, and 
quantum wells. (The only exceptions are 
\cite{ezawa1,nakamura,glazman,kirakos}. All these treatments will be
discussed later.)

It is usually accepted that the reflection is unimportant and contributes at
most a correction factor of the order of unity. However, it will be shown
here that this is not the case. Actually, there are situations when the reflection
of phonons alters the energy dependence of the electron relaxation times and
even can change the order of magnitude of the relaxation times.
\begin{figure}[htb]
\epsfxsize=14cm
\epsfysize=7cm
\mbox{\hskip 0cm}\epsfbox{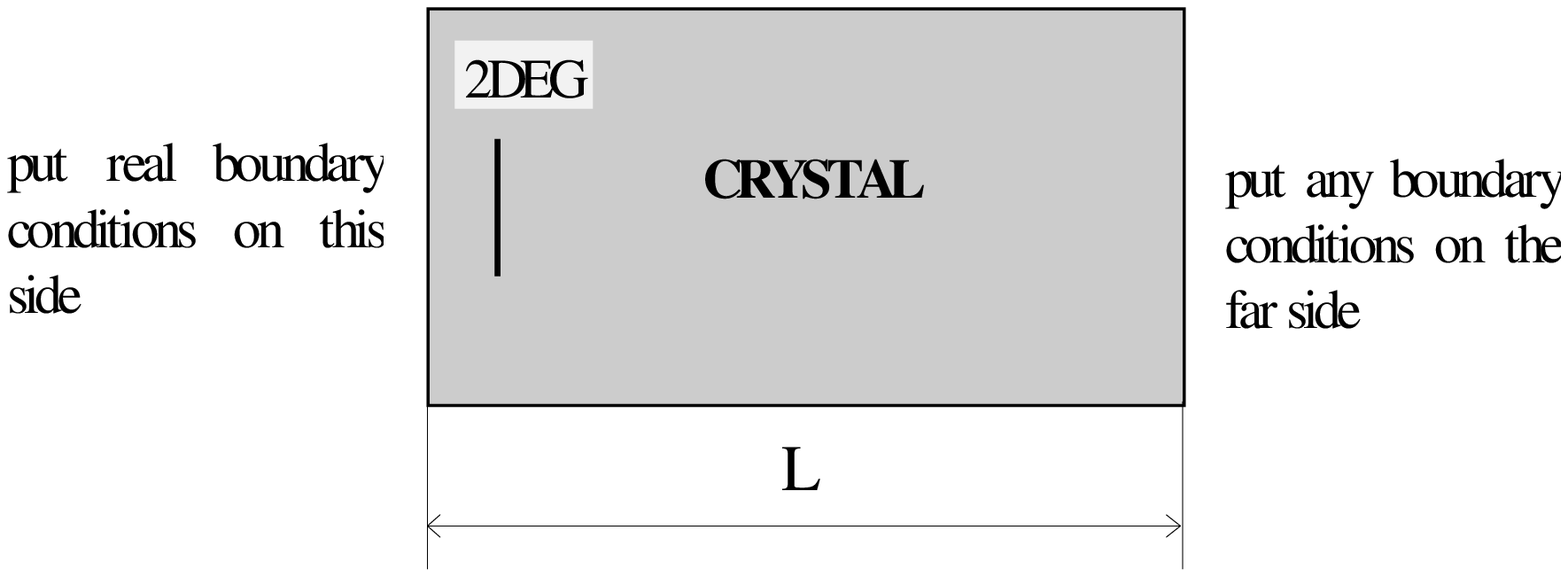}
\caption{In usual method for calculating the transition probability 
an infinite crystal is replace by finite box of length $L$.}
\label{fg12}
\end{figure}

It is very difficult to include the phonon reflection using the traditional
methods. Consider the following situation: electrons in an inversion layer
in the vicinity of a semiconductor surface. If we adopt the usual method, a
semi-infinite crystal should be replaced by a finite volume $L^3$ (see 
Fig.~\ref{fg12}),
the phonon field should be quantized, the scattering probability from each
phonon mode should be obtained, and then all these probabilities should be
combined, which is followed by going to the limit $L\rightarrow\infty$. It
is necessary in the calculation of the phonon modes to impose on the surface 
$z = 0$ near which electrons are located, the correct boundary condition, 
{\it i.g.}, the absence of a stress on the free surface. Arbitrary boundary
conditions can be applied to the far surface at $z = L$. However,
irrespective of the choice of the boundary conditions, the vector nature of
the phonon field makes it practically impossible to calculate the complex
normal modes of a plate in the limit $L\rightarrow\infty$. We can bypass
this problem by focusing our attention directly on the scattering
probability summed over all the phonon modes. {\it Such a probability 
evaluated
in the Born approximation can be expressed in terms of the phonon field
correlation function} \cite{levgant} and {\it the correlation function itself is
given in terms of the elasticity theory Green function} \cite{maradud76}. 
The limit $L\rightarrow\infty$ can be then reached for such a function 
without any difficulties.

In this chapter, the effect of interfaces separating elastic semi-spaces on
the scattering of the 2DEG from the deformation potential of acoustic
phonons is studied \cite{badalinteff,badalsemispc,badalphd}. This problem
is sufficiently multiplex, in the sense, that it contains a number of interesting
particular cases of the free and rigid surfaces, the solid-liquid interfaces 
{\it etc.}. Study of these limiting cases favors more complete understanding
of the interface effect on the electron scattering.
On the other hand, they have their own significance for device structures in
each concrete case. Par example, the case of the solid-liquid contact is
applicable directly to the electrolyte-semiconductor device structures which
are attracting increasing interest in the last years in connection with a
developing of new devices for the solar energy transformation \cite
{harout,tardella}. The model of rigid boundary can be applied to situations
when the second medium is sufficiently heavy, for example
metal-insulator-semiconductor (MIS) structures with a heavy gate and a thin
oxide layer. If in the rough approximation, differences between acoustic
properties of the dielectric and semiconductor are ignored, and a metal mass
density account for infinity, then the system can be described in the frames
of this model. The model of a free surface applies directly to inversion
layers which are formed on natural semiconductor surfaces (for example, Ge
or InAs) and to heterojunctions located at distances of the order of $10$ nm
from the surface. Our model is most applicable to systems where an
insulator in the MIS structure is not deposited on the surface, but simply
clamped ({\it Mylar foil}). The acoustic contact is then poor and phonons in
the semiconductor are reflected from the semiconductor-insulator interface
as if it were a free surface. We also wish to mention MIS structures with a
thin oxide layer and a thin (optically transparent) gate. The reflection
coefficient for the semiconductor-insulator and insulator-metal interfaces
are relatively small. Neglecting, in the first approximation, the
differences between the acoustic properties of various media, we can
describe such a system using our free surface model.

In Sec.~\ref{nmethod} a new method for calculating the probability of
electron scattering from the deformation potential of acoustic phonons is
developed \cite{badalinteff} which is applicable to an arbitrary geometry
 of the
sample. Using this method, it is possible to treat the reflection of phonons
from a crystal surface and from interfaces separating different materials.
To illustrate the method, in Secs.~\ref{rates} and \ref{trates} the energy
(and electron temperature) \cite{badalinteff,badalsemispc} and the
momentum \cite{badalphd} relaxation rates for the Fermi 2DEG in the 
vicinity
of an interface separating semi-infinite elastic spaces are calculated. Our
results and conclusions are summarized in Sec.~\ref{3results}.

\section{Interface effect on the electron-phonon interaction}
\label{nmethod} 

We consider the deformation interaction of electrons with acoustic phonons. 
Assuming a non-degenerate band, we can write the interaction Hamiltonian 
in the lowest order of atomic displacements in the form 
\begin{equation}
H_{int}={\frac{1}{2}}\Xi_{ij}\{\partial_i u_j({\bf r})+\partial_j u_i({\bf r}%
)\}.  \label{hamilton}
\end{equation}
where, $i,j=1,2,3,\cdots$ label Cartesian coordinates ($x = 1,\, y = 2,\,
z=3$), $\Xi_{ij}$ is the deformation potential tensor, $u_j({\bf r})$ are 
the displacement vector components. Summation over repeated indices is 
assumed to be carried out. It is easy to show that the probability of an 
electron transition from a state $\Psi_\upsilon({\bf r})$ with an energy $%
\varepsilon_\upsilon$ to a state $\Psi_{\upsilon^{\prime}}({\bf r})$ with 
an energy $\varepsilon_{\upsilon^{\prime}}$ due to the perturbation 
defined by Eq.~\r{hamilton} in the first Born approximation is given by the
following expression: 
\begin{eqnarray}
W_{\upsilon\rightarrow\upsilon^{\prime}}&=&\int\!d^3r_1\int\!d^3r_2\,
\Psi^*_{\upsilon^{\prime}}({\bf r_1})\Psi^*_{\upsilon}({\bf r_2})
\Psi_{\upsilon^{\prime}}({\bf r_2})\Psi_{\upsilon}({\bf r_1})  \nonumber \\
&\times& \Xi_{ii^{\prime}}\Xi_{jj^{\prime}}\partial_{1i}\partial_{2j}
<u_{i^{\prime}}({\bf r_1})u_{j^{\prime}}({\bf r_2})>_\o.  \label{transprob}
\end{eqnarray}
Here $<\cdots>_\o$ is the Fourier transform of the correlation function of
the displacement field operators in the Heisenberg representation. It is
necessary to set $\omega=\varepsilon_\upsilon-\varepsilon_{\upsilon^{\prime}}$
in the calculation of the probability.

The elasticity theory Green function $G_{ij}({\bf r_1},{\bf r_2}|\omega)$ is
defined as the displacement along the $i$ axis at a point ${\bf r_1}$ due to
a force density
\begin{equation}
-\delta({\bf r_1}-{\bf r_1})e^{-i\omega t}
 \label{forcedens}
\end{equation}
applied in the direction of the $j$ axis.

Assuming an equilibrium phonon field at temperature $ T$, we find that the
correlation function in Eq.~\r{transprob} can be expressed in terms of
the above retarded Green function \cite{maradud76} 
\begin{equation}
<u_{i^{\prime}}({\bf r_1})u_{j^{\prime}}({\bf r_2})>_\o=-2[N_T(\omega)+1]%
\mbox{Im} G_{ij}({\bf r_1},{\bf r_2}|\omega)  \label{corgreen}
\end{equation}
where $N_T$ is the Bose factor 
\begin{equation}
N_T(\omega)=[\exp({\hbar\omega/T})+1]^{-1}.  \label{bose}
\end{equation}
The Green function $G_{ij}({\bf r_1},{\bf r_2}|\omega)$ includes naturally
all the boundary conditions at the interface and at the surface of a sample.
When some of the boundary conditions are replaced by the boundary conditions
at infinity then the retarded Greet function $G_{ij}({\bf r_1},{\bf r_2}%
|\omega+i0) = G_{ret}(\omega)$ should be replaced by the Green function
corresponding to outgoing waves $G_{out}(\omega)$.

Eqs.~\r{transprob} and \r{corgreen} solve, in principle, the problem of
calculating the electron transition probability between arbitrary electron 
states in the case of deformation scattering of electrons from equilibrium
acoustic vibrations and they apply to an arbitrary geometry of the sample.
This approach is especially efficient at the calculating of scattering in
lower dimensional systems (in systems with the 2DEG, in quantum wires and
dots) when the presence of various interfaces, separating different
materials, distorts phonon modes. This method advances an easier way to
account for the phonon reflection from various interfaces and to calculate
the transition probability summed over all the phonon modes, including the
surface phonons. Recently, this method has been used to calculate the
electron scattering on surfaces of cryogen crystals (solid hydrogen,
deuterium, and neon) from lattice vibrations \cite{lev89}. It should be
noted also that the Green functions of the elasticity theory are
sufficiently well studied for several interesting sample geometries (for a
review {\it quod vide} \cite{cottam}), which makes easier their use and
notably shorts calculations of the transition probability.

\section{Scattering in a Fermi 2DEG near interfaces separating elastic
semi-spaces}
\label{semispeff} 

In most cases, the elastic properties of a sample are
those of a layered medium (for example, a semi-infinite semiconductor or a
system consisting of oxide and metal layers) and electrons move in a plane
parallel to the interface. We shall assume that the plane $z=0$ near which
the 2DEG is located in a medium $1$ represents the interface between two
elastic media and the $z$-axis is directed away from a medium $2$. Such
systems have the translation invariance in $x$ and $y$ directions so that it
is convenient to make the Fourier transformation 
\begin{equation}
G_{ij}({\bf r_1},{\bf r_2}|\omega)=\int\!{\frac{d^2q}{(2\pi)^2}}\,
g_{ij}(\omega,{\bf q}|z_1,z_2)exp[i{\bf q}\cdot({\bf R_1}-{\bf R_2})]
\label{fourier3}
\end{equation}
where ${\bf r}=({\bf R},z)$, ${\bf q}$ and ${\bf R}$ are two-dimensional
vectors in the $(x,y)$-plane. In the absence of a magnetic field, the electron
wave functions are given by 
\begin{equation}
\Psi_{n{\bf k}}({\bf r})={\frac{1}{L}}\exp[i{\bf k}\cdot{\bf R}]\psi_n(z).
\label{ewf}
\end{equation}
Here, ${\bf k}$ is the electron momentum in the $(x,y)$-plane, $n$ labels
states describing the electron motion in the direction of the $z$-axis, $L^2$ 
is the area of the structure in the $(x,y)$-plane. Substituting Eqs.~\r{fourier3}
and \r{ewf} in Eqs.~\r{transprob} and \r{corgreen} we obtain 
\begin{equation}
W^{\pm}_{n{\bf k}\rightarrow n^{\prime}{\bf k^{\prime}}}=
{\frac{2}{L^2}}[N_T(\omega)+{\frac{1}{2}}\pm {\frac{1}{2}}]
\int\!dz_1\int\!dz_2\,
\psi^*_{n^{\prime}}(z_1)\psi^*_{n}(z_2)\psi_{n^{\prime}}(z_2) \psi_{n}(z_1)
{\cal D}(\omega,{\bf q}| z_1, z_2) 
 \label{probab}
\end{equation}
where 
\begin{eqnarray}
{\cal D}(\omega,{\bf q}| z_1, z_2)&=&-\mbox{Im}[\Xi_{ii^{\prime}}\Xi_{jj^{%
\prime}}q_{i}q_{j} g_{i^{\prime}j^{\prime}}-\Xi_{33}\Xi_{jj^{\prime}}q_{j}i{%
\frac{\partial}{\partial\!{z_1}}}g_{3j^{\prime}} +
\Xi_{ii^{\prime}}\Xi_{33}q_{i}i{\frac{\partial}{\partial\!{z_2}}}%
g_{i^{\prime}3}  \nonumber \\
&+&\Xi_{33}^2{\frac{\partial^2}{\partial\!{z_1}\partial\!{z_2}}}g_{33}].
\label{dkernel}
\end{eqnarray}
Here the indices $i,j$ take only values $1,2$. For brevity, we have omitted
in Eq.~\r{dkernel} the dependence of $g_{ij}$ on their arguments. It is 
necessary to set ${\bf q}={\bf k}-{\bf k^{\prime}}$ in the
transition probability defined by Eq.~\r{probab}. The $+$ and $-$ signs
in Eq.~\r{probab} refer to transitions involving an energy loss, $%
\hbar\omega = \varepsilon_{nk}-\varepsilon_{n^{\prime}k^{\prime}} > 0$
and an energy gain, $\hbar\omega
=\varepsilon_{n^{\prime}k^{\prime}}-\varepsilon_{nk} > 0$, respectively.

We can further simplify the problem assuming that the scattering takes place
in an isotropic band with an energy minimum at the center of the Brillouin
zone 
\begin{equation}
\Xi_{ij}=\Xi\cdot\delta_{ij}.  \label{defconst}
\end{equation}
We further assume that all the layers are elastically isotropic, which means
that the dependence of $g_{ij}$ on the orientation of ${\bf q}$ is trivial
and is determined by the matrices describing rotations in the $(x,y)$ plane 
\cite{cottam}. We then can exploit this isotropy in the $(x,y)$-plane and
obtain 
\begin{equation}
{\cal D}(\omega,{\bf q}| z_1, z_2)= \Xi^2 K(\omega, q | z_1, z_2)
\label{dkkernel}
\end{equation}
where the kernel $K$ is given 
\begin{equation}
K(\omega, q | z_1, z_2)= -\mbox{Im}\left[q_{i}q_{j}g_{ij}- q_{j}i{\frac{%
\partial}{\partial\!{z_1}}}g_{3j} + q_{i}i{\frac{\partial}{\partial\!{z_2}}}%
g_{i3} + {\frac{\partial^2}{\partial\!{z_1}\partial\!{z_2}}}g_{33} \right].
\label{kernel1}
\end{equation}
Since the kernel $K$ is independent of the orientation of vector ${\bf q}$,
we may choose ${\bf q}\parallel {\bf \hat{x}}$, which yields 
\begin{equation}
K(\omega, q|z_1,z_2)= -\mbox{Im}\left[q^2g_{11}-iq{\frac{\partial}{\partial\!%
{z_1}}} g_{31}+ iq {\frac{\partial}{\partial\!{z_2}}}g_{i3} + {\frac{%
\partial^2}{\partial\!{z_1}\partial\!{z_2}}}g_{33}\right].  \label{isokernel}
\end{equation}
Recall that the notation ${\bf q}\equiv (q_x,q_y)$ is used. 

Further we
require the kernel $K$ for the contact of the semi-infinite elastic spaces.
The Green function of the elasticity theory for this case have been
calculated in \cite{djfri}. Substituting in Eq.~\r{isokernel} the obtained
results for $g_{ij}$ from \cite{djfri}, it is straightforward although somewhat 
tedious to derive an expression for $K$. We use following notations 
\begin{eqnarray}  
r_{\pm}=(\alpha_t^2+q^2)^2\pm4q^2\alpha_l\alpha_t,&
h_{\pm}=\alpha_t^2+q^2\pm2\alpha_l\alpha_t,&\gamma={\frac{%
\rho^{\prime}c^{\prime 2}}{\rho c^2}}, \label{notation1}\\
\alpha=\sqrt{q^2-{\frac{(\omega+i0)^2}{s^2}}},&  \beta=\sqrt{q^2-
{\frac{(\omega+i0)^2}{c^2}}}
\label{notation2}
\end{eqnarray}
where $s$ and $c$ are the velocities of the longitudinal LA  and the
transverse TA acoustic waves, $\rho$ is the mass density in the medium $1$
where the 2DEG is located. Quantities $r^{\prime},h^{\prime},\alpha^{\prime}$
are defined according to (\ref{notation1}) and (\ref{notation2}) with
parameters $s^{\prime},c^{\prime},\rho^{\prime}$ of the medium $2$. In Eq.~ 
\r{notation2}, the branch cut for the square root is assumed to lie along
the negative real axis. We then obtain 
\begin{equation}
K(\omega, q|z_1,z_2)={\frac{\omega^2}{\rho s^4}}\mbox{Im}
{\frac{1}{2\alpha_l}} \left \{\exp[-\alpha_l(z_1-z_2)]-{\cal R}\exp[-\alpha_t(z_1+z_2)]\right\}
\label{kernelss}
\end{equation}
where 
\begin{equation}
{\cal R}={\frac{r_{+}(\alpha_l^{\prime}\alpha_t^{\prime}-q^2)-\gamma^2
r^{\prime}_{-}(\alpha_l\alpha_t+q^2)+\gamma[2q^2h_{+}h_{-}^{\prime}+ 
{\frac{\omega^4}{c^2c^{\prime 2}}}(\alpha_l^{\prime}\alpha_t -
\alpha_l\alpha_t^{\prime})]}{r_{-}(\alpha_l^{\prime}\alpha_t^{\prime}-q^2)+%
\gamma^2 r^{\prime}_{-}(\alpha_l\alpha_t-q^2) +\gamma[2q^2h_{-}h_{-}^{%
\prime}+ {\frac{\omega^4}{c^2c^{\prime 2}}} (\alpha_l^{\prime}\alpha_t +
\alpha_l\alpha_t^{\prime})]}}.  \label{reflect}
\end{equation}
In the kernel (\ref{kernelss}), the quantity ${\cal R}$ accounts for the
contribution to the scattering by the phonon modes reflected at the
interface. In the range where $\omega>sq$ and $\omega>s^{\prime}q$,
expressing ${\cal R}$ in terms of the sound velocities and the phonon
incidence angles, we can easily show that ${\cal R}$ is the reflection
coefficient describing the LA $\to$LA  reflection at the interface separating
elastic semi-spaces. It can be easily seen that Eq.~\r{kernelss} at $%
{\cal R}=0$ yields the kernel $K$ for the case when phonons occupy the whole
infinite space. In this case, phonons of the TA type, naturally, do not
contribute to $K$ and the condition $K\neq0$ is satisfied only for $sq
<\omega$. In the range $\omega<cq$ and $\omega<c^{\prime}q$, the poles of $%
{\cal R}$ yield the dispersion law of the interface Stoneley waves \cite
{stoneley} and, therefore in this region, the kernel $K$ is nonzero only on
the line $\omega=c_Sq$ where $c_S$ is the velocity of the Stoneley waves.
This is consistent with the result that the Green function used in the
elasticity theory to describe the scattering probability has simple poles as
a function of $\omega$ at the frequencies of the normal phonon modes of the
system. Notice that for such layered systems, we have no normal modes
corresponding to the LA  and TA vibrations. The Green function of the
elasticity theory has only branch cut singularities corresponding to the 
frequencies $\omega=sq,\,s^{\prime}q$ and $\omega=cq,\,c^{\prime}q$. 
In general, the condition for
the existence of the Stoneley waves is quite complex and cannot be
investigated analytically for an arbitrary ratio of the parameters of  the two
contacted media \cite{viktorov}. It is, therefore, reasonable taking into 
account the scattering kinematics to analyze the kernel $K$ separately 
for different ranges of the electron energy typical for the scattering in the
Fermi 2DEG. This will be done in the next section. {\it Vide infra} we
consider limiting cases of the kernel $K$.

In the case of the contact between solid and liquid semi-spaces taking the 
imaginary part in Eq.~\r{kernelss}, beforehand tending $c^{\prime}\to 0$
in Eq.~\r{reflect}, we obtain for the kernel
\begin{eqnarray}
K(\omega, q|z_1,z_2)&=&{\frac{\omega^2}{\rho s^4}}{\frac{1}{2a}} 
\{\cos[a(z_1-z_2)]-{\cal R}_{s-l}\cos[a(z_1+z_2)]\}, sq<\omega,
\label{kernblk1} \\
K(\omega, q|z_1,z_2)&=&{\frac{\omega^2}{\rho s^4}} {\frac{[4bq^2+{\frac{%
\rho^{\prime}}{\rho a^{\prime}}}{\frac{\omega^4}{c^4}}](b^2-q^2)^2}{[4\alpha
bq^2+{\frac{\rho^{\prime}\alpha}{\rho a^{\prime}}}{\frac{\omega^4}{c^4}}%
]^2+(b^2-q^2)^4}} e^{-\alpha(z_1+z_2)}, cq<\omega<sq,  \label{kernblk2} \\
K(\omega, q|z_1,z_2)&=&{\frac{\omega^2}{\rho s^4}} {\frac{{\frac{%
\rho^{\prime}}{\rho a^{\prime}}}{\frac{\omega^4}{c^4}}(\beta^2+q^2)^2}{%
[4\alpha bq^2+{\frac{\rho^{\prime}\alpha}{\rho a^{\prime}}}{\frac{\omega^4}{%
c^4}}]^2+ {\frac{\rho^{\prime}\alpha}{\rho a^{\prime}}}^2{\frac{\omega^8}{c^8%
}}}} e^{-\alpha(z_1+z_2)}, s^{\prime}q<\omega<cq,  \label{kernlw} \\
K(\omega, q|z_1,z_2)&=&{\frac{\omega^2}{\rho s^4}} [4\beta q^2-{\frac{%
\rho^{\prime}}{\rho \alpha^{\prime}}}{\frac{\omega^4}{c^4}}]  \nonumber \\
&\times&\pi\delta\left(4\alpha \beta q^2-(\beta^2+q^2)^2- {\frac{%
\rho^{\prime}\alpha}{\rho \alpha^{\prime}}}{\frac{\omega^4}{c^4}}%
\right)e^{-\alpha(z_1+z_2)}, \omega<s^{\prime}q  \label{kernsw}
\end{eqnarray}
where 
\begin{equation}
{\cal R}_{s-l}=-{\frac{4abq^2-(b^2-q^2)^2+{\frac{\rho^{\prime}a}{%
\rho a^{\prime}}}{\frac{\omega^4}{c^4}}}{4abq^2+(b^2-q^2)^2+{\frac{%
\rho^{\prime}a}{\rho a^{\prime}}}{\frac{\omega^4}{c^4}}}}  
\label{reflectsl}
\end{equation}
is the LA $\to$LA  reflection coefficient at the solid-liquid interface.
Following notation is used here 
\begin{eqnarray}
a=\sqrt{{\frac{\omega^2}{s^2}}- q^2}, & b= \sqrt{{\frac{\omega^2}{c^2}}- q^2}%
,&  \label{notationa} \\
\alpha=\sqrt{q^2-{\frac{\omega^2}{s^2}}}, & \beta= \sqrt{q^2-{\frac{\omega^2%
}{c^2}}},&  \label{notationalpha}
\end{eqnarray}
and $a^{\prime}$ and $\alpha^{\prime}$ are defined using the parameters of
the liquid medium. In the majority of cases the sound velocity $s^{\prime}$ in
a liquid is less than the velocity of the TA waves in a solid, which we
assume to be true here. The bulk acoustic phonon contribution to the
scattering is determined by formulas (\ref{kernblk1}) and (\ref{kernblk2}).
Moreover, the LA phonons contribute to the scattering only in the range $%
\omega>sq$ while the TA phonons $-$ in the whole range $\omega>cq$.
Therefore, it is impossible to distinguish the LA  and TA phonon
contributions. This is natural, as far as, in strict sense, there exist no
LA  and TA normal modes in this case. They are mixed in the reflection at the
interface and form a single bulk wave. It follows from the formulas (\ref
{kernblk1})-(\ref{kernsw}) that in the range $\omega< s^{\prime}q$ the
kernel $K$ differs from zero only if the argument of the delta function in
Eq.~\r{kernsw} vanishes. In contrast to the boundary between the two
solid semi-spaces, this condition is satisfied for any ratio of the
parameters of solid and liquid media \cite{gogo} and yields the dispersion
law of the Stoneley waves \cite{brekh}. The energy of the surface Stoneley
waves is then concentrated mainly in the liquid medium on the distance of
some phonon wavelengths. In the range $s^{\prime}q <\omega<cq$ the 
value of $%
K$ differs from zero because of the interaction of electrons with surface 
{\it leaky waves} \cite{brekh}. These {\it leaky waves} are actually damped
surface Rayleigh waves modified somewhat by the response of the liquid
medium. This type of wave transfers energy continuously to the liquid
forming an inhomogeneous wave moving away from the interface. This is due
to the fact that, strictly speaking, {\it leaky waves} are not of the surface
type. It follows from Eq.~\r{kernlw} that the {\it leaky waves}
correspond to a pole on a nonphysical sheet of the Green function of the
elasticity theory for the solid-liquid contact. As the differences between
the elastic properties of the two contacted semi-spaces decrease, the {\it %
leaky wave} becomes even more strongly damped, so that we cannot regard it
as a surface wave.

The kernel $K$ for the semi-infinite space with free surface is easy to
obtain from formulas (\ref{kernblk1})-(\ref{kernsw}) taking the limit $%
\rho^{\prime}\to0$. This gives 
\begin{eqnarray}
K(\omega, q|z_1,z_2)&=&{\frac{\omega^2}{\rho s^4}}{\frac{1}{2a}} \left\{\cos[%
a(z_1-z_2)]-{\cal R}_{fr}\cos[a(z_1+z_2)]\right\}, sq<\omega,
\label{fkernblk1} \\
K(\omega, q|z_1,z_2)&=&{\frac{\omega^2}{\rho s^4}} {\frac{4bq^2(b^2-q^2)^2}{%
16\alpha^2 b^2q^4+(b^2-q^2)^4}} e^{-\alpha(z_1+z_2)}, cq<\omega<sq,
\label{fkernblk2} \\
K(\omega, q|z_1,z_2)&=&{\frac{\omega^2}{\rho s^4}} 4\beta
q^2\pi\delta\left(4\alpha\beta q^2-(\beta^2+q^2)^2\right) 
e^{-\alpha(z_1+z_2)}, \omega<cq  
\label{fkernel}
\end{eqnarray}
where 
\begin{equation}
{\cal R}_{fr}=-{\frac{4abq^2-(b^2-q^2)^2}{4abq^2+(b^2-q^2)^2}}
\label{freflect}
\end{equation}
is the LA $\to$LA  reflection coefficient at the free crystal surface.
Comparison of Eqs.~\r{kernblk1}-\r{kernsw} with Eqs.~\r{fkernblk1}-%
\r{fkernel} shows that only surface waves are qualitatively changed: the
Stoneley waves disappear while {\it leaky waves} transform into the rigorously
stationary Rayleigh waves. Therefore in the range $\omega<cq$, the kernel $K$
differs from zero only on the line $\omega=c_Rq$, $c_R$ is the velocity of
the Rayleigh waves.

In the case of the rigid boundary, the kernel $K$ can be obtained taking the
imaginary part in Eq.~\r{kernelss}, beforehand in Eq.~\r{reflect}
tending $\rho^{\prime}\to\infty$. This gives 
\begin{eqnarray}
K(\omega, q|z_1,z_2)&=&{\frac{\omega^2}{\rho s^4}}{\frac{1}{2a}} \{\cos[%
a(z_1-z_2)]-{\cal R}_{rgd}\cos[a(z_1+z_2)]\},\; sq<\omega,
\label{rkernblk1} \\
K(\omega, q|z_1,z_2)&=&{\frac{\omega^2}{\rho s^4}} {\frac{bq^2}{%
q^4+\alpha^2b^2}} \exp[-\alpha(z_1+z_2)],\: cq<\omega<sq  \label{rkernblk2}
\end{eqnarray}
where the LA $\to$LA  reflection coefficient at the rigid crystal surface is
given by 
\begin{equation}
{\cal R}_{rgd}=-{\frac{ab-q^2}{ab+q^2}}.
 \label{rreflect}
\end{equation}
As it was to be expected, the scattering is determined only by the bulk waves
in this case.

\section{Scattering probability in a Fermi gas}
\label{scatprob} 

\begin{figure}[htb]
\epsfxsize=10cm
\epsfysize=10cm
\mbox{\hskip 1.5cm}\epsfbox{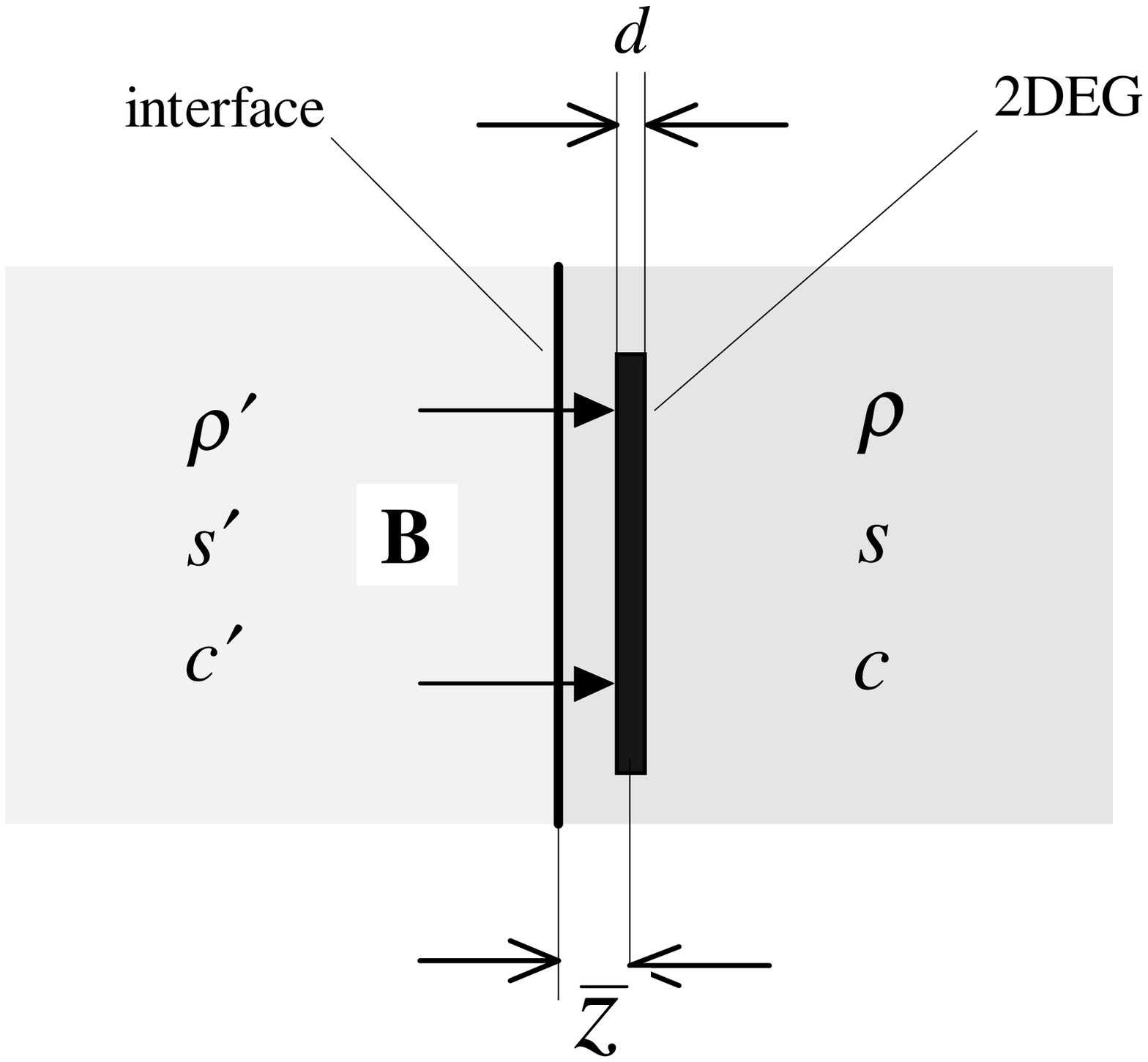}
\caption{The Fermi 2DEG located near the interface separating elastic
semi-spaces.}
\label{fg13}
\end{figure}

It is assume that all the electrons occupy a single level
of transverse motion, {\it i.e.} $\varepsilon_F\ll\hbar\omega_d,\;
\hbar\omega_d\equiv\pi^2\hbar^2/ m_cd^2$ (or $k_F\ll\pi/d$), where $d$ is
the thickness of the region of the electron transverse motion in the
direction of the $z$-axis. The scattering takes place between states with
different $k$ but all belonging to the same transverse level. The electron
gas is separated from the surface by a distance $\bar{z}$ (see Fig.~\ref{fg13}).
For clarity, we first discuss the situation when the lattice
temperature $ T$ and the electron temperature $ T_e$ are zero. 
Neglecting the reflection of phonons from the surface, it is easy to
determine the phonon momenta which dominate in the scattering. A typical
normal wave vector component $q_\perp\equiv q_z$ and a typical
tangential wave vector component ${\bf q}_\parallel\equiv(q_x,q_y)$ may
generally be different. According to \cite{karpus} there are three energy
ranges  $\varepsilon-\varepsilon_F$ for a test electron in which the
scattering is qualitatively different. The boundaries of these ranges are
determined by the following two energies: 
\be
\varepsilon_1=\hbar sk_F=\sqrt{2m_cs^2\varepsilon_F},\:
\varepsilon_2=\pi\hbar s/d=\sqrt{2m_cs^2\hbar\omega_d},\: 
\varepsilon_1/\varepsilon_2=k_Fd/\pi\ll1.  
\label{boundenerg}
\ee
We now list the characteristic phonon momenta and energies corresponding to
the scattering in the three ranges defined above:

{\it range A} -- $\varepsilon-\varepsilon_F\ll\varepsilon_1$, in which
small-angle inelastic scattering takes place 
\begin{equation}
q_\perp\sim q_\parallel\sim{\frac{\varepsilon-\varepsilon_F}{\hbar s}}\ll
k_F;\;\hbar\omega\sim\varepsilon-\varepsilon_F,  \label{arng}
\end{equation}

{\it range B} -- $\varepsilon_1\ll\varepsilon-\varepsilon_F\ll\varepsilon_2$%
, in which large-angle inelastic scattering is dominant 
\begin{equation}
q_\perp\sim {\frac{\varepsilon-\varepsilon_F}{\hbar s}}\gg q_\parallel\sim
k_F;\;\hbar\omega\sim\varepsilon-\varepsilon_F,  \label{brng}
\end{equation}

{\it range C} -- $\varepsilon_2\ll\varepsilon-\varepsilon_F$, in which
large-angle quasi-elastic scattering takes place 
\begin{equation}
q_\perp\sim \pi/d\gg q_\parallel\sim k_F;\;\hbar\omega\sim\hbar\pi
s/d\ll\varepsilon-\varepsilon_F.  \label{crng}
\end{equation}

Now we discuss the interface effect on the scattering probability $W$ in the
Fermi 2DEG. For simplicity we first consider the free surface effect. It is
clear that the boundary has no effect on the scattering probability $W$ for
$q_\perp\bar{z}\gg 1$ and should modify $W$ for $q_\perp\bar{z}\lesssim 1$.
We shall show below that a free boundary suppresses strongly the scattering in the
range B for $q_\perp\bar{z}\ll 1$.

Consider now only longitudinal LA phonons ({\it i.e.}, assume that the shear
modulus is zero). The dilatation on a free surface satisfies $u\equiv %
\mbox{{\bf div}}\;{\bf u}=0$ and scattering from the deformation potential
is strongly suppressed when electrons are located near a dilatation node, 
{\it i.e.}, for $q_\perp\bar{z}\ll 1$. Since the condition $\bar{z}\gtrsim d$
is satisfied, this cannot occur in the range C. However, we may expect 
suppression of scattering in the ranges A and B provided for 
\begin{equation}
{ \e-\e_F \over \h s}\bar{z}\ll1.
\end{equation}
Now we account for the TA phonons. We find that $u\neq 0$ holds on a free
surface and it would be incorrect to conclude quite generally that the
scattering is suppressed. We shall now consider $q_\parallel$. The
inequality $q_\parallel\ll q_\perp$ holds in the range B, {\it i.e.},
LA phonons traveling in a direction almost perpendicular to the surface
contribute mainly to the scattering. When such LA phonons are reflected from the
boundary, TA phonons do not participate in the scattering and, therefore,
suppression of the scattering takes place. A strong mixing of LA  and TA
modes occurs in the region A where the condition $q_\parallel\sim q_\perp$
is not a sufficient requirement for the scattering to be suppressed. It
follows that strong suppression of the scattering in the vicinity of a free
surface should occur only in the range B. The effect of the reflection of
phonons from a free surface in the ranges A and C, when $\bar{z}%
(\varepsilon-\varepsilon_F)/hs\lesssim 1$, should lead only to a numerical
factor of the order of unity.

Now we consider in which way situation is changed for the interface
between elastic semi-spaces. In this case because of the response of the
second medium ${\bf div}\;{\bf u}\neq0$ even if the mixing of the TA phonons
do not take into account. The measure of the second medium response at $%
q_\parallel=0$ is the parameter $\lambda=\rho s/\rho^{\prime}s^{\prime}$
where $\rho s$ is a medium normal impedance. Therefore in those cases when
the elastic semi-space is contiguous with a rarefied medium, {\it i.e.} $%
\lambda\ll1$ then one can expect that in the range B, the scattering
suppression takes place as before.

We have assumed that $ T = 0$. This condition is essentially equivalent to
the requirement $ T\ll\varepsilon-\varepsilon_F$. It is, therefore, clear
that our estimates of $q_\parallel$ and $q_\perp$ remain also valid for $%
T\neq 0$ provided the inequality $ T\lesssim\varepsilon-\varepsilon_F$ holds.
We shall now consider scattering in the range B. It is then possible to 
expand the kernel $K$ in powers of a small parameter $s^2q^2/\omega^2$. 
Such an expansion yields 
\begin{eqnarray}
&&K(\omega, q|z_1,z_2)={\frac{\omega}{2\rho s^3}} \left\{\cos\left[{\frac{%
\omega}{s}}(z_1-z_2)\right]-{\cal R}_{s-l} \cos\left[{\frac{\omega}{%
s}}(z_1+z_2)\right]\right\},  \label{kernexp1} \\
&&{\cal R}_{s-l}={\cal R}_{0}- {\frac{s^2q^2}{\omega^2}}{\cal R}%
_{1},\; {\cal R}_{0}={\frac{1-\lambda}{1+\lambda}},  \label{kernexp2} \\
&&{\cal R}_{1}={\frac{1}{(1+\sigma)(1+\lambda)^2}} \Biggl[\left({\frac{2c}{s}%
}\right)^3-\lambda\left(1-{\frac{s^{\prime 2}}{s^2}}+ {\frac{%
2s^{\prime}c^{\prime}}{s^2}}-{\frac{8c^2}{s^2}}-{\frac{8cc^{\prime}}{s^2}}+{%
\frac{16cc^{\prime 2}}{s^2s^{\prime}}}\right)  \nonumber \\
&&\quad -\lambda\sigma\left(1-{\frac{s^{\prime 2}}{s^2}}+{\frac{4s^{\prime}c%
}{s^2}} -{\frac{8cc^{\prime}}{s^2}}-{\frac{8c^{\prime 2}}{s^2}}+{\frac{%
8c^{\prime 3}}{s^2s^{\prime}}}\right) -2\lambda^2\sigma {\frac{c}{s}}\Biggr]%
, \; \sigma={\frac{\rho c }{\rho^{\prime}c^{\prime}}}.  \label{kernexp3}
\end{eqnarray}
In the region C, the arguments of the cosines in Eq.~\r{kernexp1} are of
the order of $\pi$ so that we can simplify $K$ by dropping out low-order 
terms. This gives 
\begin{equation}
K(\omega, q|z_1,z_2)={\frac{\omega}{2\rho s^3}}\left\{\cos\left[{\frac{\omega%
}{s}}(z_1-z_2) \right]-{\cal R}_{0}\cos\left[{\frac{\omega}{s}}%
(z_1+z_2)\right]\right\}.  \label{ckernexp}
\end{equation}
Substituting Eqs.~\r{ckernexp} and Eq.~\r{dkkernel} into Eq.~ \ref
{probab}, we obtain the scattering probability in the range C between 
states belonging to the same level of the electron transverse motion 
\begin{eqnarray}
W^{\pm}_{{\bf k}\rightarrow n^{\prime}{\bf k^{\prime}}}&=&{\frac{1}{%
\bar\tau_F}}\:{\frac{\pi}{L^2}}\:{\frac{\hbar\omega}{m_c (sk_F)^2}}\:
\left[N_T(\omega)+{\frac{1}{2}}\pm{\frac{1}{2}}\right] 
\int\!dz_1\!dz_2\,{\cal \rho}(z_1){\cal \rho}(z_2) \nonumber \\
&\times&  \left\{\cos\left[%
{\frac{\omega}{s}}(z_1-z_2)\right]-{\cal R}_{0}\cos\left[{\frac{\omega}{s}}%
(z_1+z_2)\right]\right\}.  \label{cprob}
\end{eqnarray}
where ${\cal \rho}(z)=|\psi(z)|^2$ is the probability to find an electron on
a level of the transverse quantization, and 
\begin{equation}
{\frac{1}{\bar\tau_F}}={1\over \pi}\:{ m_c k_F^2\over\h^2}\:
{\frac{ \Xi^2}{\rho s}}  
\label{ftime}
\end{equation}
is the characteristic scattering time in the Fermi gas of electrons by the
deformation potential of acoustic phonons \cite{gantlev}. It is seen from
Eq.~\r{cprob} that $W$ does not depend on the TA phonon velocity in the
medium $2$, therefore this expression is true for the solid-liquid
contact too. It is easy to find the scattering probability for the cases of
free and rigid surfaces taking, respectively, the limits $\rho\to 0$ and $%
\rho\to\infty$ in Eq.~\r{cprob}.

In region B, we can use the condition $\omega d/s \ll 1$ and represent the
kernel $K$ in the form 
\bea
K(\omega, q|z_1,z_2)={\frac{\omega}{2\rho s^3}} \Biggl\{\sin{\frac{\omega}{s}}%
z_1\sin{\frac{\omega}{s}}z_2 -
\nonumber\\
- \left[{\frac{\lambda}{1+\lambda}}+{\frac{s^2q^2%
}{\omega^2}}{\cal R}_{1}\right] \cos\left[{\frac{\omega}{s}}%
(z_1+z_2)\right]\Biggl\}. 
 \label{bexpkern}
\eea
Provided the scattering takes place between states of the same level of the
transverse motion $n$, it follows from Eq.~\r{probab} that this kernel $K$
should be averaged over the functions ${\cal \rho}(z)$ describing the
transverse motion. The first term then involves integrals 
\begin{equation}
\int d\!z\, {\cal \rho}(z) \sin{\frac{\omega}{s}}z.  \label{int}
\end{equation}
We set $z=\bar{z}+\zeta$, where 
\begin{equation}
\bar{z}=\int\!dz\,z{\cal \rho}(z),  \label{zbar}
\end{equation}
then the inequality $(\omega/s)\zeta\sim(\varepsilon-\varepsilon_F)d/s\ll1$
holds and the sinus in the integral (\ref{int}) can be expanded in powers of 
$\zeta$ (for any $\bar{z}$ including small values of this quantity). Bearing
in mind the definition of $\bar{z}$, we find the integral in Eq.~\r{int}
is equal to $\sin(\omega/s)\bar{z}$. As for the cosine in the second term,
we can replace $z_1+z_2$ in this function by $2\bar{z}$. As a result for the
scattering probability in the region B, we obtain 
\begin{eqnarray}
W^{\pm}_{{\bf k}\rightarrow n^{\prime}{\bf k^{\prime}}}&=&{\frac{1}{%
\bar\tau_F}}\:{\frac{2\pi}{L^2}}\:{\frac{\hbar\omega}{m_c (sk_F)^2}}\:
\left[N_T(\omega)+{\frac{1}{2}}\pm{\frac{1}{2}}\right]  \nonumber \\
&\times& \left\{{\sin^2{\frac{\omega}{s}}\bar{z}}-\left[{\frac{\lambda}{%
1+\lambda}}+{\frac{s^2q^2 }{\omega^2}}{\cal R}_{1}\right]\cos 2
{\frac{\omega}{s}}\bar{z}\right\}.  
\label{bprob}
\end{eqnarray}
The first term in the brackets in Eq.~\r{bprob} appears because electrons
are far from a node of the deformation potential which occurs at $z = 0$ for
LA phonons. The second term is due to an admixture of TA phonons at the
boundary due to the reflection of LA phonons and eliminates the deformation
potential node. If the elastic properties of contacting media differ
slightly, {\it i.e.} $\lambda\approx 1$, one can neglect the second term in
the square bracket in Eq.~\r{bprob}. Moreover, consider the 2DEG located
in the immediate vicinity from the interface, {\it i.e.} $\bar{z}\sim d$,
the first term in Eq.~\r{bprob} also can be neglected. Then, the
scattering probability $W$ only by a factor $2\lambda/(1+\lambda)$ differs
from the scattering probability $W_{3D}$ calculated without taking into
account the phonon reflection at the interface. For the latter case $W_{3D}$
can be obtained by omitting the terms in Eq.~\r{bprob} which oscillate
rapidly in the limit $\bar{z}\to\infty$, {\it i.e.} the figure bracket is
replaced by $1/2$. This fully correspond also to the rigid surface case: in
this case we have ${\cal R}_{1}\to 2c/s$ and $2\lambda/(1+\lambda)\to2$ when 
$\rho^{\prime}\to\infty$.

The situation is changed for $\lambda\ll 1$. In this case neither term in
the Eq.~\r{bprob} can be neglected and the scattering probability is
determined by "competition" of all terms. Notice, that Eq.~\r{bprob} is
valid also for the case of the solid-liquid contact for which 
\begin{equation}
{\cal R}_{1}={\frac{1}{(1+\lambda)^2}}\left[\left({\frac{2c}{s}}\right)^3-
\lambda\left(1-{\frac{s^{\prime 2}}{s^2}}-8{\frac{c^2}{s^2}}\right)\right].
\label{rsol-liq}
\end{equation}

Consider now the third range A representing inelastic low-angle scattering.
In view of the absence in this region of a corresponding small parameter,
further simplifications of Eq.~\r{kernelss} for the kernel $K$ is not
generally possible. Therefore, we restrict our analysis to two limiting
cases of a contact between a solid and a liquid and their rigidity fixed
boundary. Making use that in the region A, $ad\ll1$ and $\alpha d\ll 1$, the
kernel $K$ can be averaged over the functions ${\cal \rho}(z)$ in the way 
that the scattering probability in the region A can be represented in the form 
\begin{eqnarray}
W^{\pm}_{{\bf k}\rightarrow n^{\prime}{\bf k^{\prime}}}&=&{\frac{1}{%
\bar\tau_F}}\:{\frac{2\pi}{L^2}}\:{\frac{\rho s}{m_c k_F^2}} \:
\left[N_T(\omega)+{\frac{1}{2}}\pm{\frac{1}{2}}\right] \:
K(\omega,q|\bar{z},\bar{z}).
\label{aprob}
\end{eqnarray}
Here the kernel $K$ is given either by Eqs.~\r{kernblk1}-\r{kernsw}, \ref
{fkernblk1}-\r{fkernel} or \r{rkernblk1}-\r{rkernblk2}. The scattering
rate without allowance for the phonon reflection at the interface can be
obtained by taking the limit $\bar{z}\to\infty$ in the above equation, {\it %
i.e.} by replacing $K\ra\omega^2/2a\rho s^4$.

\section{Energy and momentum relaxation of a test electron}
\label{rates} 

We calculate the energy and momentum relaxation rates for a test electron
which can be measured accurately in a wide range of temperatures $ T$ 
and $ T_e$ \cite{sakaki,dolgopolov}. According to \cite{gantlev}, the 
energy relaxation time for the test electron in the Fermi gas is defined as 
\begin{eqnarray}
{\cal Q}_{n\rightarrow n^{\prime}}(\varepsilon)&=&{\frac{\varepsilon-%
\varepsilon_F }{\bar{\tau}(\varepsilon)}}= {\cal Q}^{+}_{n\rightarrow
n^{\prime}}(\varepsilon)-{\cal Q}^{-}_{n\rightarrow n^{\prime}}(\varepsilon),
\\
{\cal Q}^{\pm}_{n\rightarrow n^{\prime}}(\varepsilon)&=&\sum^{(\pm)}_{{\bf %
k^{\prime}}}\hbar\omega W^{\pm}_{n{\bf k} \rightarrow n^{\prime}{\bf %
k^{\prime}}} {\frac{1-f_T(\varepsilon\mp\hbar\omega)}{1-f_T(\varepsilon)}}.
\label{energrel}
\end{eqnarray}
Here ${\cal Q}_{n\rightarrow n^{\prime}}(\varepsilon)$ is the energy-loss
power, $f_T$ is the Fermi factor. The summation (+) is over the final states 
$\varepsilon(k^{\prime})< \varepsilon(k)$, the summation (-) is over the
final states $\varepsilon(k^{\prime}) >\varepsilon(k)$.

Substituting Eq.~\r{cprob} in Eq.~\r{energrel}, we find that ${\cal Q}%
_{n\rightarrow n^{\prime}}(\varepsilon)$ in the range C is described by 
\be  
{\cal Q}(x)={\frac{1}{\bar{\tau}_F}}\; J_1\;{\frac{(\pi s/d)}
{2(k_F d)^2}}\; th{\frac{x}{2}},\quad x={\frac{\varepsilon-\varepsilon_F}{T}}
\label{cerrate}
\ee
where the dimensionless integral
\be
J_1=d^3\int_0^\infty\!dz\,\left[\partial_z {\cal \rho}(z)\right]^2
\label{integ1}
\ee
is of the order of unity. Hence, it is clear that the energy relaxation rate in
the region C is just the same (apart from a numerical factor) with that of
considered without the phonon reflection \cite{karpus} and this is true for 
any parameters of the medium $2$. This result is explained by the fact that in
the region C, the interface is always distant, {\it i.e.} $\bar{z}$ is
greater or of the order of the characteristic phonon wavelength along ${\bf 
\hat{z}}$. Notice also that for hot electrons, $x\gg1$, the relaxation rate
does not depend on the electron energy 
\begin{equation}
{\cal Q}(x)= {\frac{1}{\bar{\tau}_F}}\; J_1\;{\frac{(\pi s/d)}{2(k_F d)^2}}.
\label{hotcerrate}
\end{equation}

As far as in the region C, the scattering is quasi-elastic (see Eq.~\r{crng})
then the momentum relaxation time is defined by 
\begin{equation}
{\frac{1}{\bar{\tau}_1(\varepsilon)}}=\sum_{{\bf k^{\prime}}}\,
(1-\cos\theta)W_{{\bf k}\to{\bf k^{\prime}}} {\frac{1-f_T(\varepsilon-%
\hbar\omega)}{1-f_T(\varepsilon)}},  \label{mrate}
\end{equation}
where $\theta$ is the scattering angle between ${\bf k}$ and ${\bf k^{\prime}}$.
Substituting Eq.~\r{cprob} into Eq.~\r{mrate}, we obtain 
\begin{equation}
{\frac{1}{\bar{\tau}_1(\varepsilon)}}={\frac{1}{\bar{\tau}_F}}\;J_0\;
{\frac{(\pi s/d)}{2(k_F d)^2}} T,\quad J_0=d\int_0^\infty dz{\cal \rho}^2(z).
\label{cmtime}
\end{equation}
It is easy to check that this expression for the momentum relaxation time is
in agreement with that of obtained in \cite{karpus} without allowance for
the phonon reflection. Thus in the range C, the interface effect on the
energy and momentum relaxation rates is practically absent.

Making use the scattering probability \r{bprob}, we find the following
result for the energy relaxation rate in the range B: 
\begin{equation}
{\frac{\varepsilon-\varepsilon_F}{\bar{\tau}(\varepsilon)}}={\frac{1}{\bar{%
\tau}_F}}\:\left\{ {\frac{T^3}{2(\hbar s k_F)^2}}\left[{\cal F}^{-}_2(0|x) - 
{\cal R}_0 {\cal F}^{-}_2(\xi|x)\right]+ {\cal R}_1 {\cal F}^{-}_0(\xi|x)\right\}  \label{berate}
\end{equation}
where 
\begin{equation}
{\cal F}^{-}_m(\xi|x)=\int_0^\infty\!dy\, y^m\cos\xi y \left\{{\frac{1}{%
1+\exp(y-x)}}+{\frac{1}{1+\exp(y+x)}}\right\},\; \xi=2{\frac{T}{\hbar(s/\bar{%
z})}}.  \label{finteg}
\end{equation}
For hot electrons taking the limit $ T\to 0$ in Eq.~\r{berate}, we can
obtain 
\begin{equation}
{\frac{1}{\bar{\tau}(\varepsilon)}}={\frac{1}{\bar{\tau}_F}}\:\left\{ {\frac{%
(\varepsilon-\varepsilon_F)^2}{6(\hbar s k_F)^2}}{\cal I}(\eta,\lambda)+ 
{\cal R}_1 {\frac{\sin\eta}{\eta}}\right\}  \label{betime}
\end{equation}
where 
\begin{equation}
{\cal I}(\eta,\lambda)=1-3{\frac{1-\lambda}{1+\lambda}}\left[\left(1-{\frac{2%
}{\eta^2}}\right) {\frac{\sin\eta}{\eta}}+{\frac{2}{\eta^2}}\cos\eta\right]
\label{iinteg}
\end{equation}
Here $\eta=2(\varepsilon-\varepsilon_F)\bar{z}/(\hbar s)$ is a parameter
describing the distance of a hot electron from the surface. It can be seen
that the energy relaxation rate for the test electron is an oscillatory
function of the separation $\bar{z}$ of the electron gas from the surface.
When the 2DEG is far from the surface, {\it i.e} $\bar{z}\gg d$ ($\eta\gg1$),
 we obtain 
\begin{equation}
{\frac{1}{\bar{\tau}(\varepsilon)}}={\frac{1}{\bar{\tau}_F}}\:{\frac{%
(\varepsilon-\varepsilon_F)^2}{6(\hbar s k_F)^2}} 
 \label{betimefar}
\end{equation}
The condition $\bar{z}\sim d$ is satisfied for a 2DEG near the surface and
the inequality $\eta\lesssim1
$ holds, which yields 
\begin{equation}
{\frac{1}{\bar{\tau}(\varepsilon)}}={\frac{1}{\bar{\tau}_F}}\left\{{\cal R}%
_1 + {\frac{\lambda}{1+\lambda}}{\frac{(\varepsilon-\varepsilon_F)^2}{%
3(\hbar s k_F)^2}} +{\frac{(\varepsilon-\varepsilon_F)^4}{5(\hbar s
k_F)^2(\hbar s/\bar{z})^2}}\right\}  \label{betimenear}
\end{equation}
From this equation for the free surface ($\lambda=0$ and ${\cal R}_1=(2c/s)^3
$) , it is easy to find 
\begin{equation}
{\frac{1}{\bar{\tau}(\varepsilon)}}={\frac{1}{\bar{\tau}_F}}\left\{\left(%
\frac{2c}{s}\right)^3 + {\frac{(\varepsilon-\varepsilon_F)^4}{5(\hbar s
k_F)^2(\hbar s/\bar{z})^2}}\right\}  \label{betimenearf}
\end{equation}
Both terms become comparable for $\varepsilon-\varepsilon_F\sim\sqrt{%
\varepsilon_1\varepsilon_2}$: The first term is dominant for small $%
\varepsilon-\varepsilon_F$, {\it i.e.}, mixing of LA  and TA vibrations is
important whereas, for large $\varepsilon-\varepsilon_F$, the second term
predominates, {\it i.e.}, we have the effect corresponding to large
distances of electrons from the surface. It can be seen from 
Eqs.~\r{betimefar} and \r{betimenearf} that a free surface modifies the energy
dependence of the energy relaxation rate and leads to the strong suppression
of the relaxation since 
\begin{equation}
{\frac{\bar{\tau}(\varepsilon)|_{\bar{z}\to\infty}}{\bar{\tau}(\varepsilon)
|_{\bar{z}\sim d}}}\sim {\frac{(\varepsilon-\varepsilon_F)^2}{\varepsilon_2^2%
}}+{\frac{\varepsilon_1^2}{(\varepsilon-\varepsilon_F)^2}}\ll1.
\label{suppressf}
\end{equation}
There is no suppression of the relaxation at the boundaries of the range B:
for $\varepsilon-\varepsilon_F\sim \varepsilon_1$ because of the strong
mixing of LA  and TA vibrations and for $\varepsilon-\varepsilon_F\sim%
\varepsilon_2$ because the distance from the surface becomes of the order of
the effective phonon wavelength $\pi/q_\perp$.

From comparison of Eqs.~\r{betimenear} and \r{betimenearf}, we obtain
the criterion 
\begin{equation}
\lambda\ll{\frac{\varepsilon_1}{\varepsilon_2}}  \label{criteria}
\end{equation}
which allows us to determine when the interface near which the Fermi 2DEG is
located, can be regarded as free. Actually, if inequality (\ref{criteria})
is obeyed, we can ignore the second term in parentheses in 
Eq.~\r{betimenear}. Then, Eq.~\r{betimenear} reduces to 
Eq.~\r{betimenearf} for the free surface. In the case if 
\begin{equation}
{\frac{\varepsilon_1}{\varepsilon_2}}\sim\lambda\ll 1  
\label{criteria1}
\end{equation}
the boundary can no longer be regarded as free, but we can easily see that
the reduction in the relaxation rate still occurs. If the condition (\ref
{criteria1}) is satisfied, comparison of Eqs.~\r{betimefar} and 
\r{betimenear} yields 
\begin{equation}
{\frac{\bar{\tau}(\varepsilon)|_{\bar{z}\to\infty}}{\bar{\tau}(\varepsilon)
|_{\bar{z}\sim d}}}\sim {\frac{(\varepsilon-\varepsilon_F)^2}{\varepsilon_2^2%
}}+\lambda+{\frac{\varepsilon_1^2}{(\varepsilon-\varepsilon_F)^2}}\ll1.
\label{suppress}
\end{equation}
Now it is seen, if $\lambda\gtrsim 1$, the suppression of the relaxation 
vanishes
because of the strong response of the second medium. In this case we ought
to retain only the second term in Eq.~\r{betimenear}. Then the energy
relaxation rate differs only by the factor $2\lambda(1+\lambda)$ from that 
of when the phonon reflection is not taken into account. It therefore follows
that for the rigid boundary, the relaxation of a test electron is twice as fast
as in the case when phonons are three dimensional.

For thermal electrons ($x\sim1$), $\xi$ is the parameter describing the
separation between the interfaces and 2DEG. For remote boundary, $\xi\gg1$,
integrals containing a rapidly oscillating factor $\cos\xi y$ are small , and 
according to \cite{karpus} we obtain 
\begin{equation}
{\frac{\varepsilon-\varepsilon_F}{\bar{\tau}(\varepsilon)}}={\frac{1}{\bar{%
\tau}_F}}\; {\frac{T^3}{2(\hbar s k_F)^2}}\;{\cal F}^{-}_2(0|x).
\label{berate3}
\end{equation}
Consider now a nearby boundary when the condition $\xi\ll1$ is satisfied.
Expanding $\cos\xi y$, we obtain 
\begin{equation}
{\frac{\varepsilon-\varepsilon_F}{\bar{\tau}(\varepsilon)}}={\frac{1}{\bar{%
\tau}_F}}\: \left\{{\cal R}_1T+ {\frac{\lambda}{1+\lambda}}{\frac{T}{(\hbar
s k_F)^2}}{\cal F}^{-}_2(0|x)+ {\frac{T^5}{(\hbar s k_F)^2(\hbar s/\bar{z})^2%
}}{\cal F}^{-}_4(\xi|x) \right\}  \label{beratet}
\end{equation}
We have used the result ${\cal F}^{-}_0(0|x)=x$. It can be seen that the
criteria (\ref{criteria}) holds for thermal electrons too. For $ T$ in the
range B, provided for Eq.~\r{criteria1} holds, the relaxation is
suppressed even for thermal electrons. For $x\sim1$, we obtain ${\cal F}%
^{-}_m(0|x)\sim1$ and the ratio of the results defined by Eqs.~\r{berate3}
and \r{beratet} is again determined by Eq.~\r{suppress} where $%
\varepsilon-\varepsilon_F$ should be replace by $ T$.

In the case when $\lambda\gtrsim 1$, making use Eq.~\r{berate3}, we can
write 
\begin{equation}
{\frac{1}{\bar{\tau}(\varepsilon)}}= {\frac{2\lambda}{1+\lambda}}\:
 {\frac{1}{\bar{\tau}(\varepsilon)|_{\bar{z}\to\infty}}}.  
\label{largelambda}
\end{equation}

It should be noticed that both in the ranges C and B, the electron
relaxation is mainly due to the bulk acoustic waves. In these energy ranges $%
\varepsilon-\varepsilon_F\gg\varepsilon_1$ and the scattering due to surface
phonons is always quasi-elastic: 
\begin{equation}
q_\parallel\sim k_F;\;
\omega=c_Sq\sim\varepsilon_1\ll\varepsilon-\varepsilon_F  \label{surfrange}
\end{equation}
therefore their contribution to $\bar{\tau}(\varepsilon)$ is small.

The scattering on the bulk waves is neither quasi-elastic nor small-angle,
therefore the momentum relaxation time $\bar{\tau}_1(\varepsilon)$ is of the
same order of the energy relaxation time $\bar{\tau}(\varepsilon)$ given by
Eq.~\r{berate}. To estimate the role of the surface phonons in the
momentum relaxation, we calculate $\bar{\tau}_1(\varepsilon)$ conditioned by
the Rayleigh waves in the free surface case and by the Stoneley waves for
the solid-liquid contact. Substituting Eqs.~\r{kernsw} and \r{fkernel}
into Eqs.~\r{probab}, \r{dkkernel}, and \r{mrate}, we obtain at $ T=0$ 
\begin{equation}
{\frac{1}{\bar{\tau}_1(\varepsilon)|_{R,\,S}}}={\frac{3\pi}{4}}\:{\frac{1}{%
\bar{\tau}_F}}\: \left({\frac{c_{R,S}}{s}}\right)^3\bar{\Phi}^{R,\,S}_{0}
\label{bmrates}
\end{equation}
where the dimensionless factors $\bar{\Phi}^{R,\,S}_{0}$ are of the order of
unity and determined by parameters of the contacted elastic media ({\it vide
infra} in Appendix \r{appndphi}). Indices $R,\,S$ pertain to the Rayleigh
and Stoneley waves, respectively. From Eqs.~\r{betimenear} and 
\r{bmrates}, one can see that the surface phonon contribution to momentum
relaxation is of the order of the bulk phonon contribution in the energy
interval $\varepsilon_1\ll\varepsilon-\varepsilon_F\ll\lambda^{-1/2}
\varepsilon_1$ of the range B while in the energy interval $%
\lambda^{-1/2}e_1\ll\varepsilon- \varepsilon_F\ll\varepsilon_2$, the surface
phonon contribution to $\bar{\tau}_1(\varepsilon)$ is small. Thus, in the
energy range B if the relations (\ref{criteria}) and (\ref{criteria1}) hold,
the interface suppresses both the energy and momentum relaxation rates.

Now we consider the range A. Substituting Eq.~\r{aprob} into 
Eq.~\r{energrel}, the energy relaxation rate can be represented in the form 
\begin{equation}
{\frac{\varepsilon-\varepsilon_F}{\bar{\tau}(\varepsilon)}}={\frac{1}{\bar{%
\tau}_F}}\: {\frac{T^4}{\pi(\hbar s k_F)^3}}\int_0^\infty\!dtdy\,y^3 \left\{{%
\frac{1}{1+\exp(y-x)}}-{\frac{1}{1+\exp(y+x)}}\right\}\Phi(t|\xi y)
\label{ceratet}
\end{equation}
where 
\begin{equation}
\Phi(t|\xi y)\equiv K(\omega, q|\bar{z},\bar{z}),
\;y={\frac{\hbar\omega}{T}},\; t={\frac{sq}{\omega}}. 
\label{phi}
\end{equation}
For the remote interface neglecting in the above formula rapidly oscillating
and exponentially small terms at $\xi\gg1$, we obtain in agreement with \cite
{karpus} 
\begin{equation}
{\frac{\varepsilon-\varepsilon_F}{\bar{\tau}(\varepsilon)}}={\frac{1}{\bar{%
\tau}_F}}\; {\frac{T^4}{4(\hbar s k_F)^3}}\;{\cal F}^{-}_3(0|x).
\label{cerate3}
\end{equation}
For the adjacent interface, tending $\xi\to 0$ in Eq.~\r{ceratet}, we
find 
\begin{equation}
{\frac{\varepsilon-\varepsilon_F}{\bar{\tau}(\varepsilon)}}={\frac{1}{\bar{%
\tau}_F}}\;{\frac{T^4}{4(\hbar s k_F)^3}}\;\hat{\Phi}_0\;{\cal F}^{-}_3(0|x).
\label{cerate2}
\end{equation}
where 
\begin{equation}
\hat{\Phi}_m={\frac{2^{2+m}}{\pi}}\int_0^\infty\!dt\, t^{2m}\,\Phi(t|0).
\label{phihat}
\end{equation}
For hot electrons using an asymptotics of the function ${\cal F}^{-}_3(0|x)$
for large $x$, we obtain 
\begin{equation}
{\frac{1}{\bar{\tau}(\varepsilon)}}= \hat{\Phi}_0\: {\frac{1}{\bar{\tau}%
(\varepsilon)|_{\bar{z}\to\infty}}},\; {\frac{1}{\bar{\tau}(\varepsilon)|_{
\bar{z}\to\infty}}}={\frac{1}{\bar{\tau}_F}}\: {\frac{\varepsilon-%
\varepsilon_F}{16(\hbar s k_F)^3}}.  
\label{c3and2}
\end{equation}

As far as $|\varepsilon-\varepsilon_F|\ll\varepsilon_F$, the scattering in
the Fermi gas always is kinematically quasi-elastic, $k=k^{\prime}=k_F$,
therefore in the region A too, the momentum relaxation rate can be
calculated by the formula \r{mrate}. Using the scattering probability 
\r{aprob},  for the momentum relaxation time we obtain:

for the remote interface ($\xi\gg1$) 
\begin{equation}
{\frac{1}{\bar{\tau}_1(\varepsilon)}}={\frac{1}{16}}\:{\frac{1}{\bar{\tau}_F}%
}\; {\frac{T^5}{(\hbar s k_F)^5}}\;{\cal F}^{+}_4(0|x),  \label{cmrate3}
\end{equation}

for the adjacent interface ($\xi\ll1$) 
\begin{equation}
{\frac{1}{\bar{\tau}_1(\varepsilon)}}=\hat{\Phi}_1\;{\frac{1}{\bar{\tau}%
_1(\varepsilon) |_{\bar{z}\to\infty}}}.  \label{mtime3}
\end{equation}
From comparison of the above two formulas, it is clear that in the region A,
the relaxation rates, obtained allowing and ignoring the phonon reflection
from interfaces, differ only in respect of the factor $\hat{\Phi}_m$ which
is of the order of unity. This result is due to the strong mixing of LA  and
TA waves at the reflection in this range.

The integrals $\hat{\Phi}_m$ cannot be calculated analytically. Explicit
formulas for functions $\hat{\Phi}_m$ in the case of the solid-liquid
contact, the free and rigid boundary, as well as some numerical estimates 
for particular values of parameters and asymptotic expressions are given in
Appendix~\ref{appndphi}. Notice only that in the case of the solid-liquid
contact we find that $\hat{\Phi}_m$ is governed by both the bulk and surface
wave contributions to the scattering: 
\begin{equation}
\hat{\Phi}_m=\hat{\Phi}_m^{B_1}+\hat{\Phi}_m^{B_2}+ \hat{\Phi}_m^L+\hat{\Phi}%
_m^S  \label{phisllq}
\end{equation}
where ${\hat{\Phi}}_m^B$, ${\hat{\Phi}}_m^L$ and ${\hat{\Phi}}_m^S$
correspond to the bulk phonon, {\it leaky} and Stoneley wave contributions,
respectively. Moreover, the range A is the only region where the surface
wave contribution is comparable with that of the bulk modes. In the case of
a rigid boundary we find, naturally, that $\hat{\Phi}_m$ includes only the
bulk phonon contributions.

\section{Relaxation of electron temperature}

\label{trates} When the distribution of hot electrons can be described by an
electron temperature $ T_e > T$, we can determine the energy relaxation rate
for  the whole electron gas. Following \cite{gantlev}, we find that the
relaxation rate per electron is given by 
\begin{eqnarray}
\bar{{\cal Q}}(T_e, T)&=&{\frac{1}{NL^2}}\sum_{{\bf k}}\,
f_{T_e}(\varepsilon) \Biggl \{\sum^{(+)}_{{\bf k^{\prime}}}\, \hbar\omega
W_{n{\bf k}\to n^{\prime}{\bf k^{\prime}}}[1-f_{T_e}(\varepsilon-\hbar%
\omega)]  \nonumber \\
&-& \sum^{(-)}_{{\bf k^{\prime}}}\, \hbar\omega W_{n{\bf k}\to n^{\prime}%
{\bf k^{\prime}}}[1-f_{T_e}(\varepsilon+\hbar\omega)] \Biggl\}
\label{etrate}
\end{eqnarray}
Straightforward calculations show that $\bar{{\cal Q}}(T_e, T)$ can be
written as the difference between two functions depending only on $ T_e$ 
and $ T$ \cite{price}, {\it i.e.}, 
\begin{equation}
\bar{{\cal Q}}(T_e, T)= \bar{{\cal Q}}(T_e)-\bar{{\cal Q}}(T)
\label{etratesep}
\end{equation}
Moreover for $\bar{{\cal Q}}$ in the corresponding regions, the following
expressions are obtained.

In the region C 
\begin{equation}
\bar{{\cal Q}}(T)= \bar{{\cal Q}}(T)|_{\bar{z}\to\infty}={\frac{1}{\bar{\tau}%
_F}}\; {\frac{(\pi\hbar s/d)}{2(k_F d)^2}}\;{\frac{T}{\varepsilon_F}}\;J_1.
\label{ctrate}
\end{equation}

In the region B 
\begin{equation}
\bar{{\cal Q}}(T)={\frac{1}{\bar{\tau}_F}}\:\left\{ {\frac{T^4}{2(\hbar s
k_F)^2\varepsilon_F}}\left[{\cal F}_4(0) - {\cal R}_0 {\cal F}_4(\xi)\right]+ {\cal R}_1
{\cal F}_2(\xi)\right\}  \label{btrate}
\end{equation}
where 
\begin{equation}
{\cal F}_m(\xi)={\frac{1}{2}}\Gamma(m)\left[\zeta(1,\,1+i\xi)+
\zeta(1,\,1-i\xi)\right]  \label{ffunc}
\end{equation}
and $\zeta(m,\,z)$ is the generalized Riemann zeta function. Now it is easy
to obtain the relaxation rate for the distant interface ($\xi\gg1$): 
\begin{equation}
\bar{{\cal Q}}(T)|_{\bar{z}\to\infty} = {\frac{\pi^4}{30}}\;{\frac{1}{\bar{%
\tau}_F}}\; {\frac{T^4}{(\hbar s k_F)^2\varepsilon_F}},  
\label{btrate3}
\end{equation}
while for adjacent interface, $\xi\ll1$, we obtain 
\begin{equation}
\bar{{\cal Q}}(T)={\frac{1}{\bar{\tau}_F}}\:\left\{ {\frac{\pi^2}{6}}{\cal R}%
_1 {\frac{T^2}{\varepsilon_F}}+ {\frac{\pi^4}{15}}{\frac{\lambda}{1+\lambda}}%
{\frac{T^4}{(\hbar s k_F)^2\varepsilon_F}}+ {\frac{8\pi^6}{63}}{\frac{T^6}{%
(\hbar s k_F)^2(\hbar s/ \bar{z})^2\varepsilon_F}} \right\}.
\label{btratenear}
\end{equation}
Follow to the above method, we can easily show that the criterias (\ref
{criteria}) and (\ref{criteria1}) hold for electron temperature relaxation
too. The  fulfillment of them implies that the interface separating elastic 
semi-spaces suppresses the electron temperature relaxation in the range B. 
If $\lambda \gtrsim 1$ then the interface changes the relaxation rate only 
by a numerical factor: 
\begin{equation}
\bar{{\cal Q}}(T)= {\frac{2\lambda}{1+\lambda}}\;\bar{{\cal Q}}(T) |_{\bar{z}%
\to\infty}.  \label{btrate32}
\end{equation}
Finally, in the region A, the cooling of the Fermi 2DEG for small $\xi$ is
given 
\begin{equation}
\bar{{\cal Q}}(T)= \bar{\Phi}_0 \;\bar{{\cal Q}}(T) |_{\bar{z}\to\infty}.
\label{atrate32}
\end{equation}
where 
\begin{equation}
\bar{{\cal Q}}(T) |_{\bar{z}\to\infty}={\frac{1}{\bar{\tau}_F}}\;
\Gamma(5)\zeta(5){\frac{T^4}{4(\hbar s k_F)^2\varepsilon_F}}  \label{atrate3}
\end{equation}
is the electron temperature relaxation rate without allowance for the phonon
reflection in agreement with the previous result \cite{karpus}.

\section{Discussion of results}
\label{3results} 

From the results of calculations in this chapter, it
becomes clear that the interface between elastic semi-spaces in different
ways influences the electron scattering in the various characteristic energy
ranges of the Fermi 2DEG. In the ranges of the inelastic small-angle A ($%
\varepsilon-\varepsilon_F \ll\hbar sk_F\equiv \varepsilon_1$) and of the
elastic large-angle C ($\varepsilon-\varepsilon_F\gg\pi\hbar
s/d\equiv\varepsilon_2$), the interface effect is slight. In the region A,
it reduces simply to multiplication of the energy and momentum relaxation
times by the coefficient $\bar{\Phi}$, whereas in the region C there is no
changes at all. In the range A, the weak interface effect is conditioned by
the strong conversion effect of the LA  and TA phonon modes at the interface
reflection. In the range C, the interface is always distant, {\it i.e.}, the
effective wavelength of the interacting phonons, $\pi/q_\perp$, is much
small than the distance between the 2DEG and interface. However, we can
demonstrate that the reflection of phonons affects in higher orders of the
small parameter that occurs in the region C. The range A is the only region
where the surface phonon contribution to the scattering is essential both
for the energy and momentum relaxation. The temperature dependencies obtained
for the energy and momentum relaxation have been observed experimentally
both in the low temperature range A \cite{kawaguchi,dolgopolov,hess77} and
the high temperature range C \cite{price84,kawaguchi,sakaki84}. Recall that
in both ranges, the temperature dependencies of the relaxation times do not
differ from that of for the bulk three dimensional phonons.

\begin{figure}[htb]
\epsfxsize=14cm
\epsfysize=11cm
\mbox{\hskip 0cm}\epsfbox{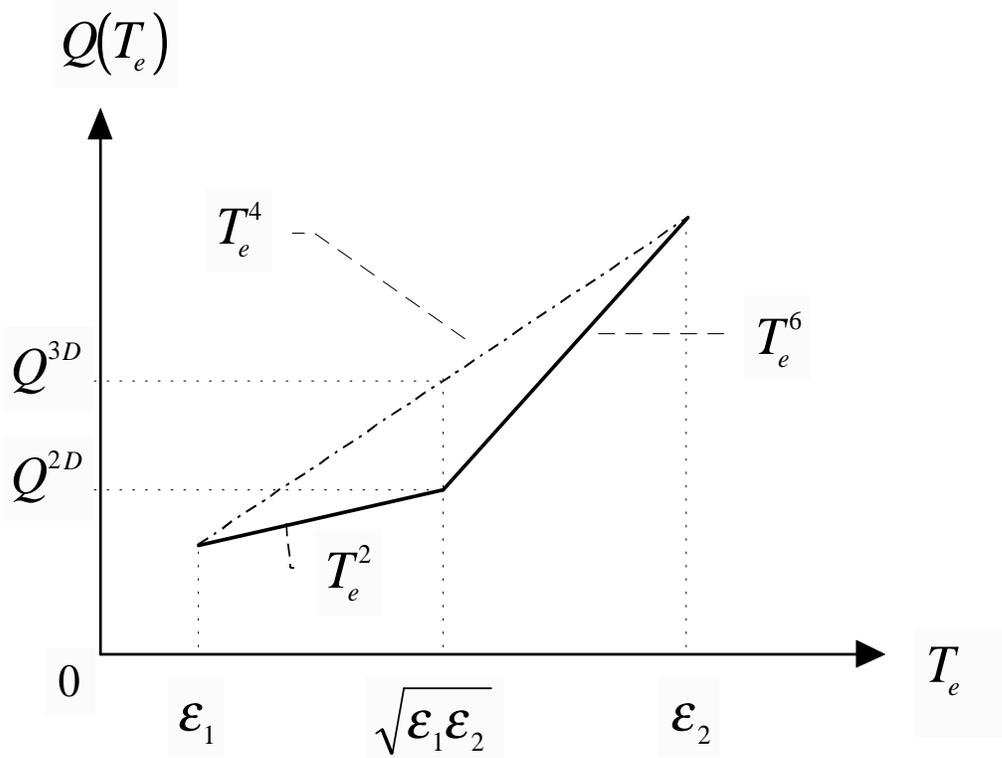}
\caption{The energy-loss power $\bar{{\cal Q}}$ of the Fermi 2DEG located
near the free crystal surface. The dashed line corresponds to the case of
the infinitely distant, ($\bar{z}\to\infty $), boundary.}
\label{fg.energy-loss}
\end{figure}

The large-angle inelastic scattering region B ($\varepsilon_1\ll\varepsilon-%
\varepsilon_F\varepsilon_2$) limited from both sides. Therefore, the
existence of the energy range B is ensured by the condition $\pi/(k_Fd)$, 
{\it i.e.} by sufficiently low electron concentrations, $N\approx10^{15}$
m$^{-2}$, and if the localization of the transverse motion of an
electrons in the 2DEG is sufficiently strong, $d\approx3$ nm. We then find
for GaAs nanostructures (see Table~\ref{tb5}) that $k_F=8\cdot10^7$
m$^{-1}$, $\varepsilon_1=3$ K and $\varepsilon_2=40$ K. It now 
follows that the criteria (\ref{criteria1}) is satisfied if $\lambda\ll0.1$, {\it i.e.} in
the situations when the elastic semi-space is in contact with a small mass
density medium, such as gas. In this case the maximum interface effect is
achieved in the middle of the energy interval $(\varepsilon_1,\varepsilon_2)$,
{\it i.e.} for $\varepsilon-\varepsilon_F,\:T_e\approx 10$ K. Therefore,
according to Eq.~\r{suppressf}, the presence of a free surface increases $%
\bar{\tau}$ and $\bar{\tau}_1$ in $\varepsilon_2/\varepsilon_1$ times, {\it %
i.e.} by a factor of the order of $10$. Moreover, for the temperature range
below $10$ K (in this range, the surface waves also contribute to the
momentum relaxation) we have $\bar{\tau}, \: \bar{\tau}_1\propto T_e^0$
while for the temperature range above $10$ K, $\bar{\tau}, \:\bar{\tau}%
_1\propto T_e^4$, {\it i.e.} the free surface also alters the dependence of
relaxation rates on the electron energy or temperature. Recall that if do
not take into account the phonon reflection then $\bar{\tau}, \:\bar{\tau}%
_1\propto T^2$. Estimates of the parameter $\lambda$ indicate that in the
majority of cases for a solid-liquid contact $\lambda\sim\varepsilon_1/%
\varepsilon_2$, therefore according to (\ref{suppress}), the solid-liquid
interface also strongly suppresses the electron relaxation rate: $%
\varepsilon_1/\varepsilon_2$ times if $\lambda\lesssim\varepsilon_1/
\varepsilon_2$ and $\lambda^{-1}$ times if $\lambda\gtrsim\varepsilon_1/
\varepsilon_2$. In the latter case, the temperature dependence
of the relaxation rate is altered in comparison with free surface case. In
the middle of the interval $(\varepsilon_1,\varepsilon_2)$, the new mini
interval $(\lambda^{-1/2}\varepsilon_1,\lambda^{1/2}\varepsilon_2)$ appears
where $\bar{\tau}, \:\bar{\tau}_1\propto T^2$. As the parameter $\lambda$
increases the interval $(\lambda^{-1/2}\varepsilon_1,\lambda^{1/2}%
\varepsilon_2)$ becomes broader and for $\lambda\gtrsim 1$ covers all the 
range B. In this case $\bar{%
\tau}, \:\bar{\tau}_1$ only by the factor $2\lambda /(1+\lambda )$ differ
from the relaxation times when the phonons are three dimensional. The
dependence of the energy-loss power $\bar{{\cal Q}}$ on the electron
temperature in the all temperature range is shown schematically in Fig.~ \ref
{fg.energy-loss}. 

Notice that the characteristic distance which is required for an interface
to have an effect on the electron relaxation is $\hbar s/T_e$. In the middle
of the range B this for GaAs gives $\hbar s/T_e\approx10$ nm. In \cite
{dolgopolov}, the dependence of the energy relaxation on the electron
temperature in inversion layers Si $(100)$ has been measured. It has been
shown that the $\bar{{\cal Q}}\propto T_e^4$ temperature dependence is
realized in the range $ T_e\gtrsim\h sk_F$. This dependence is explained
by the fact that in the
samples under the study, the interfaces separating different elastic
materials are remote, $\bar{z}\sim100$ nm, so that their effect on the
scattering is negligible. In the temperature dependence of the average
momentum relaxation time $\bar{\tau}_1$ deducted from the total mobility
measurements \cite{kawaguchi}, between low and high temperature ranges where 
$\bar{\tau}_1\propto T_e^{4-5}$ and $\bar{\tau}_1\propto T_e$, respectively,
a sharp dip has been observed in the temperature range near $10$ K. The
authors suppose that this dip can be caused by other vibration modes of the
surface layer which has been not described in the theory of surfons \cite
{ezawa1,ezawa2}. However, it is possible that this dip is connected by the
interface effect on the electron scattering which leads to the analogous
behavior of the temperature dependence in this temperature range ({\it quod
vide} in Fig.~\ref{fg.energy-loss}). 

Now we discuss the results of works 
\cite{ezawa1,nakamura,glazman,kirakos}. In \cite{ezawa1} and \cite
{nakamura}, the electron scattering in Si inversion layers has been
calculated in the high, $ T_e=200\div300$ K, and low, $ T_e= 1\div 30$ K,
temperature ranges, respectively. In both works, the scattering probability
has been expressed in terms of the phonon field correlation function.
Assuming that the scattering is quasi-elastic, the authors consider that the
static correlator can be used, {\it i.e.} the Fourrier component $\omega=0$
can be taken. However, this is not the case. From the explicit expressions 
\r{fkernblk1}-\r{fkernel} it is seen that to take $\omega=0$ in the
correlator is justified if only $\omega$ is small in comparison with $%
sq_\parallel$. Meanwhile, it follows from Eqs.~\r{arng}-\r{crng} that
such situation is never realized: in the range A, we have $\omega\sim s
q_\parallel$ while in the ranges B and C, $\omega\gg sq_\parallel$.
Therefore, we suppose that the results of these works are not correct.

Although no assumption of elastic scattering has been made in the work 
\cite{kirakos}, its results in the range B do not agree with our results.
However, the authors have also obtained that reflection of bulk phonon
modes from the interface is important in the range B and leads to a
reduction of relaxation while in the range A, the surface
phonon contribution to relaxation is important.

In the work \cite{glazman}, in particular, the conversion process LA $\to$ TA
has been assumed to take place at the sample surface. The dependence $\bar{%
{\cal Q}}\propto (T_e^2-T^2)$ has been obtained which corresponds to the
second term in Eq.~\r{btratenear}. (It appears that the approximation
adopted in this reference corresponds to $\bar{z} = 0$.) However, it follows
from Eq.~$12$ of \cite{glazman} that $\bar{{\cal Q}}$ does not vanish in the
limit $c\to 0$ when the process LA $\to$ TA is not allowed. (It is conceivable
that this is a misprint since the dimensions on the right-hand side of Eq.~$12$
after substitution of Eq.~$13$ are incorrect.)

\chapter{Phonon emission by Landau states}
\la{pemissLL}

\section{Introduction}
\label{intrdc4}

Interaction of a 2DEG with acoustic phonons in a magnetic field normal
to the electron plane is essential in a number of physical phenomena. One
may mention as examples the breakdown of the quantum Hall effect due to
phonon-assisted transitions and the steady-state power absorption for a 
GaAs heterojunction (see Refs.~\cite{eaves86,heinonen}, and also 
the review paper \cite{rashba86}), the thermalization of the heated 2DEG 
\cite{reinen,tamura,prasad}, as well as the absorption and emission of ballistic
phonon pulses by the bulk Landau states (see the reviews \cite{challis,butchallis}).

\begin{figure}[htb]
\epsfxsize=14cm
\epsfysize=11cm
\mbox{\hskip 0cm}\epsfbox{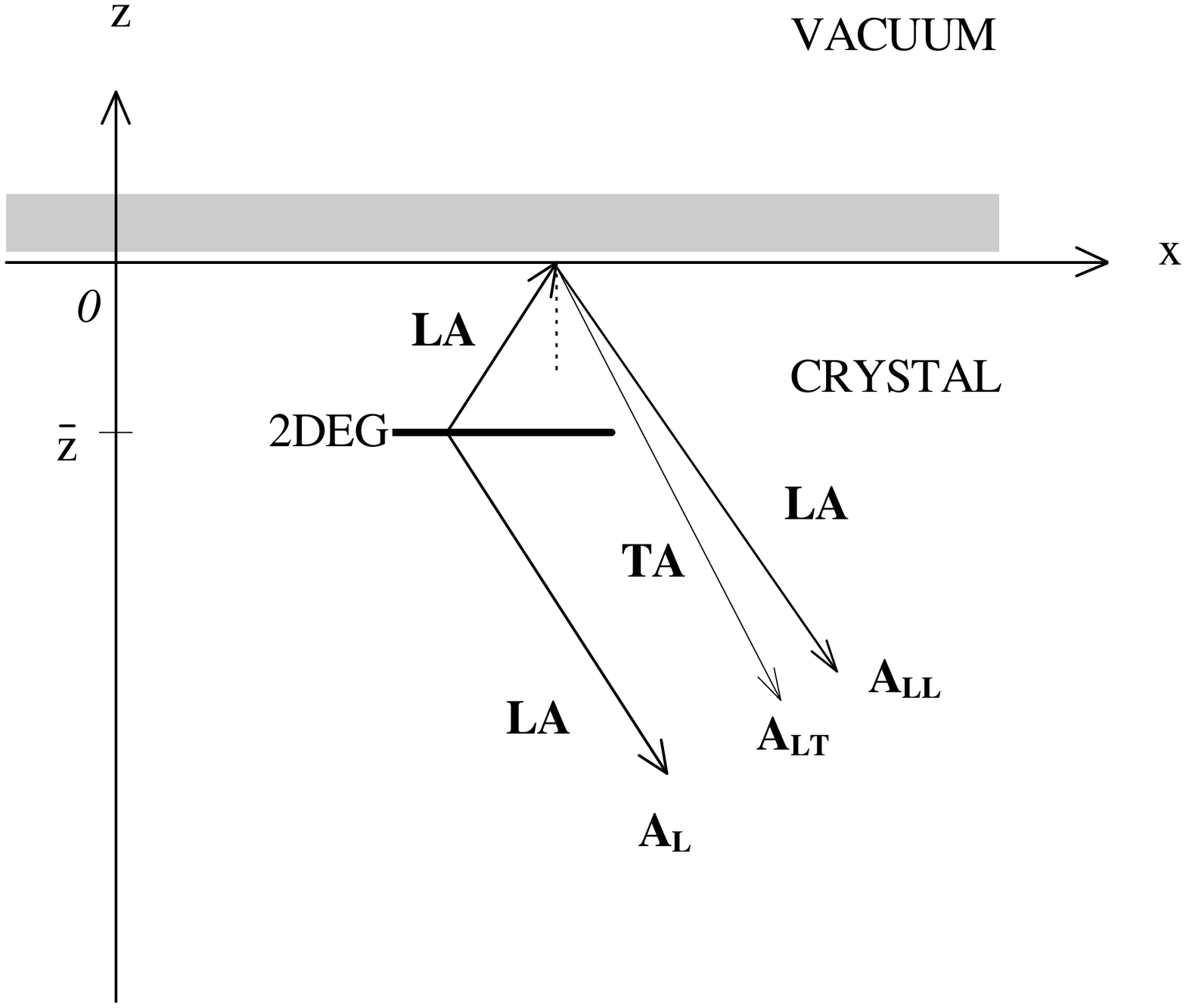}
\caption{Schematic representation of phonon emission from the 2DEG with 
account of the phonon reflection. $A_L$ and $A_{LL}$ are the amplitudes 
of  LA phonons emitted, respectively, directly into the bulk of crystal and to 
the crystal surface side then reflected from it. $A_{LT}$ is the amplitude 
of TA phonons converted from the emitted LA phonons at the reflection from 
the crystal surface.}
\label{reflection}
\end{figure}

The 2DEG is always located near a free crystal surface or near a various
interfaces separating different materials. In previous section it has been 
shown that these boundaries affect electron-acoustic-phonon interaction, 
and a free surface has the highest effect. Interaction of the 2DEG with acoustic
phonons in the magnetic field has been calculated in a number of papers
\cite{reinen,prasad,toombs}. However in all these papers it has been
assumed that crystal boundaries have no effect on interaction. 
One may easily appreciate the free surface effect if an electron 
is considered as an acoustic wave emitter. Schematically phonon emission
from the 2DEG with account of the phonon reflection is shown in 
Fig.~\ref{reflection}. Due to the reflection there appear two new effects in
phonon emission. The LA waves emitted into the crystal bulk interfere
with the LA waves reflected from the crystal surface. As a result the 
emission efficiency depends on the distance between the 2DEG and the 
surface. The phonon reflection also affects essentially the composition of the 
emitted phonon field.  Detector records the TA phonons converted from the
emitted LA phonons at interfaces separating different materials. Thus, for full 
interpretation of experimental results, the conversion of various phonon modes
and interference between reflected and non-reflected phonons of the same 
mode should be taken into account.

The 2DEG emits acoustic phonons with wave-vectors predominantly within
a narrow cone near the normal to the plane of electrons. 
At $B=0$ the frequencies of these phonons are $\omega\sim T_e/\hbar$
where $ T_e$ is the electron temperature. The frequencies are spread in a
broad band, the frequency dispersion being of the order of the frequency
itself, {\it i.e.} $\Delta\omega\sim\o\sim T_e/\hbar$. Such a wave
packet with momentum dispersion $\Delta q\sim T_e/\hbar s$. (where $s$ is
the sound velocity) attenuates at a distance o the order of $\hbar s/T_e$. If 
the distance $\bar{z}$ between the 2DEG and the surface is large ($\b{z}
\gg \h s/T_e$), the emitted phonon field will not reach the surface and the 
latter has no effect on electron-phonon interaction. However, the situation is
drastically changed if the electrons are in a strong magnetic field $B$
normal to the 2DEG plane. In this case the frequencies of the emitted
phonons are close to the cyclotron frequency ($\omega\approx \omega_B$)
while the width of the frequency band is of the order of the cyclotron
resonance line width, {\it i.e.} $\Delta\omega\sim 1/\tau$ where $\tau$
is the momentum relaxation time \cite{badalfree}. This corresponds to the 
phonon wave packet attenuation length equal to $s\tau$. In high-quality 
heterostructures even at liquid helium temperatures, the length $s\tau$ is 
larger than $\hbar s/T_e$. Therefore, in quantizing magnetic fields 
electron-phonon interaction can be modified by a free surface far more 
distant from the 2DEG.

Study of  emission and absorption processes for ballistic phonons in
systems with the 2DEG gives the most detailed information about the
character and specific features of electron-phonon interaction. In this 
case one may easily trace contributions of separate phonon modes to
\begin{figure}[htb]
\epsfxsize=14cm
\epsfysize=9cm
\mbox{\hskip 0cm}\epsfbox{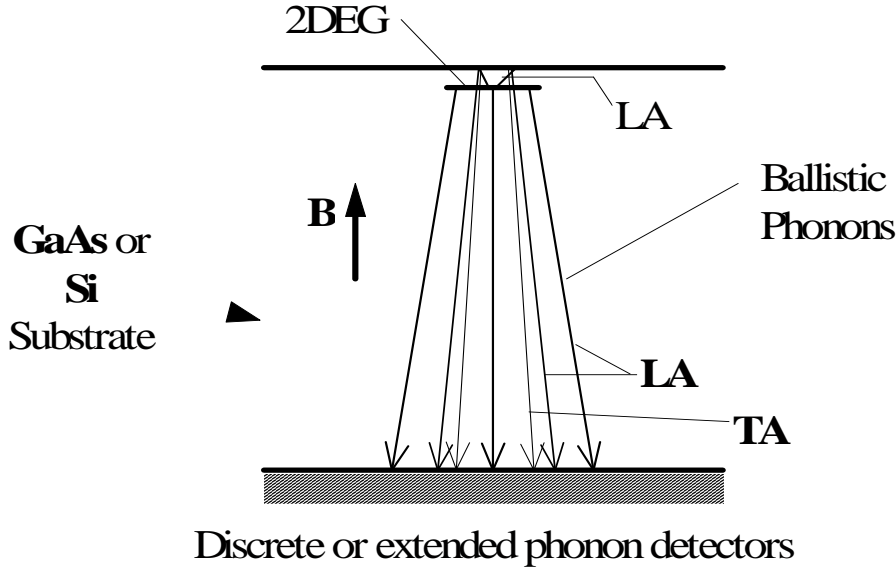}
\caption{Schematic view of the ballistic acoustic phonon emission.}
\label{ballemission}
\end{figure}
the interaction characteristics. An experiment of ballistic phonon emission is
schematically shown in Fig.~\ref{ballemission}. The 2DEG is heated by 
passing a current along it. Then emitted phonons travel ballistically from the 
2DEG until they reach the opposite face of the sample. Usually it is measured: 
the intensity distribution on the reverse face of the sample, {\it i.e.} the angular
distribution (travel is ballistic), and a relative proportion of LA and TA 
phonons using time-of-flight techniques.
However, previous works \cite{toombs,rothenfus,hensel,narayanamurti,vass}, 
analyzing these effects, both in the presence and absence of the quantizing 
magnetic field, did not take into account the ballistic phonon reflection from
 the interfaces near which the 2DEG is placed. The exception is \cite{hensel} 
in which it has been shown that the absorption of ballistic thermal phonons by
the 2DEG has the correct order of magnitude in the absence of the magnetic 
field only when the reflection of these phonons from the interfaces is taken 
into account. The results obtained in the previous chapter also indicate directly
the importance of the interface effect. 
The energy and momentum relaxation of the 2DEG have been
calculated \cite{badalinteff,badalsemispc,badalphd} and it has been shown 
that there are situations when interfaces strongly suppress the relaxation rates.
Recently the acoustic phonon emission from the heated 2DEG in 
Si-metal-oxide nanostructure has been studied theoretically by taking into 
account of the phonon reflection from Si-SiO$_2$ interface and it has been 
shown that the interface  plays an important role in the observed phonon
intensity.
Below it will be shown that the interface effect is highest in the case when
the 2DEG is exposed to the quantizing magnetic field. This is conditioned by
the fact that in this case, the interface interacts with almost monochromatic
phonons.

The previous experimental results (see review \cite{challis,butchallis}) show
that the angular distribution of phonon emission has a distinctly expressed 
peak at small angles and the signal detected on the reverse face of the 
sample consists of both longitudinal acoustic (LA) and transverse acoustic 
(TA) phonons. Giving their interpretation of these experimental data for the
GaAs/AlGaAs heterostructure, the authors suggest that the presence of
TA phonons in the detected signal is only due to a piezoelectric
electron-phonon interaction. They proceed from the assumption that in a
parabolic energy band, emission of TA phonons is forbidden in the case of
deformation electron-phonon interaction. However, this is not the case. As
it can be seen later, the reflection of phonons emitted by the 2DEG only 
due to the deformation coupling results in the presence of both LA and TA 
phonons in the detected signal. The position and height of the peak 
in the angular distribution of phonon emission are also significantly different
from those obtained in the case when the phonon reflection from interfaces 
is not taken into account.

In recent years there has been much success in using of electron-optical 
phonon interaction to probe the various electronic properties of the 2DEG 
(see \cite{moriando} and references cited therein).
The electron-longitudinal optical (LO) phonon scattering rates of the 2DEG
in the magnetic field free case have been calculated both for usual bulk LO 
phonons \cite{hess79,price81,price82,price84prb,ridley82,ridley83,shah,mason}
as well as for bulk-like confined LO and surface optical (SO) modes 
\cite{moriando,lassnig,ridley85,ridley89,swk,deg87,deg88,rudin}. If the 
magnetic field is applied normal to the electron layer, the electron motion is fully
quantized. In this quantum Hall effect (QHE) geometry, there have been
considerable experimental and theoretical interests in the various LO phonon
assisted effects of 2D-(resonant)-megneto-polarons, cyclotron resonances 
\cite{horst83,sarmamason,peet85,sigg85,zawad85,horstmerkt86,wu86,%
wu87,wu88,mason}, 2D-electron-phonon bound states \cite{badalbs}, 
cyclotron-phonon resonances \cite{badalcpr}, magneto-phonon \cite
{englert,laszwd1,laszwd2,vasil,mori} and energy loss 
\cite{reinen,hawker,warmenbol} effects.

In various recent experiments of light radiation \cite{galvao1,galvao2,knap}
or Auger \cite{potemski1} processes in the QHE geometry, phonon emission
remains the basic mechanism which ensures the relaxation of electrons
firstly created in higher Landau states. However, in such effectively zero
dimensional (0D) systems with rather thin electron layers and subjected to 
rather strong magnetic fields, a large separation between Landau levels 
cannot be covered by an acoustical LA phonon \cite{badalfree} so that the 
multiphonon 2LA phonon emission processes become more efficient 
\cite{falko}. Meanwhile an optical phonon emission requires a precise 
resonance $\Delta l\o_{B}=\omega_{LO},\, \Delta l=1,2,3,\cdots$ 
in this regime ($\omega_{B}$ and $\omega_{LO}$ are the cyclotron and 
the LO phonon frequencies). As far as it moves off from the resonance, 
the efficiency of this process is steeply falls so that out of resonance,
LO phonon emission becomes possible in accompaniment of LA phonon 
emission via the two-phonon emission mechanism. Possibly for this reason, 
except for the work \cite{prasad}, calculations of the electron-optical 
phonon relaxation rates have been restricted thus far to the zero magnetic 
field case where LO phonon emission is the dominant relaxation mechanism
with subpicosecond emission time. In Ref.~\cite{prasad} the direct calculation 
of the electron-phonon scattering rates have been carried out for a quantum 
well structure with infinitely high confining wells. It has been obtained, however,
that in the case of polar optical (PO) phonon scattering, the relaxation rate 
diverges.
This divergence has been physically attributed to the complete quantization
of the electron spectrum although it does not make clear why the relaxation
rate diverges only for polar interaction. We suppose that this divergence can 
be caused by use of the momentum-conservation-approximation 
\cite{ridley82,ridley83} in the extreme quantum limit. Whereas this
approximation fails in this limit \cite{ridley82,ridley83}. It can be seen
later that calculations without using of the momentum-conservation-approximation 
allow us to avoid this divergence.

Recently, to estimate the intensity of infrared cyclotron radiation from the
2DEG in a double barrier quantum well structure, the LA, 2LA phonon, photon
emission and Auger processes happening in the QHE geometry have been
analyzed \cite{levknap}. Although, the LO+LA phonon emission processes have
not been considered, one may expect that in some experimental situations,
electron relaxation from a higher Landau level directly into the lowest
one can turn out crucial either due to LO phonon emission or, if it is
far from the resonance, due to LO+LA phonon emission.

The LO phonon assisted electron relaxation in quantum dots has been studied 
in \cite{bockelmann,inoshita}. The multiphonon relaxation rate calculation in a
GaAs quantum dot \cite{inoshita} has indicated the significance of LO+LA
processes in such 0D systems which create a window of rapid subnanosecond
relaxation. 

Up to now, however, the study of the LO phonon assisted multiphonon 
processes in effectively 0D systems of the QHE geometry is missing from 
the literature.

The aim of the present chapter is to study theoretically the deformation
acoustic DA phonon and the polar optical PO phonon interaction with 
the 2DEG in the QHE geometry. The calculations incorporate one DA 
or PO phonon and two DA+PO phonon emission processes for electron
relaxation between bulk Landau states. In Sec.~ \ref{trans} the transition 
probability between two Landau levels with acoustic phonon emission 
via a deformation potential is calculated taking into account phonon 
reflection from the free crystal surface near which the 2DEG is located 
\cite{badalfree}. This probability is shown to be an oscillating function of 
the magnetic field $B$ and of the distance $\bar{z}$ from the 2DEG to 
the free surface. The oscillation period is given by the condition that 
$\bar{z}$ is a multiple of the half-wavelength of the emitted phonons 
$\lambda/2=\pi s/\omega_B$. In Sec.~\ref{ballistic} the emission 
spectrum of ballistic acoustic phonons in quantizing magnetic fields 
normal to the plane of the 2DEG is calculated when the reflection of 
these phonons from a GaAs/AlGaAs type interface is taken into
account \cite{badalbal}. Phonon emission only due to the deformation
electron-phonon interaction is considered. It is shown that the interface
affects essentially the intensity and the composition of the emitted phonon
field. The emission spectrum of surface acoustic phonons is calculated in
Sec.~\ref{surfacebal} \cite{badalsurf}. Strong exponential suppression of
emission of surface acoustic phonons occurs in a wide range of the
magnetic field variation, $\omega_B\gg 2mc^2_R$ ($m_c$ is the electron
mass and $c_R$ is the surface wave velocity). Therefore, cooling of the 
heated 2DEG is only at the expense of bulk longitudinal and
transverse acoustic phonons. In Sec.~ \ref{popt} the longitudinal optical
phonon assisted electron relaxation is investigated in the 2DEG in the
quantum Hall effect geometry \cite{badalpoda}. The phonon emission 
rates versus inter-Landau-level separations are calculated. Electron
interaction with bulk PO and interface SO phonons are considered. In
quantizing magnetic fields,  emission of LA phonons via piezoelectric
interaction is suppressed in comparison with deformation interaction 
\cite{badalrspt}. Therefore, the calculations of LO+LA phonon emission
are carried out for the deformation potential of DA phonons and for the
Fr\"olich coupling of PO phonons. To obtain a finite relaxation rate
associated with one-phonon emission, the allowance for the Landau level
broadening or for the LO phonon dispersion is made. Below the LO phonon
energy, $\h\omega_{LO}$, within an energy range of the order of $\h\sqrt{%
\o_{B}/\tau}$, the one-phonon relaxation rate exceeds $1$ ps$^{-1}$, 
$\tau$ is the relaxation time deduced from the mobility. In GaAs/AlGaAs
heterostructure with the mobility $\mu=25$ V$^{-1}$ s$^{-1}$ m$^2$, 
this range makes up $0.7$ meV. The two-phonon emission has a significant
contribution to relaxation above $\h\o_{LO}$. At energy separations of
the order of $sa^{-1}_{B}$ ($a_{B}$ is the magnetic length), LO+LA
phonon emission provides a mechanism of subpicosecond relaxation while
in a wide energy range of the order of $\hbar\omega_{B}$, subnanosecond
relaxation can be achieved.

Particular attention is given to the comparison of  electron relaxation going
on in two different ways: first, relaxation in two consecutive emission acts 
(emission of a PO phonon with subsequent emission of a LA or 2LA 
phonons) and, second, PO+DA phonon relaxation via the multiphonon 
emission mechanism between the same Landau levels. Particularly, emission
processes in magnetic fields up to $10$ T are considered. On the one hand 
in such fields, the resonance $\Delta l\omega_{B} =\omega_{PO}$ can
be achieved for small values of $\Delta l=1,2,3$. On the other hand,
acoustical phonon emission at electron transitions between such distant
Landau levels is still a rather efficient effect \cite{badalfree,falko,toombs}
observable in experiment \cite{galvao1,galvao2,knap,potemski1,potemski2}. 
Numerical results are illustrated for PO phonon assisted relaxation between
Landau levels $l=3$ and $l=0$. The last section contains concluding remarks.

\section{Electron transition probability between Landau levels: Free 
surface effect on electron-acoustic phonon interaction}
\label{trans}

In what follows we evaluate the probability of  electron transitions between
two Landau levels due to electron-acoustic-phonon interaction via the 
deformation potential. Implying the 2DEG in a GaAs/AlGaAs
heterostructure, the electron effective mass is considered to be isotropic. 
In this case, according to \cite{badalinteff}, the transition probability from the
state $\Psi_i({\bf r})$ with energy $\varepsilon_i$ to the state $\Psi_i(%
{\bf r})$ with energy $\varepsilon_i$ with lattice temperature $ T=0$ is
given by 
\begin{eqnarray}
W_{i\rightarrow f}&=&{\frac{2\Xi^2}{\hbar L^2}}\sum_{{\bf q_\perp}}
\int d^3r_1 \int d^3 r_2 \Psi^*_f({\bf r_1})\Psi_f({\bf r_2})\Psi_i({\bf r_1})
\Psi^*_i({\bf r_2})  
\nonumber \\&\times& 
\exp[i {\bf q_\perp\cdot(R_1-R_2)}]\,K(\omega,{\bf q_\perp}| z_1,z_2).
\label{transprb}
\end{eqnarray}
Here we have used following notations: $\Xi$ is the deformation potential
constant, $L^2$ is the normalization area in the $(x,y)$-plane occupied by 
the 2DEG, ${\bf q_\perp}$ and ${\bf R}$ are vectors in the $(x,y)$-plane.
The kernel $K$ is given in terms of the force-displacement Green function, 
calculated for elastic media, composed from GaAs and AlGaAs layers (and
a metallic layer, if present). The frequency $\omega$ entering this Green 
function satisfies $\hbar\omega=\varepsilon_f-\varepsilon_i$. When the 
temperature $ T\neq 0$, the probabilities of the downward transitions
$\varepsilon_f<\varepsilon_i$ are obtained from expression (\ref{transprb})
by multiplying the latter by a Bose factor $N_\omega+1$ and a Fermi factor
$f_i(1-f_f)$ while for the upward transitions $\varepsilon_f>\varepsilon_i$ 
the probabilities are
calculated from the detailed balance principle. The elastic properties of
the crystals GaAs and AlGaAs differ very little. Therefore. in the absence
of a metallic gate the Green function of interest is the half-space Green
function. In the isotropic approximation the latter has been calculated in
Ref.~ \cite{maradud76}, and the corresponding kernel $K$ was calculated
in Ref.~ \cite{badalinteff}. A metallic gate changes the kernel $K$: however,
in a rough approximation this fact can be ignored (since the reflection
factor for longitudinal acoustic wave energy at the Ni/GaAs interface is
less than $10\% $). The electron wave functions $\Psi$ in (\ref{transprb}) 
are of the following form: 
\begin{equation}
\Psi_{nlk_x}({\bf r})=\chi_{lk_x}({\bf R})\psi_n(z)  
\label{wfmag}
\end{equation}
where $\chi$ is the Landau wave function ($l$ is the Landau level number, $%
k_x$ is the $x$-component of the electron momentum), and $\psi$ is the
spatial quantization wave function ($n$ is the subband quantum number).
Inserting the wave functions (\ref{wfmag}) into (\ref{transprb}) and
performing the integration over ${\bf R}$, we obtain 
\begin{eqnarray}
W_{nlk_x\rightarrow n^{\prime}l^{\prime}k^{\prime}_x}&=&{\frac{2\Xi^2}{\hbar
L^2}} \sum_{{\bf q_\perp}}\delta_{k_x,
k^{\prime}_x-q_x}Q^2_{ll^{\prime}}(q_\perp)  \nonumber \\
&\times& \int dz_1\int d z_2 \psi_f({z_1})^*\psi_f({z_2})\psi_i({z_1})
\psi_i({z_2})^*K(\omega,{\bf q_\perp}|z_1,z_2).  
\label{magprob}
\end{eqnarray}
where the modulus squared of the form factor $Q_{ll^{\prime}}$ is given by
Eq.~\r{ffmagnetic} \cite{kubo}. The form factor $Q_{ll^{\prime}}$ does not
depend on momenta $k$ and $k^{\prime}$ apart but depends only on $q_{\perp}$
which is bound up with the axial symmetry of the magnetic field. We shall
consider only transitions between Landau levels corresponding to the ground
state of spatial quantization omitting the subband index $n$. Summation over
the final values $k^{\prime}_x$ and averaging over the initial values $k_x$
gives 
\begin{eqnarray}
W_{l\rightarrow l^{\prime}}&=& \mbox{Av}_{\phantom{|}_{\phantom{|}_{%
\mbox{\hspace{-3.7 ex}
\scriptsize $k_x$}}}}\sum_{k^{\prime}_x}W_{nlk_x\rightarrow
n^{\prime}l^{\prime}k^{\prime}_x}  \nonumber \\
&=& {\frac{\Xi^2}{\pi \hbar}}\int_0^\infty\! dq_\perp q_\perp
Q^2_{ll^{\prime}}(q_\perp) \int\! dz_1\int\! d z_2 |\psi({z_1})|^2|\psi({z_2}%
)|^2K(\omega, q_\perp |z_1,z_2).  \label{avprob}
\end{eqnarray}
Here we have made use of the fact that in the case of the isotropic
approximation the kernel $K$ is independent of the direction of the vector 
${\bf q_\perp}$. It follows from (\ref{ffmagnetic}) that the integration in
(\ref{avprob}) is limited by values $q_\perp\lesssim a_B^{-1}$. On the 
other hand, $\o=\o_B,\,2\o_B ,\,3\o_B,\cdots $. Therefore for the most 
important transitions with $\o\sim\o_B$ we have $(sq_\perp/\o)^2\lesssim 
ms^2/\hbar\o_B$. Since in GaAs $ms^2=0.11$ K, we may conclude that 
in the quantizing magnetic fields always the $sq_\perp/\omega\ll1$ inequality
takes place. Thus, we can expand the kernel $K$ in $q_\perp$ and use the
simplified expression 
\begin{eqnarray}
K(\omega, q_\perp |z_1,z_2)={\frac{\omega}{2\rho s^3}}
\left\{\cos\left[a(z_1-z_2)\right]- \cos\left[a(z_1+z_2)\right]\right\}, \; a=\sqrt{{\frac{\omega^2}{s^2}}-q_\perp^2}.
\label{expkern}
\end{eqnarray}
Recall that here $\rho$ is the crystal mass density, $s$ is the longitudinal
sound velocity. The value $z=0$ refers to the free surface position. The
quantity $a$, which represents the component of the phonon wave vector
normal to the surface, is not expanded in $q_\perp$, since $z_1$ and $z_2$
can be large if the 2DEG is far from the surface. Now let us choose a
certain point $\bar{z}$ inside the 2DEG (e.g. the "center of gravity" of the
distribution $|\psi(z)|^2$ or the GaAs/AlGaAs interface) and substitute: $z=%
\bar{z}+\zeta$ into the integral (\ref{avprob}) and the kernel (\ref{expkern}%
). Then carrying the intergration over $\zeta_1$ and $\zeta_2$, we obtain 
\begin{eqnarray}
W_{l\rightarrow l^{\prime}}={\frac{\Xi^2\omega}{2\pi \hbar\rho s^3}} %
\mbox{Re}\int_0^\infty\! dq_\perp q_\perp Q^2_{ll^{\prime}}(q_\perp)
\left[|I(a)|^2-\exp(2ia\bar{z})I(a)^2\right].  \label{1intprobmag}
\end{eqnarray}
The form factor $I(a)$ is given by 
\begin{eqnarray}
I(a)=\int\! d\zeta|\psi(z)|^2\exp(ia\zeta)\equiv|I(a)|\exp\{i\varphi(a)\}.
\label{ffi}
\end{eqnarray}
Since in the latter integral the integration range is limited due to the
localization of the distribution $|\psi(z)|^2$, we can put $q_\perp=0$, {\it %
i.e.} take $a=\omega/s$. However, this approximation can be invalid in the
exponential factor $\exp(2ia\bar{z})$, since $\bar{z}$ can be large. But
still, for values $\bar{z}$ not too large we may use the expansion 
\begin{equation}
a={\frac{\omega}{s}}\left[1-{\frac{1}{2}}{\frac{s^2q^2}{\omega^2}}%
-\dots\right] .
\label{expmomz}
\end{equation}
Inserting this expansion in the exponential factor, we obtain 
\begin{eqnarray}
W_{l\rightarrow l^{\prime}}&=&{\frac{\Xi^2\omega}{2\pi a^2_B \hbar\rho s^3}} %
\mbox{Re}\left\{1-\exp\left[i\left({\frac{2\omega}{s}}\bar{z}+ \varphi\left({%
\frac{\omega }{s}}\right)\right)\right]F_{ll^{\prime}}(\xi)\right\}, \\
F_{ll^{\prime}}(\xi)&=&\int_0^\infty\! dt Q^2_{ll^{\prime}}(t)\exp(-i\xi t) 
\nonumber \\
&=&{\frac{l!}{l^{\prime}!(l-l^{\prime})!}}{\frac{(i\xi)^{2l^{\prime}}}{%
(1+i\xi)^{l+l^{\prime}+1}}} {_{\phantom{|}{\scriptstyle \scriptsize{2}}}}
F_1(-l^{\prime},-l^{\prime}; l-l^{\prime}+1;-\xi^{-2}), \\
&&
\xi={\frac{\bar{z}}{\Delta l\, \Lambda_1}}, \: 
\Lambda_1={\frac{\hbar}{2ms}}, \: \Delta l=l-l^{\prime}.  
\label{finalprobmag}
\end{eqnarray}
Here ${_{\phantom{|}{\scriptstyle \scriptsize{2}}}}F_1$ is the Gauss 
hypergeometric function.
The integrals $F_{ll^{\prime}}\rightarrow 0$ when $\bar{z}\rightarrow 0$
so that the probability $W$ has a definite limit corresponding to the infinitely
distant free surface, as one should expect. We list explicit formulae for $F$
for the most important transitions, 
\begin{eqnarray}
F_{10}(\xi)=(1+i\xi)^{-2},\: F_{20}(\xi)=(1+i\xi)^{-3}, \\
F_{21}(\xi)=2 {\frac{(i\xi)^2}{(1+i\xi)^4}} 
{_{\phantom{|}{\scriptstyle \scriptsize{2}}}}F_1(-1,-1,2;-\xi^{-2}). 
 \label{funcf}
\end{eqnarray}
It follows from the integrals (\ref{finalprobmag}) that one can neglect the
free surface effects when $\bar{z}\gg\Lambda_1$. For GaAs 
$\Lambda_1=170$ nm while typical values of $\b{z}$ are in the interval 
$50$ to $150$ nm. The expansion (\ref{expmomz}) can be used in 
$\exp(2ia\bar{z})$ if the quantity 
\begin{eqnarray}
\bar{z}{\frac{\omega}{s}}\left({\frac{sq_\perp}{\omega}}\right)^4\approx 
\b{z}{\frac{a^2_B}{\Lambda^3_1}} 
\label{approx}
\end{eqnarray}
is small. This condition is usually satisfied in the quantizing fields.
Equally with the momentum spread, there exists another reason why the effect
of the free surface becomes weaker with distance, namely, the spread of the
frequencies of the emitted phonons due to the Landau level broadening. To
take this broadening into account, we average the transition probability (%
\ref{finalprobmag}) over $\omega$ using a Lorentzian distribution with its
center at the transition frequency $\Delta l\omega_B$ and width $1/\tau$.
Only the fast oscillating factor containing $\bar{z}$ should be averaged,
and as a result this factor is substituted as follows 
\bea
\exp\left(i{\frac{2\omega}{s}}\bar{z}\right)\rightarrow \exp\left(-{\frac{%
\bar{z}}{\Lambda_2}}+i{\frac{2\omega}{s}}\bar{z}\right),  
\label{broadening}
\\
\Lambda_2={\frac{1}{2}}{\frac{\hbar s}{\Gamma}}  
\label{lambda2}
\end{eqnarray}
where $\Gamma$ is the Landau level width. Before comparing the two
attenuation lengths $\Lambda_1$ and $\Lambda_2$, we discuss what sort of
Landau level width is to be used in averaging. In high-quality
heterostructures the Landau level width being determined from the de
Haas-van Alphen effect, is a factor of about $10$ larger than the width $%
\Gamma=\hbar/\tau$, with $\tau$ determined from mobility or from cyclotron
resonance \cite{fangsmith}. This is because the scattering potential has a
smooth component with a large amplitude. The smooth potential produces an
inhomogeneous broadening of the Landau levels seen in the de Haas-van Alphen
effect. However, the smooth potential gives no contribution to scattering,
since its correlation length exceeds the electron Fermi wavelength \cite
{sarmastern}. Neither does the smooth component contribute to the cyclotron
resonance linewidth, since the absorption occurs without change of $k_x$, 
{\it i.e.} with a fixed position of the center of the Landau oscillator.

It follows from (\ref{magprob}) that phonon emission is accompanied by a
change of $k_x$, the change being of the order of $a^{-1}_B$. In other words,
when a phonon is emitted, the center of the Landau oscillator is shifted by
a distance of the order of $a_B$. Since in quantizing magnetic fields $%
a_B\ll\Lambda$, no contribution is made by the smooth potential component to
the frequency spread of the emitted phonons. Therefore the $\Gamma=
\hbar/\tau$ in (\ref{lambda2}) should be deduced from the momentum
relaxation time.

Even in high-quality heterojunctions with mobility $\mu=100$ m$^2$ 
V$^{-1}$ s$^{-1}$ the length $\Lambda_2=100$ nm is comparable 
with the distance $\b{z}$. Therefore the Landau level broadening is 
essential.

With the model wave function \cite{fanghow} 
\begin{eqnarray}
|\psi(\zeta)|^2={\frac{1}{2d^3}}\zeta^2\exp(-\zeta/d),  \label{modelwf}
\end{eqnarray}
where $\zeta=0$ corresponds to the interface position, one obtains 
\begin{equation}
\left|I\left({\frac{\omega}{s}}\right)\right|^2={\frac{1}{(1 +\eta)^3}},\:
\tan\varphi\left({\frac{\omega}{s}}\right)={\frac{\eta(3-\eta)^2}{(3\eta-1)^2%
}},\:\eta={\frac{\omega^2}{(s/d)^2}}.  \label{modelff}
\end{equation}
For a GaAs/AlGaAs heterojunction the typical value is $d=3$ nm (see Ref.~%
\cite{stern}), and $\omega_B=s/d$ when $B=0.66$ T. With the wave function (%
\ref{modelwf}), the probability of the most important transition from $l=1$
to $l=0$ is 
\begin{eqnarray}
W_{1\rightarrow 0}&=&{\frac{1}{\bar{\tau}}}\left({\frac{\omega_B}{s/d}}%
\right)^2 \left[1+\left({\frac{\omega_B}{s/d}}\right)^2\right]^{-3} 
\label{10probmag}
\nonumber \\
&\times& \left\{1-{\frac{\exp(-\bar{z}/\Lambda_2)}{1+(\bar{z}/\Lambda_1)^2}}
\cos\left(2{\frac{\omega_B}{s}}\bar{z} +6\tan^{-1} {\frac{\omega_B}{s/d}}
-2\tan^{-1}{\frac{\bar{z}}{\Lambda_1}}\right)\right\}, \\
&&{\frac{1}{\bar{\tau}}}={\frac{m\Xi^2}{2\pi\hbar^2\rho s d^2}}.
\la{nomtau}
\end{eqnarray}
Here $\bar{\tau}$ is a nominal relaxation time, depending only on the
heterojunction. For a GaAs/AlGaAs heterojunction with $d=3$ nm we 
have $\bar{\tau}=0.20$ ns. 
\begin{figure}[htb]
\epsfxsize=13cm
\epsfysize=13cm
\mbox{\hskip 0.5cm}\epsfbox{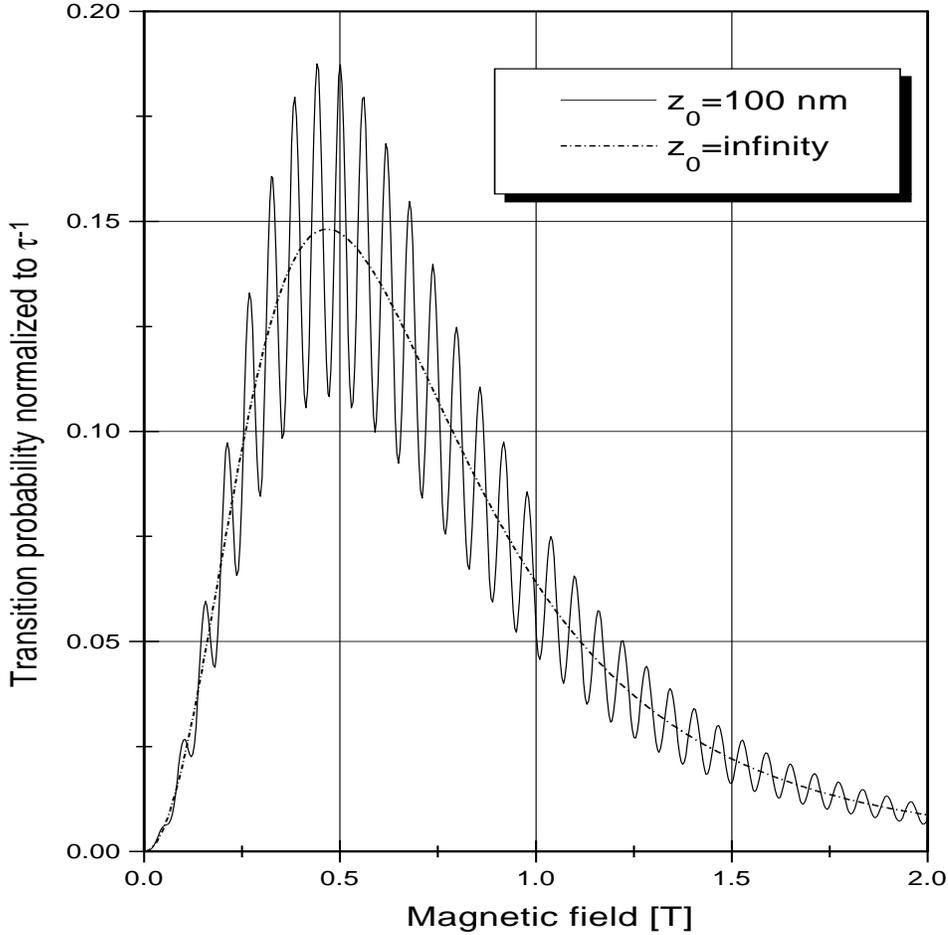}
\caption{Transition probability $W_{1\rightarrow 0}$ between Landau levels 
$l_1=1$ and $l_2=0$ versus magnetic field $B$ for the distance $\b{z}=100$
nm and $\b{z}=\infty$ from the 2DEG to the free surface.}
\label{fgprobmag10}
\end{figure}
One may see from (\ref{10probmag}) that when the 2DEG is not too far from
the free surface, the transition probability $W$ has an oscillatory
dependence on the magnetic field $B$ and on the distance $\bar{z}$ from the
free surface. For illustration we calculate from (\ref{10probmag}) $%
W_{1\rightarrow 0}$ (normalized to $1/\bar{\tau}$) versus $B$ for $\bar{z}%
=100$ nm and $\bar{z}=\infty$ (Fig.~ \ref{fgprobmag10}) 
\begin{figure}[htb]
\epsfxsize=13cm
\epsfysize=12cm
\mbox{\hskip 0.5cm}\epsfbox{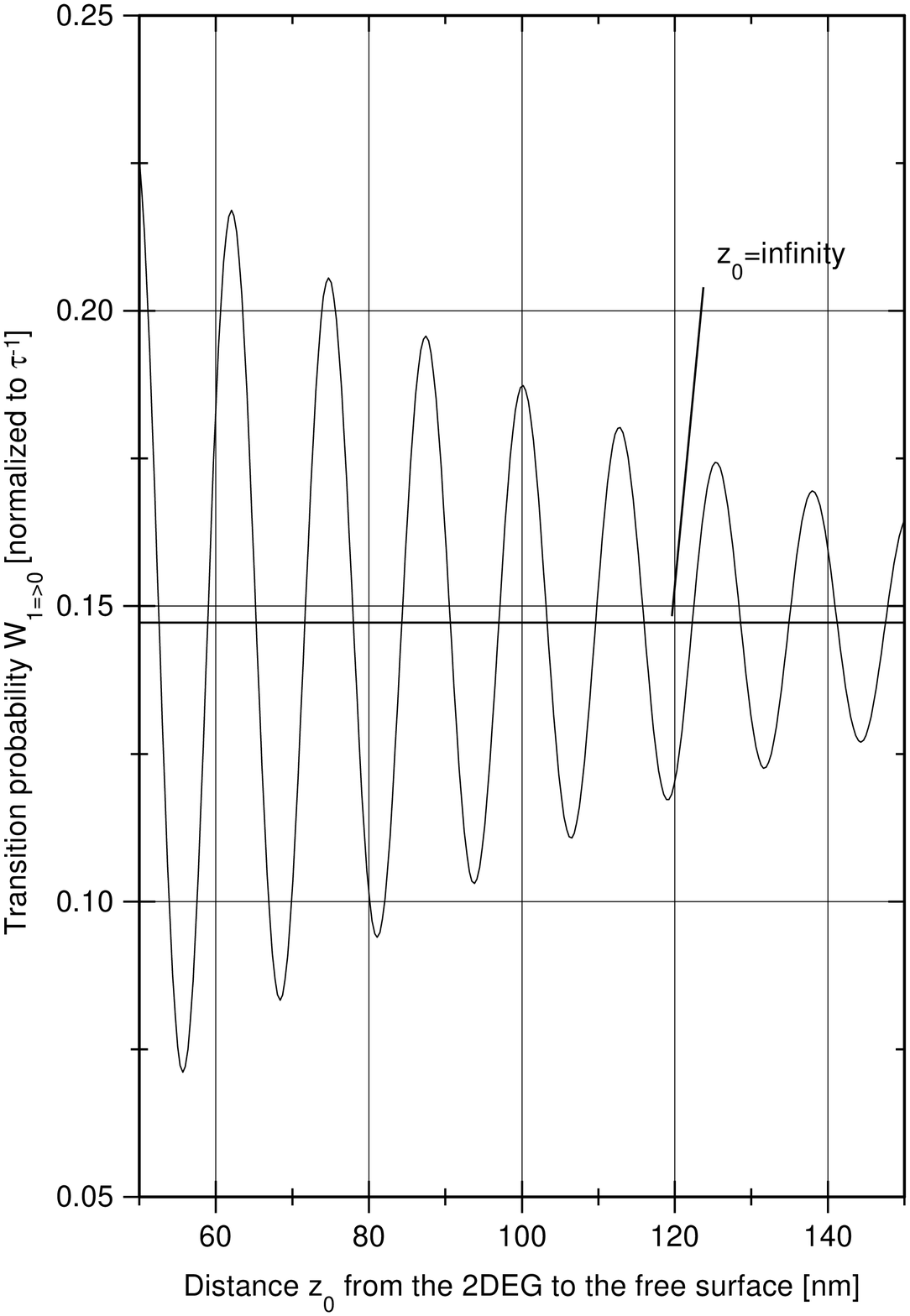}
\caption{Transition probability $W_{1\rightarrow 0}$ between Landau levels 
$l_1=1$ and $l_2=0$ versus distance from the 2DEG to the free surface
for the magnetic field $B=0.5$ T. The transition probability is also shown
when no phonon reflection is taken into account.}
\label{fgprobz10}
\end{figure}
and also versus $\bar{z}$ for $B=0.5$ T (Fig.~ \ref{fgprobz10}).

\section{Ballistic acoustic phonon emission by Landau states}
\label{ballistic}

\subsection{Acoustic energy flux density}

For deformation electron-phonon interaction in the non-degenerate
isotropic energy band, with the energy minimum in the Brillouin zone center,
the interaction Hamiltonian density can be represented in the following form
\begin{equation}
H_{int}=\Xi\:|\Psi({\bf r}, t)|^2 \:{\bf div}\:{\bf u}({\bf r}, t).  
\label{hint}
\end{equation}
Here $\Xi$ is the deformational potential constant, $\Psi(r, t)$ and ${\bf u}%
({\bf r}, t)$ are respectively the electron and phonon field operators in
the Heisenberg representation. It can be shown by a straightforward
calculation that the spectral density of the acoustic energy flux at the
frequency $\omega$ has the form 
\begin{eqnarray}
{\cal W}_\alpha&=&\mbox{Re}\left[-i{\frac{\omega}{2\pi}}
 <\sigma^{+}_{\alpha%
\beta}({\bf r_0})u_{\beta}({\bf r_0})>_\o\right],  \label{flux} \\
\sigma_{\alpha\beta}({\bf r_0})&=&2\mu u_{\alpha\beta}+ \lambda
u_{\gamma\gamma}\delta_{\alpha\beta},\: u_{\alpha\beta}={\frac{1}{2}}%
(\partial_\alpha u_{\beta}+ \partial_\beta u_{\alpha}).  \label{stress}
\end{eqnarray}
Here $\sigma_{\alpha\beta}$ and $u_{\alpha\beta}$ are the stress and
deformation tensors of the elastic medium, $\mu$ and $\lambda$ are the Lame
coefficients, $<\cdots >_\o$ is the Fourier transform of the correlator of
the operators $\sigma$ and $u$. Repeated Cartesian indices imply summation
over the values $\alpha,\beta=1, 2, 3$, and $\partial_\alpha \equiv\partial
/ \partial_\alpha$ is the partial derivative operator.

As follows from expressions (\ref{hint}) and (\ref{flux}), in order to obtain 
the flux density ${\cal W}_\alpha$ we have to calculate the correlator for
the phonon field operators 
\begin{equation}
K_{\alpha\beta}({\bf r, r^{\prime}})=<u^{+}_{\alpha}({\bf r})u_{\beta}({\bf %
r^{\prime}})>_\o.  \label{corel}
\end{equation}
With that purpose we note that the deformational displacement $u_\alpha(r, t)
$ is caused by the force 
\begin{equation}
F_\alpha({\bf r},t)=-{\frac{\delta}{\delta u_\alpha}}H_{int} =
\Xi\,\partial_\alpha|\Psi({\bf r}, t)|^2.  \label{force}
\end{equation}
Therefore, making use of the Green function $G_{\alpha\beta}$ from
elasticity theory, for the displacement under the force (\ref{force}) we can
write 
\begin{equation}
u_{\alpha}({\bf r})=\int\!d^3r_1dt_1 G_{\alpha\beta}({\bf r},{\bf r_1}%
|t-t_1) F_\beta({\bf r_1},t_1)  \label{displ}
\end{equation}
The electron field operators in the occupation number representation are
expressed through the anti-commutative Fermi operators $a_s$ in the
following form, 
\begin{equation}
\Psi({\bf r}, t)=\sum_s a_s \psi_s({\bf r})\exp( -i\varepsilon_st)
\label{fermiop}
\end{equation}
where $\varepsilon_s$ and $\psi_s$ are, respectively,  energies and 
wavefunctions of electrons in the single-particle states which are 
characterized by a set of quantum number $s$. Substituting expressions 
(\ref{force})-(\ref{fermiop}) into (\ref{corel}) we can represent the 
correlator $K$ in the following form
\begin{eqnarray}
K_{\alpha\beta}({\bf r, r^{\prime}})&=&
2\pi\Xi^2\sum_{s,s^{\prime}}f_{s}(1-f_{s^{\prime}})\delta\left(\e_{s}-%
\e_{s^{\prime}}-\h\o\right) \: \int\!d^3r_1 d^3r_2  \nonumber \\
&\times& \psi^*_{s}({\bf r_1})\psi_{s^{\prime}}({\bf r_1}%
) \psi_{s}({\bf r_2})\psi^*_{s^{\prime}}({\bf r_2}) D^*_\alpha({\bf r},{\bf %
r_1}|\omega)D_\beta({\bf r^{\prime}},{\bf r_2}|\omega)  \label{correl}
\end{eqnarray}
in which 
\begin{equation}
D_\alpha({\bf r},{\bf r^{\prime}}|\omega)=\partial^{\prime}_\beta
G_{\alpha\beta}({\bf r},{\bf r^{\prime}}|\omega)  \label{divgreen}
\end{equation}
and the $f_s$ are the electronic occupation numbers.

Now, in order to obtain the energy flux, we have to find the Green function
for a layered elastic medium consisting of GaAs and AlGaAs, and a metallic
gate. As was shown in previous section, such a layered medium can be
described for our purposes in the framework of a semi-infinite elastic
medium, {\it i.e.} we can consider the elastic waves to be reflected only 
from a surface adjacent to the vacuum. In this case it is assumed that the 2DEG is
at a distance $\bar{z}$ from the free surface.

The Green function for an isotropic semi-infinite elastic medium in the
momentum representation in the plane of the translation invariance has been
obtained in \cite{cottam}. Using these results, we have calculated the Green
function in the coordinate representation. As far as phonon emission is 
detected in an infinitely distant point on the reverse face of the sample, we 
are interested in obtaining the Green function $G({\bf r_0},{\bf r_1}|\omega)$
when ${\bf r_0} \to {\bf n}\infty$, ${\bf n}$ is a unit vector directed
towards the detector.

The result of calculations shows that the acoustic energy flux density
is represented as a sum of the flux densities of the bulk LA and TA phonons: 
\begin{eqnarray}
{\cal W}^{Bulk}_\alpha&=&{\cal W}_\alpha^{L}+{\cal W}^{T}_\alpha
\label{bulkflux}
\end{eqnarray}
where 
\begin{eqnarray}
{\cal W}_\alpha^{L}&=&{\frac{1}{\pi}}\rho s^2q^L_\alpha\hbar\omega
K^L_{\beta\beta}({\bf r_0, r_0}),  \label{fluxl} \\
{\cal W}_\alpha^{T}&=&{\frac{1}{\pi}}\rho c^2q^T_\alpha\hbar\omega
K^T_{\beta\beta}({\bf r_0, r_0}),  \label{fluxt}
\end{eqnarray}
in which $s$ ($c$) and $q^L_\alpha=n_\alpha \omega/s$ ($q^T_\alpha=n_\alpha
\omega/c$) are the velocities and momenta for the LA (TA) phonons,
respectively. Further, the kernel of the correlator $K^L$ in expression 
(\ref{correl}) is given by
\begin{equation}
D^*_\alpha({\bf r_0},{\bf r_1}|\omega)D_\beta({\bf r_0},{\bf r_2}|\omega)= {%
\frac{\omega^2}{16\pi^2 r_0^2\rho^2s^6}} {\cal H}^{L*}({\bf r_1}){\cal H}%
^{L}({\bf r_2})  \label{lkernel}
\end{equation}
where 
\begin{equation}
{\cal H}^{L}({\bf r})=\exp(-i{\bf q_\perp^L R}) \left[\exp(-iq^L_z z)+{\cal R%
}_{LA\to LA}\exp(iq^L_z z)\right].  \label{lcalh}
\end{equation}
Similarly, in calculating $K^T$ we have exploited 
\begin{equation}
D^*_\alpha({\bf r_0},{\bf r_1}|\omega)D_\beta({\bf r_0},{\bf r_2}|\omega)= {%
\frac{\omega^2}{16\pi^2 r_0^2\rho^2 c^4s^2}} {\cal H}^{T*}({\bf r_1}){\cal H}%
^{T}({\bf r_2})  \label{tkernel}
\end{equation}
where 
\begin{equation}
{\cal H}^{L}({\bf r})={\cal R}_{LA\to TA} \exp(-i{\bf q_\perp^L R}
-i\alpha^L z).
  \label{tcalh}
\end{equation}
The following notations were used in expressions (\ref{lcalh}) and (\ref
{tcalh}): ${\bf r}=({\bf R}, z)$ and ${\bf q}=({\bf q_\perp},q_z)$, where $%
{\bf R}$ and ${\bf q_\perp}$ are two dimensional vectors in the $(x, y)$%
-plane. The reflected-to-incident LA wave amplitude ratio ${\cal R}_{LA\to
LA}$ is given by the following expression 
\begin{eqnarray}
{\cal R}_{LA\to LA}= {\frac{c^2\sin(2\theta)\sin(2\theta_T)-s^2\cos^2(2%
\theta_T) }{c^2\sin(2\theta)\sin(2\theta_T)+s^2\cos^2(2\theta_T)}}, \;
\sin\theta_T={\frac{c}{s}}\sin\theta  \label{llampltd}
\end{eqnarray}
where $\theta$ is the angle between the vector ${\bf n}$ and the normal to
the plane of incidence.

The amplitude ratio of the reflected TA wave to the incident LA wave as well
as the momentum $\alpha^L$ are determined differently in various ranges of
the angle $\theta$, $\theta<\theta<\theta_c$ and $\theta_c<\theta<\pi/2$,
where $\theta_c=\arcsin(c/s)$ is the critical angle of the TA wave total
reflection.

When $\theta<\theta<\theta_c$ then the momentum $\alpha^L=-i(\omega/s)\cos%
\theta_L$ where $\sin\theta_L=(s/c) sin\theta$ and 
\begin{equation}
{\cal R}_{LA\to TA}= {\frac{2cs\sin(2\theta)\cos(2\theta)}{%
c^2\sin(2\theta)\sin(2\theta_L)+s^2\cos^2(2\theta)}}.  \label{ltampltd1}
\end{equation}

When $\theta_c<\theta<\pi/2$ then the momentum $\alpha^L=(\omega/ s)
sh\vartheta_L$ where $ch \vartheta_L = (s/c) sin\theta$ and 
\begin{equation}
|{\cal R}_{LA\to TA}|^2= {\frac{4c^2s^2\sin^2(2\theta)\cos^2(2\theta) }{%
c^4\sin^2(2\theta)sh(2\vartheta_L)+s^4\cos^4(2\theta)}}.  \label{ltampltd2}
\end{equation}
Since we are interested in emission taking place in the magnetic field,
we have to take in the corellator (\ref{correl}) the electron wave functions
given by Eq.~\r{wfmag}. Substituting expressions (\ref{correl}), (\ref
{lkernel})-(\ref{tcalh}) and (\ref{wfmag}) into the relation (\ref{fluxl}),
we finally obtain the following expression for the LA phonon energy flux
density emitted from a unit area of the 2DEG into a solid angle element d$o$
around the direction $\theta$: 
\begin{equation}
d{\cal W}^L(\theta)=\left[1+{\cal R}^2_{LA\to LA}+2{\cal R}_{LA\to LA}
\cos2\varphi\right]{\cal W}^{3D}(\theta)do  
\label{l2dflux}
\end{equation}
where 
\begin{equation}
{\cal W}^{3D}(\theta)= {\cal W}_0\,Q^2_{l_1l_2}\left({\frac{\omega}{s}}%
\sin\theta\right)\, \left|I\left({\frac{\omega}{s}}\cos\theta
\right)\right|^2,\: {\cal W}_0={\frac{\Xi^2\omega_B^4}{8\pi^3\rho s^5 a_B^2}}%
f_{l_1}(1- f_{l_2})
\label{flux3d}
\end{equation}
is the LA phonon acoustic energy flux when the phonon reflection is not
taken into account. The form factors $Q_{l_1l_2}$ and $I(q_z)$ are given by
Eq.~\r{ffmagnetic} and Eq.~\r{ffi}. The phase $\varphi$ in Eq.~\r{ffi}
can be represented in the form 
\begin{equation}
\varphi(\theta,\bar{z})={\frac{\omega}{s}}\bar{z}\sin\theta+\varphi_0,
\label{phase}
\end{equation}
$a_B$ is the magnetic length. It is assumed that the interacting electrons
stay at the same spatial quantization level. It is also taken into account
the spin degeneration factor $g_s=2$.

For TA phonons in the same way, we obtain in the angular range $%
0<\theta<\theta_c$ 
\begin{equation}
d{\cal W}^T(\theta)={\frac{s^3}{c^3}}|{\cal R}_{LA\to TA}|^2{\cal W}%
^{3D}(\theta_L)do  \label{t2dflux1}
\end{equation}
while in the range $\theta_c<\theta<\pi/2$ 
\begin{equation}
d{\cal W}^T(\theta)={\frac{s^3}{c^3}}|{\cal R}_{LA\to TA}|^2 
{\cal W}^{3D}(\pi/2-i\vartheta_L)do  \label{t2dflux2}
\end{equation}
Notice that in the range $0<\theta<\theta_c$, ${\cal W}^T$ does not depend
on the distance $\bar{z}$ while in the range $\theta_c<\theta<\pi/2$, there
appears a strong exponential dependence via a factor $\exp[-2(\omega/s)\bar{z}
sh \vartheta_L]$.

\subsection{Suppression of surface acoustic phonon emission}
\label{surfacebal} 

In the previous section, the emission spectrum for the bulk ballistic acoustic
phonons has been calculated when the reflection of these phonons from the 
interface near the 2DEG has been taken into account.
It is clear that, in spite of the considered deformational electron-phonon
coupling, the detected signal consists of both longitudinal LA and
transverse TA phonons. However, the emitted phonon field, in general, must
also include surface acoustic SA phonons. Over the last few years the
interaction between the 2DEG and surface acoustic waves in the 
quantum Hall regime have been widely investigated 
\cite{wixforth,efr3,rampton,tam,knaebchen}.

In order to find the emission spectrum of the ballistic surface phonons, we
have to complement our calculations of the previous section also for the
energy range $\omega<c q_\perp$, where $c$ is the TA phonon velocity, $\omega
$ and $q_\perp$ are the energy and in-plane momentum of the phonons. Recall
that the elasticity theory Green function has the pole in this energy range
corresponding to the normal surface mode of the elastic layered medium. 
Straightforward calculations show that the spectral density of the acoustic
energy flux ${\cal W}$ at frequency $\omega$ is represented as a sum of the
flux densities of the bulk and surface SA phonons, 
\begin{equation}
{\cal W}= {\cal W}^{Bulk} + {\cal W}^S.  \label{bsaflux}
\end{equation}
Moreover, if one describes the layered medium of the GaAs/AlGaAs  
heterostructure in the framework of a semi-infinite elastic medium which is
an acceptable approximation for our purposes \cite{badalfree} then for the
flux density of the Rayleigh waves emitting from a unit area of the 2DEG 
we obtain 
\begin{eqnarray}
{\cal W}^R|_{1\to 0}=g\,{\cal W}_0\,\kappa_R^2\exp(-\kappa_R^2) (1+
{\frac{\eta_R d}{\bar{z}}})^{-6}\exp(-2\eta_R), \\
\kappa_R^2={\frac{1}{2}}(q^R a_B)^2={\frac{\omega_B}{2m_c c_R^2}},\:
\eta_R=\gamma_1{\frac{\omega_B}{c_R/\bar{z}}}.  \label{sflux}
\end{eqnarray}
where the dimensionless quantity $g$ is given 
\begin{eqnarray}
g(\xi_1,\xi_2)&=&2\pi\left[{\frac{1}{2\gamma_1}}\left(1-{\frac{\gamma_1}{%
\gamma_1}}(1+\gamma_2^2)+{\frac{\gamma_1^2}{\gamma_2^2}}\right)
-(\gamma_1-\gamma_2){\frac{1+\gamma_2^2}{1-\gamma_2^2}}\right]  \nonumber \\
&\times& \left\{\sqrt{1-\gamma_2}\left[2(\gamma_1-\gamma_2)+ \gamma_1\left({%
\frac{1}{\gamma_1}}-{\frac{1}{\gamma_2}}\right)^2\right]^2 \right\}^{-1}, \\
&& \gamma_1=\sqrt{1-\xi_1^2},\: \gamma_2=\sqrt{1-\xi_2^2},\: \xi_1={\frac{c_R%
}{s}},\: \xi_2={\frac{c_R}{c}}.  \label{g}
\end{eqnarray}
Here $c_R$ is the velocity of the Rayleigh waves. For GaAs with $c/s=0.59$
and $c_R/c=0.92$ (see Table~ \ref{tb5}), we obtain $g=2.51$. This surface
acoustic energy flux isotropic and concentrated in the 2DEG plane. The factor 
${\cal W}_0$ is given by \ref{flux3d} and depends on the parameters of the 
problem which are the same for both bulk and surface phonons. Moreover, 
we have ${\cal W}^B\sim {\cal W}_0$ at the peak of the bulk phonon 
emission ({\it cf.} Eq.~\r{flux3d}). However, in magnetic fields $B=1$ T 
taking $d= 3$ nm and $\b{z}= 100$ nm one can be convinced that 
$\kappa_R^2 = 303$ and $2\eta_R= 157$. This means that the
emission of SA phonons is extremely weak. The first exponent in 
Eq.~\r{sflux} is also present in the corresponding expression for the bulk 
phonon
emission intensity. However, in contrast with the bulk phonon case where for
the same transition energy $\h\o_B$, the emission is almost normal to the 
2DEG plane, the momentum of the Rayleigh waves completely lies in the
2DEG plane. That is why in this case $\kappa_R^2 \gg 1$ and the interaction
matrix element rapidly decreases. This is a direct consequence of the fact
that the electron states in this magnetic field regime constitute a plane
wave packet, and in this packet there is an exponentially small number of
waves for which the momentum conservation law is fulfilled for interaction 
with surface phonons with $q=\o_B/c_R\gg p_B\equiv a_B^{-1}$. The 
second exponent appears because the SA phonons propagate along the 
free surface and exponentially attenuate if departing from this surface. It is 
easy to see that for typical values of $\b{z} =50-150$ nm the surface is 
always distant: $\eta_R\gg1$. Thus one can confirm that the strong 
exponential suppression of the SA phonon emission occurs in the wide range
of the magnetic field variation, $\h\o_B\gg 2m_c c^2_R$ so the cooling of
the heated 2DEG is only at the expense of emission of the bulk phonons.

\subsection{Interference and conversion effects}
\label{conversion}

The relations obtained in previous section show that in spite of deformation
 electron-phonon interaction, and though the energy band has a parabolic 
shape, the phonon field emitted by the 2DEG consists of both bulk LA and
TA phonons. On the other hand, it is clear that the electrons interact directly 
only with LA phonons ({\it cf.} Eq.~\r{hint}). This situation can be explained
as follows (see Fig.~\ref{reflection}). The electrons of the 2DEG emit only
LA phonons in all directions. Therefore the detector at the point ${\bf r_0}$ 
will record all the LA phonons which are emitted in the direction of negative 
$z$-axes at the angle of $\theta$. Only such phonons were taken into account
in the 3D case.
However, the detector will also respond to all those phonons which have 
been primary emitted in the direction of positive $z$-axes at the angles of 
$\theta$ and $\theta_L$ to the free surface. Such LA phonons, upon their
reflection and conversion at the free surface, propagate backwards in the
form of LA and TA phonons, respectively, in the negative $z$-direction 
both at the same $\theta$ angle. The reflected LA phonons interfere with the
initial LA phonons. Thus the detector will respond to the {\it interference}
field of the LA phonons and to the {\it conversion} field of the TA phonons.
Now, since the phase difference of the initial and reflected LA phonons is
the function of the angle $\theta$ and the distance $\bar{z}$ ({\it cf.} Eq.~%
\ref{phase}), the intensity of the detected interference of the LA phonons
should also depend on $\theta$ and $\bar{z}$ (see expressions (\ref{l2dflux}%
)). At the same time, one should expect that the intensity of the conversion
field of the TA phonons should be independent of the distance $\bar{z}$.
This, indeed, is the case when $\theta<\theta_c$. According to Eq.~\r
{t2dflux1}, the flux of the converted TA phonons depends only on $\theta$
but not on the distance $\bar{z}$. When $\theta$ exceeds $\theta_c$, the
intensity of the converted TA phonons begins to depend strongly on $\bar{z}$
via the factor $\exp[-2(\omega/s)\bar{z}sh\vartheta_L]$ ({\it cf.} Eq.~\r
{t2dflux2}). The appearance of such a dependence can be most easily
understood if we consider the inverse process of the TA$\rightarrow$LA
conversion at the phonon absorption by the 2DEG. When the TA phonons 
are incident on the free surface at the angle of $\theta>\theta_c$ due to the
total reflection effect, the LA phonons are being created in the $(x,y)$-plane
(one should not confuse them with surface phonons). The amplitude of such
phonons decreases exponentially with the distance from the surface. 
Therefore, their
interaction with the 2DEG is effective and the absorption of the TA phonons
incident at angles of $\theta>\theta_c$ takes place when the distance $\b{%
z}$ is of the order of the LA phonon wavelength $\lambda$. Usually in real
structures this is not the case. Therefore, in terms of the phonon emission,
one can state that for $\bar{z}\gg\lambda$, the conversion field of the TA
phonons is concentrated inside a cone $\theta<\theta_c$. Recall that if the
reflection of phonons from the free surface is not taken into account then
according to relation (\ref{flux3d}), the emission of the phonon field is
concentrated in a narrow cone around the magnetic field direction. As 
can be seen from expression (\ref{l2dflux}) , the presence of the free surface
changes the LA phonon flux magnitude by a factor of 
$\left[1+{\cal R}^2_{LA\to LA}+ 2{\cal R}_{LA\to LA}\cos2\varphi\right]$. 
This factor varies between the values $0$ and $4$ (see insets in 
Figs.~ \ref{angdstr1}-\ref{angdstrb}). Therefore, there can be situations when 
the effect of the free surface on the magnitude of the acoustic energy flux can 
be very strong. Due to oscillations of this factor, new peaks can originate
in the dependence of ${\cal W}^L$ on $\theta$, {\it i.e.} the initial and
reflected LA phonons can cancel each other in the interference when their
phase difference is close to $\pi$. This can take place, for example, at small 
angles $\theta$ and at distances $\bar{z}\sim d$ (Fig.~\ref{angdstr2}) where 
$d$ is the electron layer thickness. One should observe, however, that the 
Landau level broadening, the account of which requires the following substitution
in (\ref{l2dflux}) 
\begin{equation}
\cos2\varphi \rightarrow\exp\left[-{\frac{\bar{z}}{\Lambda_2}}\right]
\cos2\varphi  \label{broad}
\end{equation}
where the characteristic damping length $\Lambda_2$ is given by Eq.~\r
{lambda2}, will somewhat spoil the monochromatic character of the emitted
phonons and, thus, smooth the oscillations (Fig.~\ref{angdstrb2}). Therefore it 
will be difficult to observe the interference peaks on the background of the main
peak for large values of $\bar{z}\gtrsim\Lambda_2$.
Note also that the factor $\exp(-\bar{z}/\Lambda_2)$ ensure that the Landau
level broadening diminishes the effect of the free surface when $\bar{z}%
\to\infty$. This effect, however, is not eliminated {\it completely}. When $%
\bar{z}\to\infty$ the interference between the LA phonons disappears, and so
does the TA phonons in the range $\theta_c<\theta< \pi/2$ while in the range 
$0<\theta<\theta_c$, the conversion effect remains active. If we take into
account that the main emission is concentrated near small angles $\theta$,
then ${\cal R}_{L\to L}\approx1$ and, therefore, ${\cal W}^L|_{\b{z}
\to\infty}$ is
nearly twice as large as ${\cal W}^{3D}$. This result is natural, since the
detector in this case will respond to the LA phonons emitted both in
positive and negative $z$-directions. However, the most intriguing for
experiment is that due to the conversion effect of the LA phonons into the
TA, it becomes possible under the deformational electron-phonon interaction
to detect the TA phonons on the reverse side of the sample. Moreover, the
intensity of these TA phonons is independent on $\bar{z}$ and by its order
of magnitude is comparable with ${\cal W}^{3D}$.

\subsection{Angular distribution of the emitted phonon field}

In this section the angular distribution of the emitted phonon field is
calculated under the condition that the electrons occupy only the ground
level of the size quantization. Therefore, to obtain the form-factor $I$ we
use the model function (\ref{modelwf}) in which $\bar{z}=0$ corresponds to
the free surface position. Usually for a GaAs/AlGaAs heterojunction the
characteristic distance $d= 3$ nm while typical values of $\bar{z}$ vary
between $50$ and $150$ nm. Calculations carried out for the electron
transitions between the Landau levels $l = 1$ and $l=0$ in the magnetic
field with $B=0.5$ T. It is assumed the Fermi level is to be midway between
these Landau levels and the electron temperature $ T_e = 7$ K. 

\begin{figure}[htb]
\epsfxsize=12cm
\epsfysize=13cm
\mbox{\hskip 1cm}\epsfbox{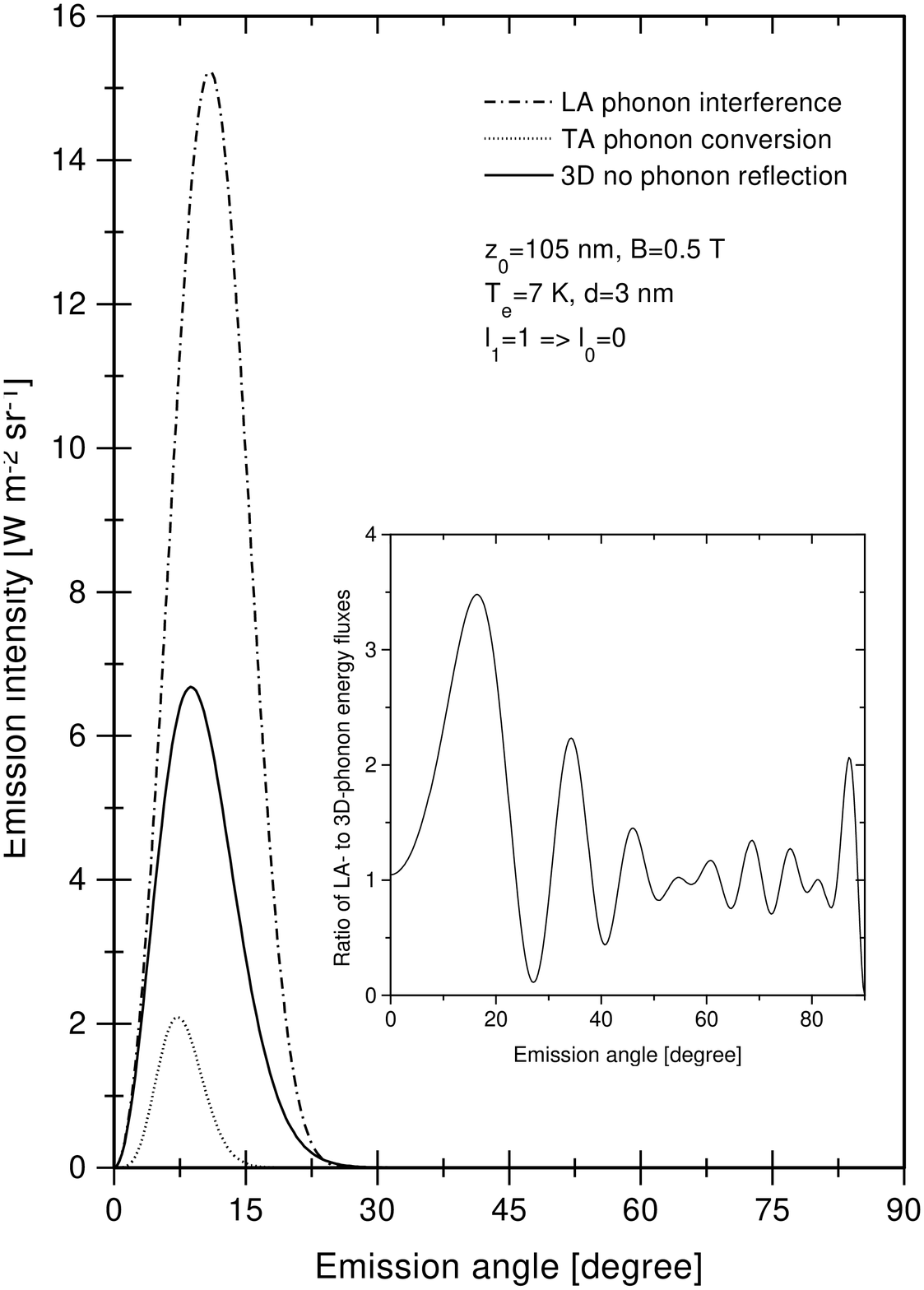}
\caption{Angular distribution of the phonon emission intensity at $\bar{z}%
=105$ nm when the Landau level broadening is not taken into account.
Inset shows dependence of the ratio ${\cal W}^L/ {\cal W}^{3D}$ on the
emission angle.}
\label{angdstr1}
\end{figure}

\begin{figure}[htb]
\epsfxsize=12cm
\epsfysize=13cm
\mbox{\hskip 1cm}\epsfbox{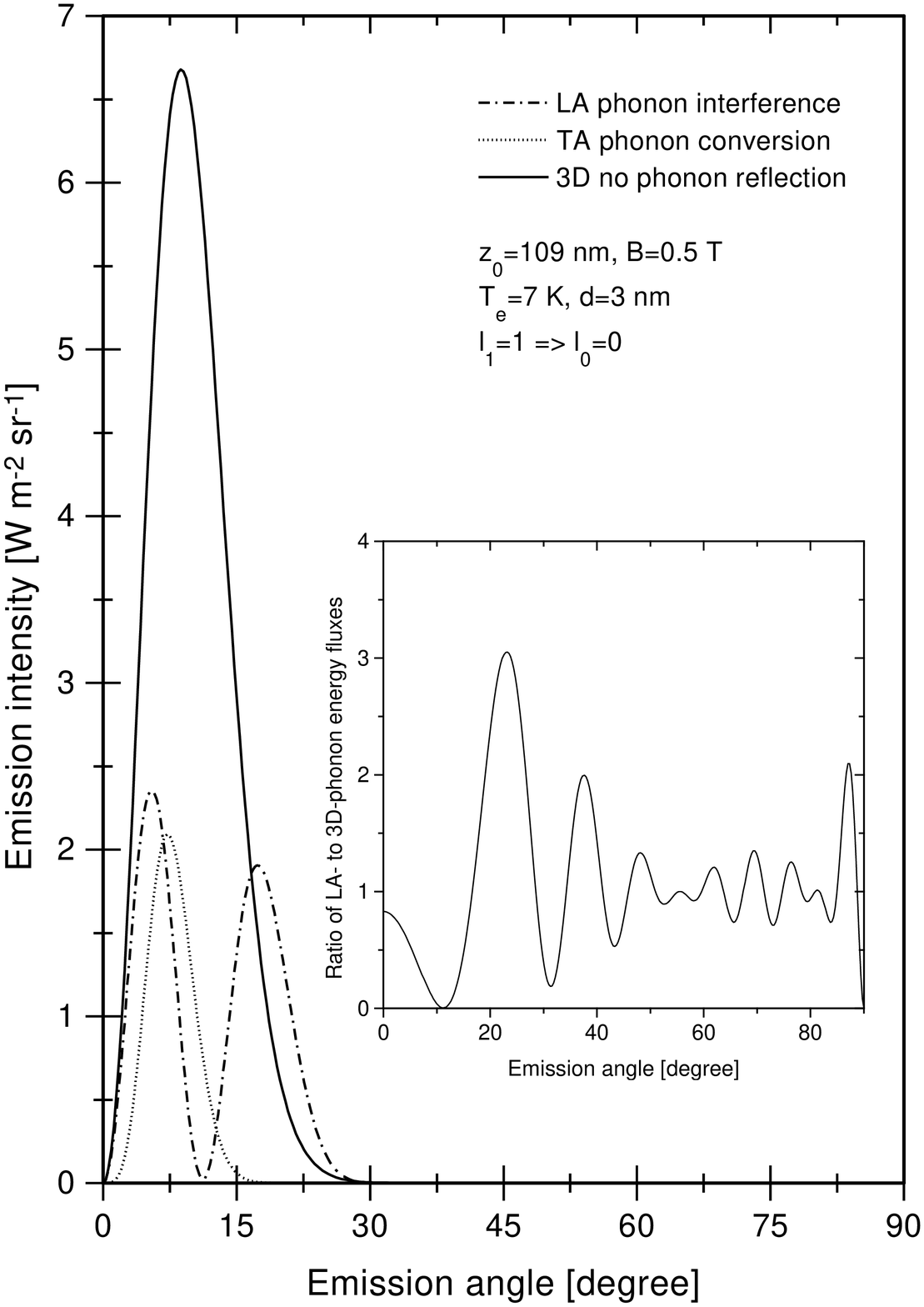}
\caption{ Angular distribution of the phonon emission intensity at $\bar{z}%
=109$ nm when the Landau level broadening is not taken into account.
Inset shows the interference oscillations.}
\label{angdstr2}
\end{figure}

\begin{figure}[htb]
\epsfxsize=12cm
\epsfysize=13cm
\mbox{\hskip 1cm}\epsfbox{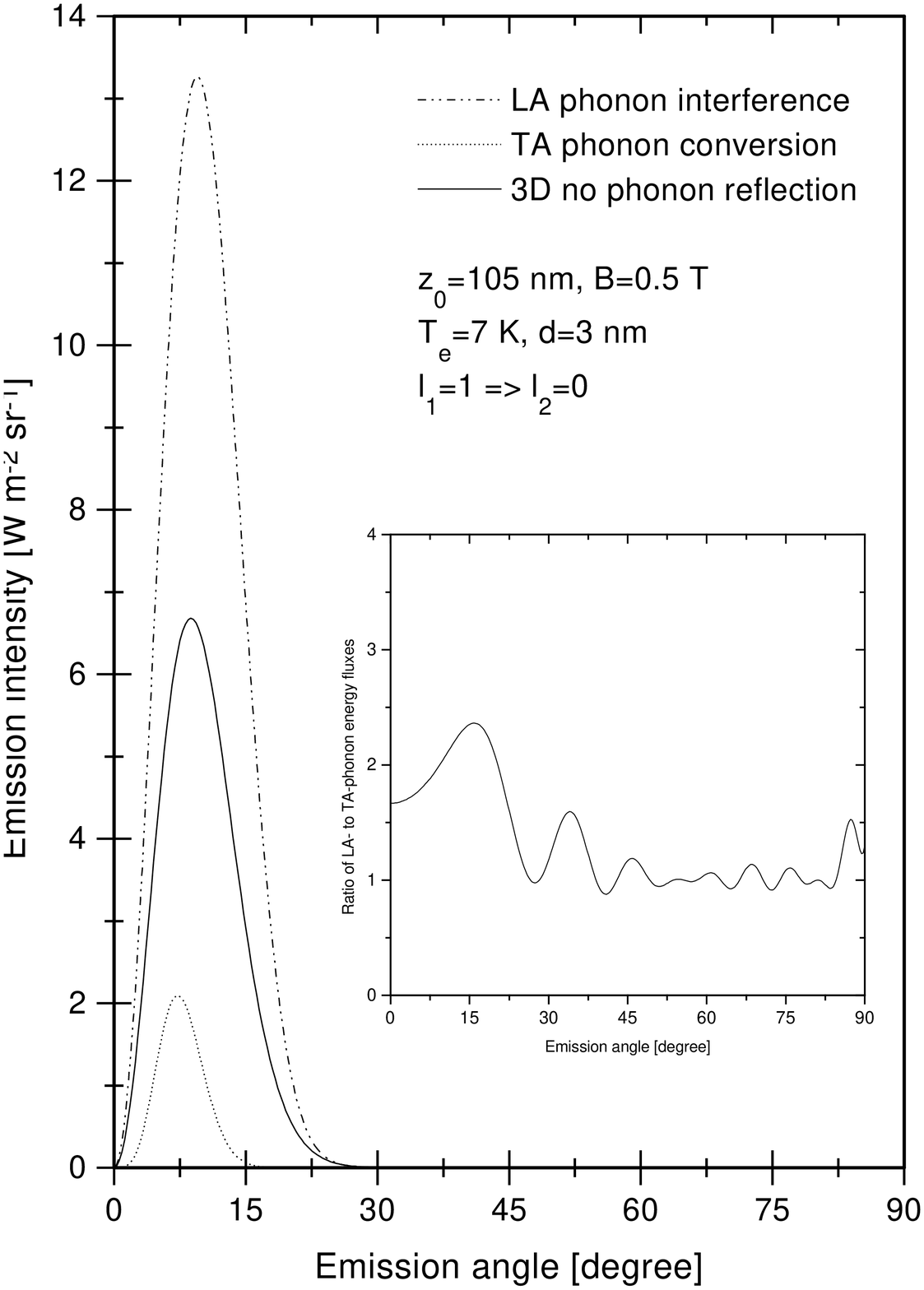}
\caption{ Angular distribution of the phonon emission intensity at $\bar{z}%
=105$ nm when the Landau level broadening is taken into account.}
\label{angdstrb1}
\end{figure}

\begin{figure}[htb]
\epsfxsize=12cm
\epsfysize=13cm
\mbox{\hskip 1cm}\epsfbox{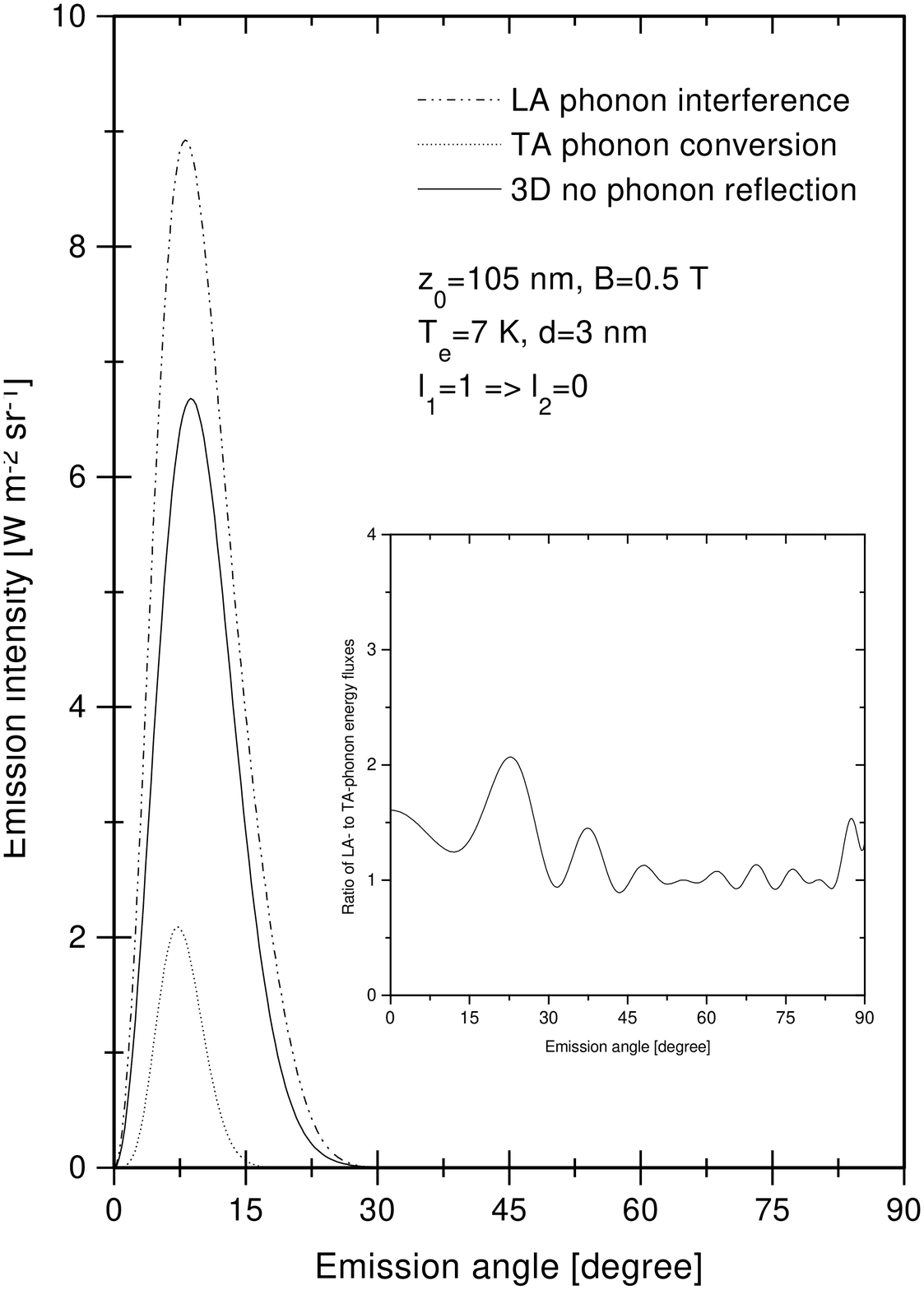}
\caption{Angular distribution of the phonon emission intensity at $\bar{z}%
=109$ nm when the Landau level broadening is taken into account.}
\label{angdstrb2}
\end{figure}

In Figs.~\ref{angdstr1}-\ref{angdstrb2} the angular dependence of phonon
emission is shown for two values of the distance $\bar{z}$. The thin
curve shows the emission intensity when the phonon reflection is not taken
into account. From Figs.~ \ref{angdstr1}-\ref{angdstrb2} one can see that
the phonon emission is practically absent for angles larger than $30^\circ$.
The TA phonon field is concentrated in a narrower cone around the magnetic
field than the LA phonon field. Note that the total inner reflection angle 
$\theta_c=36^\circ$. As one can see from the diagrams in Figs.~\ref{angdstr1}
and \ref{angdstr2}, the angular pattern of the LA phonon emission is very
much different for $\b{z}=105$ nm and $\b{z}=109$ nm when the Landau
level broadening is not taken into account. Namely, for $\b{z}=105$ nm, LA
phonon emission has a peak at $\theta\approx 11^\circ$ while for 
$\b{z}=109$ nm, emission is strongly suppressed at the same angle so that 
there appear two other peaks. This is due to the interference oscillations 
discussed above. These oscillations are more distinctly demonstrated on insets
of Figs.~\ref{angdstr1} and \ref{angdstr2} where the dependence of the ratio
 ${\cal W}^L/{\cal W}^{3D}$ on the angle $\theta$ is shown. The 
oscillations almost preserve their form with varying $\b{z}$, they are simply 
shifted along the $\theta$-axis, so that emission can be strongly suppressed 
by choosing a proper $\b{z}$. However, for large values 
$\b{z}\gtrsim\Lambda_2$ (even for high-quality heterostructures with 
mobility $\mu= 100$ m$^2$ V$^{-1}$ s$^{-1}$ we have 
$\Lambda_2\approx100$ nm), the Landau level broadening smoothes these
oscillations, as it is shown on inset of Figs.~\ref{angdstrb1} and \ref{angdstrb2}. 
So the emission character will not be impressively different for different values
of $\b{z}$ ({\it quod vide} Figs.~ \ref{angdstr1} and \ref{angdstr2}). Finally,
note that the intensity of TA phonon emission is nearly $3-6$ times smaller than
that of LA phonons but still remains a quite measurable quantity for experiment.

\section{Optical phonon emission: One phonon processes}
\label{popt}

The 2DEG embedded in an elastic medium of a single heterostructure and
subjected to the quantizing magnetic field in $z$-direction normal to the
electron layer is considered in this section. The scattering probability at 
which one phonon of a given mode $\Upsilon$ and a 3D-wavevector 
${\bf q}=({\bf q_{\perp}}, q_{z})$ is emitted by an electron of the 2DEG
is given by Fermi's golden rule as 
\begin{equation}
W^{\Upsilon{\bf q}}_{nlk_x \rightarrow n^{\prime}l^{\prime}k_x^{\prime}} ={%
\frac{2\pi}{\hbar}}|M^{\Upsilon{\bf q}}_{nlk_x \rightarrow
n^{\prime}l^{\prime}k_x^{\prime}}|^2 \delta\left(\varepsilon_{nl}-
\varepsilon_{n^{\prime}l^{\prime}}-\hbar \omega_{\Upsilon}(q)\right)
\end{equation}
The electron eigenstate $|nlk_x>$ is labeled by a subband index $n$
corresponding to the quantization of electron motion in $z$-direction, by a
Landau index $l$ and an electron momentum $x$-component $k_x$. In the case
of the bulk Landau states the eigenenergy $\varepsilon_{nl}$ does not depend
on the quantum number $k_x$ which counts the degeneracy of a Landau level.
The frequency of a phonon in a given mode is $\omega_{\Upsilon}({\bf q})$.
Because corresponding wave functions $|nlk_x>$ are factored into a subband
function $|n>$ and a Landau oscillator function $|lk_x>$ the matrix element
can be represented in the form 
\begin{equation}
M^\U_{nlk_x\rightarrow n^{\prime}l^{\prime}k_x^{\prime}} =B^\U(q)
Q_{ll^{\prime}}(q_{\perp})I^\U_{nn^{\prime}}(q_{\U})
\delta_{k_x^{\prime},k_x-q_{x}}  \label{matrixelement}
\end{equation}
where the form factor $Q_{ll^{\prime}}$ is given by Eq.~\r{ffmagnetic}.
In the case of $n=n^{\prime}=0$, using the model wave function (\ref{modelwf}%
), one can obtain the following explicit forms for the subband part of the
matrix element 
\begin{equation}
|I^{PO, DA}_{00}(q_z)|^2=(1+q^2_z d^2)^{-3},\: |I^{SO}_{00}
(q_\perp)|^2=(1+q_\perp d)^{-6}  \label{subbandff}
\end{equation}
where $d$ is the length scale of the 2DEG in $z$-direction. It is assumed
that $d$ is the smallest parameter of the problem so that the electron
relaxation between Landau levels of the lowest subband should be considered.

In Eq.~\r{matrixelement} the factors $B^\U (q)$ characterize the
electron-phonon interaction and are given by 
\begin{eqnarray}
&&B^{DA}(q)=iq^{1/2}{\frac{B_{DA}^{1/2}}{L^{3/2}}}, \: {\frac{1}{\bar{\tau}%
_{DA}}}={\frac{m_cB_{DA}p_{PO}^3}{\pi\hbar^2 s}} ={\frac{\Xi^2 p^3_{PO}}{%
2\pi\hbar\rho s^2}}={\frac{1}{4\;\mbox{ps}}},  \label{bda} \\
&&B^{PO}(q)={\frac{1}{iq}}{\frac{B_{PO}^{1/2}}{L^{3/2}}}, \: {\frac{1}{\bar{%
\tau}_{PO}}}={\frac{m_cB_{PO}}{\pi\hbar^3p_{PO}}}= 2\alpha_{PO}\omega_{PO}={%
\frac{1}{0.14\;\mbox{ps}}},  \label{bpo} \\
&&B^{SO}(q_{\perp})={\frac{1}{iq_{\perp}^{1/2}}}{\frac{B_{SO}^{1/2}}{L}}, \: 
{\frac{1}{\bar{\tau}_{SO}}}={\frac{m_cB_{SO}}{\pi\hbar^3p_{SO}}}={\frac{%
\alpha_{SO} \omega_{SO}}{\alpha_{PO}\omega_{PO}}}{\frac{1}{\bar{\tau}_{PO}}}
\nonumber \\
&&\quad\quad\quad =\cases{(0.50\;\mbox{ps})^{-1}\quad \mbox{for\quad
$x=0.3$},\cr (0.34\;\mbox{ps})^{-1}\quad \mbox{for\quad $x=1$}.}  \label{bso}
\end{eqnarray}
Here $L$ is the normalization length, $m_c$ is the electron effective mass, $%
s$ is the sound velocity and $\hbar p_{\Upsilon}=\sqrt{2m_c\hbar\omega_{%
\Upsilon}}$. The constants $B_{DA}$, $B_{PO}$ \cite{gantlev} and $B_{SO}$
are bound up, respectively, with the deformation potential constant $\Xi$,
the usual Fr\"olich constant $\alpha_{PO}$ and the electron-SO phonon
coupling $\alpha_{SO}$. The constant $\alpha_{SO}$ for interface modes in
a single heterostructure is defined according to Eq.~\r{scoupling}.

In Eqs.~\r{bpo}-\r{bso} instead of the constants $B_{\Upsilon}$, the
nominal times $\bar{\tau}_{\Upsilon}$ \cite{gantlev} are defined which give
a visual view of scattering rates. It can be seen that $\bar{\tau}_{SO}$ is $%
2.5\sim 3.5$ times larger than $\bar{\tau}_{PO}$, {\it i.e.}, generally
speaking, the interaction with interface modes should be weaker than with
bulk modes.

\subsection{Polar optical phonon scattering}
\label{pophon}

The scattering rate of an electron between two Landau levels $l$ and $%
l^{\prime}$ interacting with PO phonon can be represented in the form 
\begin{eqnarray}
{\frac{1}{\tau^{PO}_{l\rightarrow l^{\prime}}}}&=&\sum_{{\bf q},
k_x^{\prime}} W^{PO {\bf q}}_{nlk_x \rightarrow
n^{\prime}l^{\prime}k_x^{\prime}}={\frac{2\pi^2}{L^3}}{\frac{v_{PO} }{\bar{%
\tau}_{PO}}} \sum_{q}{\frac{1}{q^2}} Q^2_{ll^{\prime}}(q_{\perp})  \nonumber
\\
&\times& |I_{00}(q_z)|^2\delta(\Delta l\omega_B-\omega_{PO}(q))= {\frac{%
\sqrt{\omega_B \omega_{PO}}}{\bar{\tau}_{PO}}}D^{PO}_{ll^{\prime}}(\gamma)
\label{scrt}
\end{eqnarray}
where the overlap integral $D^{PO}_{ll^{\prime}}(\gamma)$ is given as 
\begin{eqnarray}
D^{PO}_{ll^{\prime}}(\gamma)&=&\int_0^\infty dxdt{\frac{Q^2_{ll^{\prime}}(t)%
}{(t+x^2)(1+\gamma^2 x^2)^3}}  \nonumber \\
&\times& \delta\left(\Delta l\omega_B-\omega_{PO}({\bf q})\right),\,%
\gamma^2= 2{\frac{d^2}{a_B^2}}.  \label{ovrlp}
\end{eqnarray}
The PO phonon emission is governed by the density of final states of a
two-particle system: an electron at the level $l^{\prime}$ and a PO phonon.
If the PO phonon dispersion be ignored then both particles have an infinite
mass so that this system does not have a continuous spectrum. Thus, to 
obtain the finite relaxation rate the Landau level broadening or the PO phonon
dispersion is to be taken into account.

\subsubsection{Allowance for the Landau level broadening}
\label{sec:level3}

For actual calculations of the scattering rate (\ref{scrt}) we have to smear
the $\delta$-function in Eq.~\r{ovrlp}. Replacing the $\delta$-function
by a Lorentzian 
\begin{equation}
\delta(\cdots)\rightarrow{\frac{\tau}{\pi[1+\tau^2(\omega_{PO}-\Delta l
\omega_{B})^2]}}  \label{lorenzian}
\end{equation}
characterized by the total width $\hbar/\tau$ where $\tau$ is the
relaxation time deduced from the mobility ({\it e.g.} $\mu=25$ V$^{-1}$ 
s$^{-1}$ m$^2$ corresponds to $\tau=10$ ps), it is
easy to obtain 
\begin{eqnarray}
D^{PO}_{ll^{\prime}}(\gamma)&=&{\frac{\tau}{1+\tau^2(\omega_{PO}- \Delta
l\omega_{B})^2}}d^{PO}_{ll^{\prime}}(\gamma), \\
d^{PO}_{ll^{\prime}}(\gamma)&=&{\frac{1}{16}}\int_0^\infty dt
Q^2_{ll^{\prime}}(t){\frac{8+9\gamma \sqrt{t}+3\gamma^2t}{\sqrt{t}(1+\gamma%
\sqrt{t})^3}}, \\
d^{PO}_{l0}(0)&=&\sqrt{\pi}{\frac{(2l-1)!!}{2^{l+1}l!}}.  \label{ovrlp2}
\end{eqnarray}
One can see that $x^2\simeq t\simeq 1$, {\it i.e.} $q_{z}\simeq
q_{\perp}\simeq a_{B}^{-1}$, have the main contribution to the integral (\ref
{ovrlp}). This isotropic distribution in momenta of emitted phonons is
conditioned by the long-range nature of the polar interaction. It forces the
scattering time to be governed by the minimum scale in the momentum space,
$min\{a_{B}^ {-1},d^{-1}\}$. Recall that in the case of DA phonon
scattering, phonon emission is anisotropic in the momentum space: 
$q_{z}\simeq d^{-1}\gg q_{\perp}\simeq a_{B}^{-1}$ 
\cite{toombs,badalfree,kitagawa}.

In magnetic fields below $10$ T using $d=3$ nm as a typical value for a
GaAs/AlGaAs heterostructures we have $\gamma^2\ll 1$. Therefore, in rough
estimates, the value of $d^{PO}_{ll^{\prime}}$ at $\gamma=0$ can be used. 
\begin{figure}[htb]
\epsfxsize=12cm
\epsfysize=12cm
\mbox{\hskip 1cm}\epsfbox{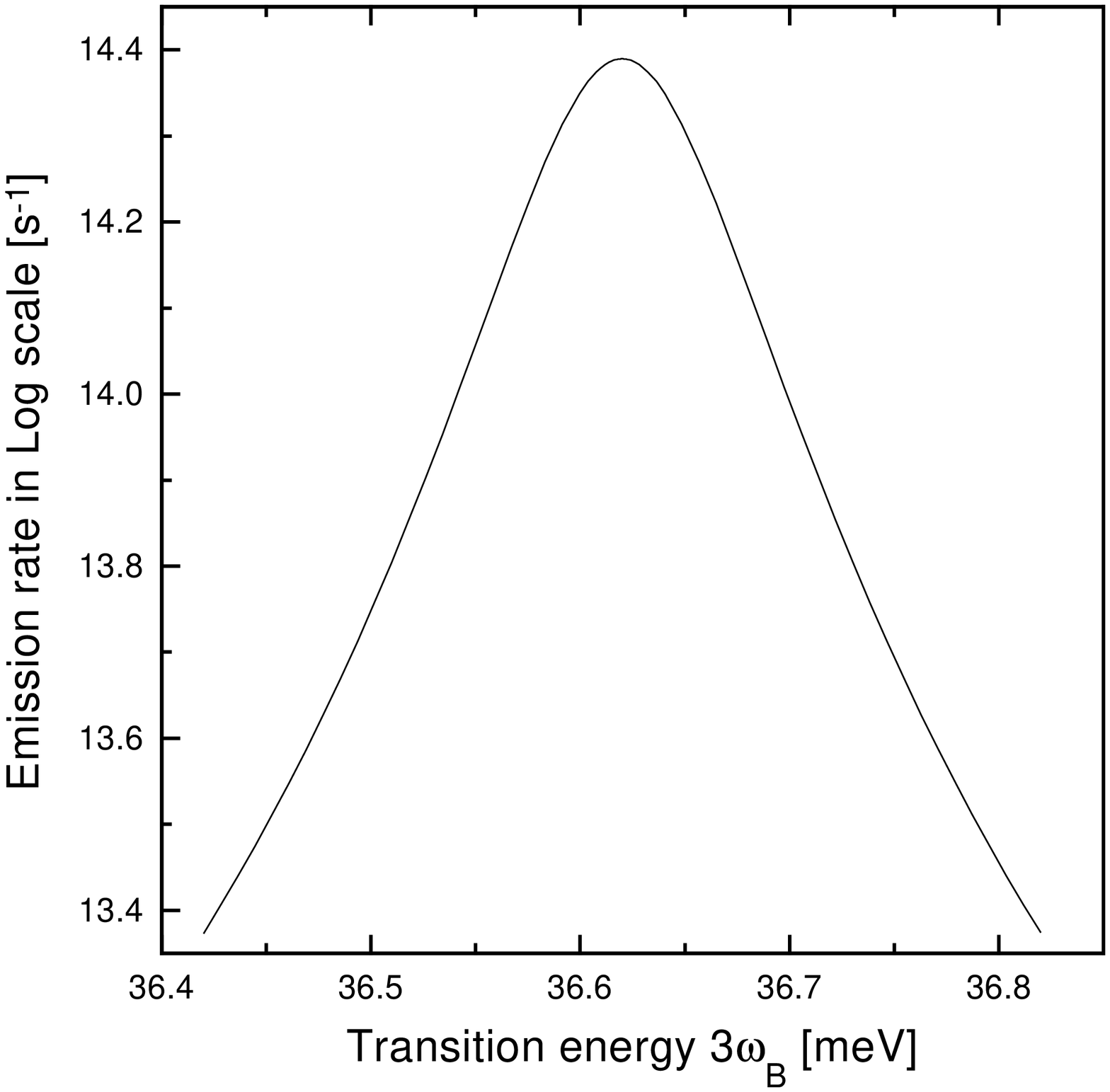}
\caption{The Landau level broadening contribution to the
PO phonon emission rate versus on the inter-level-spacing 
$\Delta l\omega_{B}$ for the electron transitions between $l=3$ and 
$l^{\prime}=0$ levels in
the vicinity of the PO phonon energy $\omega_{PO}$.}
\label{fg23}
\end{figure}

The result of numerical calculations according to Eqs.~\r{scrt} and \ref
{ovrlp2} at $\gamma\neq 0$ and $\tau=10$ ps is shown in Fig.~ \ref{fg23}. The
diagram of the scattering rate dependence on the inter-Landau-level
separation for $l=3$ and $l^{\prime}=0$ represents a narrow peak at the
PO phonon energy with peak value exceeding $10^2$ ps$^{-1}$. One may see
that for detuning $0.2$ meV the scattering time $\tau_{3\rightarrow 0}^{PO}$
increases by an order.

Note that in the case of the relaxation between Landau levels $l=1$ and $%
l^{\prime}=0$ the scattering rate dependence on the inter-level spacing is
also represented as a Lorentzian with approximately the same width and with
the peak value $1.17$ times larger than that of for the relaxation between
levels $l=3$ and $l^{\prime}=0$.

\subsubsection{Allowance for the PO phonon dispersion}

\begin{figure}[htb]
\epsfxsize=12cm
\epsfysize=12cm
\mbox{\hskip 1cm}\epsfbox{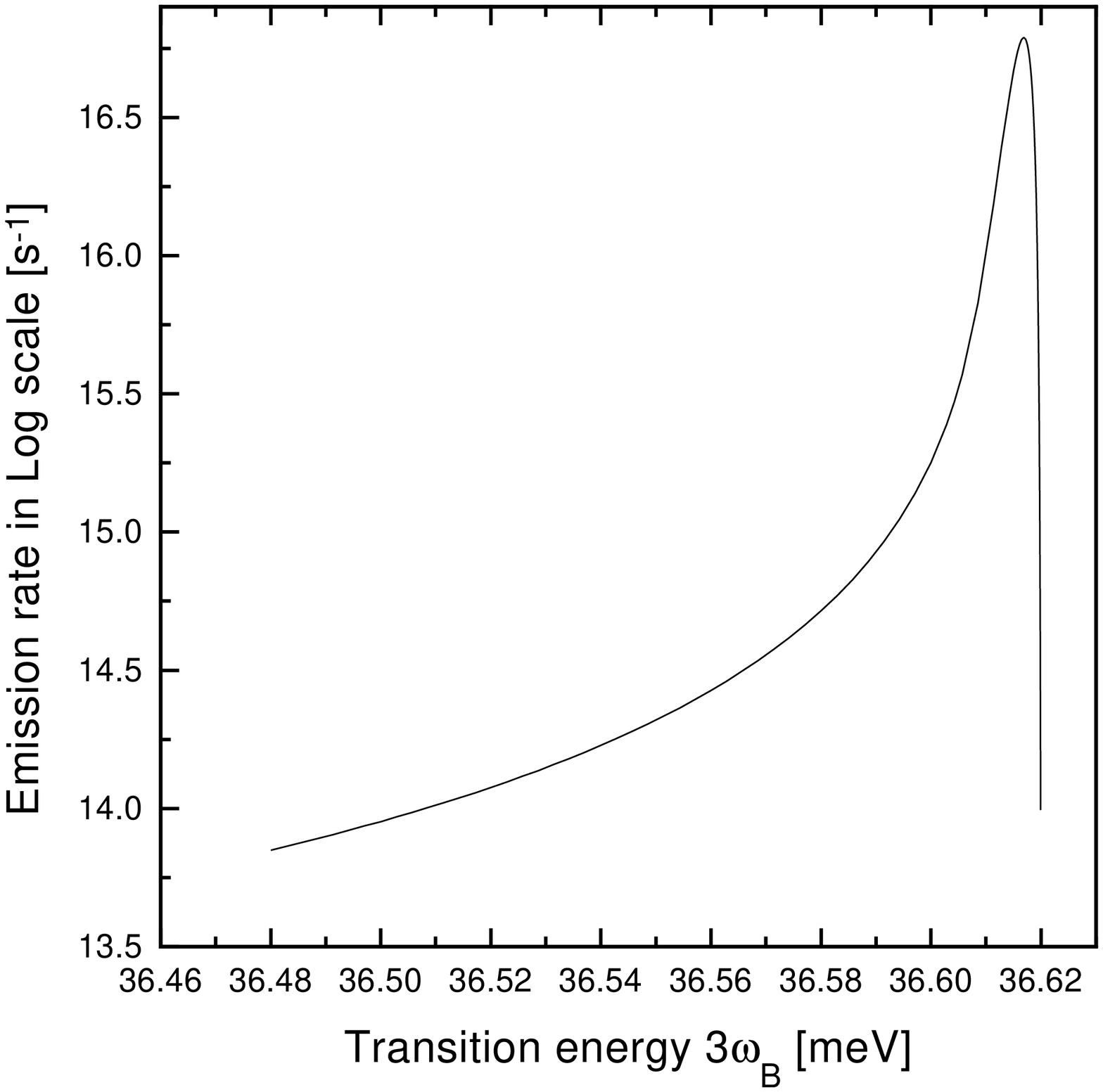}
\caption{The PO phonon dispersion contribution to the
PO phonon emission rate as a function on the inter-level-spacing $\Delta l
\omega_{B}$ for the electron transitions between $l=3$ and $l^{\prime}=0$
levels.}
\label{fg24}
\end{figure}

In this case taking $\omega_{PO}(q)=\o_{PO}(1-{q^2}/{2q_{0}^2})$ with 
$\o_{PO}=36.62$ meV and $q_{0}=18.5$ nm$^{-1}$ the overlap integral 
(\ref{ovrlp}) can be reduced to a one-dimensional integral of the form 
\begin{eqnarray}
D^{PO}_{ll^{\prime}}(\gamma)&=&\int_0^\infty dxdt{\frac{Q^2_{ll^{\prime}}(t)%
}{(t+x^2)(1+ \gamma^2 x^2)^3}}\delta\left(\Delta l\omega_B-\omega_{PO}+ {%
\frac{t+x^2}{2x^2_{0}}}\omega_{PO}\right) 
\label{phdsp1}
 \nonumber \\
&=&{\frac{a}{\omega_{PO}-\Delta l \omega_{B}}} \int^1_0 dx {\frac{%
Q^2_{ll^{\prime}}(y) }{(1+b^2 x^2)^3}},  
\: a^2=(1-{\frac{\Delta l \omega_{B}}{\omega_{PO}}})q_{0}^2a^2_{B},
\label{phdsp}
\nonumber \\
&&x_{0}^2={\frac{1}{2}}q^2_{0}a_{B}^2,\: y=a^2(1-x^2),\: b^2=a^2 \gamma^2
\label{phdsp2}
\end{eqnarray}
which is evaluated numerically for the electron transitions between $l=3$
and $l^{\prime}=0$ Landau levels (Fig.~ \ref{fg24}). Because of the energy
conservation, electron relaxation via this mechanism is possible only on
the low magnetic field side, $3\omega_{B}<\omega_{PO}$. The scattering rate
has a sharp and strongly asymmetric peak immediately below $\omega_{PO}$
with the peak value exceeding $10$ fs$^{-1}$. It can be seen from Esq.~ 
\ref{scrt} and \ref{phdsp} that the scattering rate decreases exponentially
for energies below the peak 
value while it drops more strongly for energies above the peak value (Fig.~ 
\ref{fg24}). As following from the comparison of diagrams of Figs.~ \ref
{fg23} and \ref{fg24}, the PO phonon dispersion contribution at the
relaxation peak by an order of magnitude is greater than the Landau level
broadening contribution. The former remains significant even for samples of
exceptional quality ($\mu>100$ V$^{-1}$ s$^{-1}$ m$^2$). On the
other hand the allowance for the Landau level broadening gives rise to a
symmetric peak at $\omega_{PO}$ thereby providing the one-phonon relaxation
on the upper side of the resonance.

Note that the electron-PO phonon relaxation rate calculated for a GaAs
quantum dot \cite{inoshita} as a function of a dot diameter or, that is the
same, of an inter-level-spacing has an approximately $3\sim 4$ times narrow
peak than the peak of the PO phonon dispersion contribution obtained here in
the QHE geometry. Both have approximately the same peak value.

\subsection{Surface optical phonon scattering}

\label{sophon} In the case of a single heterostructure, the surface optical
SO phonons do not exhibit dispersion. Therefore for SO phonons we consider
only the Landau level broadening contribution to the relaxation. To obtain
the scattering rate we have to replace all indexes $PO$ by $SO$ and to take
the overlap integral $D_{ll^{\prime}}^{SO}$ in the form 
\begin{eqnarray}
D_{ll^{\prime}}^{SO}(\gamma)&=&{\frac{\tau}{1+\tau^2(\omega_{SO}-\Delta l
\omega_{B})^2}} d^{SO}_{ll^{\prime}}(\gamma), 
\label{ovrlp30}\\
d^{SO}_{ll^{\prime}}(\gamma)&=&\int_0^\infty dt{\frac{Q^2_{ll^{\prime}}(t)}{%
\sqrt{t}(1+\gamma \sqrt{t})^6}}, 
\label{ovrlp3}\\
d^{SO}_{l0}(0)&=&\sqrt{\pi}{\frac{(2l-1)!!}{2^l l!}}.  
\label{ovrlp300}
\end{eqnarray}
Note that $t\simeq 1$, {\it i.e.} $q_{\perp}\simeq a_{B}^{-1}$ have
contributed heavily to both integrals in Eqs.~\r{ovrlp2} and \ref{ovrlp3}.
It is clear that the SO phonon relaxation peak is shifted on the low
energy side by $\o_{PO}-\o_{SO}\approx 2$ meV. Taking $d=3$ nm 
and $B=7$ T it is easy to obtain $d^{SO}_{30}=0.023$ and 
$d^{PO}_{30}=0.1$. Therefore at the relaxation peak, the PO phonon 
scattering rate is to the SO phonon scattering rate as 
\begin{eqnarray}
\max{\frac{\tau^{SO}_{3\rightarrow 0}}{\tau^{PO}_{3\rightarrow 0}}}= {\frac{%
\bar{\tau}_{SO}}{\bar{\tau}_{PO}}}\sqrt{\frac{\omega_{PO}}{\omega_{SO}}}{%
\frac{d^{PO}_{30}}{d^{SO}_{30}}}= \cases{15.9\quad \mbox{if\quad
$x=0.3$},\cr 10.8\quad \mbox{if\quad $x=1$}.}
\end{eqnarray} 
Thus SO phonon relaxation at least by an order is weaker than 
relaxation via PO phonon emission.

\section{Phonon emission: Two-phonon processes}
\label{two}

To calculate the electron transition probability due to the two-phonon 
emission mechanism, the quasiparticle approach has been exploited. 
Therefore, the two-phonon contribution to electron relaxation in the first 
order of the perturbation theory arising from the interaction Hamiltonian 
expanded up to second order in the phonon displacement operators can 
be neglected \cite{gantlev}. The second order contribution to the probability
of an electron transition from a bulk Landau state $|lk_x>$ into a state $%
|l^{\prime}k_x^{\prime}>$ of the same lowest subband (for brevity the
subband index will be omitted) at which one PO phonon with a 3D-wavevector $%
{\bf q}$ and one DA phonon with a 3D-wavevector ${\bf q^{\prime}}$ are
emitted is given by 
\begin{eqnarray}
W^{{\bf +q, +q^{\prime}}}_{lk_x \rightarrow l^{\prime}k_x^{\prime}}&=&{\frac{%
2\pi}{\hbar^4}}{\frac{B_{PO}B_{DA} }{L^6}}{\frac{q^{\prime}}{q^2}}%
\delta_{k_x^{\prime},k_x-q_{x}-q^{\prime}_{x}}\delta(\Delta l \omega_{B}-
\omega_{PO} -sq^{\prime})  \nonumber \\
&\times& \Biggl|\sum_{\bar{l},\bar{k_x}}\biggl\{{\frac{<l^{\prime}k_x^{%
\prime}|\exp(-i{\bf q^{\prime}r})|\bar{l}\bar{k_x}> <\bar{l}\bar{k_x}|\exp(-i%
{\bf qr})|lk_x>}{(l-\bar{l})\omega_{B}-\omega_{PO}}}  \nonumber \\
&+&{\frac{<l^{\prime}k_x^{\prime}|\exp(-i{\bf qr})|\bar{l}\bar{k_x}><\bar{l}%
\bar{k_x}|\exp(-i{\bf q^{\prime}r})|lk_x> }{(l-\bar{l})\omega_{B}-sq^{\prime}%
}}\biggl\}\Biggl|^2.  \label{poda}
\end{eqnarray}
The PO+DA phonon emission rate at electron transitions between Landau levels 
$l$ and $l^{\prime}$ can be obtained after summing up over the phonon and
the final electron momenta, ${\bf q},{\bf q^{\prime}}$ and $k_x^{\prime}$,
\begin{equation}
{\frac{1}{\tau^{PO+DA}_{l\rightarrow l^{\prime}}}}=\sum_{k_x^{\prime}, {\bf %
q, q^{\prime}}} W^{{\bf +q, +q^{\prime}}}_{lk_x \rightarrow
l^{\prime}k_x^{\prime}}.
\end{equation}
Note that after summation over $k_x^{\prime}$ the result cannot depend on
gauge non-invariant quantum number $k_x$. The explicit calculation of the
emission rate can be carried out by consideration separately the following
two situations: $(\Delta l\omega_{B} -\omega_{PO})\ll \omega_{B},\omega_{PO}$
and $(\Delta l\omega_{B}- \omega_{PO})\lesssim \omega_{B},\omega_{PO}$, 
where the relaxation is qualitatively different.

\subsection{Low magnetic fields}
\label{prgr1}

\paragraph{ $(\Delta l\omega_{B}-\omega_{PO})\ll\omega_{B},\omega_{PO}$}
In this energy range the main contribution to the sum over intermediate
states $\bar{l}$ in Eq.~\ref{poda} have the state $\bar{l}=l^{\prime}$ in
the first term and the state $\bar{l}=l$ in the second term. Therefore, the
emission rate can be rewritten as 
\begin{eqnarray}
{\frac{1}{\tau^{PO+DA}_{l\rightarrow l}}}&=&{\frac{2\pi}{\hbar^4}}{\frac{%
B_{PO} B_{DA}}{L^6c^2}}\sum_{{\bf q, q^{\prime}}}\delta(\Delta l\omega_{B}-
\omega_{PO}-sq^{\prime})  \nonumber \\
&\times& {\frac{Q^2_{ll^{\prime}}(q_{\perp})}{q^2q^{\prime}}}%
|Q_{l^{\prime}l^{\prime}}(q^{\prime}_{\perp})-
Q_{ll}(q^{\prime}_{\perp})|^2|I_{00}(q_z)|^2|I_{00}(q^{\prime}_z)|^2
\end{eqnarray}
which again can be reduced to a one-dimensional integral of the form 
\begin{eqnarray}
{\frac{1}{\tau^{PO+DA}_{l\rightarrow l}}}&=& {\frac{1}{\bar{\tau}_{PO+DA}}}{%
\frac{\omega_{B}}{s p_{PO}}} d^{PO}_{ll^{\prime}}(\gamma)\int_0^{%
\beta^2}dt^{\prime}\exp(-t^{\prime})  \nonumber \\
&\times& {\frac{[L_{l^{\prime}}(t^{\prime})-L_l(t^{\prime})]^2}{\sqrt{%
\beta^2-t^{\prime}}[1+\gamma^2 (\beta^2-t^{\prime})]^3}},\beta={\frac{\Delta
l\omega_{B}-\omega_{PO}}{\sqrt{2}s}}a_{B}
\end{eqnarray}
where a nominal relaxation time is introduced for the DA+PO phonon emission 
mechanism
\begin{equation}
{\frac{1}{\bar{\tau}_{PO+DA}}}={\frac{1}{4 s p_{PO}\bar{\tau}_{PO}
\bar{\tau}_{DA}}} 
\end{equation}
which depends only on a heterojunction parameters. For a GaAs/AlGaAs
heterojunction with $d=3$ nm we have $\bar{\tau}_{PO+DA}\approx
2.9$ ps.

Actually in this energy range of $\Delta l\omega_{B}$, the electron
transitions take place in following two ways. (i) Remaining on the level $l$,
an electron emits a DA phonon thereby the electron-phonon system is forced
to transit into a virtual intermediate state. Then the electron emits a
second PO phonon so that the electron-phonon system turns out in the final state
with the real electron on the Landau level $l^{\prime}$ and with two real
DA and PO phonons. (ii) In the second way, the electron emits firstly the
PO phonon and simultaneously makes a transition to the level $l^{\prime}$.
By emission of the second DA phonon, the created virtual intermediate state is
forced to transit into the same final state. In both cases 3D-wavevectors of
emitted PO- and DA phonons are not correlated. The 3D-wavevectors of emitted
PO phonons have the same isotropic distribution in the momentum space as in
the one-phonon emission case. While the momentum distribution of the emitted
DA phonons is different in different ranges of $\Delta l\omega_{B}$.
Immediately above the PO phonon energy for $\beta\ll1$
(this corresponds to the energies $(\Delta l\omega_{B}-\omega_{PO})\ll
s/a_{B}$), electrons emit DA phonons with $q^{\prime}_{\perp}\simeq
q^{\prime}_{z}\simeq a^{-1}_{B}$. In this case the following asymptotic
expression is obtained for the emission rate 
\begin{equation}
{\frac{1}{\tau^{PO+DA}_{l\rightarrow l^{\prime}}}}={\frac{16(l-l^{\prime})^2%
}{15}}{\frac{\beta^5}{\bar{\tau}_{PO+DA}}}{\frac{\omega_{B}}{sp_{PO}}}%
d^{PO}_{ll^{\prime}}(\gamma).
\end{equation}
The essential part of the magnetic field dependence in this range is given
by 
\begin{equation}
{\frac{1}{\tau^{PO+DA}_{l\rightarrow l^{\prime}}}}\propto (B-B_{\Delta l})^5,
\end{equation}
{\it i.e.}, relaxation is enhanced as a fifth power of $B-B_{\Delta l}$ with 
increase in the magnetic field. Here $B_{\Delta l}$ is the magnetic field
for which $\Delta l\omega_{B}=\omega_{PO}$.

In opposite limiting case of $\beta\gg 1$, actually for energies $s/d
\lesssim (\Delta l\omega_{B}-\omega_{PO})\ll\omega_{B},\omega_{PO}$, 
electrons emit DA phonons with $q^{\prime}_{\perp}\simeq a^{-1}_{B}\ll 
q^{\prime}_{z}\simeq d^{-1}$, {\it i.e.}, DA phonon emission is heavily
concentrated in a narrow cone around the magnetic field. In this case for 
the emission rate we obtain 
\begin{equation}
{\frac{1}{\tau^{PO+DA}_{l\rightarrow l^{\prime}}}}={\frac{1}{\bar{\tau}%
_{PO+DA}}} {\frac{2}{\beta(1+\gamma^2\beta^2)^3}}{\frac{\omega_{B}}{sp_{PO}}}%
d^{PO}_{ll^{\prime}} (\gamma).
\end{equation}
The essential part of the magnetic field dependence is given by 
\begin{equation}
{\frac{1}{\tau^{PO+DA}_{l\rightarrow l^{\prime}}}}\propto (B-B_{\Delta
l})^{-1},
\end{equation}
{\it i.e.}, in this range relaxation becomes linearly weaker with
increase in the magnetic field.

\begin{figure}[htb]
\epsfxsize=12cm
\epsfysize=12cm
\mbox{\hskip 1cm}\epsfbox{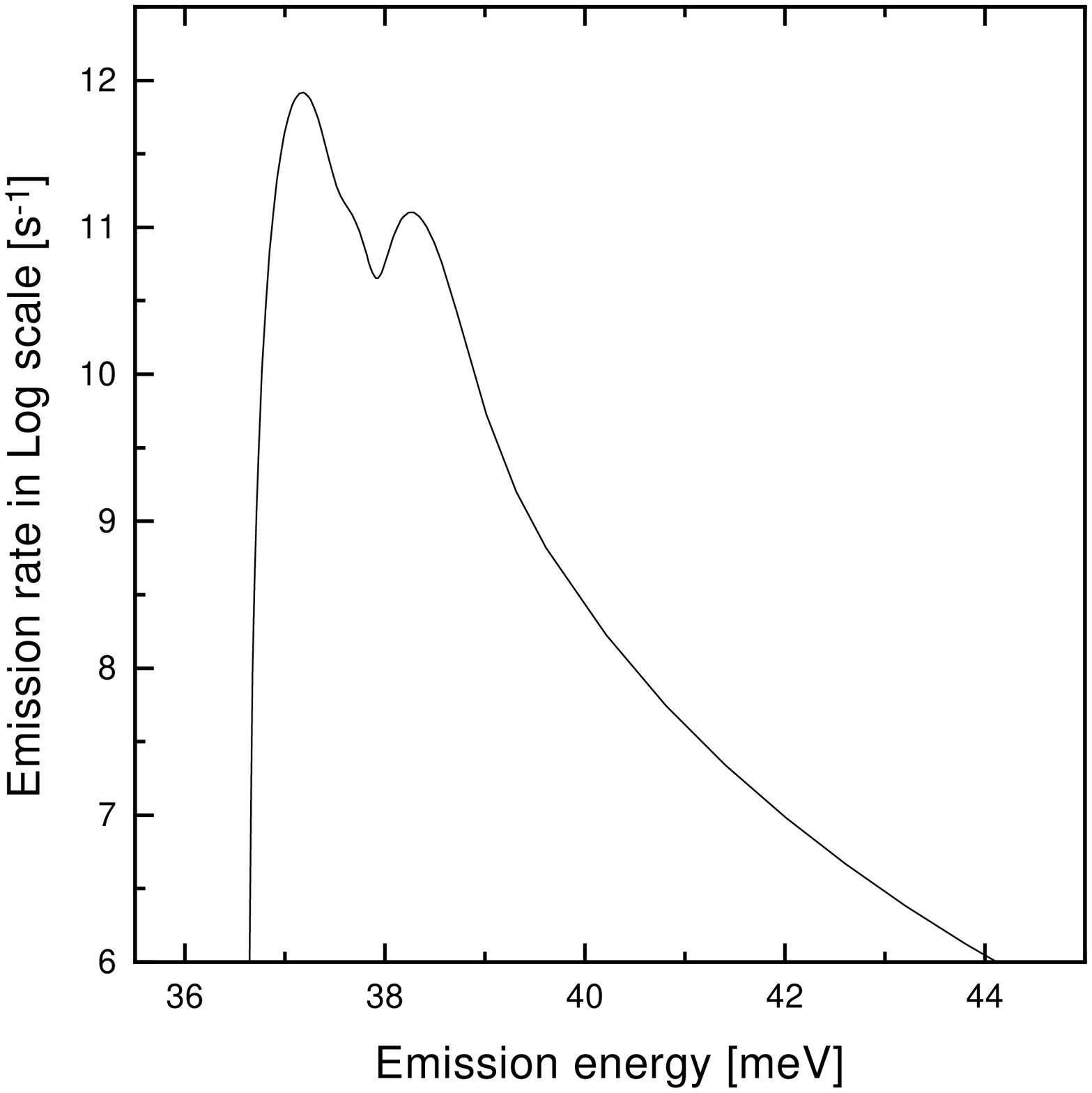}
\caption{The PO+DA phonon emission rate versus on the inter-Landau-level
separation $\Delta l\omega_{B}$ for the electron transitions between $l=3$
and $l^{\prime}=0$ levels in the energy range $(3\omega_{B}-
\omega_{PO})\ll\omega_{B},\omega_{PO}$.}
\label{fg25}
\end{figure}

Results of numerical evaluation of the emission rate as a function of the
inter-Landau-level separation $\Delta l\omega_{B}$ in the whole energy range
(\ref{prgr1}) are illustrated for transitions between Landau levels $l=3$
and $l^{\prime}=0$ in Fig.~\ref{fg25}. So far as the two-phonon processes
contain the small electron-phonon coupling in the second order,
PO+DA emission gives rise to a peak which is lower than that of for 
one-phonon emission. PO+DA  phonon relaxation has a sharp onset at low 
magnetic
fields corresponding to energies immediately above $\hbar\omega_{PO}$ where
the peak increases as a fifth power in the magnetic field achieving the peak
value exceeding $1$ ps$^{-1}$. At high magnetic fields the peak
decreases much slowly, linearly in $B-B_{\Delta l}$, so that the
PO+DA phonon emission mechanism gives rise to a rather broad peak than the
PO phonon emission peak.

\subsection{High magnetic fields}
\paragraph{$(\Delta l\omega_{B}-\omega_{PO})\simeq\omega_{B}, \omega_{PO}$.}
\label{prgr2}

As it follows from the energy conservation, the energy of an emitted
DA phonon is $s q^{\prime}\simeq \omega_{B}$ so that $q^{\prime}\simeq
\omega_{B}/s \gg a^{-1}_{B}$. On the other hand, electrons in the states with
Landau indexes $l\simeq 1$ have momenta of the order of $a_{B}^{-1}$.
Therefore for more important electron transitions with $l,l^{\prime}\simeq 1$
the momentum transmission to the phonon system is also of the same order of 
$a_{B}^{-1}$. Hence in this energy range, electrons should emit phonons 
with almost
oppositely directed momenta of approximately equal absolute values to avoid
an additional suppression of the two-phonon emission. The large momentum
transferred to each phonon results in the large Landau index $\bar{l}$ for
intermediate states in Eq.~\r{poda}. As far as the quasiclassic
description takes place for $\bar{l}\gg 1$, the intermediate state energies
are 
\begin{equation}
\bar{l}\omega_{B}\approx{\frac{(\Delta\bar{l}\omega_{B}-\omega_{PO})^2 }{%
2ms^2}}\gg l\omega_{B},\omega_{PO},sq^{\prime}.
\end{equation}
Therefore for the emission rate in the range (\ref{prgr2}) we obtain 
\begin{equation}
{\frac{1}{\tau_{l\rightarrow l^{\prime}}^{PO+DA}}}={\frac{3}{2\b{\tau}%
_{PO+DA}}} {\frac{(2ms^2/\hbar)^2\omega_{B}}{(\Delta
l\omega_{B}-\omega_{PO})^3}} {\frac{1}{p_{PO}d}}.  
\label{sctmb}
\end{equation}
The magnetic field dependence is given by 
\begin{equation}
{\frac{1}{\tau_{3\rightarrow 0}^{PO+DA}}}\propto (B-B_{\Delta l})^{-3}.
\end{equation}
Because in the range (\ref{prgr2}), $B$ is of the order of $B_{\Delta l}$ one
can replace $B-B_{\Delta l}$ by $B_{\Delta l}$ so that the emission rate
dependence on the magnetic field is weak in a rather wide energy range above 
$\omega_{PO}$.

It is interesting also to compare the PO+DA phonon emission rate for
transitions between Landau levels $l=3$ and $l^{\prime}=0$ with the
2LA phonon emission rate at transitions between levels $l=1$ and $%
l^{\prime}=0$. Using the result obtained in \cite{falko} it is easy to
obtain 
\begin{equation}
{\frac{1}{\tau_{3\rightarrow 0}^{PO+DA}}}=8{\frac{\tau_{DA}}{\tau_{PO}}} {%
\frac{\omega_{PO}^2 (2mc^2/\hbar)}{(3\omega_{B}-\omega_{PO})^3}} {\frac{1}{%
\tau^{2DA}_{1\rightarrow 0}}}.
\end{equation}
Taking $B=8.4$ T and $d=3$ nm for a GaAs/AlGaAs heterojunction we have 
$2ms^2\approx0.02$ meV and $\omega_{B}=2\omega_{PO}/5 \approx14.65$ 
meV. So as it follows from Eq.~\r{sctmb} at transitions between Landau levels 
$l=3$ and $l^{\prime}=0$ one can use $\tau_{3\rightarrow 0}^{PO+DA}=100$ 
ns as a characteristic relaxation time in the range (\ref{prgr2}). Under the same
conditions at transitions between levels $l=1$ and $l^{\prime}=0$ for the
LA and 2LA phonon emission times we have, respectively, $\tau_{1\rightarrow
0}^{DA}=5.4$ $\mu$s \cite{badalfree} and $\tau_{1 \rightarrow
0}^{2DA}=15.6\tau_{3\rightarrow 0}^{PO+DA}$. Thus comparison of these times
shows the importance of the PO+DA emission processes in the QHE geometry. In
some experimental arrangements the PO+DA phonon emission mechanism is much
more efficient than relaxation in the following two consecutive emission acts: 
PO phonon emission (even under the sharp resonance) + either LA or
2LA phonon emission.

\section{Summary}
\label{four}

In conclusion, PO phonon assisted electron relaxation is calculated as a
function of the inter-Landau-level spacing in the 2DEG in the QHE geometry.
The PO, SO and PO+DA phonon emission processes via polar optical and
deformation acoustical interactions are considered. The interface SO phonon
relaxation is at least by an order weaker than the relaxation via PO phonon
emission. To obtain a finite relaxation rate associated with one-phonon
emission, the allowance for the Landau level broadening and for the PO phonon
dispersion is made. Immediately below the phonon energy $\hbar\omega_{PO}$,
the PO phonon dispersion contribution gives rise to a sharp peak with the
peak value approximately $0.17$ fs$^{-1}$. The Landau level broadening
contribution has a rather broad peak with the relatively lower peak value. 
Below $\o_{PO}$ within an energy range of the order of 
$\hbar\sqrt{\o_{B}/\tau}$, the one-phonon relaxation rate exceeds $1$ ps$^{-1}$
($\tau$ is the relaxation time deduced from the mobility). In GaAs/AlGaAs
heterostructures with the mobility $\mu=25$ V$^{-1}$ s$^{-1}$ m$^2$ this 
range makes up $0.7$ meV.

Two-phonon emission is a controlling relaxation mechanism above $\h\o_{PO}$. 
For $\Delta l\o_{B}$ immediately above $\o_{PO}$, PO+DA phonon relaxation
has a sharp onset. The relaxation rate increases as a fifth power in the magnetic 
field achieving to the peak value exceeding $1$ ps$^{-1}$ at energy separations 
of the order of $s/a_{B}$ (in GaAs at $B=7$ T we have $\h(s/a_{B})\approx 0.4$
meV). At higher magnetic fields in the energy range $sa^{-1}_{B}\lesssim\Delta 
l\omega_{B}- \omega_{PO}\lesssim sd^{-1}$ (in GaAs with $d=3$ nm we have 
$\hbar(s/d) \approx 1.2$ meV), the two-phonon peak decreases linearly in the 
magnetic field. Above $\omega_{PO}$ within the wide energy range (in GaAs this 
range makes up to $5$ meV), the magnetic field dependence of the relaxation rate 
is rather weak and the subnanosecond relaxation between Landau levels $l=3$ and 
$l^{\prime}=0$ can be achieved via two-phonon emission mechanism.

Our analysis demonstrates also that in some experimental situations, the
PO+DA phonon emission mechanism is more efficient than relaxation in two
consecutive emission acts: PO phonon emission (even under the sharp
resonance) with the subsequent emission of either LA- or 2LA phonon.

\chapter{Edge state scattering}
\label{sec:intro}

\section{Introduction}

\begin{figure}[htb]
\epsfxsize=7cm
\epsfysize=8cm
\mbox{\hskip 0cm}\epsfbox{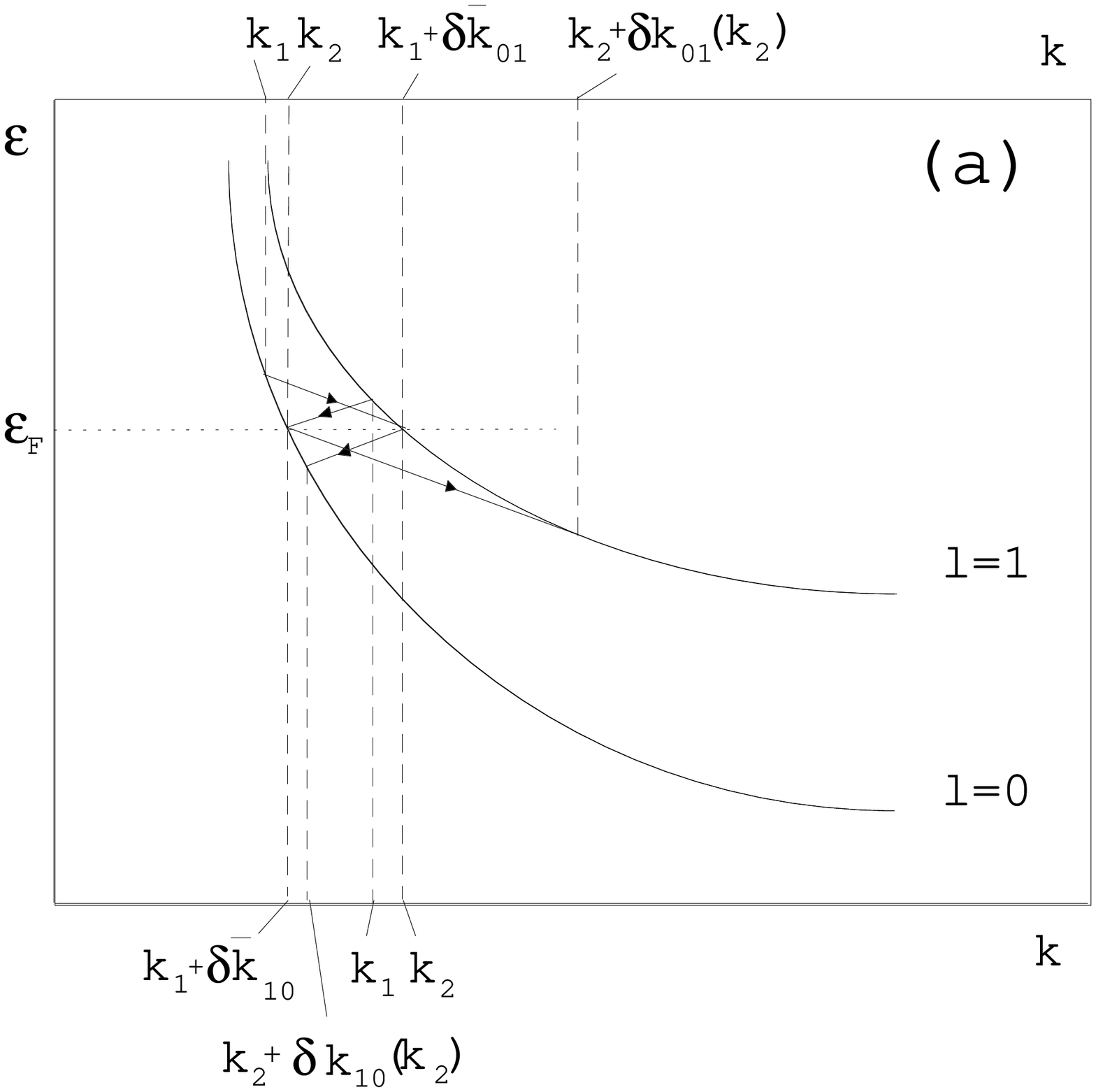}
\epsfxsize=7cm
\epsfysize=8cm
\mbox{\hskip 0cm}\epsfbox{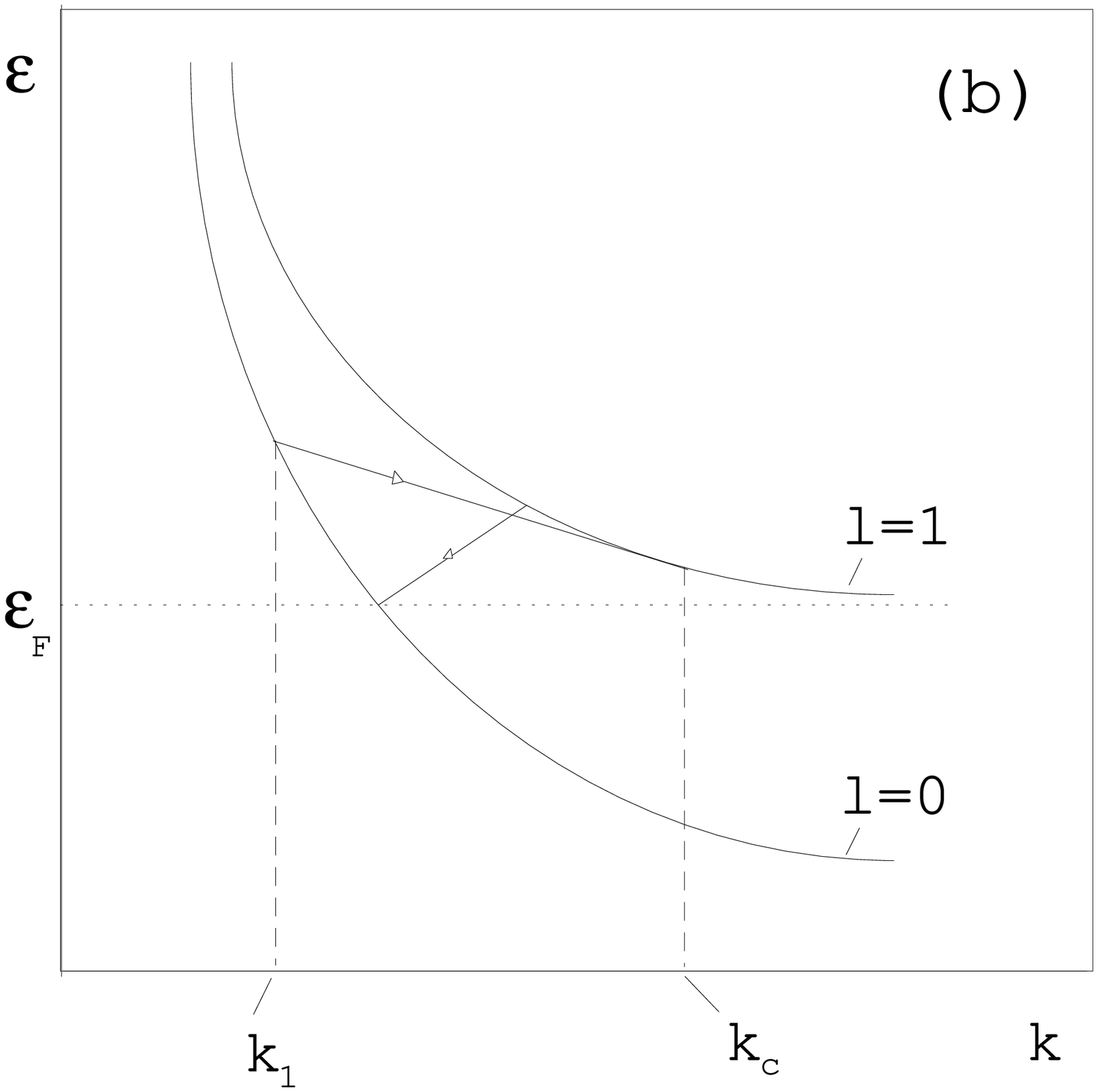}
\caption{A schematic diagram of the edge state energy spectrum. (a)
Transitions $0\rightarrow1$ and $1\rightarrow0$ due to acoustic phonon 
emission are shown. At low
temperatures, states close to the $0k_1$ are important. (b) If the Fermi
level is close to the bulk Landau level, the transitions $0\rightarrow1$ 
involving states above the threshold $k_c$ are only possible. At high
temperatures, states disposed at the separations of $ T_e$ above and 
below $\varepsilon_F$ are important.}
\label{spect}
\end{figure}

The single particle energy spectrum in a 2DEG exposed to a homogeneous
magnetic field normal to the electron plane is separated into the quasibulk
Landau and edge states. The edge states correspond to the classical skipping
orbits. Their location with respect to the boundary of the sample depends on
the wave vector $k$. When the $k\rightarrow\infty$, the edge state number $l$
transforms into a quasibulk Landau state (see Fig.~\ref{spect}). 
The confining potential in the plane of electrons removes the degeneracy of
the Landau levels so that, in contrast to the bulk Landau states, the edge
state energy exhibits a finite dispersion $\varepsilon_{l}(k)$ with $%
\varepsilon_{l}(k)\rightarrow\varepsilon_{l}=\hbar\omega_{B}(l+ 1/2)$ if 
$k\rightarrow\infty$ ($\omega_B$ is the cyclotron frequency). There 
exist edge states also in quantum wires (QWs) in the magnetic field
normal to the wires. If wire width $L$ is much greater than the magnetic
length, $L\gg a_B$, ($a_B=\sqrt{{\hbar}/{m_{c}\omega_B}}$ is the magnetic
length, $m_{c}$ is the electron effective mass), then the backscattering
between edge states moving in opposite directions on the opposite sides of
the wire is strongly suppressed. So in sufficiently strong magnetic fields,
the edge states in the 2DEG and QW can be treated in the same way as 
one dimensional system structures. Currently semiconductor nanostructures
with such one dimensional electron systems attract considerable interest 
both to study in them novel physical phenomena as well as for possible 
device applications such as QW lasers (see review \cite{haug}).

To explain the quantum Hall effect (QHE) in high mobility samples with a
2DEG of submicron dimensions, since shortly after its discovery \cite
{klitzing}, a new concept based on the B\"uttiker-Landauer quantum edge
state transport formalism has been advanced \cite{but,laughlin,halperin,streda}. 
In this description the conducting properties of the 2DEG are determined by 
the edge states at the Fermi level, $\e_{F}$, which are propagated along the 
sample boundaries. Within
the sample, electrons occupy $N$ quasibulk Landau levels below $%
\varepsilon_{F}$, but near the boundary of the sample, the confining
potential bent up these energy levels so that they rapidly increase and
intersect $\varepsilon_{F}$. Thus each quasibulk Landau level below $%
\varepsilon_{F}$ gives rise to a quasi-one dimensional transport edge
channel at $\varepsilon_{F}$ on either side of the sample. If the 2DEG is
coupled to two large electron reservoirs with applied potential $\Delta \mu$
and with ideal contacts, {\it i.e.}, they transmit every electron without
reflection, then the edge states are equally populated and each edge channel
carries an equal fraction of the current on a given side of the sample. On
opposite sides of the sample, the current flows in opposite directions
resulting in a total current $I=N(2e/h)\Delta \mu $ and a Hall voltage $%
V_{H}= \Delta \mu /e$ so that we have the QHE with the Hall conductance $%
G_{H}=N(2e^{2}/h)$. Under these conditions $G_{H}$ does not depend on the
properties of the contacts and is determined by the number $N$ of the
quasibulk Landau levels. Kent {\it et al.} in their recent experiment \cite
{kent92} have obtained qualitative agreement with such theoretical
description of the QHE.

On the other hand, electrons can be selectively injected into different edge
channels and selectively detected using non-ideal probes. As a result we can
have a non-equal population of the edge states and significant deviation
from the normal integer QHE. In this case, as it has been theoretically
shown by B\"uttiker \cite{but}, $G_{H}$ depends on the properties of the
current and voltage contacts and is not related to the number of bulk Landau
states below $\varepsilon_F$. While flowing along an edge of samples, the
current tends to redistribute so that an equal fraction of the current is to
be carried by each edge channel. Such a re-establishment of the current
equilibrium can be achieved via electron scattering processes by phonons and
impurities. If spatial separation between edge channels is not much greater
than the magnetic length then the equilibrium is expected to occur after the
current travels a distance of the order of the zero-field inelastic and
elastic scattering length, $10$ $\mu$m. However, recently van Wees {\it et al.}
\cite{wees1} and Komiyama {\it et al.} \cite{kom1}, involving high mobility
2DEG systems with non-ideal current probes separated more than $200$ $\mu$m
from one another, have observed that a long distance in excess of $100$ $ \mu$m
is necessary to equilibrate the populations between different edge channels.
It has been shown in these works that under the conditions of strong
magnetic fields and low temperatures, the quantization of the Hall
conductance is determined {\it only} by the number of the quantum edge
channels which are coupled to the non-ideal probes and is independent of the
total number of the occupied quasibulk edge states. This effect, usually
referred to as the anomalous integer QHE, has also been revealed in a number
of recent experiments \cite{klitz1,klitz2,kom1,kom2,kom3,wees1,wees2}. Data
of these experiments, being analyzed with the help of Landauer-B\"uttiker
formalism, provide some information about the inter-edge-relaxation that
leads to the re-establishment of the equilibrium. In the course of such
analyses two important features were revealed which indicate the importance
of the analogous theoretical investigations. Firstly, the characteristic
scattering length ${\cal L}_{l\rightarrow l^{\prime}}$ between states $l$
and $l^{\prime}$ exceeds significantly the transport scattering length
obtained from zero-field mobility. It points out the significant suppression
of the scattering compared to the case of zero magnetic field. Secondly, a
strong temperature dependence of ${\cal L}_{l\rightarrow l{\prime}}$ was
revealed \cite{klitz2,kom3}.

In their experiment Alphenaar {\it et al.} \cite{alph} have further clarified 
the long range of non-equilibrium distribution of electrons. The carrier
source contact has been adjusted in such a way that the current has been
injected into the sample from the non-ideal current probe reservoir through
only the lowest of edge channels. The authors have studied in detail the
dependence of the inter-edge-state equilibration length on different pairs
of edge states, on the magnetic field and filling factors. The result
indicates that the $N-1$ outermost edge states equilibrate but these states
are decoupled from the innermost edge state. A clear understanding of the
reason for this effect has not been achieved yet. In principle, it is impossible
to do this in the framework of the theories with a parabolic confining
potential. Such potential has no flat domain corresponding to the interior
of the sample. Therefore, there are no quasibulk edge states and one cannot
reveal properties of the scattering which appear when the Fermi level is
close to a bulk Landau level. Thus, it is important to carry out
calculations for the scattering length for transitions between outermost and
the innermost quasibulk edge states for an arbitrary shape of the confining
potential. The efficiency of such an interaction is chiefly determined by
the form of the edge state spectrum. As it follows from the energy and momentum
conservation laws, there are two threshold-like points on the innermost
quasibulk Landau level so that transitions from outermost to innermost edge
state after emission of acoustic phonons are restricted by the final states
between of these two points. As we can see below, these states, the velocity 
of which is approximately equal to the sound velocity, play an important role in the
scattering processes.

Theoretically, inter-edge-state relaxation has been addressed in several works 
\cite{kom3,glazm89,lev90,martin,khaet}. Scattering by irregularities of the
boundary is discussed in Refs.~\cite{glazm89,lev90} and scattering by
delta-function impurities and phonons in \cite{kom3,martin}. A two-phonon
scattering mechanism has been suggested recently \cite{khaet}. Although
significant progress has been achieved the problem still cannot be considered as
solved. Results of Refs.~\cite{glazm89,lev90} are difficult to compare with
experiment because little is known about the actual profile of the
boundaries. The application of the delta-function impurity model to the
GaAs/Ga$_{1-x }$Al$_{x}$As heterostructures which have been used 
in the experiments \cite{wees1,kom1,klitz1,klitz2,kom2,kom3,wees2,alph}
is somewhat questionable since it is known ({\it cf., e.g.,} 
Refs.~\cite{sarma,fangsmith,col}) that in these systems at least zero-field 
scattering is determined by the long-range potential fluctuations due to 
the remote ionized donors.

One of the aims of the present chapter is to calculate the inter-edge-state
scattering length due to phonons and impurities under more general
assumptions than used previously, for realistic models of the
electron-phonon and electron-impurity interaction and for a realistic
confining potential.

In contrast to the above conventional transport measurement experiments, in
this chapter we study the inter-edge-state relaxation also in the phonon emission
experimental technique. The phonon emission or absorption experiments (see
review papers \cite{challis,butchallis} measure directly the total energy flux from
the 2DEG interacting
with different phonon modes and, in some cases, are more powerful tools to
investigate the electron-phonon interaction in the 2DEG systems. The basic
difference between the conventional energy-loss experiments and the phonon
emission experiments is that in the latter case, the phonon signal is
measured in the certain phonon emission angle and at the fixed excitation
energy. Therefore, to have a knowledge of the frequency and angular distributions
of the emitted phonon field is very important for the detection of the phonon signal.
These distributions are determined both by the energy and
momentum conservation laws during a scattering act. At inter-Landau-state 
transitions, the phonon emission energy is fixed by the cyclotron energy, therefore
the angular dependence of the emitted phonons remains as the sole
important characteristic of emission in this case.

Till recently theoretical investigations of the phonon emission in the 2DEG
under the quantizing magnetic field have been limited only to the
considerations of the bulk Landau states which have been the main subject of
the previous chapter. However, equally with the bulk Landau states, the
quantum edge states also give contribution to the emission and absorption of
ballistic phonons by the 2DEG. In the recent experiment \cite{kent92}, the
absorption of nonequilibrium phonon pulses has been already used to
investigate a backscattering of electrons between edge states on opposite
sides of the sample. Possibly, absorption of ballistic acoustic phonons by edge
states play an important role in the phonon-drag effect in the QHE regime
observed in experiments \cite{dietzel,dietzelphd}. The frequency spectrum and
angular distribution of the total energy-loss rate due to LA phonon emission
have been obtained in a QW \cite{shik}.

Electron-optical phonon interaction in a polar semiconductor controls such
phenomena as the cooling of photoexcited hot carriers on the picosecond time
scale as well as transport and optical properties at relatively high
temperatures. To date, however, this aspect of electron-phonon scattering
remains the least investigated when either the 2DEG or QW subjected to the
quantizing magnetic field. The polar optical PO phonon assisted edge state
relaxation in the 2DEG has not been given. Recent studies have been reported
for cylindrical QWs where the axial magnetic field effect on the
electron-PO phonon interaction has been discussed \cite{constantinou}. In
rectangular QWs, scattering processes involving only one transverse subband
have been considered \cite{telang} using the delta-function approximation
for electron wave functions along the magnetic field \cite{danish}.

The second goal of present chapter is to calculate the ballistic acoustic
(both deformation and piezoelectric interaction) and polar optical phonon
emission by quantum edge states.

This chapter is organized as follows. First in Sec.~\ref{sec:edgerel} we
consider the inter edge state relaxation due to electron-phonon and electron
impurity interactions. Results of this section have been reported in \cite
{bad,maslov}. Then in Secs.~\ref{sec:edstemiss} and \ref{edge-lo} 
we discuss emission of acoustic and optic phonons by quantum edge states 
\cite{badaledbal,badaledbal1}.

Sect.~\ref{sec:imp} is devoted to the impurity scattering. We use the
standard model of the heterostructures \cite{afs,ando} which takes into
account long-range potential fluctuations due to the layer of ionized donors
as well as short-range fluctuations due to the uniformly distributed
acceptors. Analytical expressions for the scattering length are derived for
an arbitrary confining potential. In Sect.~\ref{sec:phon} phonon scattering
is discussed. We consider deformation acoustic (DA) and piezoelectric (PA)
interactions and again derive analytical expressions for an arbitrary
confining potential. As follows from energy and momentum conservation, only
phonons with frequencies above some threshold can participate in the
transitions between edge states. As a result, phonon scattering is
exponentially suppressed at low temperatures. According to our numerical
evaluations, the observed temperature dependence of the scattering length
cannot be attributed to phonon scattering.

The impurity-assisted phonon scattering suggested by Y B Levinson and 
D L Maslov \cite{maslov} is discussed in Sect.~\ref{sec:assist}. The difficulty
with the exponential suppression of the scattering does not exist in this
mechanism, since  the impurity  takes up the momentum and thermal phonons
are able to participate in the process. Thus, one can expect a significant
enhancement of the scattering. Actually, it is found that impurity-assisted
phonon scattering can dominate over ordinary phonon scattering only at very
low temperatures (in the hundred mK range). The temperature dependence
associated with this kind of scattering is non-exponential: the scattering
rate goes like $ T^{2}$ below some crossover point and after that like $ T$. A
similar temperature dependence has been observed recently in the hundred mK
range \cite{klitz2}.
In Sect.~\ref{sec:comp} 
it is found that the experimental temperature dependence of the scattering
can be explained solely by the thermal averaging of the long-range impurity
scattering rate while short-range and phonon scattering play a minor role.

In the Sec.~\ref{edgeflux} an analytic expression for the ballistic acoustic
energy flux emitted by quantum edge states is derived. Detailed
analysis of the phonon emission intensity distribution is made in the low and
high regimes of electron temperature $ T_{e}$ as well as different positions of
the Fermi level $\varepsilon_{F}$ are considered in Sec.~\ref{edgetemlev}.
It is shown that at low temperatures the phonon emission is predominantly
concentrated within a narrow cone around the direction of the edge state
propagation while at high temperatures $-$ around the magnetic field normal to
the electron plane. The emission intensity decreases when the Fermi level
falls down. This diminution is exponential at low temperatures. In Sec.~\ref
{edgepiez} phonon emission due to the piezoelectric interaction is discussed.
In contrast to the case of bulk Landau states where piezoelectric
interaction is always suppressed in comparison with deformation interaction,
in the edge state case, the relative contributions of the piezoelectric and
deformation interactions depend on the magnetic field. The angular
distribution of emitted phonons is presented in Sec.~\ref{angdistrb}.

In Sec.~\ref{edge-lo} the LO-phonon assisted edge state relaxation is
investigated in QWs with rectangular cross sections exposed to the normal
magnetic field. The energy and momentum relaxation rates for a test electron
and electron temperature are calculated. Detailed result are given for the
intrasubband and intersubband scattering rates as a function of the initial
electron energy and of the magnetic field. By considering different limiting
cases of the ratio of the cyclotron frequency to the strength of the lateral
confinement, it is obtained results for the edge state relaxation both in
the 2DEG and QW as well as for the magnetic field free case.

\section{Inter-edge-state relaxation}
\label{sec:edgerel}

\subsection{Impurity scattering}
\label{sec:imp}

Consider a 2DEG in a uniform perpendicular magnetic field $B$. Let the $z$%
-axis be parallel to the magnetic field and the x-axis be parallel to the
sample boundary. Positive values of $y$-coordinate correspond to the
interior of the sample. In the following we assume that the edge-states are
spin-degenerate.

The wave function and the spectrum of the edge state is 
\be
\Psi_{nlk}({\bf r})=\exp(ikx)\chi_{lk}({y})\psi_{n}(z) \;
\varepsilon _{nl}(k)=\varepsilon _{l}(k)+\varepsilon_{n}.  
\label{edstspect}
\ee
Here $\varepsilon_{l}(k)$ and $\chi_{lk}$ are the energy and wave function
of the quantum edge state in the plane of electrons specified by the Landau
index $l$ and momentum $k$ (Fig. \ref{spect}). The energy and wave function
of the spatial quantazation of the 2DEG are $\varepsilon_{n}$ and $\psi_{n}$.

In the Born approximation one can calculate the elastic scattering length
for transition $l\rightarrow l^{\prime} $ according to this formula 
\begin{equation}
{\frac{1}{{\cal L}_{l\rightarrow l^{\prime}}}}={\frac{1}{%
\hbar^{2}v_{l}v_{l^{\prime}}}} \int\!{\frac{dp_{y}}{2\pi }}%
|Q_{ll^{\prime}}|^{2}<UU>_{p},  \label{2}
\end{equation}
where $<UU>_{p}$ is the 2D Fourier component of the scattering potential
correlation function taken at the 2DEG plane, ${\bf p}=(p_{x}, p_{y})$, $%
v_{l}$ is the group velocity of the edge state $l$ and the form factor 
\begin{equation}
Q_{ll^{\prime}}(p_{y}, k, k^{\prime})=\intop dy
\chi_{l^{\prime}k^{\prime}}(y)e^{-ip_{y}y}\chi_{lk}(y)  \label{3}
\end{equation}
In (\ref{2}) one should put $p_{x}=\delta
k_{ll^{\prime}}=|k_{l}(\varepsilon)-k_{l^{\prime}}(\varepsilon)|$ with $%
\varepsilon$ being the energy of the initial state.
Here the velocities $v$ and functions $\chi$ for states $l$ and $l^{\prime}$
correspond to the energy $\varepsilon$. Note that ${\cal L}_{l\rightarrow
l^{\prime}}={\cal L}_{l^{\prime}\rightarrow l}$.

We assume that the ionized donors are situated in a narrow layer separated
from the 2DEG plane by the undoped spacer of the thickness $z_{0}$. Due to
the electroneutrality the density of donors (per cm$^{-2}$) equals $%
N_{s}+N_{d}$, where $N_{s}$ is the density of the 2DEG and $N_{d}$ is the
density of the depletion charge layer on the GaAs side, the latter is
assumed to be uniformly doped by the acceptors with the net density ( per cm$%
^{3}$) $N_{AC}$. Then the correlation function $<UU>_{q}$ can be written in
the form \cite{afs,ando} 
\begin{equation}
<UU>_{p}=\left(\frac{2\pi e^{2}}{\epsilon \epsilon_{s}(p)p}%
\right)^{2}[(N_{s}+ N_{d}) \exp(-2pz_{0})+{\frac{1}{2p}}N_{AC}],  \label{4}
\end{equation}
where $\epsilon$ is the lattice dielectric constant taken to be the same for
GaAs and GaAlAs, $\epsilon_{s}(p)= 1+q_{s}/p$ is the dielectric function of
the 2DEG, $q_{s}$ is the screening parameter. To determine $q_{s}$ in a
quantizing magnetic field constitutes a special problem, and we will discuss
it later (Sect.~\ref{sec:comp}). The first term in (\ref{4}) corresponds to
the long-range part of the scattering, while the second does to the
short-range part.

For the long-range scattering one can proceed further assuming that $\delta
k_{ll^{\prime}}z_{0}\gg 1$. In this case only small values of $p_{y}\simeq
(\delta k_{ll^{\prime}}/z_{0})^{1/2}$ contribute to the integral (\ref{2}).
Using this simplification one can calculate the scattering length due to the
long-range potential 
\begin{equation}
{\frac{1}{{\cal L}_{l\rightarrow l^{\prime}}^{L}}}= {\frac{1}{\bar{{\cal L}}%
_{L}}}\exp(-2\delta k_{ll^{\prime}}z_{0})A^{2}_{ll^{\prime}},  \label{5}
\end{equation}
where $A_{ll^{\prime}}=Q_{ll^{\prime}}(0)$ and the nominal scattering length
is defined as 
\begin{equation}
{\frac{1}{\bar{{\cal L}}_{L}}}=2\pi^{3/2}\left[{\frac{2\pi e^{2}}{%
\hbar\epsilon (v_{l}v_{l^{\prime}})^{1/2}}}\right]^{2}(N_{s}+N_{d}) {\frac{1%
}{(\delta k_{ll^{\prime}}+q_{s})^2}}\left({\frac{\delta k_{ll^{\prime}}}{%
z_{0}}}\right)^{1/2}.  \label{6}
\end{equation}
Note that due to the small factor $\exp(-2\delta k_{ll^{\prime}}z_{0})$ in (%
\ref{5}), scattering can be strongly suppressed compared to the case of
zero magnetic field. It can happen even in the case when the spatial
separation between edge channels $\delta y_{ll^{\prime}}= a^2_{B}\delta
k_{ll^{\prime}}$, where $a_{B}$ is the magnetic length, is not large
compared to $a_{B}$. For the short- range scattering Eq.~\r{2} cannot be
reduced to a more simple form without any assumption on the confining
potential.

Now we consider the case of the smooth potential $V(y) $, with $%
V^{\prime}(y)a_{B}\ll\hbar\omega_B$. In this case 
\begin{equation}
\chi_{lk}(y)=\Phi_{l}(y-ka_{B}^{2}),  \label{7}
\end{equation}
\[
\varepsilon_{lk}=\varepsilon_{l}+V(ka_{B}^{2}),
\]
where $\Phi_{l} $ is the harmonic oscillator wave function. In the smooth
potential the overlap integral (\ref{3}) can be calculated explicitly. Using
(\ref{7}) we have 
\begin{equation}
A_{ll^{\prime}}^{2}=(2^{l+l^{\prime}}l!l^{\prime}!)^{-1}
\sigma^{2l+2l^{\prime}}\exp(-\sigma^{2}/2),  
\label{8}
\end{equation}
where $\sigma=(\delta y_{ll^{\prime}}/a_{B})^{2}\gg1$. The short-range
scattering length ${\cal L}_{l\rightarrow l^{\prime}}^{S}$ becomes 
\begin{equation}
{\frac{1}{{\cal L}_{l\rightarrow l^{\prime}}^{S}}}=(2\pi)^{-1/2}\left[{\frac{%
2\pi e^{2}}{\hbar\epsilon(v_{l}v_{l^{\prime}})^{1/2}}}\right]^{2} {\frac{%
N_{AC}}{a_{B}\delta k_{ll^{\prime}}(\delta k_{ll^{\prime}}+q_{s})^{2}}}
A^{2}_{ll^{\prime}}.  \label{9}
\end{equation}
The group velocity in the smooth potential is 
\begin{equation}
v_{lk}=a_{B}^{2}V^{\prime}(ka_{B}^{2}).  \label{10}
\end{equation}
It follows from (\ref{8}) that in the smooth potential the dominant
transitions are $l\rightarrow l+1$.

The inverse scattering lengths (\ref{5}), (\ref{9}) are to be averaged near
the Fermi energy 
\begin{equation}
\left\langle\frac{1}{{\cal L}_{l\rightarrow n\prime}}\right\rangle=\int
d\varepsilon \left(-\frac{\partial f_{0}}{\partial \varepsilon}\right) \frac{%
1}{{\cal L}_{l\rightarrow n^{\prime}}(\varepsilon)}.  
\label{11}
\end{equation}
where $f_{0}$ is the equilibrium Fermi distribution. Averaging (\ref{11})
plays a minor role for the impurity scattering of free electrons when $ T\ll
\varepsilon_{F}$ and reduces usually to the substitution $\e=\e_{F}$ in 
scattering lengths. However, for inter-edge-state relaxation, the averaging
 (\ref{11}) results in a pronounced temperature dependence appearing in a 
comparatively low temperature range.
This can be seen from the following arguments. When integrating in (\ref{11}%
), each value of $\varepsilon$ corresponds to some value of $\delta
k_{ll^{\prime}}(\varepsilon)$ entering scattering lengths (\ref{5}), (\ref{9}%
). The gap between any two branches of the spectrum corresponding to
different Landau levels, say $\varepsilon_{lk}$ and $\varepsilon_{l^{\prime}k}$
decreases as $k\rightarrow\infty$. This means that $\delta k_{ll^{\prime}}
(\varepsilon)$ decreases with $\varepsilon$ (cf. Fig.~\ref{spect}). Therefore, 
the main contribution to (\ref{11}) stems from $\varepsilon\simeq \e_{F}+T$
or $\delta k_{ll^{\prime}}(\varepsilon)=\delta k_{ll^{\prime}}(\e_{F})+\delta
{\h v}^{-1}_{ll^{\prime}}T$ where ${\delta v_{ll^{\prime}}}^{-1}=
|v_{l}^{-1}-v_{l^\prime}^{-1}|$. 
Due to the exponential factors in (\ref{5}) and (\ref{9}), corrections 
to $\delta k_{ll^{\prime}}$ can become significant even if $\delta
{\h v_{ll^{\prime}}}^{-1}T \ll\delta k_{ll^{\prime}}(\varepsilon_{F}).$

If the confining potential is smooth, the temperature dependence is
determined mainly by the $\delta y_{ll^{\prime}}$ dependence of the overlap
integral (\ref{8}). The onset temperature is 
\begin{equation}
T_{1}\simeq {\frac{\hbar \delta v_{ll^{\prime}}}{a_{B}^{2}\delta
k_{ll^{\prime}}(\varepsilon_{F})}}.  \label{12}
\end{equation}
For smooth potentials $ T_{1}\ll\hbar \omega _{B}$, while in the typical
experimental conditions $\hbar \omega _{B}\simeq \varepsilon_{F}$. However,
even if $V(y)$ is not smooth, the long-range scattering length (\ref{5}) can
strongly depend on $ T$ due to the exponential factor containing thickness of
the spacer $z_{0}$. The onset temperature for this kind of dependence is 
\begin{equation}
T_{2}={\frac{\hbar \delta v_{ll^{\prime}}}{z_{0}}}.  \label{13}
\end{equation}
Again, if the impurity potential is smooth, {\it i.e.} $z_{0}\gg a_{B}$,
then $ T_{2}\ll\hbar \omega_B$. Results of the numerical calculation of (\ref
{11}) will be given in Sect.~\ref{sec:comp}.

To learn what is the dominant mechanism of impurity scattering,, it is
sufficient to put all of the characteristic lengths entering the
pre-exponential factors in (\ref{5}) and (\ref{9}) to be the same: $%
a_{B}=z_{0}=k^{-1}_{F}=q_{s}^{-1}\equiv\bar{k}^{-1}$  and 
$N_{d}=N_{s}=\bar
{k}^{2}$. Comparing (\ref{5}) and (\ref{9}) one can see that the long-range
scattering dominates, if the thickness of the spacer $z_{0}$ is less than 
\begin{equation}
z_{0}^{*}={\frac{1}{2\delta k_{ll^{\prime}}}}\ln{\frac{\bar{k}^{3}}{N_{AC}}}.
\label{14}
\end{equation}
For instance, taking $k_{ll^{\prime}}=1.5/a_{B}$ at $B=2$ T, $\bar {k}^{-1}=10
$ nm and $N_{AC}=10^{14}$ cm$^{-2}$ we get $z_{0}^{*}\approx 60$ nm.

\subsection{Acoustic phonon scattering}
\label{sec:phon}

The scattering length due to phonon scattering is 
\begin{equation}
\frac{1}{{\cal L}_{l\rightarrow l^{\prime}}}=\frac{2\pi}{\hbar v_{l}}%
\sum_{k^{\prime}} \sum_{{\bf q}} [|M^{{\bf +q}}_{lk\rightarrow
l^{\prime}k^{\prime}}|^{2}\delta
(\varepsilon_{lk}-\varepsilon_{l^{\prime}k^{\prime}}- \hbar\omega_{q}) (N_{%
{\bf q}}+1)(1-f_{0}(\varepsilon_{l^{\prime}k^{\prime}}))+  \label{15}
\end{equation}
\[
|M^{{\bf -q}}_{lk\rightarrow l^{\prime}k^{\prime}}|^{2}\delta
(\varepsilon_{lk}-\varepsilon_{l^{\prime}k^{\prime}} +\hbar\omega_{q})N_{%
{\bf q}}(1-f_{0}(\varepsilon_{l^{\prime}k^{\prime}}))],
\]
where ${\bf q}$ is the phonon wavevector, $\omega_{q}=sq$, with $s$ being
the sound velocity, $N_{{\bf q}}$ and $f_{0}(\varepsilon)$ are the
equilibrium Bose and Fermi distributions, respectively.

The emission matrix element $M^{{\bf +q}}_{lk\rightarrow
l^{\prime}k^{\prime}}$ is calculated with the phonon perturbation potential 
\cite{gantlev} 
\begin{equation}
W^{+}({\bf r})=B(q)e^{-i{\bf qr}},  \label{16}
\end{equation}
where for the deformation potential scattering (DA), $B^{DA}$ is given by Eq.~\r{}
while for the piezoacoustic scattering (PA) 
\begin{equation}
B^{PA}(q)=e\beta\left (\frac{\hbar}{2\rho \omega_{q} L^{3}}\right)^{1/2}.
\label{18}
\end{equation}
Here $L^{3}$ is the normalization volume, $\beta$ is an effective piezoelectric
modulus, $\rho$ is the crystal mass density. For absorption processes
$W^{-}= (W^{+})^{*}$.

Since $s\ll v_{l},v_{l^{\prime}}$, the scattering is quasielastic, {\it i.e.},
the energy of the emitted or absorbed phonons $\hbar\omega_{q}\ll\hbar
\omega_B$. Hence, in the transition $l\rightarrow l^{\prime}$ (cf. Fig.~\ref
{spect}) the change of $k$ is $\delta k_{ll^{\prime}}$. The minimal energy
of the phonon is $\Delta_{ll^{\prime}}=\hbar s \delta k_{ll^{\prime}}$. In
the following we consider low temperatures $ T \ll \Delta_{ll^{\prime}}$. In
this case, due to the phonon Bose factor and the Pauli exclusion principle,
the phonon energy $\hbar\omega_{q}$ is close to the threshold $%
\Delta_{ll\prime}$. As a result the calculations of the scattering length is
greatly simplified. In the Born approximation we get for DA scattering 
\begin{equation}
\frac{1}{{\cal L}_{l\rightarrow n^{\prime}}} =\frac{\Xi^{2}}{2\pi\hbar\rho
s^{2}}A_{ll^{\prime}}^{2} \frac{\delta k_{ll^{\prime}}^{2}}{\hbar v_{l}
v_{l^{\prime}}}T F\left(\frac{\varepsilon-\varepsilon_{F}}{T},\frac{%
\Delta_{ll^{\prime}}}{T}\right),  \label{19}
\end{equation}
where 
\begin{equation}
F(\xi,\eta)=(1/2)[\ln(1+\exp(\xi-\eta)+\exp(\xi)\ln(1+ \exp(-\xi\eta)]
\label{20}
\end{equation}
The velocity $v_{l^{\prime}}$ and function $\chi_{l^{\prime}}$ of the final
state correspond to the energy of this state $\varepsilon=\varepsilon^{%
\prime}$. Equation (\ref{19}) is valid if $\exp(\varepsilon-\varepsilon_F-%
\Delta_{ll^\prime}/T)\ll\exp(\Delta_{ll^\prime}/T)$ or, in other words, if $|\varepsilon-%
\varepsilon_F|< \Delta_{ll^{\prime}}$ and if $\varepsilon-\varepsilon_F
-\Delta_{ll^{\prime}}\ll\Delta_{ll^{\prime}}$. In the first case $%
\hbar\omega-\Delta_{ll^{\prime}}\simeq T $, while in the second case $%
\hbar\omega-\Delta_{ll^{\prime}}\simeq \varepsilon-\varepsilon_F -
\Delta_{ll^{\prime}}$. It has been also assumed that $ T\gg ms^{2}$ and $%
\Delta_{ll^{\prime}}\ll \hbar s/d $, where $d $ is the scale of the electron
wave function $\psi(z) $. For GaAs/AlGaAs heterostructure $d=3$ nm, $%
s=5\times10^{3} $ m s$^{-1}$, $ms^{2}=0.1$ K and $\hbar s/d = 13$ K. 
Taking $%
\delta k_{ll^{\prime}}=a_{B}^{-1} $, we have for $B=2$ T: $a_{B} =18$ nm, $%
\Delta_{ll^{\prime}}=2$ K and $\hbar \omega_B=39$ K. The inverse scattering
length (\ref{19}) is to be averaged near Fermi energy according to (\ref{11}%
). Since the function $F$ grows exponentially with $\varepsilon-\varepsilon_F
$ for $\varepsilon-\varepsilon_F > 0$, the average value is rather due to
hot electrons $(\varepsilon-\varepsilon_F\simeq\Delta_{ll^{\prime}}\gg T)$
than to thermal ones $(\varepsilon-\varepsilon_F\simeq T)$. Until $%
\Delta_{ll^{\prime}} \ll\hbar\omega_B $, one can put $\varepsilon=%
\varepsilon_{F}$. With the above mentioned assumptions for DA scattering we
get 
\begin{equation}
\left\langle\frac{1}{{\cal L}_{l\rightarrow n^{\prime}}}\right\rangle_{DA}= 
\frac{1}{{\cal L}_{DA}}A_{ll^{\prime}}^{2}(\delta k_{ll^{\prime}}a_{B})^{3} 
\frac{s^{2}}{v_{l}v_{l^{\prime}}}\exp\left(-\frac{\Delta_{ll^{\prime}}}{T}%
\right).  \label{21}
\end{equation}
For piezoelectric PA scattering calculations are similar: 
\begin{equation}
\left\langle\frac{1}{{\cal L}_{l\rightarrow n^{\prime}}}\right\rangle_{PA}= 
\frac{1}{{\cal L}_{PA}}A_{ll^{\prime}}^{2}(\delta k_{ll^{\prime}}a_{B}) 
\frac{s^{2}}{v_{l}v_{l^{\prime}}}\exp\left(-\frac{\Delta_{ll^{\prime}}}
{T}\right).  \label{22}
\end{equation}
Here we defined nominal scattering lengths 
\begin{eqnarray}
{\frac{1}{\bar{{\cal L}}_{DA}}}={\frac{\Xi^{2}}{4\pi\hbar\rho s^{3}a_{B}^{3}}%
}, \\
{\frac{1}{\bar{{\cal L}}_{PA}}}={\frac{(e\beta)^{2}}{4\pi\hbar\rho s^{3}a_{B}%
}}.  \label{23}
\end{eqnarray}
For GaAs at $B=2$ T we have $(\bar{\tau}_{DA})_{B}=800$ ps and $(\bar{\tau}%
_{PA})_{B}=75$ ps ($\Xi^{2}$ and $\beta^{2}$ taken from Ref.~\cite{gantlev}).
The exponential suppression of the scattering rate is because of the deficit
of the superthermal phonons for absorption and deficit of the free final
states below $\varepsilon_{F}$ for emission.

To consider the case of the smooth potential one has to substitute Eq.~(\ref
{8}) for $A_{ll^{\prime}}$ into (\ref{21}) and (\ref{22}). Comparing (\ref
{21}) or (\ref{22}) with Eq.~(7) in Ref.~\cite{martin} one can see that the
corresponding equations agree only in the exponential factors from the
overlap integral and from the deficit of the final states.

The results presented above are valid provided that the chemical potential
difference $|\Delta \mu|$ between edge levels is small compared to
temperature. If this condition is not satisfied, then Eqs. (\ref{21}) and (%
\ref{22}) should be multiplied by the additional factor $\exp(|\mu |/T)$ 
\cite{khaet}.

\subsection{Impurity-assisted phonon scattering}
\label{sec:assist}

According to Eqs.~(\ref{21}) and (\ref{22}), phonon scattering is
strongly suppressed at $ T\ll \Delta_{ll^{\prime}}$ due to the exponential
factors. The origin of the threshold $\Delta_{ll^{\prime}}$ entering this
factors is momentum conservation. If one considers a three-body collision,
namely, impurity-assisted phonon scattering, the momentum $\delta
k_{ll^{\prime}}$ is taken up by the impurity, and no exponential suppression
of the scattering appears.

The matrix element of the impurity-assisted transition of the electron from
state $(n.k)$ to the state $(l^{\prime},k^{\prime})$ accompanied by the
emission of a phonon with momentum $q$ can be written in the form 
\begin{equation}
M^{+q}_{lk\rightarrow
l^{\prime}k^{\prime}}=\sum_{l^{\prime\prime}k^{\prime\prime}}\frac
{<l^{\prime}k^{\prime}q
|U|l^{\prime\prime}k^{\prime\prime}q><l^{\prime\prime}k^{\prime\prime}q |W
|nk>}{\varepsilon_{lk}-\varepsilon_{l^{\prime\prime}k^{\prime\prime}}-\hbar
\omega_{q}+i\gamma}+  \label{24}
\end{equation}
\[
+\frac {<l^{\prime}k^{\prime}q |W|
l^{\prime\prime}k^{\prime\prime}><l^{\prime\prime}k^{\prime\prime}|U| nk>}
{\varepsilon_{lk}-\varepsilon_{l^{\prime\prime}k^{\prime\prime}}+i\gamma}
\]
where $1/2\gamma$ is the decay time of the intermediate state $%
(l^{\prime\prime},k^{\prime\prime})$ due to all possible scattering
processes. The absorption matrix element $M^{-q}_{lk\rightarrow
l^{\prime}k^{\prime}}$ is obtained in a similar way. The calculations in (%
\ref{24}) are greatly simplified by using the arguments given below.

Although the summation in (\ref{24}) is presumed to be over all intermediate
values of $l^{\prime\prime}$, the main contribution to (\ref{24}) comes from
the transition $l=l^{\prime\prime}$ corresponding to the smallest value of
the denominator. In the first term of (\ref{24}) $k^{\prime\prime}=k-q_{x}$,
in the second $k^{\prime\prime}=k+q_{x}$, thus, in general, the impurity
first-order matrix elements $<...| U |...>$ entering (\ref{24}) depends on $%
q_{x}$. But, if $q_{x}$ is small enough, one can neglect it in $<...| U |...>
$. The condition for this simplification is different for short- and
long-range impurity potentials. For the former case this condition is 
\begin{equation}
q_{x}\ll \delta k_{ll^{\prime}},  \label{25}
\end{equation}
while for the latter 
\begin{equation}
q_{x}z_{0}\ll 1.  \label{26}
\end{equation}
If 
\begin{equation}
max\{q_{x}a_{B}, kq_{x}a^{2}_{B}\}\ll1
\label{27}
\end{equation}
and 
\begin{equation}
q_{y}a_{B}\ll 1  \label{28}
\end{equation}
one can put the overlap integrals (\ref{3}) stemming from the first-order
phonon matrix elements to be equal to unity. When substituting (\ref{24})
into (\ref{15}), it is convenient to express $<...| U |...>$ through the
first-order impurity scattering length ${\cal L}^{i}_{l\rightarrow
l^{\prime}}$ calculated in Sect.~\ref{sec:imp} ({\it cf.} Eq.~(\ref{2})).
This can be done by using the relation 
\begin{equation}
{\frac{1}{{\cal L}^{i}_{l\rightarrow l^{\prime}}}}={\frac{L}{%
\hbar^{2}v_{l}v_{l^{\prime}}}} |<l^{\prime}k^{\prime}| U |lk>^{2}.
\label{29}
\end{equation}
Using (\ref{16}) and (\ref{17}), we have for DA phonon scattering after some
manipulations 
\begin{equation}
\frac{1}{{\cal L}_{l}\rightarrow l^{\prime}}=\frac {1}{2(2\pi)^{3}} \frac{1}{%
{\cal L}^{i}_{l\rightarrow l^{\prime}}}\frac {\Xi^{2}}{\hbar\rho s} \int
d^{3}qq |a^{+}+a^{-}|^{2}[(N_{q}+1)(1-f_{0}(\varepsilon- \hbar\omega_{q}))+
\label{30}
\end{equation}
\[
+N_{q}(1-f_{0}(\varepsilon+\hbar\omega_{q}))],
\]
where $a^{+}=(v_{l}q_{x}-sq+i\gamma)^{-1}$ and $a^{-}=
(qs-v_{l^{\prime}} q_{x}+i\gamma)^{-1}$. Eq.~(\ref{30}) includes 
contributions from two types of
transitions: from real and from virtual ones \cite{gantlev}. For real
transitions the energy conservation takes place: $\varepsilon_{l^{\prime%
\prime}k^{\prime\prime}}=\varepsilon_{lk}$. These transitions can be
imagined as proceeding through the successive scatterings $-$ first, by
impurities and, second, by phonons or vice versa $-$ and are not of interest
here. The contribution of the real transitions to (\ref{30}) comes from the
terms $|a^{\pm}|^{2}$ which reduce to the $\delta$-functions reflecting
energy conservation at $\gamma\rightarrow +0$. Hence, these terms should be
omitted. The true second-order transitions proceed through the virtual
states with $\varepsilon_{l^{\prime\prime}k^{\prime\prime}}\not=
\varepsilon_{lk}$. Virtual transitions are related to the terms $%
a^{+}(a^{-})^{*}+c.c.$. The principal value of the integral over $d\Omega_{q}
$ is 
\begin{equation}
J=v.p.\int d\Omega_{q}\left(a^{+}(a^{-})^{*}+c.c.\right)= \frac
{4\pi}{v_{l}v_{l^{\prime}}q^{2}} \frac {1}{x-x^{\prime}}\ln\left(\frac{1-x}{%
1+x}\frac{1+x^{\prime}}{1-x^{\prime}}\right),  \label{31}
\end{equation}
where $x=s/v_{l}$, $x^{\prime}=s/v_{l^{\prime}}$. As in the Sect.~\ref
{sec:phon}, we assume that $s\ll v_{l}, v_{l^{\prime}}$. Then (\ref{31})
reduces to $J=8\pi/(v_{l}v_{l^{\prime}}q^{2})$. Finally, substituting (\ref{31}%
) into (\ref{30}) and integrating over $dq$, we have 
\begin{equation}
\frac {1}{{\cal L}_{l\rightarrow l^{\prime}}}=\frac{1}{{\cal L}^{i}_{DA}} 
\frac{s^{2}}{v_{l}v_{l^{\prime}}}\Phi \left(\frac{\varepsilon-\varepsilon_{F}%
}{T}\right),  \label{32}
\end{equation}
where 
\begin{equation}
{\frac{1}{{\cal L}^{i}_{DA}}}={\frac{1}{{\cal L}^{i}_{l\rightarrow
l^{\prime}}}}{\frac{\pi^{2}\Xi^{2} }{\hbar^3\rho s^{3}}}{\frac{T^{2}}{%
\hbar^2s^2}}  \label{33}
\end{equation}
and 
\begin{equation}
\Phi(x)=\int^{\infty}_{0}\!\! dyy\left((N(y)+1)(1-f_{0}(y-x))+
N(y)(1-f_{0}(y+x))\right).  \label{34}
\end{equation}

To make the conditions (\ref{26})-(\ref{28}) more explicit we assume that
the confining potential is not smooth, {\it i.e.} $\delta k\simeq a_{B}^{-1}$%
, $k\simeq a_{B}^{-1}$ and $v_{l}\simeq v_{l^{\prime}}$. One can see that
the main contribution to (\ref{32}) comes from $q\simeq q_{y}\simeq
q_{z}\simeq q_{T} \equiv T/\hbar s$ and $q_{x}\approx q_{T}s/v_{l}$. Then,
instead of (\ref{26})-(\ref{28}), we have $ T\ll \hbar s/a_{B}$, if the
short-range impurity scattering dominates, and $ T\ll min\{\hbar
s/a_{B},\hbar v_{l}/z_{0}\}$, if the long-range one does.

For PA scattering the calculations proceed in a quite similar way up to the
integration over $dq$. At this point we faced the problem: if $B^{PA}(q)$ is
taken in the form (\ref{18}), the integral over $dq$ diverges at $%
q\rightarrow 0$ as $\int dq/q^{2}$. Since PA interaction is interaction of 
an electron with the electrostatic potential induced by the
phonon waves, the divergence at $q\rightarrow 0$ means that the screening of
this interaction by free electrons has not been taken into account. In our
case the screening problem is somewhat specific, since we must take into
account the screening of three-dimensional phonons by two-dimensional
electrons. The solution of this problem is given in the Appendix \ref
{screening}. The result is quite obvious: $B^{PA}(q)$ entering (\ref{16}) and
given by (\ref{18}) should be replaced by $B^{PA}(q)/\epsilon_{s}(q_{||})$ where $%
q_{||}$ is the component of the phonon vector parallel to the 2DEG plane.
The remaining calculations are straightforward. Explicit results can be
obtained in two limiting cases: $q_{T}\ll q_{s}$ or $q_{T}\gg q_{s}$. In the
former case 
\begin{equation}
\frac {1}{{\cal L}_{l\rightarrow l^{\prime}}}=\frac {1}{4\pi^{2}} \frac {1}{%
{\cal L}_{PA}}\left (\frac {T}{\hbar sq_{s}}\right)^{2} \frac
{s^{2}}{v_{l}v_{l^{\prime}}}\Phi \left(\frac
{\varepsilon-\varepsilon_{F}}{T}\right).  \label{35}
\end{equation}
In the latter 
\begin{equation}
\frac {1}{{\cal L}_{l\rightarrow l^{\prime}}}=\frac {2}{\pi^{3}} \frac {1}{%
{\cal L}^{i}_{PA}}\frac {T}{\hbar sq_{s}},  \label{36}
\end{equation}
where 
\begin{equation}
\frac {1}{{\cal L}^{i}_{PA}}=\frac {1}{{\cal L}^{i}_{l\rightarrow
l^{\prime}}} \frac {(e\beta)^{2}}{\hbar\rho s^{3}}.  \label{37}
\end{equation}
Averaging of the inverse scattering lengths (\ref{32}) and (\ref{35})
according to Eq.~(\ref{11}) reduces to the averaging of function $\Phi(x)$.
Numerical calculation gives $\langle\Phi \rangle \approx 5.2$.

\subsection{Discussion}
\label{sec:comp}

As far as we are aware, the temperature dependence of the
inter-edge-state relaxation rate has been measured only in three works \cite
{alph,klitz2,kom3}. Measurements in Ref.~\cite{alph} were carried out only
at two values of temperature. High magnetic fields and low temperatures were
used in Ref.~\cite{klitz2} and, as a result, spin-splitting of the edge
states was observed. Thus, results of Ref.~\cite{klitz2} cannot be compared
directly to the present theory because we assume that the edge states are
spin-degenerate. Fortunately, the experimental conditions of Ref.~\cite{kom3}%
, lower magnetic fields, higher temperatures, and the Fermi level lying far
away from the adjacent Landau levels, allow us to make such a comparison. To
obtain some numerical evaluations using the results of Secs.~\ref{sec:imp}-~%
\ref{sec:assist} one needs to choose the confining potential $V(y)$ in some
specific form. ${\em Ab \ initio}$ calculation of $V(y)$ constitutes a very
complicated problem and is beyond the framework of the present work. We take 
$V(y)$ in the model form 
\begin{equation}
V(y)=V_{0}exp\left(-{\frac{y}{a_{0}}}\right)  \label{38}
\end{equation}
where $V_{0}$ and $a_{0}$ are free parameters to be determined from the fit
to the experimental data. To make our heterostructure model consistent with
the experimental value of the zero-field mobility we found it necessary to
consider the spacer thickness $z_{0}$ as free parameter also.


Another subtle question is the choice of the screening parameter $q_{s}$
entering the expressions for impurity scattering length (\ref{5}) and (\ref
{9}) and for impurity-assisted phonon scattering length (\ref{35}) and (\ref
{36}). For impurity-assisted phonon scattering we assume that 
$q_{T}\ll a^{-1}_{B}$. 
If the confining potential is not smooth, the edge states are located at
the distance $\bar {y}\simeq a_{B}$ near the sample boundary. For typical
experimental conditions, the width of the sample channel is much larger than $%
\bar {y}$. This means that the screening of the electron-phonon interaction
potential is provided rather by the wide domain of bulk electrons filling
the Landau levels than by the narrow strips of width $\bar {y}$ where the
electrons filling the edge states are situated. Thus, to learn what is $q_{s}
$ entering Eqs.~(\ref{35}) and (\ref{36}) one can consider the infinite 2DEG
in a quantizing magnetic field. The screening problem for these conditions
is very complicated and is not well understood. It is known \cite{afs,hor}
that in a quantizing magnetic field the usual Thomas-Fermi expression
connecting $q_{s}$ with the density of states (DOS) at the Fermi level $%
\nu_{F}$ holds 
\begin{equation}
q_{s}={\frac{2\pi e^{2}}{\epsilon}}\nu_{F},  \label{39}
\end{equation}
but $\nu_{F}$ differs from the DOS for zero-field case $\nu_{F}^{0}$. At $ T=0
$ and in the absence of any scattering, $\nu_{F}$ is a sum of $\delta$%
-functions centered at Landau levels. If a short-range scattering is taken
into account, then each $\delta$-function is smoothed and becomes
Gaussian-shaped \cite{afs,hor}. In this case $\nu_{F}$ is extremely small
compared with $\nu_{F}^{0}$ unless the Fermi level is very close to one of
the adjacent Landau levels. As a result, for short-range scattering,
screening is very weak. However, as was shown in a number of experimental
works \cite{gorn,eis,stahl,kuk}, the observed behavior $\nu_{F}$ has no
relation to the theory that assumes short-range scattering. For GaAs/GaAlAs
heterostructures the value of $\nu_{F}$ in the center of the gap between two
Landau levels where DOS has its minimum was found to be $(0.1-0.2)\times
\nu_{F}^{0}$ \cite{gorn,eis,stahl}. Such high values of DOS unequivocally
point at the long-range character of the scattering, as was confirmed also
by independent experiments on the determination of the quantum decay
lifetime and its comparison with the transport time entering the mobility 
\cite{fangsmith,col}. Unfortunately, no quantitative theory for DOS under
the long-range scattering conditions exists at present. This is why to
estimate roughly $q_{s}$ entering (\ref{35}) and (\ref{36}) we use the
above-mentioned experimental values of $\nu_{F}$ in (\ref{39}).

For impurity scattering the momentum transfer is $\delta k_{ll^{\prime}}$.
If the potential is not smooth, $\delta k_{ll^{\prime}}^{-1}\simeq
a_{B}\simeq \bar {y}$. Thus, the screening of electron-impurity interaction
is provided by the same electrons that fill the edge states. The proper
accounting of the screening in this situation constitutes a very complicated
problem
We do not know the solution of this problem. Nevertheless, one
can assume that the screening properties of edge states do not differ
drastically from those of the 2DEG without the magnetic field. This
assumption is based on the fact that the specific features of the screening
by Landau electrons are caused by the gaps in the DOS, while the DOS of edge
states is gapless. Thus, to estimate roughly $q_{s}$ entering Eqs.~(\ref{5})
and (\ref{9}) we use $\nu_{F}^{0}$ in instead of $\nu_{F}$ in (\ref{39}).

According to the results of numerical calculations, the main contribution to
the total inverse scattering lengths due to phonons (Eqs.~(\ref{21}) and (%
\ref{22})) and impurities (Eqs.~(\ref{5})and (\ref{9}) averaged according to
Eq.~(\ref{11})) is given by the long-range impurity scattering. The
scattering rates due to the short-range impurities and both the DA and PA 
phonons underestimate the experimental data, taken from Ref.~\cite{kom3}
and corresponding to the transition $0\rightarrow 1$ at $B=3.7$ T, by at least
a factor of $50$. According to calculations by Komiyama {\it et al.} \cite{kom3},
phonon scattering dominates while impurity scattering plays a minor role.
Thus, using a more realistic model than the delta-function impurity model
used in Ref.~\cite{kom3} leads to different conclusions. One can note also
that the DA phonon scattering length estimated in Ref.~\cite{martin} is much
shorter than ours. The reason of the difference is mainly attributed to the
another choice of the electron-phonon interaction constant. The constant
used in Ref.~\cite{martin} is not related to the deformation potential
constant $\Xi$.

The contribution of the impurity-assisted phonon scattering essentially
depends on temperature. The impurity-assisted phonon scattering lengths (\ref
{32}), (\ref{35}) and (\ref{36}) do not contain the small exponential
factors that the first-order scattering lengths (\ref{21}) and (\ref{22})
do, since only thermal phonons are involved in the process. However, there
are two other reasons that do not allow the impurity-assisted phonon
scattering rate to be very high. Firstly, it is a second-order process: the
scattering rate is proportional to the product of the electron-impurity and
of the electron-phonon interaction constants. Secondly, the phase volume of
the thermal phonons is lower than the phase volume of the superthermal
phonons involved in the first-order scattering. As a result, gaining in the
exponent we yield the smaller pre-factors than before. The impurity-assisted
phonon scattering rate exceeds the first-order one if the temperature is low
enough, so that the exponential factors in (\ref{21}), (\ref{22}) become
very small. Using the list of parameters corresponding to the experiment 
\cite{kom3} and taking $\nu_{F}=0.1\nu_{F}^{0}$ to estimate the screening
parameter entering Eqs.~(\ref{35}) and (\ref{36}), one can see that the
impurity-assisted scattering is stronger than the first-order phonon
scattering if $ T\lesssim T_{DA}\approx 0.4$ K for DA interaction and at 
$ T\lesssim T_{PA}\approx 0.5$ K for PA interaction, while
the measurements in Ref.~\cite{kom3} were carried out at significantly
higher temperatures ($ T=1.5-12$ K). Since, according to the calculations,
even the DA  and PA scattering is much weaker than the impurity scattering
at these temperatures we did not consider the impurity-assisted phonon
scattering when comparing to the results of Ref.~\cite{kom3}. However, at
lower temperatures: $ T<min(T_{1},T_{2},T_{DA},T_{PA})$, where $ T_{1}$ and $%
T_{2}$ are defined by Eqs.~(\ref{12}) and (\ref{13}) respectively, the
impurity-assisted phonon scattering can dominate. The temperature dependence
similar to that given by Eqs.~(\ref{32})-(\ref{36}) has been observed
recently in Ref.~\cite{klitz2}, although one must bear in mind that present
theory cannot be applied directly to this experiment ({\it cf.} beginning of
the present section).

The strong suppression of the inter-edge-state relaxation scattering compared to
the zero-field case was observed in the majority of the experiments . For
example, in Ref.~\cite{kom3} the zero-field transport scattering length is $%
4.3$ $\mu$m while at $ T=1.5$ K the inter-edge-relaxation length is $500$ $\mu$m.
Usually this suppression is attributed to the smoothness of the confining
potential and, consequently, to the exponential smallness of the overlap
integrals. On the other hand, the potential (\ref{38}) with the parameter
values given above cannot be considered as being smooth, but we also have
strong suppression of the scattering. According to our calculations, the
suppression is caused mainly by the exponential smallness of the correlation
function (\ref{4}) and not by the smallness of the overlap integrals. As has been
shown in  Refs.~\cite{glazm91}, $V(y)$ decreases as $1/y$ at $%
y>>a_{d}$, where $a_{d}$ is the width of the depletion region. 
%

In order to avoid the
contradiction with Refs.~\cite{glazm91} we have to assume that Eq. (%
\ref{38}) is referred to the intermediate region where the confining
potential changes drastically whereas long-decaying tail of $V(y)$ is
described by asymptotic law obtained in Refs.~\cite{glazm91}.
However, as follows from the comparison with the experiment, edge channels
are situated in the intermediate region where presence of the long-decaying
tail is of minor importance. Notice that recent study of the inter edge
relaxation due to the multiple impurity scattering by Martin and Feng \cite
{martinfeng} similarly shows that the impurity assisted scattering gives
rise to the linear temperature dependence.

\section{Phonon Emission by Quantum Edge States}
\label{sec:edstemiss}

\subsection{Acoustic energy flux: Deformation Potential}
\label{edgeflux} 

When the 2DEG is embedded in an elastic medium and the
phonon displacement is caused by the deformation electron-phonon interaction
then the spectral density of the acoustic energy flux ${\cal P}_{\alpha}$ in
a point ${\bf r_{0}}$ and at a frequency $\omega$ is given by Eq.~(\ref{flux}). 
To obtain ${\cal P}_{\alpha}$ explicitly it is necessary to calculate the correlator
of the phonon field operators $K_{\alpha\beta}({\bf r}, {\bf r^{\prime}})=
<u_{\alpha}^{*}({\bf r})u_{\beta}({\bf r^{\prime}})>_{\omega}$ which can
be represented in the form (\ref{correl}). If do not take into account the
phonon reflection from interfaces separating different materials then it
follows from Eq.~(\ref{lcalh}) that the following function should be used as
a kernel of the correlator (\ref{correl}): 
\begin{equation}
D^{*}_{\alpha}({\bf r_{0}}, {\bf r_{1}} \mid \omega)D_{\alpha}({\bf r_{0}}, 
{\bf r_{2}} \mid \omega)={\frac{\omega^2}{16\pi^2r^2_{0}\rho^2s^6}} \exp[-i%
{\bf q(r_{2}-r_{1})}]  \label{bkernel}
\end{equation}
where $\rho$ is the mass density of the elastic medium, $s$ and ${\bf q}= 
{\bf n}\omega/s$ are the velocity and momentum of LA-phonons. It is clear
from experimental situation that the acoustic energy flux emitted from the
2DEG located near one side of the sample is detected in an infinitely
distant point on the reverse side of the sample. Therefore, in obtaining the
kernel (\ref{bkernel}) we have assumed ${\bf r_{0}}\to{\bf n}\infty$, ${\bf n}$
is an unit vector towards the detector. This kernel is averaged in (\ref
{correl}) by electron wave functions. In this case $\varepsilon$ and $\Psi$
represent, respectively, the electron energy and wave
function for edge states. We assume that all electrons occupy a single level
$n$ of the spatial quantization and the electron transitions take place
between edge states with different Landau index $l$ and momentum $k$.
Substituting Eqs. (\ref{edstspect}) and (\ref{bkernel}) into Eq. (\ref
{correl}) we obtain the acoustic energy flux density in a frequency range $%
d\omega$ emitted from an unit length (towards ${\bf x}$) of the 2DEG into a
solid angle $do$ around the direction of ${\bf n}$ 
\begin{equation}
{\cal P}^{DA}_{l\rightarrow l^{\prime}}({\bf q})= {\frac{\Xi^2\omega^4}{%
8\pi^3\rho s^5\delta v_{ll^{\prime}}}}f(\varepsilon _{l}(k_{0}))
(1-f(\varepsilon
_{l^{\prime}}(k_{0}+q_{x})))|Q_{ll^{\prime}}(q_{x},q_{y})|^2
|I_{00}(q_{z})|^2  \label{edstflux}
\end{equation}
where the form factors in directions ${\bf y}$ and ${\bf z}$ are 
\begin{equation}
Q_{ll^{\prime}}(q_{x},q_{y})=\intop dy \chi_{lk_{0}}(y)e^{-iq_{y}y}
\chi_{l^{\prime}k_{0}+q_{x}}(y)  \label{qff}
\end{equation}
and 
\begin{equation}
I_{nn}(q_{z})=\int dz |\psi_{n}(z)|^2 e^{-iq_{z}z}  \label{iff}
\end{equation}
Notice that the confining potential in the plane of the 2DEG violate the
axial symmetry of the magnetic field so that the form factor (\ref{qff})
depends on the momenta $q_x$ and $q_y$ apart. In Eq.~(\ref{edstflux}) $%
\delta v_{ll^{\prime}}=|v_{l}(k_{0})-v_{l^{\prime}}(k_{0}+q_{x})|$, $v_{l}(k)
$ is the group velocity of the edge state $l$ with the momentum $k$, $%
k_{0}=k_{0}({\bf q)}$ is determined from the energy and momentum (in the $%
{\bf x}$-direction) conservation laws, {\it i.e.} from the following
equation 
\begin{equation}
\varepsilon_{l}(k)-\varepsilon_{l^{\prime}}(k+q_{x})=\omega=s\sqrt{q_{x}^2+
q_{y}^2+q_{z}^2}.  \label{conserv}
\end{equation}
Actually, despite lack of the momentum conservation in ${\bf y}$- and ${\bf z%
}$-directions, the phonon momentum ${\bf q}$ uniquely determines the initial 
$l, k_{0}({\bf q})$ and final $l^{\prime}, k_{0}({\bf q})+q_{x}$ edge states
in an act of the phonon emission. Thus, equation (\ref{edstflux}) gives the
distribution of the emission intensity in phonon momenta.

\subsection{Low and High Temperature Regimes}
\label{edgetemlev} 

In this section we will discuss the situation in low and
high temperature regimes in which the emission is qualitatively different.
From Eq. (\ref{conserv}) one can see that for a given $k$, $q_{x}$ cannot be
less than some value $\delta k_{ll^{\prime}}(k)$ determined from Eq.~(\ref
{conserv}) at $q_{y}=q_{z}=0$. Hence, only phonons with frequencies $%
\omega\geq s\delta k_{ll^{\prime}}(k)$ can be emitted from the edge state $%
l, k$. On the other hand, the effectiveness of each emission act depends on
the position of the Fermi level and electron temperature. Due to the deficit
of the hot electrons with energies above $\varepsilon_{l}(k_{1}) =
\varepsilon_{F}+s\delta k_{ll^{\prime}}(k_{1})$ and deficit of the free
final states with energies below $\varepsilon_{l^{\prime}}(k_{2}+s\delta
k_{ll^{\prime}}(k_{2}))=\varepsilon_{F}- s\delta k_{ll^{\prime}}(k_{2})$,
emission acts from states $l, k$ with $k$ between $k_{1}$ and $k_{2}$
are comparatively more efficient (see Fig.~(\ref{spect})(a)). The energy 
spectrum of the edge states is arranged so that for a fixed energy level we
always have $v_{l}>v_{l^{\prime}}$ if $l<l^{\prime}$. Therefore 
$\delta k_{ll^{\prime}}(k)$ achieves its minimum 
$\delta {\bar k}_{ll^{\prime}}=\delta
k_{ll^{\prime}}(k_{1})$ at upper edge of the interval $(k_{1},k_{2})$ .
Thus from the Fermi factors as well as form factors (\ref{qff})
and (\ref{iff}) we have following obvious restrictions on the emission
processes: 
\begin{eqnarray}
|q_{x}-\delta {\bar k}_{ll^{\prime}}|\leq T_{e}/s,& |q_{y}|\leq
min\{a_{B}^{-1}, T_{e}/s\}, &|q_{z}|\leq min\{d^{-1},T_{e}/s\},
\label{restric}
\end{eqnarray}
where $a_{B}$ is the magnetic length, $d$ is the characteristic length of
electron motion in ${\bf z}$-direction. (So far as the electron transitions
take place between edge states of one spatial quantization, $d$ is the
minimum length scale of the problem). Here $q_{x}$ is not a free parameter
but it should firstly satisfy the momentum conservation in the 
${\bf x}$-direction.
\paragraph{{\bf At low temperatures, $ T_{e}\ll s\delta {\bar k}_{ll^{\prime}}
$}},
we have $q_{x}\approx \delta {\bar k}_{ll^{\prime}}$, $q_{z}\sim T_{e}/s$.
If the confining potential is not smooth, $\delta {\bar k}_{ll^{\prime}}\sim
a_{B}\gg T_{e}/s$, so that $q_{y}\sim T_{e}/s$ and $q_{x}\gg q_{y}$. In the
case of the smooth potential $\delta {\bar k}_{ll^{\prime}}\gg a_{B}^{-1}$
so that $ T_{e}/s$ and $a_{B}^{-1}$ can be of the same order of magnitude.
But in any case, the relation $q_{x}\gg q_{y}$ remains true. Thus at low
temperatures $q_{x}\gg q_{y},q_{z}$, {\it i.e.} phonons are mainly emitted
in ${\bf x}$-direction. Therefore one can substitute $q_{x}=\delta {\bar k}%
_{ll^{\prime}}$ and $q_{y}=q_{z}=0$ in the prefactor of (\ref{edstflux}) and
form factors (\ref{qff}), (\ref{iff}). Taking into account that $I_{nn}(0)=1$, 
one can find for the acoustic energy flux distribution in the momentum space
\begin{equation}
{\cal P}^{DA}_{l\rightarrow l^{\prime}}({\bf q})= {\frac{\Xi^2{\delta {\bar k%
}_{ll^{\prime}}}^4}{8\pi^3\rho s\delta {\bar v}_{ll^{\prime}}}}
Q^2_{ll^{\prime}}(q_{x},0)\exp\left\{-{\frac{s\delta {\bar k}_{ll^{\prime}}}{%
T_{e}}}- {\frac{s}{T_{e}}}\left[(q_{x}-\delta {\bar k}_{ll^{\prime}})+{\frac{%
q_{y}^2+q_{z}^2}{2\delta {\bar k}_{ll^{\prime}}}}\right]\right\}
\label{lowemiss}
\end{equation}
where $\delta {\bar v}_{ll^{\prime}}=\delta v_{ll^{\prime}}$ at $k=k_{1}$
and $q_{x}=\delta {\bar k}_{ll^{\prime}}$. In the case of the non-smooth
potential we have only one length scale $a_{B}$ in the $(x,y)$-plane and so $%
Q_{ll^{\prime}}\sim 1$ if $q_{y}\sim a^{-1}_{B}$. In the case of the smooth
potential, $Q_{ll^{\prime}}$ can be calculated explicitly \cite{kom1} and is 
given by
\begin{equation}
Q^2_{ll^{\prime}}(q_{x},0)=(2^{l+l^{\prime}}l!l^{\prime}!)^{-1}(q_{x}
a_{B})^{2l+2l^{\prime}}\exp\{- (q_{x} a_{B})^2/2\}.  \label{sqff}
\end{equation}
Now one can see that at low temperatures electrons at inter-edge-state
transitions emit almost monochromatic acoustic phonons with frequencies $%
\omega \approx s\delta {\bar k}_{ll^{\prime}}$. The emission predominantly
concentrated within a narrow cone around the edge state propagation. The
emission intensity exponentially drops out of the cone. For the non-smooth
potential, the emission cone is isotropic in the $(y,z)$-plane and the cone
angle is determined by the length scales 
$\sqrt{\delta {\bar k}_{ll^{\prime}}T_{e}/s}$ in ${\bf x}$-direction 
(in the case of the smooth potential also by the length scale
$max \{a_{B}^{-1}, \sqrt{\delta {\bar k}_{ll^{\prime}}T_{e}/s}\}$ in the 
${\bf y}$-direction) and 
$\sqrt{\delta {\bar k}_{ll^{\prime}} T_{e}/s}$ in ${\bf z}$-direction. 
A momentum spread in ${\bf x}$-direction is $\Delta q_{x}\equiv q_{x}-
\delta {\bar k}_{ll^{\prime}}\sim T_{e}/s$. It should be noticed that only
a part of
states from the interval between $k_{1}$ and $k_{2}$ gives an essential
contribution to the phonon emission processes. When $k$ varies in this
interval (see Fig.~\ref{spect}), $\delta {k}_{ll^{\prime}}(k)$ is changed by 
$\delta {\bar k}_{ll^{\prime}}s\delta {\bar v}_{ll^{\prime}}/v_{l}v_{l^{%
\prime}}$ which is much less than $\delta {\bar k}_{ll^{\prime}}$ even if $%
\delta {\bar v}_{ll^{\prime}}\sim v_{l},v_{l^{\prime}}$. However, only
states $l,k$ for which $s\delta k_{ll^{\prime}}(k)s\delta {\bar v}%
_{ll^{\prime}}/v_{l}v_{l^{\prime}}\sim T_{e}$, are effective in the emission
processes. Because of the velocity asymmetry $v_{l}>v_{l^{\prime}}$ if $%
l<l^{\prime}$, the number of these states decreases when the Fermi level
falls down. Simultaneously $\delta {\bar k}_{ll^{\prime}}$ increases which
leads to the suppression of the phonon emission ({\it cf.} \cite{alph}). If
the Fermi level is very close to the topmost bulk Landau level (Fig.~\ref{spect}%
(b)), $v_{l^{\prime}}\sim s$ and $\delta v_{ll^{\prime}}\sim v_{l}$ so that $%
s\delta {k}_{ll^{\prime}}(k)$ is changed by $s\delta \bar{k}%
_{ll^{\prime}}s\delta \bar{v}_{ll^{\prime}}/v_{l}v_{l^{\prime}}\sim s\delta%
\bar{k}_{ll^{\prime}}\gg T_{e}$ in the interval $(k_{1},k_{2})$. Therefore
the main contribution to the phonon emission comes from the single edge
state $l, k_{c}$ where $k_{c}$ is defined from $v_{l^{\prime}}(k_{c})=s$. 
The emission processes from states lying below $%
k_{c}$ are forbidden by conservation laws, while from states above $k_{c}$:
due to the strong deficit of the hot electrons. Such critical point $k_{c}$
exists only for transitions $l\rightarrow l^{\prime}$ with $l<l^{\prime}$.
In the case of $l>l^{\prime}$ there is no critical points in the edge state
spectrum and it is possible smooth transition to the case of the phonon
emission from the bulk Landau levels \cite{badalfree,badalsurf,badalbal}.
Because of $\delta {\bar k}_{ll^{\prime}}$ for $l>l^{\prime}$ is always more
than for $l<l^{\prime}$, there is an asymmetry between emission processes $%
l\rightarrow l^{\prime}$ and $l^{\prime}\rightarrow l$. This asymmetry
depends on the Fermi level position and is pronounced when the Fermi level
is close to the bulk Landau level: ${\cal P}_{l\rightarrow l^{\prime}}\gg 
{\cal P}_{l^{\prime}\rightarrow l}$ if $l>l^{\prime}$.

\paragraph{{\bf At high temperatures, $ T_{e}\gg s\delta {\bar k}_{ll^{\prime}}$}}, 
the states, which are more efficient in the emission processes, are
disposed above and below the Fermi level at the separation of the order of $%
T_{e}$. Therefore it is clear that $q_{x}\sim \delta {\bar k}_{ll^{\prime}}$
and $q_{y}\sim a_{B}^{-1}$ are determined, respectively, only by the
momentum conservation and by the magnetic length according to (\ref{restric}).
Correlation between $q_{x}$ and $q_{y}$ is determined by the shape of the
confining potential just as at the low temperatures. In ${\bf z}$-direction
we have $q_{z}\sim min\{d^{-1},T_{e}/s\}$ so that the relation $q_{z}\gg
q_{x}, q_{y}$ takes place for any $d$ and $ T_{e}$ as well as for any shape
of the confining potential. Thus in contrast to the low temperature regime,
at high temperatures the phonon emission is concentrated within a narrow
cone around magnetic field normal to the plane of the 2DEG. Using the variation
wave function for the lowest subband $n=0$ we have 
$|I_{00}|^2=[1+(q_{z}d)^2]^{-3}$. Taking into account that in this regime 
$f(\varepsilon _{l}(k_{0}))(1-f(\varepsilon _{l^{\prime}}(k_{0}+ q_{x})))
\sim\exp\{-sq_{z}/T_{e}\}$ we obtain for the emission intensity 
\begin{equation}
{\cal P}^{DA}_{l\rightarrow l^{\prime}}={\frac{\Xi^2{q_{z}}^4}{8\pi^3\rho s
\delta {\bar v}_{ll^{\prime}}}}{\frac{Q^2_{ll^{\prime}}(q_{x},0)}{%
[1+(q_{z}d)^2]^3}} \exp\left\{- {\frac{sq_{z}}{T_{e}}}\right\}.  
\label{highemiss}
\end{equation}
By comparing Eqs.~(\ref{lowemiss}) and (\ref{highemiss}) one may see that at
low temperatures the phonon emission is exponentially suppressed. At high
temperatures electrons emit phonons with frequencies $\omega\sim T_{e}$ and
the emission is much intense than at low temperatures when $\omega \sim s
\delta{\bar k}_{ll^{\prime}}\gg T_{e}$. In the high temperature regime,
phonon emission is not so sensible with respect to the Fermi level position.

\subsection{Acoustic Phonon Emission: Piezoelectric Potential}

\label{edgepiez} Up to now we considered only the deformation
electron-phonon interaction. Direct calculation shows that to find the
phonon emission intensity due to the piezoelectric coupling it should be
done the following replacement \cite{badalrspt} in Eq.~(\ref{edstflux}): 
\begin{equation}
\Xi^{2}{\frac{\omega ^{2}}{s^{2}}}\rightarrow (e\beta)^{2}  \label{rplc}
\end{equation}
where $\beta$ is the piezoelectric modulus of the crystal averaged over
directions of a phonon propagation and its polarizations \cite{kogan,zook}.
Therefore, taking the values of $\Xi$ and $\beta$ from \cite{gantlev}, for
GaAs we find 
\begin{equation}
{\frac{{\cal P}^{DA}}{{\cal P}^{PA}}}={\frac{\Xi^{2}(\omega/s)^2 }{%
(e\beta)^{2}}}=\left(5.6{\frac{\omega}{s}}[nm^{-1}]\right)^2.
\label{datopa1}
\end{equation}
At low temperatures $\omega\sim s\delta {\bar k}_{ll^{\prime}}$. Therefore
for the non-smooth confining potential 
\begin{equation}
{\frac{{\cal P}^{DA} }{{\cal P}^{PA}}} \sim\left({\frac{5.6}{a_{B}[nm]}}%
\right) ^{2}={\frac{B[T]}{20.9}}  \label{non-smdatopa}
\end{equation}
and DA interaction is suppressed with respect to PA interaction for not so
high magnetic fields. In the case of the smooth potential 
\begin{equation}
{\frac{{\cal P}^{DA} }{{\cal P}^{PA}}}\sim\left({\frac{\delta {\bar k}%
_{ll^{\prime}}a_{B} }{4.6}}\right)^{2}B[T]  \label{smdatopa2}
\end{equation}
and because of $\delta {\bar k}_{ll^{\prime}}a_{B}\gg 1$, even at 
$B\sim 1T$, the DA  and PA interaction give roughly the same contribution 
to the phonon emission.\\
At high temperatures we have $\omega\sim T_{e}$ so that 
\begin{equation}
{\frac{{\cal P}^{DA} }{{\cal P}^{PA}}}\sim\left({\frac{T_{e}[K]}{6.9}}%
\right)^2,  \label{highdatopa}
\end{equation}
{\it i.e.} at actual temperatures $ T_{e}\sim 10$ K, contributions of the DA 
and PA interaction are approximately of the same order of magnitude.

To compare the contributions of edge and bulk Landau states to the phonon
emission from the 2DEG one has to average Eq.~(\ref{edstflux}) in $k_{0}$
and to make a substitution $\pi\delta(\cdots )\rightarrow \tau$ in Eq. 8 of
Ref. 20 where $\tau$ determines the Landau level broadening. Proceed in this
way, in the high temperature regime one may obtain for magnetic fields $%
\omega_{B} \sim T_{e}$ and for $\delta {\bar v}_{ll^{\prime}} \sim
v_{l},v_{l^{\prime}}$, even in heterojunctions not so high quality with the
mobility $\mu =10$ $m^2$ V$^{-1}$ s$^{-1}$, the contribution of the edge states to
the cooling of the 2DEG at least is not less than the contribution of the
bulk Landau states. In the regime of low temperatures, because of $%
\omega_{B}\gg s \delta {\bar k}_{ll^{\prime}}\gg T_{e}$, it is clear that
the phonon emission is only due to inter-edge-state transitions while the
emission is practically absent at electron transitions between bulk Landau
states.

\subsection{Angular distribution of emitted acoustic phonons}
\label{angdistrb} 

In this section we estimate the acoustic energy flux power emitted by
the edge states at the peak of emission for the case of the non-smooth 
confining potential. In the low temperature regime the emission goes 
mainly via piezoelectric coupling. At the peak of emission, $q_{x}=
\delta {\bar k}_{ll^{\prime}}$ and  $q_{y}=q_{z}=0$. Therefore,
the emission intensity at the emission peak can be represented in the form 
\begin{equation}
{\cal P}^{PA}_{0}={\frac{1}{(2\pi)^2}}{\frac{ms}{{\bar
\tau}_{PA}}}{\frac{v_{B}}{\delta {\bar v}_{ll^{\prime}}}} \left({\frac{%
\delta {\bar k}_{ll^{\prime}}}{p_{B}}}\right)^2 \exp\left(-{\frac{s\delta {%
\bar k}_{ll^{\prime}}}{T_{e}}}\right)  \label{paflux}
\end{equation}
where the nominal time of the piezoelectric interaction is defined as 
\begin{equation}
{\frac{1}{{\bar \tau}_{PA}}}={\frac{(e\beta)^2p_{B}}{2\pi\hbar\rho s^2}}
\label{nomtime1}
\end{equation}
\begin{figure}[htb]
\epsfxsize=14cm
\epsfysize=9.334cm
\mbox{\hskip 0cm}\epsfbox{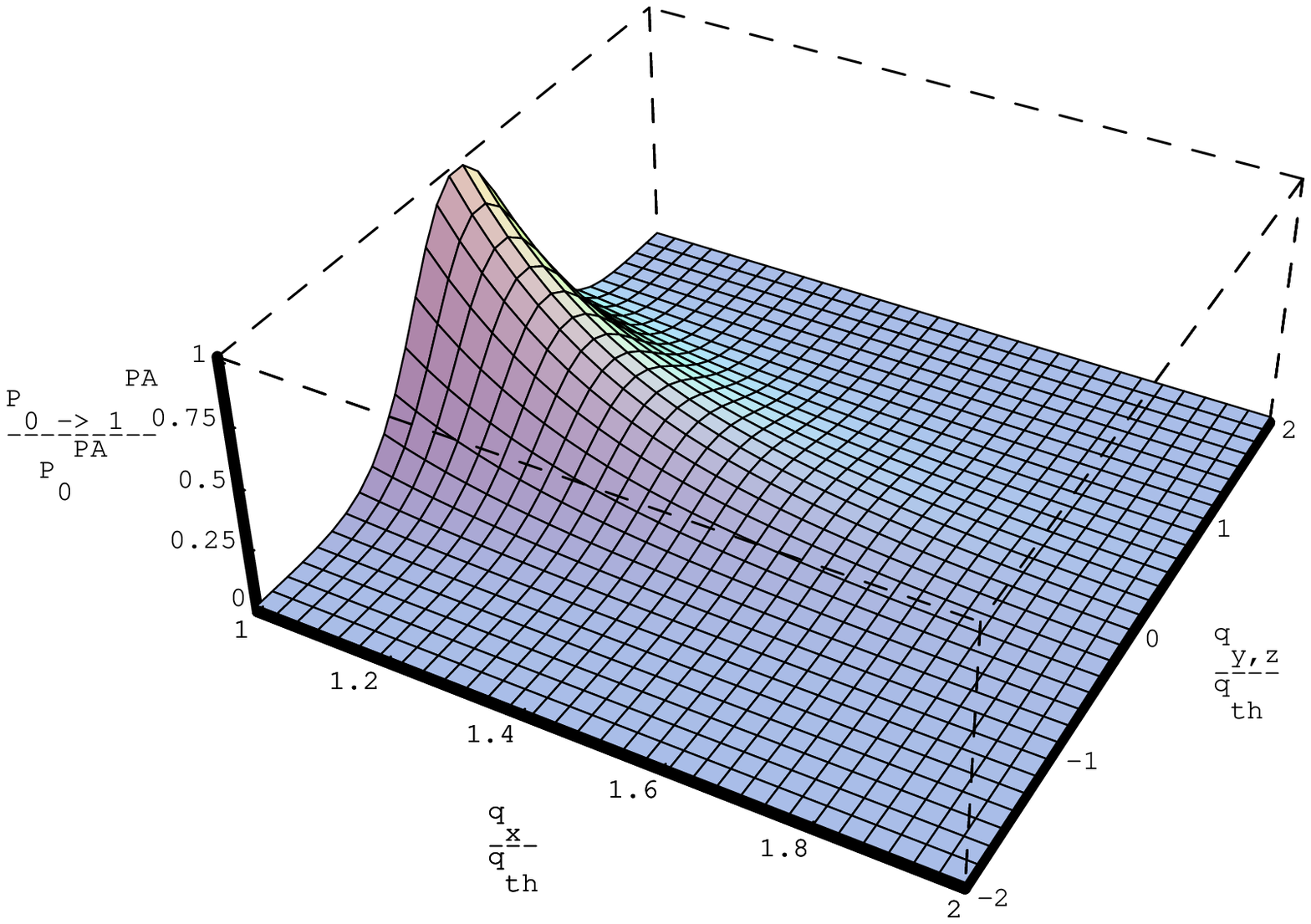}
\caption{Emission intensity distribution (piezoelectric interaction) in phonon 
momenta at low temperatures $ T_e\ll\h\o_{th}\equiv s q_{th}\equiv s\delta 
k_{10}$ and for the non-smooth confining potential.}
\label{fgpaflux}
\end{figure}
and $p_{B}=a_{B}^{-1}=mv_{B}$ is the magnetic momentum. For 
GaAs we have $ms=3.1\cdot10^{-28}$ J s m$^{-1}$ and at $B=2$ T, 
$\tau_{PA}=36.4$ ps. Taking $\delta {\bar v}_{10}=v_{B}$ and 
$\delta {\bar k}_{10}=p_{B}$ we have $s\delta {\bar k}_{10}=2.2$ K
for $B=2$ T so that at $ T_{e}=0.5$ K we obtain ${\cal P}_0^{PA}=
3.1\cdot 10^{-21}$ W s m$^{-1}$ for the electron transitions 
between $l=1$ and $l=0$ edge states. For the emission cone angle we
find $\theta =\tan^{-1}q_{x}/ \sqrt{q_{y}^2+q_{z}^2}=25^\circ$ 
(see Fig.~\ref{fgpaflux}).

At high temperatures ${\cal P}_{l\rightarrow l^{\prime}}^{DA}$ and 
${\cal P}_{l\rightarrow l^{\prime}}^{PA}$ can be represented in the 
forms 
\begin{equation}
{\cal P}^{DA}_{l\rightarrow l^{\prime}}={\frac{1}{(2\pi)^2}}{\frac{ms}{{\bar
\tau}_{DA}}} {\frac{v_{B}}{\delta {\bar v}_{ll^{\prime}}}}\left({\frac{T_{e}%
}{sp_{B}}}\right)^4 {\frac{x^4 e^{-x}}{[1+\eta^2 x^2]^3}}  \label{highdaflux}
\end{equation}
where the nominal time of the interaction is defined as 
\begin{equation}
{\frac{1}{{\bar \tau}_{DA}}}={\frac{\Xi^2p_{B}^3}{2\pi\hbar\rho s^2}}
\label{highnomtime}
\end{equation}
and 
\begin{equation}
{\cal P}^{PA}_{l\rightarrow l^{\prime}}={\frac{1}{(2\pi)^2}}{\frac{ms}{{\bar
\tau}_{PA}}} {\frac{v_{B}}{\delta {\bar v}_{ll^{\prime}}}}\left({\frac{T_{e}%
}{sp_{B}}}\right)^2 {\frac{x^2 e^{-x}}{[1+\eta^2 x^2]^3}}.
\label{highpaflux}
\end{equation}
Here $x=\frac{\h sq_{z}}{T_{e}}$ and $\eta =\frac{T_{e}}{\h s/d}$. 
The emission peaks in this temperature regime are defined from conditions 
that the last factor in Eqs. (\ref{highdaflux}) and (\ref{highpaflux}) should be 
maximum. For the 
\begin{figure}[htb]
\epsfxsize=14cm
\epsfysize=9.334cm
\mbox{\hskip 0cm}\epsfbox{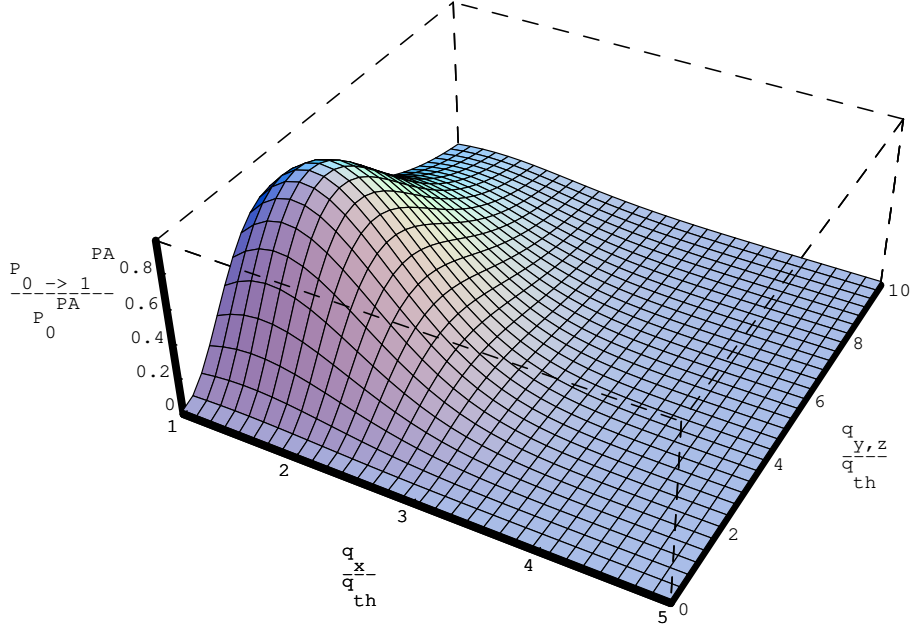}
\caption{Emission intensity distribution (deformation interaction) in phonon 
momenta at high temperatures $ T_e\gg\h\o_{th}\equiv s q_{th}\equiv s\delta 
k_{10}$ and for the non-smooth confining potential.}
\label{fghighdaflux}
\end{figure}
GaAs/AlGaAs heterojunction $d=3$ nm, $s=5\cdot 10^3$ m s$^{-1}$ and
so $\hbar s/d=13$ K. Taking $ T=10$ K we obtain that the emission peaks 
for DA  and PA interaction are determined, respectively, from $x=1.21$ and 
$x=0.85$. This means that at the emission peak, frequencies of phonons 
emitted due to the deformation coupling are approximately $1.4$ times 
larger than frequencies of phonons emitted due to the piezoelectric coupling. 
At $B=2$ T we have $\bar\tau_{DA}=382$ ps so that taking again 
$\delta {\bar v}_{10}=v_{B}$ we obtain  ${\cal P}_0^{DA}=1.03\cdot 
10^{-18}$ for the electron transitions between edge states $l=1$ and $l=0$.
W s m$^{-1}$ and ${\cal P}_0^{PA}=0.53\cdot 10^{-18}$ W s m$%
^{-1}$. For the emission cone angle we find $\theta =tan^{-1}q_{z}/ \sqrt{%
q_{x}^2+q_{y}^2}=10^\circ$ for DA phonons (see Fig.~\ref{fghighdaflux}) and 
\begin{figure}[htb]
\epsfxsize=14cm
\epsfysize=9.33cm
\mbox{\hskip 0cm}\epsfbox{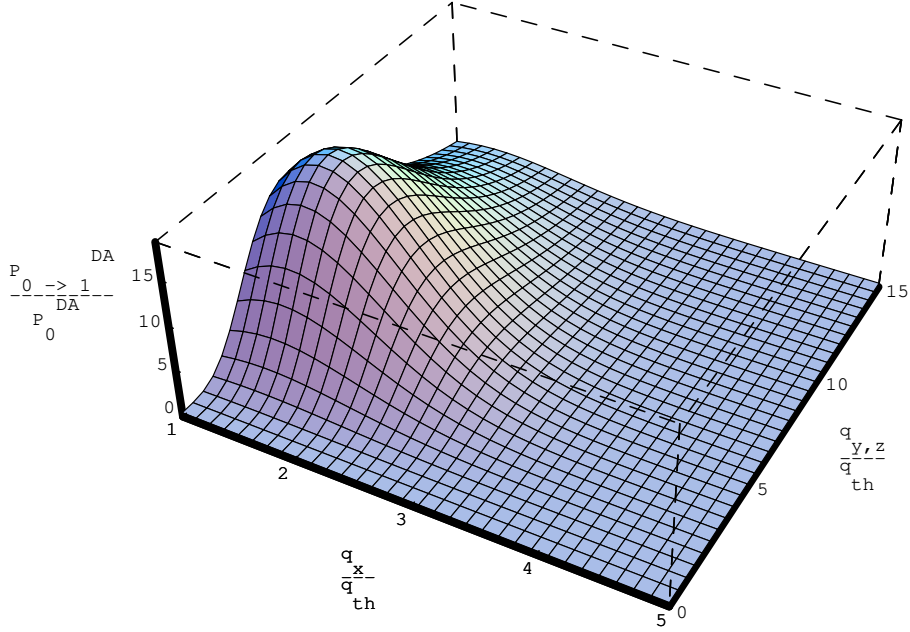}
\caption{Emission intensity distribution (piezoelectric interaction) in phonon 
momenta at high temperatures $ T_e\gg\h\o_{th}\equiv s q_{th}\equiv s\delta 
k_{10}$ and for the non-smooth confining potential.}
\label{fghighpaflux}
\end{figure}
$\theta =14^\circ$ for PA phonons (see Fig.~\ref{fghighpaflux}).

It should be observed that in the smooth confining potential, phonon
emission is suppressed exponentially. At low temperatures suppression 
takes place for two reasons. First, because of the threshold nature of 
emission, electrons are forced to emit phonons with frequencies larger than
$\o_{th}\equiv s\delta {\bar k}_{ll^{\prime}}\gg sp_{B}$. Second, 
because of the exponential smallness of the overlap integral $Q_{ll^{\prime}}$. 
While at high temperatures suppression takes place only for the last reason.

\section{LO-phonon assisted edge state relaxation}
\label{edge-lo}

\subsection{Inter edge state transition probability}

In this section we consider the edge states in the $z=0$ plane created by a
normal quantizing magnetic field ${\bf B}$ applied in $z$-direction in a
rectangular QW formed in $x$-direction by additional lateral confinement $%
V(y)$ in $y$-direction of the 2DEG. The lateral confining potential is
assumed to be parabolic, $V(y)=m\omega^2_0 y^2/2$, $\omega_0$ is a
characteristic frequency defining the strength of the lateral confinement.
Such type of the potential is sufficiently realistic one \cite{laux} and allows to
carry out more analytical calculation. In such one-dimensional electron
system, motion of particles is described by eigenfunctions $|nlk>$ labeled
by a subband index $n$ corresponding to the heterojunction quantization in
the $z$-direction, by hybrid index $l$ corresponding to the twofold
quantization due to the lateral and magnetic confinement on electrons, and
by wave vector $k$ corresponding to the free translation electron motion
along the $x$-direction. Corresponding eigenstates $|nlk>=|n>|lk>$ are
factored into a subband function $|n>=\psi(z)$ and a Landau oscillator
function $|lk>$. The latter $| l k>=e^{ikx}\chi_{lk}(y)$ are products of
plane waves and harmonic oscillator functions centered at $y_k= \lambda ka^2$
($a=\sqrt{\hbar/{m_{0} \omega}}$ is the characteristic length of the hybrid
quantization). The dependence of $y_k$ on the quantum number $k$ creates
spatial separation of the edge states with different $k$. Therefore, in
order to insure sufficient overlap between the electron wave functions, a
transferred momentum in an act of phonon emission or absorption should be
not so large. 
\begin{figure}[htb]
\epsfxsize=14cm
\epsfysize=11cm
\mbox{\hskip 0cm}\epsfbox{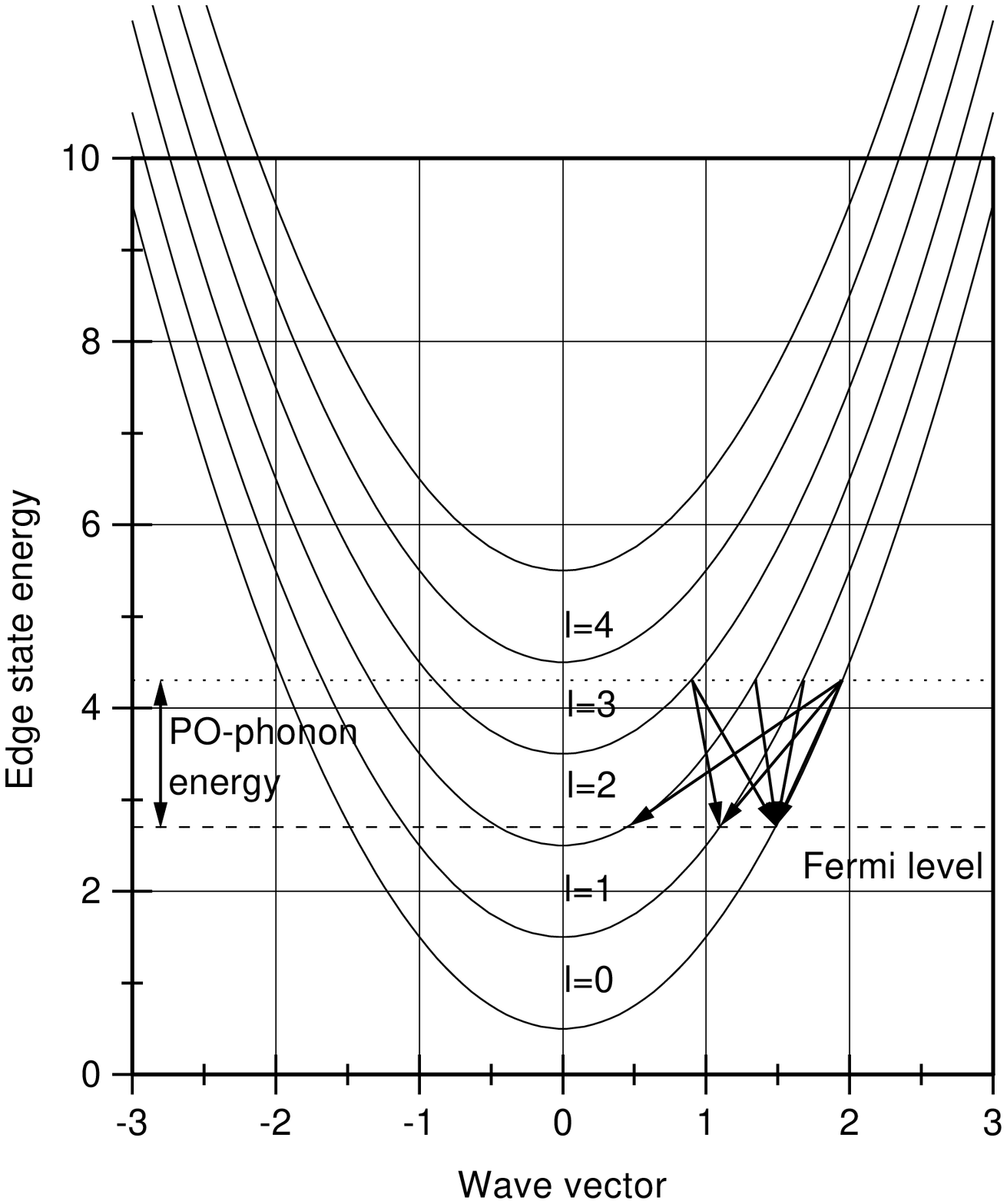}
\caption{Hybrid edge state spectrum in the parabolic lateral confinement. All
possible transitions due to polar optical phonon emission are shown.}
\label{fghybsub}
\end{figure}
The single-particle energy spectrum in the parabolic lateral confinement is
given by 
\begin{eqnarray}
\varepsilon_{nl}(k)&=&\varepsilon(k)+\varepsilon_{nl},\, \varepsilon(k)={%
\frac{\hbar^2k^2}{2m_{B}}},\, \varepsilon_{nl}=\varepsilon_{n}+\hbar\omega(l+%
{\frac{1}{2}}),\, m_{B}={\frac{\omega^2}{\omega^2_0}}m_{c}  \label{eq1}
\end{eqnarray}
where $\varepsilon_n$ is the subband energy, the frequency $\omega$ and
effective renormalized mass $m_{B}$ determine the hybrid subbands (see Fig.\ 
\ref{fghybsub}). The corresponding DOS for the subband $nl$
per unit length is given by 
\begin{equation}
g_{nl}(\varepsilon)={\frac{1}{\pi\hbar}}\:\sqrt{\frac{2m_{B}}{\varepsilon-
\varepsilon_{nl}}}\:\Theta(\varepsilon-\varepsilon_{nl})  \label{eq2}
\end{equation}
where $\Theta(x)=1$ if $x\geq 1$ and zero otherwise is the Heaviside step
function. The DOS has a square-root singularity at the bottom
of any subband $nl$. We recall that the DOS exhibits an
analogous behavior in the bulk samples exposed to the quantizing magnetic
field. The position of the Fermi level $\varepsilon_F$ and the linear
electron density $\nu$ are related in the following way 
\begin{equation}
\nu={\frac{2}{\pi\hbar}}\:\sum_{n,l}\sqrt{2m_B(\varepsilon_{F}-
\varepsilon_{nl})}\:\Theta(\varepsilon_F-\varepsilon_{nl}).  \label{eq3}
\end{equation}
Different physical situations corresponding to $\nu=const$ and $%
\varepsilon_F=const$, as well as to the mixture situations when both $\nu$
and $\varepsilon_F$ vary with field ({\it e.g.}, the situation when the ratio of
$\varepsilon_F/\omega$ remains constant) can be realized experimentally.
Notice that in these different situations the magnetic field dependence of
various physical quantities can be differ.

The scattering probability at which one polar optical PO phonon with a
wavevector ${\bf q}=(q_{x},q_y, q_{z})$ and frequency $\omega_{PO}$ (the
PO phonon dispersion is neglected, $\omega_q=\omega_{PO}=36.62$ meV) is
emitted or absorbed by an edge state can be represented as \cite{gantlev} 
\begin{eqnarray}
W^{\pm {\bf q}}_{nlk \rightarrow l^{\prime}l^{\prime}k^{\prime}}& =&{\frac{%
2\pi^2}{L_xL_yL_z}} {\frac{\hbar v_{PO}}{\bar{\tau}_{PO}}}{\frac{1}{q^2}}
|Q_{ll^{\prime}}(k^{\prime},k;q_{y})|^2|I^{PO}_{nl^{\prime}}(q_z)|^2 
\nonumber \\
&\times&\left(N_{T}(\omega_{PO})+1/2 \pm 1/2 \right) \delta_{k^{\prime},k\mp
q_{x}}\delta\left(\varepsilon_{nlk}-\varepsilon_{n^{\prime}l^{\prime}k^{%
\prime}} \mp\hbar\omega_{PO}\right).  \label{eq4}
\end{eqnarray}
In the above, $\pm$ refer to the emission and absorption of a PO phonon, 
$L_x, L_y,$ and $L_z$ are wire length, width, and height, respectively, 
$v_{PO}=\h p_{PO}/m_{c},\, \h p_{PO}=\sqrt{2m_{c}\h \o_{PO}}$.
The nominal time $\b{\tau}^{-1}_{PO} = 2\alpha_{PO}\omega_{PO}=(0.14$
ps $)^{-1}$ characterizes electron-PO phonon interaction ($\alpha_{PO}=0.07$
is the Fr\"olich coupling constant). 
$N_{T}(\omega_{PO})$ is the Bose-Einstein factor at temperature $ T$. If
phonon confinement effects are neglected, the form factor $Q_{ll^{\prime}}$
is given by Eq.~\ref{ffmagnetic} and, in the parabolic confinement, does not
depend on momenta $k$ and $k^{\prime}$ apart but depends only on the
combination $q_{\lambda}=\sqrt{q_y^2+\lambda^2q_x^2}$ ($\lambda\equiv%
\omega_B/ \omega$). To treat the electron relaxation between edge states of
the lowest $n=0$ subband, the variational wave function $\phi_0(z)= (z/{\sqrt{%
2d^3}}) \exp(- z/{2d})$ \cite{fangsmith} is used. Then the form factor in
the $z$-direction, $I^{PO}_{00}$, is given in the following explicit form 
\begin{equation}
|I^{PO}_{00}(q_z)|^2=(1+q^2_z d^2)^{-3}  
\label{eq6}
\end{equation}
where $d$ is the length scale characterizing confinement of the electrons in 
$z$-direction. By considering the inter-subband relaxation it is assumed
that the electrons in the $z$-direction are confined in a rectangular
potential with infinite high walls so that the wave functions are $\phi_n(z)=%
\sqrt{2/d}\sin{\pi nz}/d$ which give for the form factor $%
I^{PO}_{nn^{\prime}}$ following expression 
\begin{eqnarray}
&&|I^{PO}_{nn^{\prime}}(q_z)|^2={\frac{n^2n^{\prime 2}\pi^4\zeta^2}{%
(\zeta^2-\alpha^2)^2 (\zeta^2-\beta^2)^2}} \cases{\cos^2\zeta\quad
\mbox{if\quad $n+n'=3,5,7,...$}\cr \sin^2\zeta\quad \mbox{if\quad
$n'+n=2,4,6,...$}},  \nonumber \\
&&\zeta={\frac{q_z d}{2}},\,\alpha={\frac{\pi}{2}}(n+n^{\prime}),\,\beta= {%
\frac{\pi}{2}}(n^{\prime}-n)  \label{eq7}
\end{eqnarray}
Eqs.~\r{eq6} and \r{eq7} show that the form factor in the $z$-direction
decreases in both cases as a sixth power of $\zeta$ for small wavelengths, $%
q_z\gg d^{-1}$.

\subsection{Kinematics of PO phonon emission}
\label{kinematPO}

The kinematics of the PO phonon emission and absorption is determined
firstly by the conservation laws. In this system due to quantization of
electron motion, the momentum conservation remains only in $x$-direction so
that we have 
\begin{eqnarray}
\varepsilon(k)=\varepsilon(k^{\prime})+\Delta^{\pm}_{nln^{\prime}l^{%
\prime}},\,k=k^{\prime}\pm q_x,  \nonumber \\
\Delta^{\pm}_{nln^{\prime}l^{\prime}}=\varepsilon_{n^{\prime}l^{\prime}}-%
\varepsilon_{nl}\pm\hbar\omega_{PO}.  \label{eq8}
\end{eqnarray} 
These relationships are shown in Fig.\ \ref{fg1}. Electron transitions are
possible between edge states with different $n, l, k$ lying on two
horizontal lines separated by $\hbar\omega_{PO}$. If $\Delta^{\pm}_{nln^{%
\prime}l^{\prime}}>0$ ($\Delta^{\pm}_{nln^{\prime}l^{\prime}}<0$) electrons
gain (loose) the momentum, $q^+_x>0$ ($q^+_x<0$), at the phonon emission and
loose (gain) the momentum, $q^+_x<0$ ($q^-_x>0$), at the phonon absorption
processes. All emission (absorption) processes with gain (loose) of the
momentum have a threshold nature, {\it i.e.} only electrons with the energy $%
\varepsilon(k)\geq\Delta^{\pm}_{nln^{\prime}l^{\prime}}$ ($%
\varepsilon(k)\leq \Delta^{\pm}_ {nln^{\prime}l^{\prime}}$) can emit
(absorb) PO phonons. The number of all possible electron transitions
increases in Fermi level height. Notice that as it follows from Eqs.\ \ref
{eq8} at fixed electron subband quantum numbers, the electron initial energy
uniquely determines the $x$-component of the emitted or absorbed phonon
momentum. 
\begin{eqnarray}
q^{\pm}_{x}=\pm k\left(1-\sqrt{1-\Delta^{\pm}_{nln^{\prime}l^{\prime}}/%
\varepsilon(k)}\right).  \label{eq9}
\end{eqnarray}

Exact momentum conservation in the $y$- and $z$-directions is replaced by
the form factors $Q_{ll^{\prime}}$ and $I_{nn^{\prime}}$ which require
following restrictions for $l,l^{\prime}$ and $n,n^{\prime}$ of the order
unity 
\begin{eqnarray}
\lambda q_{x}a \lesssim 1,
\label{eq10} \\
q_y a \lesssim 1,
\label{eq11} \\
q_z d\lesssim 1.
\label{eq12}
\end{eqnarray}
Actually, condition (\ref{eq10}) is the restriction upon the electron
initial energy and magnetic fields for which emission and absorption of
PO phonons provide a subpicosecond relaxation between given subbands with
quantum numbers $nl$ and $n^{\prime}l^{\prime}$. By virtue of the
relationship (\ref{eq9}), the condition (\ref{eq10}) can be rewritten in the
following explicit form 
\begin{eqnarray}
2{\frac{\omega_B}{\omega_0}}\left(\sqrt{{\frac{\varepsilon(k)}{\omega}}}- 
\sqrt{{\frac{\varepsilon(k^{\prime})}{\omega}}}\right)\lesssim 1. 
\label{eq13}
\end{eqnarray}
The parameter $\omega_B/\omega_0$ determines an extend degree of the inter
edge state spatial separation at electron transitions. If $%
\omega_0/\omega_B\ll 1$ then a slight shift in the electron momentum $%
k-k^{\prime}=q_x$ or energy $\varepsilon-\varepsilon^{\prime}=\Delta^%
\pm_{lnl^{\prime}n"}$ causes a large inter edge state separation and weak
overlap between electron wave functions. Therefore in this case we restrict
the present analysis to the magnetic fields for which $\Delta^\pm_{lnl^{%
\prime}n"}\sim \omega_0\ll\omega_B$ while electron energy $\varepsilon$ can
be of the order both $\Delta^\pm_{lnl^{\prime}n"}$ and $\hbar\omega_B$. In
the limit of a non-smooth confinement, $\omega_0/\omega_B\gtrsim 1$, even 
for $\e-\e^{\prime}=\Delta^\pm_{lnl^{\prime}n'}\sim\o_B$
there exists sufficient overlap between electron wave functions. Therefore,
in this case we consider magnetic fields and energies for which 
$\e\sim \o_B\sim\Delta^\pm_{lnl^{\prime}n'}\sim\o_0$.

Due to the long-range nature of the PO interaction, phonons with minimum
momenta $q\equiv\sqrt{q_x^2+q_y^2+q_z^2}\sim\min\{q_x,q_y,q_z\}$ are emitted
or absorbed at the relaxation processes. Since for the given electron
discrete quantum numbers and initial energy, $q_x$ is a fixed quantity then
only phonons are essential for which 
\begin{eqnarray}
q_y,q_z\lesssim q_x.
\label{eq14}
\end{eqnarray}
In the limit of high magnetic fields, $d\gg a$, the distribution in phonon
momenta is isotropic in the (x,y)-plane of in the inverse space while in 
$z$-direction, the restriction (\ref{eq12}) is more severe than that of 
(\ref{eq14}) so that we have $q_z\ll q_y\lesssim q_x$. In the opposite 
limit of weak magnetic fields, $\o_B\ll\o_0$, ({\it i.e.} $\lambda\ll 1$), 
the distribution in phonon momenta is isotropic in the (x,z)-plane while in
 $y$-direction, $q_y\sim\lambda q_x$, therefore we have $q_y\ll q_z
\lesssim q_x$. In the intermediate range of magnetic fields we have 
$\lambda\sim1$ and $d\ll a$ therefore the distribution in phonon 
momenta is determined only by inequality (\ref{eq14}).

\subsection{Energy relaxation of a test electron}

The energy relaxation rate for a test electron between two subbands 
$nl$ and $n^{\prime}l^{\prime}$ due to PO phonons can be represented 
in the form \cite{gantlev} 
\begin{eqnarray}
{\cal Q}^{PO}_{nl\rightarrow n^{\prime}l^{\prime}}(k)={\cal Q}%
^{+}_{nl\rightarrow n^{\prime}l^{\prime}}(k)- {\cal Q}^{-}_{nl\rightarrow
n^{\prime}l^{\prime}}(k)  \label{eq115}
\end{eqnarray}
where 
\begin{eqnarray}
{\cal Q}^{\pm}_{nl\rightarrow n^{\prime}l^{\prime}}(k)={\frac{%
\hbar\omega_{PO}}{\bar{\tau}_{PO}}} {\frac{v_{PO}}{v(k\mp q^\pm_x)}}%
\Phi^\pm_{nl}(k) G^\pm_{nl\rightarrow n^{\prime}l^{\prime}}(k)\equiv{\frac{%
\hbar\omega_{PO}}{\tau^\pm_{PO}(\varepsilon)}}  \label{eq116}
\end{eqnarray}
is the energy transferred to the lattice (obtained from the lattice) at the
PO phonon emission (absorption) processes, ${1}/{\tau^\pm_{PO}(\varepsilon)}$
is the PO phonon emission (absorption) rate. Here $v(k)=1/\hbar
\partial\varepsilon(k)/\partial k$ is the group velocity of the edge state
which depends only on the momentum $k$ for the parabolic confinement. The
overlap integral $G^{\pm}_{nl\rightarrow n^{\prime}l^{\prime}}(k)$ is given
as 
\begin{eqnarray}
G^\pm_{nl\rightarrow n^{\prime}l^{\prime}}(k)={\frac{1}{4\pi}}%
\int^{+\infty}_{-\infty}{\frac{dq_ydq_z }{q^2_x+q^2_y+q^2_z}}%
|Q_{ll^{\prime}}(q_\lambda)|^2|I^{PO}_{nn^{\prime}}(q_z)|^2 \Biggl|%
_{q_x=q^\pm_x}.  \label{eq17}
\end{eqnarray}
Integration over phonon momenta does not catch on the temperature factor $%
\Phi^\pm_{nl}(k)$ which is a function of the electron initial quantum
numbers, 
\begin{equation}
\Phi^\pm_{nl}(k)=(N_{T}(\omega_{PO})+1/2 \pm1/2) {\frac{1-f_{T}(%
\varepsilon_{nl}(k)\mp\hbar\omega_{PO}) }{1-f_{T}(\varepsilon_{nl}(k)}},
\label{eq18}
\end{equation}
$f_{T}(\varepsilon)$ is the Fermi factor. Therefore, any variation of
temperature will change only intensity of the phonon emission and absorption
while momentum distribution of emitted and absorbed PO phonons by the test
electrons does not depend on temperature.

In the case of relaxation between edge states of the lowest subband for
actual calculations of the overlap integral $G^\pm_{nl\rightarrow
n^{\prime}l^{\prime}}(k)$ we use expression (\ref{eq6}) for the form factor $%
I_{nn^{\prime}}$. Then directly taking integration over $q_z$, Eq.~\r{eq17}
can be reduced to the following one-dimensional integral 
\begin{eqnarray}
G^\pm_{0l\rightarrow 0l^{\prime}}(k)={\frac{1}{16}}\int_0^\infty dy
Q^2_{ll^{\prime}}(t) {\frac{8+9r+3r^2}{r(1+r)^3}},  \nonumber \\
t={\frac{a^2}{2d^2}}(y^2+\lambda^2x^2),\,r=\sqrt{x^2+y^2},\,x=q_xd,\,y=q_yd.
\label{eq19}
\end{eqnarray}
which we evaluate numerically. 
\begin{figure}[htb]
\epsfxsize=14cm
\epsfysize=11cm
\mbox{\hskip 0cm}\epsfbox{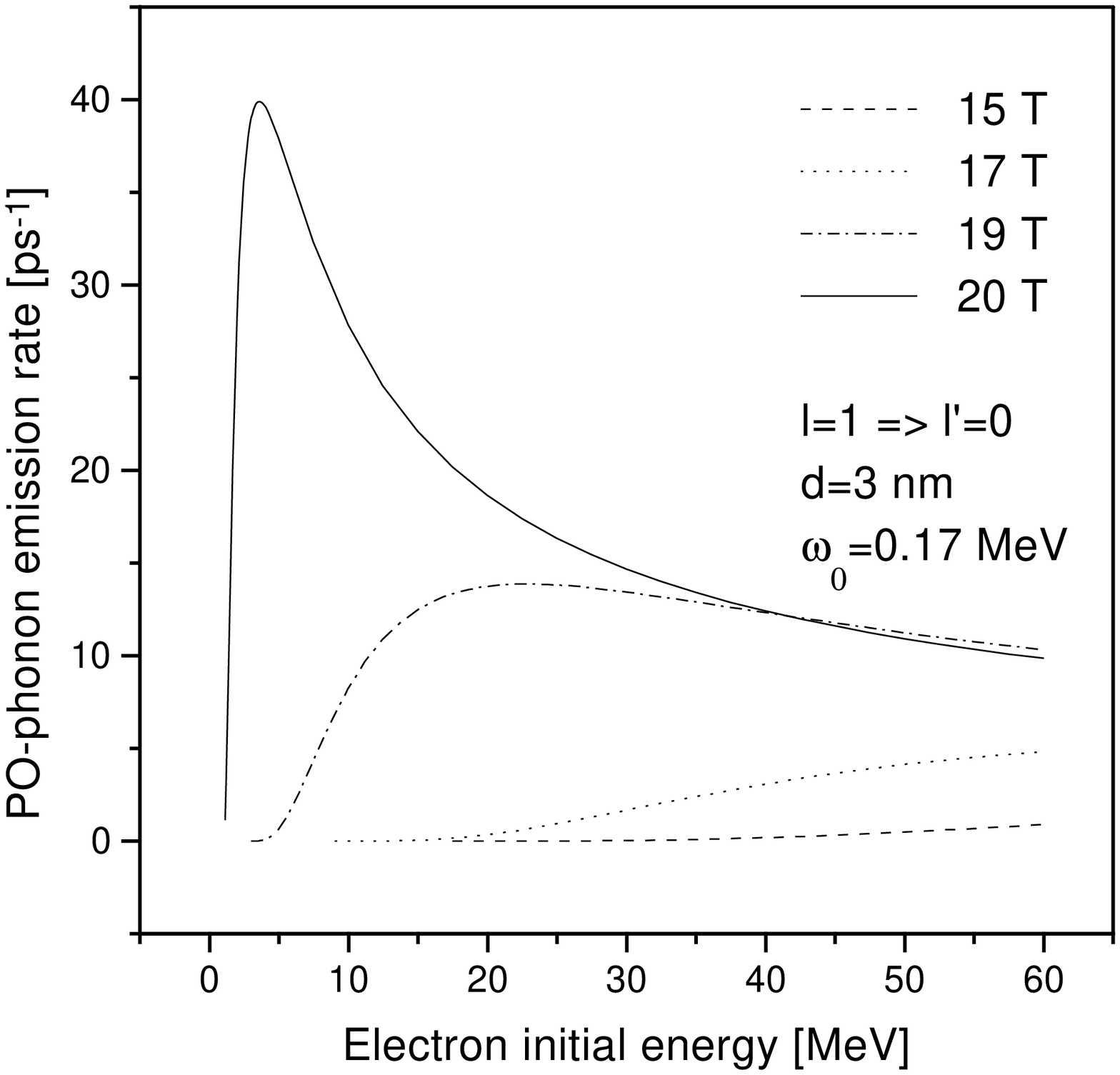}
\caption{The PO phonon emission rate versus the electron initial 
energy for the $l=1 => l=0$ transition and for different values of 
the magnetic field.}
\label{fgloemiss1}
\end{figure}
The results of numerical calculations in the limit of the smooth confinement
(we take $\omega_0=1.754$ meV) for transitions between hybrid subbands $%
l=1\rightarrow l^{\prime}=0$ and $l=2\rightarrow l^{\prime}=0$ are shown in
Figs.~\ref{fgloemiss1} and 
\begin{figure}[htb]
\epsfxsize=14cm
\epsfysize=11cm
\mbox{\hskip 0cm}\epsfbox{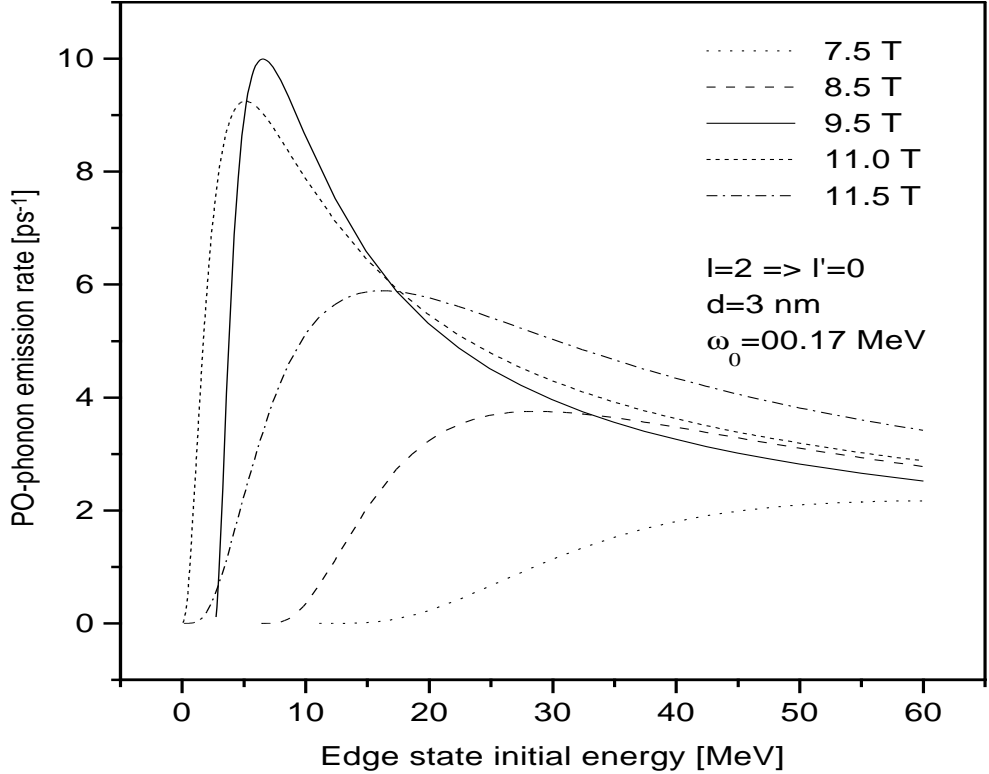}
\caption{The PO phonon emission rate versus the electron initial 
energy for the $l=2 => l=0$ transition and for different values of 
the magnetic field.}
\label{fgloemiss2}
\end{figure}
\ref{fgloemiss2}, respectively. We use $d=3$ nm as a typical value for a
GaAs/AlGaAs heterostructure. The diagrams represent the PO phonon emission
rate dependencies on the electron initial energy at room temperatures for
several values of the magnetic field. One can see from Fig.~\ref{fgloemiss1}
that for the transition $l=1\rightarrow l^{\prime}=0$ at low magnetic fields 
$B=15,\; 17$ T corresponding to the detuning $\Delta^+_{0100}=9.89,\; 
6.38$ meV,
the emission rate increases slowly but monotonously thought it is remaining
sufficiently small in the whole range of the electron initial energy variation. 
It is exceeding the value of $1$ ps$^{-1}$ only near the upper
edge of the energy variation. Such behavior is conditioned by the
sufficiently large value of the transferred momentum $q_x$ (so that the
inequality (\ref{eq10}) is not satisfied) and by its monotonous decrease in
the same energy range. The curves of the emission rate for the magnetic
fields $B=19,\; 20$ T corresponding to the detuning $\Delta^+_{0100}=2.87, \;
1.12$ meV represent peaks at the energies for which the inequality (\ref{eq10})
starts to take place. The peak values exceeding $40$ ps$^{-1}$ and $10$
ps$^{-1}$, respectively. At higher energies, the inequality (\ref{eq10}) 
takes place in the strong sense, $\lambda q_x a \ll 1$, therefore the
overlap integral (\ref{eq17}) becomes weakly depending on the energy and the
energy dependence of the emission rate is mainly determined by the behavior 
of the DOS, {\it i.e.}, it slowly decreases with an energy increase.
On the low-energy side we have $\lambda q_x a \gg 1$ causing an exponential
increase of the emission rate with energy increase. The curve with $B=20.64$ T
represents the emission rates corresponding to the very sharp detuning ($%
\Delta^+_{0100}<0.5$ meV). In this case we have $\lambda q_x a \ll 1$ in the
whole energy range and features of the emission rate are mainly determined
by an energy dependence of the DOS. Particularly, the sharp
increase of the emission rate at low energies arises from the divergence of
the DOS at the bottom of the electronic subband.

Such kinds of curves with analogous behavior we find also for the
electronic transitions between subbands $n=0,l=2$ and $n^{\prime}=0,l^{%
\prime}=0$ (Fig.\ \ref{fgloemiss2}).

\chapter{Theory of Auger-upconversion}
\label{upconversion}

\section{Introduction}
\label{augint}

Auger scattering in semiconductors is well known from investigations of
nonradiative recombination 
\cite{landsberg,haugaug,smith,takeshima,bockel}. Free electrons and holes are a
prerequisite for this process: the energy obtained in the recombination of
an electron-hole pair is taken to excite another electron. The latter
electron may loose its excess energy by electron-lattice relaxation; thus
the recombination energy is converted into heat. More recently two-particle
correlation effects in the Auger process \cite{stolz} have been demonstrated
experimentally for $p$-GaAs and $n$-Si \cite{hangleiter,hangleit}. In
quantum well structures, Auger processes become possible between different
subbands and have been investigated both in theory and experiment (see \cite
{borenstain} and references therein). The reduction of Auger scattering rate
in quantum dots due to the discreteness of the electronic states has been
used as an argument to propose quantum dots lasers \cite{pan}. Except for an
early study of transport in crossed electric and magnetic fields \cite
{zlobin}, the theory of Auger scattering was restricted so far to the
magnetic field free case.

\begin{figure}[htb]
\epsfxsize=11cm
\epsfysize=13cm
\mbox{\hskip 1.5cm}\epsfbox{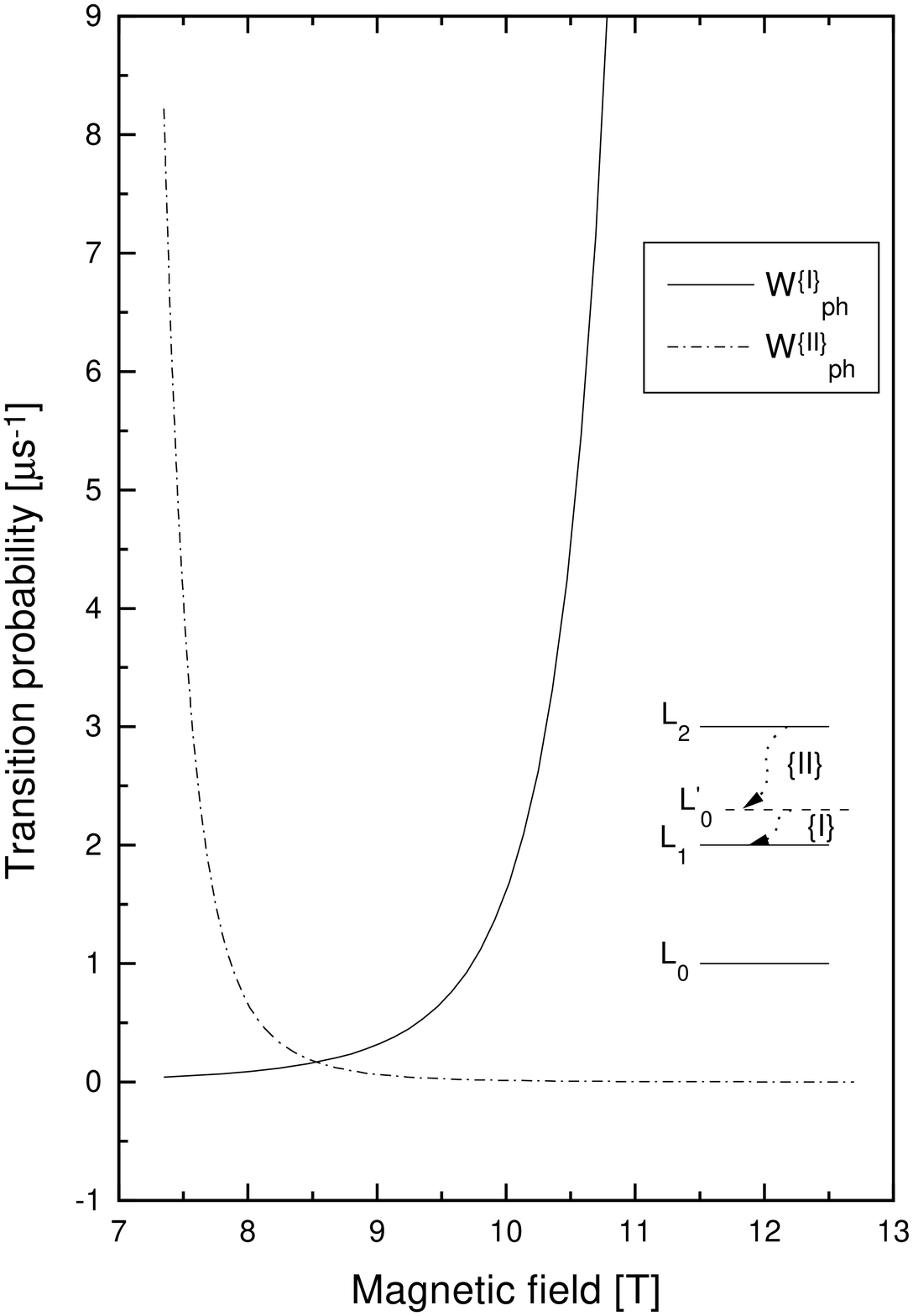}
\caption{Magnetic field dependence of the probabilities for the transitions
between Landau levels $L_{0}^{\prime}\rightarrow L_{1} (\{I\})$ and 
$L_{2}\rightarrow L_{0}^{\prime} (\{II\})$ caused by acoustic phonon 
emission.}
\label{phmagdep}
\end{figure}
In recent magneto-luminescence experiments by Potemski {\it et al.} 
\cite{potemski1,potemski2} on one-side modulation doped GaAs/AlGaAs 
quantum wells
an up-conversion has been observed and interpreted as being due to an Auger
process. These authors studied in photoluminescence and photoluminescence
under excitation an asymmetric GaAs/AlGaAs single quantum well of width 
$d=25$ nm with an electron density of $N_{s}=7.6\cdot 10^{11}$ cm$^{-2}$ 
with a magnetic field ${\bf B}$ applied in growth direction. 
The characteristic energy level scheme for $7.9$ T $\leq B\leq 12.9$ T is
depicted as inset in Fig.~\ref{phmagdep}: the lowest Landau level of the
second electric subband $L_{0}^{\prime}$ lies between the second $L_{1}$
and third $L_{2}$ Landau levels of the lowest electric subband (the index
refers to the Landau quantum number), while due to the doping concentration
the Fermi energy is pinned to the level $L_{1}$. 
\begin{figure}[htb]
\epsfxsize=11cm
\epsfysize=13cm
\mbox{\hskip 1.5cm}\epsfbox{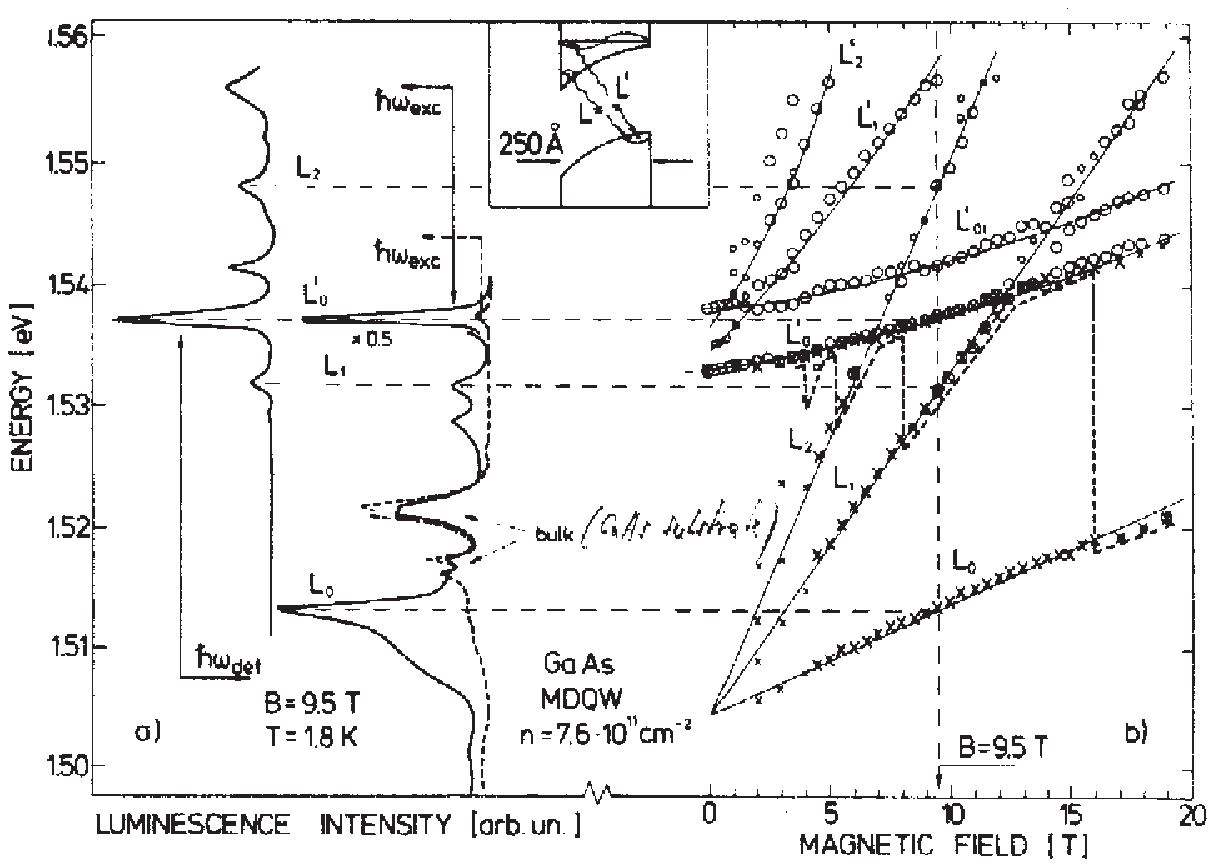}
\caption{(a) Intensity of the luminescence detected at $\h\o_{det}$ as
a function of the exciting energy (left) and luminescence spectra at 
different excitation energies $\h\o_{exc}$ (right). Two-dimensional
and bulk structures are observed when the excitation energy corresponds
to the peak of the two-dimensional density of states (solid line), and
mainly bulk-related luminescence is visible when exciting in the gap 
between the two-dimensional levels (dashed line). (b) The Landau level 
fan chart of the optically active transitions observed in luminescence
(crosses) and luminescence-excitation (open circles) spectra. The size
of the symbol reflects the transition intensity. The $L^{\prime}_{01}$
absorption line involves the light-hole level. The Fermi-level position
is shown with the dashed line.}
\label{lum}
\end{figure}
Changing the magnetic field in this interval allows to tune $L_{0}^{\prime}$
between $L_{1}$ and $L_{2}$. The luminescence spectrum under interband
excitation into $L_{1}$ at low temperatures ($ T=1.8$ K) and for low excitation
power ($P_{exc}\leq 10$ W cm$^{-2}$) shows two peaks 
(see Fig.~\ref{lum} taken from Ref.~\cite{potemski1}):
besides the luminescence due to recombination of an electron from $L_{0}$
with a hole in a valence band, a second peak is observed above the exciting
laser energy and is related to $L_{0}^{\prime}$, {\it i.e.} recombination of
an up-converted electron with a hole. In order to explain this second peak,
the following processes have been supposed: after an interband excitation of 
electrons into the partially filled level $L_{1}$ (i), a recombination takes place 
between electrons from $L_{0}$ and photo-induced holes (ii), 
then in an Auger process two electrons in $L_{1}$ are scattered to $L_{2}$ 
and $L_{0}$ (iii), and a relaxation process brings the electron from $L_{2}$ 
to $L_{0}^{^{\prime}}$ (iv), from where it recombines with a photo-induced
hole (v) to give the up-converted luminescence or, emitting a phonon, relaxes 
into the level $L_{1}$ (vi). 
\begin{figure}[htb]
\epsfxsize=11cm
\epsfysize=13cm
\mbox{\hskip 1.5cm}\epsfbox{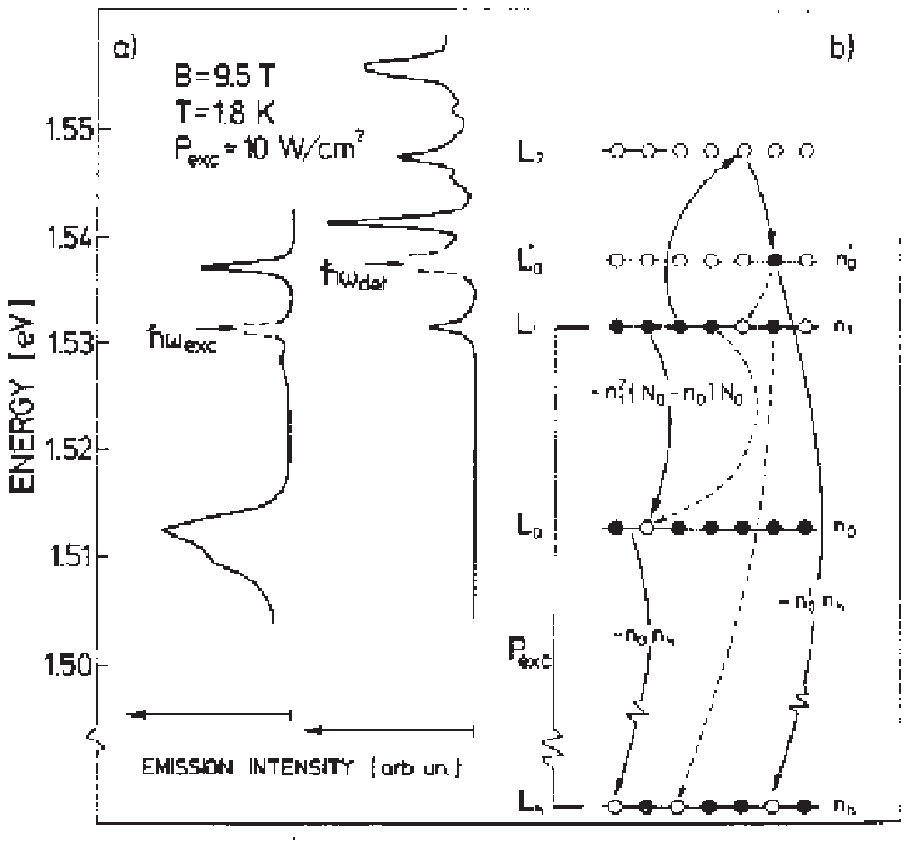}
\caption{(a) Solid circles: variation of the peak intensity ($I_0^{\prime}$)
of the above-laser emission as a function off the magnetic-field-dependent 
filling factor $\nu$ or the separation $\Delta$ between the excitation- 
and the emission-peak energies. The peak intensity of the $L_0$-related 
luminescence ($I_0$) is shown with open circles. Solid lines are guides for
the eye. (b) Power dependence (in relative units) of the $I_0^\prime$ and
$I_0$ when exciting into the $L_1$ level. Solid lines represent quadratic
(for $I_0^\prime$) and linear (for $I_0$) variations.}
\label{deplum}
\end{figure}
Potemski {\it et al.} \cite{potemski1,potemski2} present the dependence
of the luminescence intensity related to $L_{0}$ and $L_{0}^{\prime}$ 
on the magnetic field and the excitation power (see Fig.~\ref{deplum} taken
from Ref.~\cite{potemski1}). The 
most surprising result is the high intensity of the $L_{0}^{\prime}$ 
luminescence which can be of the same order as the $L_{0}^{\prime}$ 
luminescence.

Experimentally, electron-electron interaction has been studied in various
phenomena. Besides already mentioned luminescence experiment
\cite{potemski1,potemski2}, inter- and intra-Landau level Auger transitions 
have been observed also in the cyclotron resonance in the 2DEG \cite{maran}.
As a dephasing mechanism electron-electron interaction is observed in 
interference of two electron beams in the 2DEG \cite{yacoby}. It can be 
responsible for the signal decay in four wave mixing experiment \cite{bar-ad}
and consider as a possible mechanism \cite{price83,solomon} of the mutual
drag between two electron gases, 2D-3D, and 2D-2D \cite{gramila,trugman}.

Electron-electron scattering is one of the main mechanisms controlling relaxation,
 transport and optical properties of electrons. The electron scattering has been 
firstly calculated by Landau and Pomeranchuk \cite{pomeranchuk} in massive 
samples. In recent years,
intensive effort of many authors has been exhausted to develop theory of 
electron-electron scattering in low-dimensional nanostructures 
\cite{afs,esipov,schmid1}. In the 2DEG, electron scattering has been 
calculated by Chaplik \cite{chaplik} at low and by Gulianni and Quinn 
\cite{quinn} at high temperatures for free Fermi-liquid electrons. 
When electrons are strongly scattered by random
short range potential of impurities, the Fermi-liquid picture can be no
longer valid. In this situation electron-electron scattering has been
discussed by Schmid \cite{schmid2} and by Altshuler and Aronov 
\cite{aronov2} for the magnetic field free case.

As already mentioned, theoretical studies of Auger scattering between Landau
levels in a 2DEG are missing in the literature. In this section phenomenological
calculations of the electron-electron scattering rate for free electrons of the 
2DEG exposed to the normal quantizing field are presented 
\cite{badalauger,badalrspt}.

Quite recently these calculations have been generalized by Levinson \cite
{lev95} for a smooth random potential depending on only one coordinate which
allow to calculate Landau level broadening corresponding to heterostructures
and narrow quantum wells with a large spacer \cite{gornik87,dassarma88}.

In the last few years electron-acoustic phonon interaction in a 2DEG in a
quantizing magnetic field has attracted attention because of its role in the
breakdown of the dissipationless quantum Hall effect \cite{heinonen,eaves86}
and the cooling of a 2DEG at low temperatures $ T\leq 40K$ \cite
{prasad,reinen,tamura}. The emission and absorption of the ballistic phonon
pulses by a 2DEG in MOS-structures and GaAs-heterostructures have been
studied in theoretical works \cite{toombs,badalfree,badalbal,badalsurf} as
well as in many experiments (see the review paper \cite{challiskntrmp}). In
all calculations, however, only acoustic phonon-assisted transitions within
Landau levels of the lowest subband are considered.

Therefore in this chapter, we calculate the characteristic times of
processes (iii) and (iv), (vi), {\it i.e.} Auger scattering between Landau
levels of the lowest electric subband and electron-acoustic phonon
scattering between Landau levels of the two lowest electric subbands ($%
L_{2}\rightarrow L_{0}^{\prime}$ and $L_{0}^{\prime}\rightarrow L_{1}$), as
well as the lifetime of a test hole in level $L_{0}$ with respect to both
the Auger process and the phonon emission. By analyzing rate equations for
the processes (i)-(vi) we find an estimate for the time of the Auger process
as well as magnetic field and excitation power dependencies of the two
luminescence peaks which are consistent with the experimental findings.

\section{Auger scattering between Landau levels}
\label{aug} 

Due to the combined effect of quantum well confinement and Landau
quantization, the energy spectrum of electrons in a quantum well with
magnetic field in growth direction (parallel $z$) is discrete. The single
particle energy $\varepsilon_{nl}$ is characterized by a subband index $n$
and a Landau level index $l$ (the spin degeneracy is not removed and is
taken into account in the occupation factors). The corresponding wave
functions $\Psi({\bf r})$ (${\bf r}$=(${\bf R},z)$) are factored into an
oscillator function $\chi _{lk}({\bf R})$ for the Landau oscillator and a
subband function $\psi _{n}(z)$. The single particle energy does not depend
on the quantum number $k$ which results from the asymmetric gauge of the
vector potential and counts the degeneracy of the Landau levels.

The scattering time for a single electron due to the Auger process, in which
two electrons are scattered from single particle states $1,2$ into states $%
1^{\prime},2^{\prime}$, is given by 
\begin{equation}
{\frac{1 }{\tau ^{e}_{Auger}}}= \sum _{k_{2},k_{1}^{\prime},k_{2}^{\prime}}
W_{1,2 \rightarrow
1^{\prime},2^{\prime}}f_{2}(1-f_{1^{\prime}})(1-f_{2^{\prime}})
\label{augerscat}
\end{equation}
where the occupation probabilities $f$ take into account the Pauli exclusion
principle. In detail the transition probability is given by 
\begin{equation}
W_{1,2 \rightarrow 1^{\prime},2^{\prime}}={\frac{2\pi }{\hbar}} \left|
M_{1,2 \rightarrow 1^{\prime},2^{\prime}} \right|^{2} \delta
(\varepsilon_{n_{1}l_{1}}+\varepsilon_{n_{2}l_{2}}-\varepsilon_{n_{1}^{%
\prime}l_{1}^{\prime}}-\varepsilon_{n_{2}^{\prime}l_{2}^{\prime}}).
\label{augerprob}
\end{equation}
In Eq.~(\ref{augerprob}) we have taken the sum over the momenta $k$ to
account for all equivalent scattering processes, which are possible due to
the degeneracy for a given set of Landau levels. After the summation the
result does not depend on $k_{1}$ (and thus on the choice of the gauge, as
it should be). Thus $\tau ^{e}_{Auger}$ is the relaxation time of a test
electron from the level $L_{1}$ to $L_{0}$ with respect to the Auger
process. For later considerations we introduce also the lifetime of a test
hole in the level $L_{0}$ with respect to the Auger process from the level 
$L_{1}$ as 
\begin{equation}
{\frac{1 }{\tau _{Auger}^{h}}} = \sum _{k_{1},k_{2},k_{2}^{\prime}} W_{1,2
\rightarrow 1^{\prime},2^{\prime}}f_{1}f_{2}(1-f_{2^{\prime}})
\label{hauger}
\end{equation}
where $1-f_{2}^{\prime}$ is the occupation probability of a hole in $L_{2}$.

The matrix element $M_{1,2 \rightarrow 1^{\prime},2^{\prime}}$ to be
calculated with functions of Eq.~\r{} is that of the Coulomb interaction
potential after 2D Fourier transformation 
\begin{equation}
V_{ee}({\bf r_{1}},{\bf r_{2}})={\frac{1 }{A}} \sum_{{\bf q}} {\frac{2\pi
e^{2} }{\kappa _{0} q}} \exp\left\{i {\bf q} ({\bf R_{1}}- {\bf R_{2}})-q
\mid z_{1}-z_{2}\mid \right\}  \label{coulombfour}
\end{equation}
where $A$ is the normalization area in the $(x,y)$-plane, $\kappa _{0}$ is
the low-frequency dielectric constant. Here the screening due to free
carriers in the electron subband is not considered.

In our special case we are interested in transitions only within the first
electric subband $(n=1)$. Therefore, using the form factor 
\begin{equation}
F(q)=\int dz_{1}dz_{2} \psi _{1} ^{2} (z_{1}) \psi _{1} ^{2} (z_{2})
\exp\left\{-q\mid z_{1}-z_{2} \mid\right\}  \label{coulff}
\end{equation}
the matrix element can be represented in the form 
\begin{eqnarray}
M_{1,2 \rightarrow 1^{\prime},2^{\prime}}& =& {\frac{1 }{A}} \sum_{{\bf q}}{%
\frac{2 \pi e ^{2} }{\kappa _{0} q}} F(q)
Q_{l_{1}l_{1}^{\prime}}(q)Q_{l_{2}l_{2}^{\prime}}(q)\delta_{k_{1}^{%
\prime},k_{1}+q_{x}} \delta_{k_{2}^{\prime},k_{2}-q_{x}}  \nonumber \\
&\times& \exp\left[{\frac{1 }{2}} i a^{2}_{B} q_{y}
(k_{1}+k_{1}^{\prime}-k_{2}-k_{2}^{\prime})-i\phi(l_{1}^{\prime}+l_{2}^{%
\prime}-l_{1}-l_{2})\right].  \label{eematrix}
\end{eqnarray}
Here $a_{B}=(\hbar /\mid e \mid B)^{1/2}$ is the magnetic length, $\phi$ is
the polar angle of the vector ${\bf q}$. The functions $Q_{ll^{\prime}}$ are
given by Eq.~(\ref{ffmagnetic}).   
Using  the model wavefunction (\ref{wfmodel}) \cite{fanghow},
the form factor \r{coulff} can be calculated explicitly
\be
F(q)={{{10\,{b^6}\,{e^{-b + q}}}\over 
         {{{\left( b + q \right)}^6}}} + 
      {b\left({8\,{b^2} + 9\,b\,q + 3\,{q^2}}\right)\over 
         {16\,{{\left( b + q \right) }^3}}}}.
\la{eeff}
\ee

It is easy to see from Eqs.~(\ref{augerscat})-(\ref{hauger}) that the
momenta $k$ appear only in the matrix element so that using (\ref{eematrix})
one can find 
\begin{equation}
\sum_{k_{2},k_{1}^{\prime},k_{2}^{\prime}} \left| M_{1,2 \rightarrow
1^{\prime},2^{\prime}} \right| ^{2} = {\frac{1 }{A}} {\frac{1}{2\pi a_{B}^{2}%
}} {\frac{4\pi ^{2}e^{4} }{\kappa _{0}^{2}}} \sum_{{\bf q}} {\frac{1}{q^{2}}}
F^{2}(q)Q^{2}_{l_{1}l_{1}^{\prime}}(q) Q^{2}_{l_{2}l_{2}^{\prime}}(q).
\label{aveematrix}
\end{equation}
It should be noted that after performing the summation of the modulus
squared of the matrix element over the momenta $k$ no interference between
different Fourier components of the Coulomb potential occurs.

So far we have not considered the $\delta$-function in Eq.~(\ref{augerprob})
which would give a factor of infinity because of energy conservation. For
a realistic 2DEG, impurity scattering results in a broadening of the
Landau levels \cite{ando74}. The same mechanism also limits the mobility of
the carriers in the system (see Chapter \ref{pemissLL}). Therefore, we replace 
the $\delta$-function by a Lorentzian with a width corresponding to the
scattering time $\tau \approx 10$ ps for the mobility $\mu =25$ V$^{-1}$ 
s$^{-1}$ m$^{2}$ which means $\pi \delta(\e_{n_{1}l_{1}}+\e_{n_{2}l_{2}}-
\e_{n_{1}^{\prime}l_{1}^{\prime}}-\e_{n_{2}^{\prime}l_{2}^{\prime}})
\rightarrow \tau$. This approach is inconsistent in so far as impurity
scattering, in principle, gives $k$-depended energies, however, we perform
the $k$-sum without taking this into account.

\begin{figure}[htb]
\epsfxsize=14.5cm
\epsfysize=12.7cm
\mbox{\hskip 0cm}\epsfbox{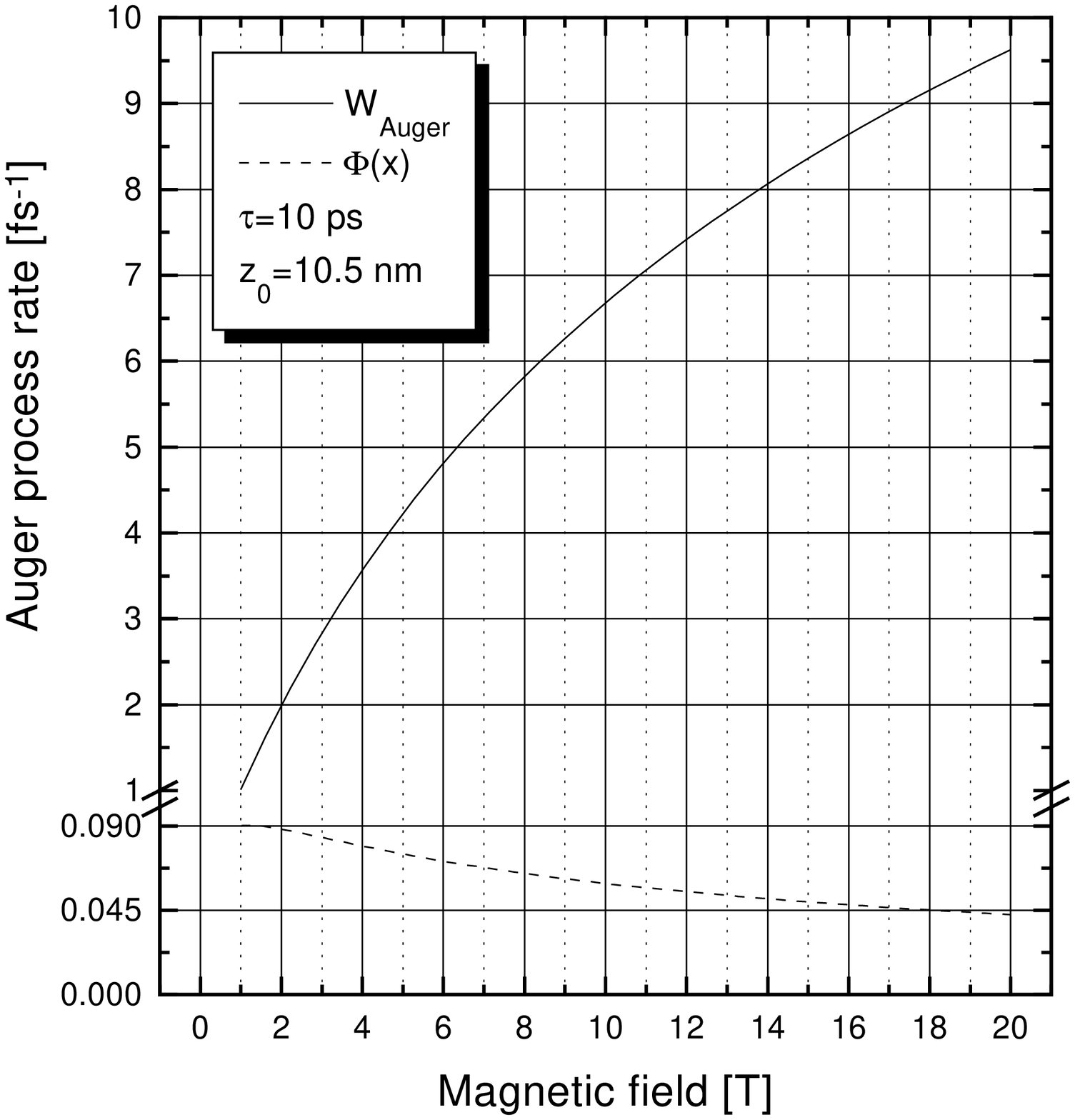}
\caption{Probability of the Auger process $W_{Auger}$ from the level $L_{1}$
into the levels $L_{0}$ and $L_{2}$. Dashed line shows the overlap integral
$\Phi(a_B/z_0)$.}
\label{ovlplot}
\end{figure}

Now from (\ref{augerscat}), (\ref{augerprob}), and (\ref{aveematrix}) we
obtain 
\begin{equation}
{\frac{1}{\tau ^{e}_{Auger}}}=
W_{Auger}f_{2}(1-f_{1^{\prime}})(1-f_{2^{\prime}}) \bigg|%
_{l_{1}+l_{2}=l_{1^{\prime}}+l_{2^{\prime}}}  \label{augertime}
\end{equation}
and the probability of the Auger process is 
\begin{equation}
W_{Auger}=2\left ({\frac{e^{2} }{\hbar}}\right)^{2}{\frac{\tau }{\kappa
_{0}^{2}}} {\frac{1}{a_{B}^{2}}}\Phi\left({\frac{a_{B}}{z_{0}}}\right).
\label{augprob}
\end{equation}
The overlap integral 
\begin{equation}
\Phi\left({\frac{a_{B}}{z_{0}}}\right)=\int_{0}^{\infty} {\frac{dq }{q}}
F^{2}(q)Q^{2}_{l_{1}l_{1}^{\prime}}(q)Q^{2}_{l_{2}l_{2}^{\prime}}(q) \bigg|%
_{l_{1}+l_{2}=l_{1^{\prime}}+l_{2^{\prime}}}  \label{overlap}
\end{equation}
depends on magnetic field via the dimensionless parameter ${a_{B}/ z_{0}}$
where $z_{0}$ is a characterizing length parameter of the lowest electric
subband. We have calculated the overlap integral $\Phi({a_{B}/ z_{0}})$ for
the Auger process involving Landau levels $l_{1}=l_{2}=1$ and 
$l_{1}^{\prime}=0$, $l_{2}^{\prime}=2$ (Auger process from the level $L_{1}$ 
into the levels $L_{0}$ and $L_{2}$) using the form factor \r{eeff} (see the lower
part of Fig.~\ref{ovlplot}). The value of the parameter $z_{0}=3/b=10.5$ nm has been
chosen to reproduce the separation of the two lowest subbands of the actual
quantum well in \cite{potemski1} and \cite{potemski2} by a triangular potential 
model. 
The magnetic field dependence of the probability of Auger process $W_{Auger}$
is also plotted in Fig.~\ref{ovlplot}.
One can see that in the range of magnetic fields between $7.9$ T and $12.9$ T, 
the function $\Phi$ is slowly varying function in the magnetic field so that the 
probability $W_{Auger}$ is a {\it linearly} increasing function in $B$ in the same 
range.
When the magnetic field is varied from $1$ to $20$ T, the probability
$W_{Auger}$ increases approximately from $1$ to $10$ fs. Notice that,
recently, such a fast electron-electron thermalization (faster than $10$ fs)
has been observed in modulation-doped GaAs quantum wells \cite{knox}.

However, the occupation factors, which have to be included in order to
obtain $\tau ^{e}_{Auger}$, will drastically increase this value. For the
case under consideration in the experiment of \cite{potemski1} and \cite
{potemski2} (with $N_{s}=7.6\cdot10^15$ cm$^{-2}$, $ T=1.8$ K and 
$B=9.5$ T ) but without pumping, we find by including the occupation factors 
$\tau^{e}_{Auger} \simeq 10^{63}W_{Auger}^{-1}$, {\it i.e.} the Auger
process is not possible at all because the lower Landau level $L_{0}$ is 
almost completely filled. It becomes possible only by optical pumping into the
level $L_{1}$ and subsequent recombination from $L_{0}$, thus creating the
empty states required for the Auger process. Without pumping, the Auger
process is possible only at much lower magnetic fields (for lower carrier
density) or much higher temperatures.

Applying the same considerations to $\tau _{Auger}^{h}$ we obtain an
expression as (\ref{augertime}) but with the factor $f_{1}$ instead of the
factor $1-f_{1}^{\prime}$. This gives for half-filled $L_{1}$ level $\tau
_{Auger}^{h} \approx 2$ fs. In contrast to $\tau _{Auger}^{e}$ we see that $%
\tau _{Auger}^{h}$ does not depend on available free places in the level $%
L_{0}$ and shows the efficiency of the Auger process in comparison with
other processes which add (by emission of phonons) or remove (by interband
recombination) electrons in $L_{0}$.

\section{Electron-acoustic phonon scattering between Landau levels}

Transitions between Landau levels of the different electric subbands are
possible by emission of acoustic phonons via the deformation (DA) potential. 
The corresponding relaxation time of a test electron is given by 
\begin{equation}
{\frac{1}{\tau _{nl\rightarrow n^{\prime}l^{\prime}}^{DA}}} =
(1-f_{n^{\prime}l^{\prime}})\sum_{k^{\prime}}W_{nlk \rightarrow
n^{\prime}l^{\prime}k^{\prime}}^{DA}  \label{ptime}
\end{equation}
where the transition probability from the state $\Psi _{nlk}$ to the state $%
\Psi _{n^{\prime}l^{\prime}k^{\prime}}$ is given by \cite{badalfree} 
\begin{equation}
W_{nlk \rightarrow n^{\prime}l^{\prime}k^{\prime}}^{DA} ={\frac{1}{A}}{\frac{%
\Xi ^{2}}{\hbar \rho s^{2}}}{\frac{\omega ^{2}}{s^{2}}}\sum_{{\bf q}}{\frac{1%
}{a}}\delta_{k^{\prime},k+q_{x}} Q^{2}_{ll^{\prime}}(q) \left|
I_{nn^{\prime}}(a) \right|^{2},\: a=\sqrt{{\frac{\omega^{2}}{s^{2}}}-q^{2}}.
\label{ptrprob}
\end{equation}
with the form factors $Q$ and $I$ given by \r{qff} and (\ref{iff}), respectively. 
Recall that $\Xi $ is the deformation potential constant, $\rho $ is the mass density
of the crystal, $s$ is the velocity of sound and the transferred energy 
$\hbar\omega=\varepsilon_{nl}-\varepsilon_{n^{\prime}l^{\prime}}$.

Direct calculation shows that the probability for the piezoelectric
potential (PA) can be obtained from (\ref{ptrprob}) by replacing $%
\Xi^{2}(\omega ^{2}/ s^{2})$ by $(e\beta)^{2}$ where $\beta$ is the
piezoelectric modulus of the crystal averaged over directions of propagation
of phonons and its polarizations. Therefore, for GaAs one can find 
\begin{equation}
{\frac{W^{DA} }{W^{PA}}}= {\frac{\Xi^{2}(\omega ^{2}/s^{2}) }
{(e\beta)^{2}}}=\left({\frac{\omega[meV]}{0.42}}\right)^{2}.  
\label{datopa3}
\end{equation}
This means that in strong quantizing magnetic fields ($B\sim 10$ T) when $%
L_{0}^{\prime}$ is sufficiently far away from $L_{1}$ and $L_{2}$, {\it i.e.}
it is far from the crossing point of the magnetic field when $\omega \ll
\omega _{B}$, the PA-interaction is suppressed with respect to the
DA-interaction.

From (\ref{ptrprob}), taking a sum over the momentum $k^{\prime}$ of the
final states, it is easy to find 
\begin{eqnarray}
W_{nl \rightarrow n^{\prime}l^{\prime}}&=&\sum_{k^{\prime}}W_{nlk
\rightarrow n^{\prime}l^{\prime}k^{\prime}}^{DA}= {\frac{\Xi ^{2}}{2\pi
\hbar \rho s^{2}}}{\frac{\omega ^{2}}{s^{2}}}  \nonumber \\
&\times& \int dz_{1}dz_{2} \psi _{n_{1}} (z_{1}) \psi _{n_{2}} (z_{1}) \psi
_{n_{1}}(z_{2})\psi _{n_{2}}(z_{2}) K\left({\frac{\omega }{s}}%
|z_{1}-z_{2}\right)  \label{pprob}
\end{eqnarray}
where the kernel is 
\begin{equation}
K\left({\frac{\omega }{s}}|z_{1}-z_{2}\right)=\int_{0}^{\omega /s} \!dq{%
\frac{q}{a}}Q^{2}_{ll^{\prime}}(q)\exp\left\{ia(z_{1}-z_{2})\right\}.
\label{ker}
\end{equation}
The following two transitions will be considered below: 
$$
n=2,\,l=0 \rightarrow n^{\prime}=1,\,l^{\prime}=1\quad
(L_{0}^{\prime}\rightarrow L_{1})\eqno\{I\} 
$$
with 
\begin{equation}
\hbar\omega=\hbar\omega_{1}=\varepsilon_{20}-\varepsilon_{11}=\omega_{E}-%
\omega_{B},\quad Q_{10}^{2}(q)=t\exp(-t)  \label{1tr}
\end{equation}
and

$$
n=1,\,l=2 \rightarrow n^{\prime}=2,\,l^{\prime}=0\quad (L_{2}\rightarrow
L_{0}^{\prime})\eqno{\{II\}} 
$$
with 
\begin{equation}
\hbar\omega=\hbar\omega_{1}=\varepsilon_{12}-\varepsilon_{20}=2\hbar%
\omega_{B}-\hbar\omega_{E},\quad Q_{20}^{2}(q)=(t^{2}/2) \exp(-t).
\label{2tr}
\end{equation}
Here $\omega_{B}=eB/mc$ is the cyclotron energy and $\hbar\omega_{E}=%
\varepsilon_{20}- \varepsilon_{10}$ is the energy separation between first
and second electric subband.

The kernel (\ref{ker}) for these two transitions can be rewritten 
\begin{equation}
K\left({\frac{\omega }{s}}| z_{1}-z_{2}\right)\bigg|_{\{I; II\}} ={\frac{1}{%
a_{B}^{2}}}\int_{0}^{\alpha}\!dt\{t; (t^{2}/2)\}{\frac{1}{a}}
\exp\left[ia(z_{1}-z_{2})-t\right].  \label{ker1}
\end{equation}
where $\alpha=(\omega ^{2}/2s^{2})a_{B}^{2}$ is an integration parameter. 
This integration has been evaluated by means of the steepest descent method using
the fact that in quantizing magnetic fields the parameter $\alpha=(\omega 
^{2}/2s^{2})a_{B}^{2}\simeq\omega_{B}/2ms^{2}\gg 1$ is always large. Then
we find 
\begin{equation}
K\left({\frac{\omega }{s}}| z_{1}-z_{2}\right)\bigg|_{\{I; II\}}= {\frac{%
\omega }{s}}{\frac{1}{2\alpha}}{\frac{(1-i\beta)^{2}}{(1+\beta^{2})^{2}}}
\left\{1; {\frac{1-i\beta }{1+\beta^{2}}}\right\}\exp\left[i{\frac{\omega}{s}%
}(z_{1}- z_{2})\right]+o\left({\frac{1}{\alpha}}\right)  \label{ker2}
\end{equation}
where the parameter $\beta =(\omega /s)(z_{1}-z_{2})/2\alpha$. For $\omega
\sim \omega_{B}$ and for $z_{1},z_{2} \sim d$ we have $\beta \sim
ms(z_{1}-z_{2}) \sim msd $. For GaAs the length $\hbar /ms = 340$ nm so that
at $d=25$ nm we have $\beta\ll 1$. Hence, neglecting the terms of the higher
order in $\beta$ we obtain 
\begin{equation}
K\left({\frac{\omega }{s}}|z_{1}-z_{2}\right)\bigg|_{\{I; II\}} \approx {%
\frac{\omega }{s}}{\frac{1}{2\alpha}}\exp\left[i{\frac{\omega}{s}}
(z_{1}-z_{2})\right].  \label{ker3}
\end{equation}
One can see that in this approximation the kernel differs for the two
transitions only by the different values of the transferred energy $\omega$.
This result for $K$ is easy to find from (\ref{ker}) by replacing $a$ by $%
\omega /s$. This means that in a 2DEG in a quantizing magnetic field normal
to the electron sheet, electrons due to the DA interact mainly with phonons
which propagate along the magnetic field direction ${\bf B}$. It should be 
noticed that this statement is true not only for these particular transitions but 
also in general.

Substituting the kernel (\ref{ker3}) in (\ref{pprob}) we obtain the
probability for the transitions $\{I\}$ and $\{II\}$ 
\begin{equation}
W_{ph}^{\{I; II\}}={\frac{1}{\bar{\tau} _{B}}}{\frac{\omega }{s}}a_{B}
\left|I_{12}\left({\frac{\omega }{s}}\right)\right| ^{2},\quad
\omega=\{\omega_{1};\omega _{2}\}.  
\label{prob1,2}
\end{equation}
Here we define a nominal interaction time 
\begin{equation}
{\frac{1}{\bar {\tau} _{B}}}={\frac{\Xi ^{2} }{2\pi \hbar \rho s^{2}
a_{B}^{3}}}
\label{nomtime2}
\end{equation}
which depends on the magnetic field.
Since the parameter $(\omega/s)d \gg 1$ (actually, this inequality defines
the energy range in which we are interested), the form factor $I_{12}$ can
be calculated by the method of a stationary phase \cite{erdelyi,erdelyi1}.
Using Airy functions instead of wave functions $\psi_{n}(z)$ one can find 
\begin{equation}
\left| I_{12}({\frac{\omega}{s}})\right|^{2}=4\left ({\frac{\omega}{s}}
\bar{z}\right)^{-6}, \quad \bar{z} ={\frac{\hbar }{\sqrt{2m\omega_{E}/
(\alpha_{1}-\alpha_{0})}}}  
\label{iffairy}
\end{equation}
where $\alpha_{i}$ is the $i$th zero of the Airy function.

Substitute (\ref{iffairy}) in (\ref{prob1,2}) we find 
\begin{eqnarray}
W_{ph}^{\{I\}}= {\frac{1}{\bar {\tau}_{B}}}{\frac{a_{B} }{\bar{z}}}{\frac{%
4(s/\bar{z})^{5} }{(\omega_{E}-\omega_{B})^{5}}} \propto B(B_{E}-B)^{-5},
\label{ph1} \\
W_{ph}^{\{II\}}= {\frac{1}{\bar {\tau} _{B}}}{\frac{a_{B} }{\bar{z}}}
{\frac{%
4(s/\bar{z})^{5} }{(2\omega_{B}-\omega_{E})^{5}}}\propto B(2B-B_{E})^{-5}
\label{ph2}
\end{eqnarray}
where $B_{E}$ is the magnetic field when $\omega_{B}=\omega_{E}$. For GaAs
at $B=9.5$ T and $\bar{z}=6.7$ nm (this corresponds to $\o_{E}=22.4$ 
meV taken from \cite{potemski1} and \cite{potemski2}), we find 
$W_{ph}^{\{II\}}=(32.3\; \mu\mbox{s})^{-1}$ and $W_{ph}^{\{I\}}
=(1.5\; \mu \mbox{s})^{-1}$. Note, that the relaxation probability 
calculated in \cite{badalfree} for magnetic fields $B\simeq 1T$ is much 
greater. Such a
suppression of the electron-phonon interaction at $\omega \sim \omega_{B}$
and $(\omega_{B}/s)\bar{z} \gg 1$ follows from the conservation laws. The
electron states in this regime constitute a wave packet so that the states
with $l, n\sim 1$ have momenta of the order of $a_{B}^{-1}$ in the $(x,y)$%
-plane and of the order of $\bar{z}^{-1}$ in $z$-direction. Therefore, only
for a small number of electron states in this packet, the momentum
conservation law is fulfilled at the interaction with acoustic phonons with
momenta $\omega /s\gg a_{B}^{-1},\bar{z}^{-1}$.

The dependence of the probabilities (\ref{ph1}) and (\ref{ph2}) on the
magnetic fields is plotted in Fig.~\ref{phmagdep}. At low fields $B\simeq 8$
T the transition $\{II\}$ ($L_{2} \rightarrow L_{0}^{\prime}$) is
predominant. As $B$ increases, the probability $W_{ph}^{\{II\}}$ rapidly
falls while $W_{ph}^{\{I\}}$ (this corresponds to $L_{0}^{\prime}\rightarrow
L_{1}$) slowly increases so that already at $B=8.6$ T these two transitions
are equally probable. At high fields transition $L_{2} \rightarrow
L_{0}^{\prime}$ is suppressed with respect to $L_{0}^{\prime}\rightarrow
L_{1}$ and $W _{ph}^{\{I\}}$ rapidly increases with increasing $B$ so that
at fields near $B_{E}$ achieves to values corresponding to times less than $1
$ ns.

In order to obtain the relaxation times $\tau _{ph}^{\{I\},\{II\}}$, the
occupation factors have to be included according to Eq.~(\ref{augertime}).
Noting that for the case under consideration in experiment \cite{potemski1}
and \cite{potemski2} the $L_{1}$ and $L_{0}^{\prime}$ levels are not fully
occupied, we obtain 
\begin{equation}
\tau _{ph}^{\{I\}, \{II\}} =(1-f_{11, 20})W_{ph}^{\{I\}, \{II\}} \sim
W_{ph}^{\{I\}, \{II\}}.  \label{phtime1,2}
\end{equation}
To find out which mechanism (the Auger scattering or the phonon emission) is
responsible for filling of holes in the level $L_{0}$ with electrons, we
estimate also the lifetime of a test hole in $L_{0}$ with respect to the
phonon emission from the higher level $L_{1}$ 
\begin{equation}
{\frac{1}{\tau ^{h}_{ph}}} =\sum_{k^{\prime}}W_{11k \rightarrow
10k^{\prime}}^{DA} f_{11}\equiv W_{1 \rightarrow 0}f_{11}.  \label{phhole}
\end{equation}
Here the probability $W_{11k \rightarrow 10k^{\prime}}^{DA}$ is given by
Eq.~(\ref{augerprob}) for the transferred energy $\omega =\omega_{B}$. Since
always $\omega_{B} >\omega_{1},\omega_{2}$, the approximations, which have
been made above to calculate the kernel $K$ and form factor $I_{nn^{\prime}}$%
, are also justified for this case. Therefore one can obtain $%
W_{1\rightarrow 0}$ from Eq.~(\ref{ker3}) by substituting $\omega =\omega_{B}
$. It is clear, that the lifetime of a test hole $\tau ^{h}_{ph}$ is larger
than the relaxation times $\tau _{ph}^{\{I\},\{II\}}\sim 1\; \mu$s and thus
much larger than ($\tau^{h}_{Auger}\sim 1fs$). Hence, the Auger process is
much more efficient to fill holes in the level $L_{0}$ than the phonon
emission. This fact makes possible the observation of the Auger-upconversion
by interband optical pumping \cite{potemski1} and \cite{potemski2}.

\section{Analysis of the rate equations.}

\label{rateeqs}

So far we have considered only the processes (iii), (iv) and (vi). In order
to correctly describe the experimental situation of \cite{potemski1} and \cite
{potemski2}, we have to take into account also the processes (i), (ii) and
(v), {\it i.e.} the pumping by interband excitation to the level $L_{1}$ and
the recombination of electrons from $L_{0}$ and $L_{0}^{\prime}$ with the
photo-induced holes. As it has been discussed already in Sec.~\ref{aug}, the
Auger process becomes possible only after processes (i) and (ii).
Intensities of the emissions from the levels $L_{0}$ and $L_{0}^{\prime}$
are determined by the characteristic times of the processes (i)-(vi) from
the following set of rate equations 
\begin{eqnarray}
{\frac{\partial n_{0} }{\partial t}}&=&{\frac{n_{0}}{\tau_{0}}}-{\frac{n_{1}%
}{\tau^{e}_{Auger}}}  \label{reqs1} \\
{\frac{\partial n_{1} }{\partial t}}&=&\tilde P_{exc}+{\frac{n_{0}^{\prime}}{%
\tau_{ph}^{\{I\}}}}-{\frac{2n_{1}}{\tau^{e}_{Auger}}}  \label{reqs2} \\
{\frac{\partial n_{0}^{\prime}}{\partial t}}&=&{\frac{n_{2}}{%
\tau_{ph}^{\{II\}}}}- {\frac{n_{0}^{\prime}}{\tau_{ph}^{\{I\}}}}-{\frac{%
n_{0}^{\prime}}{\tau_{0}^{\prime}}}  \label{req3} \\
{\frac{\partial n_{2}}{\partial t}}&=&{\frac{n_{1}}{\tau_{Auger}^{e}}}- {%
\frac{n_{2}}{\tau_{ph}^{\{II\}}}}  \label{reqs4}
\end{eqnarray}
Here $n_{l}$ $(n_{l}^{\prime})$ is the areal number density of electrons in
the Landau level with index $l$ of the first (second) electric subband
including spin degeneracy. 
The characteristic times of the recombination of electrons from level $L_{0}$
and $L_{0}^{\prime}$ with photo-induced holes are $\tau_{0}$ and $%
\tau_{0}^{\prime}$ which depend on $B$ in the same way and are of same order
of magnitude. In Eq.~(\ref{reqs2}) we have also defined the flux of
electrons created by interband excitation into the level $L_{1}$ 
\begin{equation}
\tilde P_{exc}= {\frac{P_{exc}}{\omega_{exc}}} (1-f_{1})  \label{interexc}
\end{equation}
where $P_{exc}$ and $\hbar\omega_{exc}$ are the excitation power and energy.
The factor $1-f_{1}$ takes into account the availability of free states in
the level $L_{1}$. In a stationary case we have ${\partial n/\partial t}=0$
and the rate equations reduce to 
\begin{eqnarray}
I_{0}&\equiv& {\frac{n_{0}}{\tau_{0}}}={\frac{n_{1}}{\tau^{e}_{Auger}}} ={%
\frac{n_{2}}{\tau_{ph}^{\{II\}}}}  \label{intens1} \\
I_{0}^{\prime}&\equiv &{\frac{n_{0}^{\prime}}{\tau_{0}^{\prime}}}
\label{intens2} \\
I_{0}-I_{0}^{\prime}&=& {\frac{n_{0}^{\prime}}{\tau_{ph}^{\{I\}}}}
\label{intens3} \\
I_{0}+I_{0}^{\prime}&=& \tilde P_{exc}  \label{intens4} \\
\end{eqnarray}
where $I_{0}$ and $I_{0}^{\prime}$ are proportional to the luminescence
intensities for recombinations from $L_{0}$ and $L_{0}^{\prime}$,
respectively. According to Eq.~(\ref{intens1}), the luminescence intensity $%
I_{0}$ is determined both by $\tau_{Auger}^{e}$ and $\tau_{ph}^{\{II\}}$.
Because of $\tau^{e}_{Auger}=(n_{1}/n_{2})\tau_{ph}^{\{II\}}>
\tau_{ph}^{\{II\}}$ we find for $B=9.5$ T that $\tau ^{e}_{Auger}> 30\; \mu %
\mbox{s}$. This means that the relaxation time of a test electron in the
level $L_{1}$ with respect to the Auger scattering is strongly enhanced due
to the occupation factors over the characteristic time $W^{-1}_{Auger}$ and
the lifetime of a test hole $\tau_{Auger}^{h}$ in the level $L_{0}$ which
are of the order of $1$ fs.

Because $\tau_{Auger}^{h}$ is much smaller than all other characteristic
times of this system, the Auger-upconversion mechanism immediately fills all
arising holes in $L_{0}$ due to the recombination of electrons from $L_{0}$
and creates electrons in the level $L_{2}$. Thus the number of the
up-converted electrons is always equal to the number of recombining electrons
from $L_{0}$. However, the intensity of the up-converted luminescence $%
I_{0}^{\prime}$ is determined by the competition of the processes (v) and
(iv), {\it i.e.} by the ratio of $\tau_{0}^{\prime}$ and $\tau _{ph}^{\{I\}}$
(see Eqs.~(\ref{intens2}) and (\ref{intens3})).

In order to obtain the dependencies of $I_{0}$ and $I_{0}^{\prime}$ on the
magnetic field and the excitation power we consider two different cases. In
a case when process (v) is dominant over the process (vi), {\it i.e.} $\tau
_{ph}^{\{I\}}\gg \tau_{0}^{\prime}$, we find from equations (\ref{intens1})-(%
\ref{intens4}) that 
\begin{equation}
I_{0}\approx I_{0}^{\prime}\approx {\frac{\tilde P_{exc}}{2}}\gg {\frac{N_{0}%
}{\tau_{ph}^{\{I\}}}}, \quad n_{0}\sim n_{0}^{\prime}\sim N_{0}
\label{intrel1}
\end{equation}
which means 
\begin{equation}
I_{0},\: I_{0}^{\prime}\propto P_{exc},\; \left(1-{\frac{B_{1}}{B}}\right).
\label{intrel2}
\end{equation}
Here $N_{0}$ is the capacity of one Landau level and $B_{1}$ corresponds to
the full occupation of two Landau levels $L_{0}$ and $L_{1}$. With the help
of Eqs.~(\ref{ph1}), (\ref{phtime1,2}), and (\ref{interexc}) it is seen that
the inequality in Eq.~(\ref{intrel1}), which is determined by $\tau
_{ph}^{\{I\}} \gg\tau_{0}^{\prime}$, corresponds to the situation when the
magnetic field is close to the lower bound of the considered interval and at
the same time the excitation power is high. In opposite case when $\tau
_{ph}^{\{I\}}\ll \tau_{0}^{\prime}$, which corresponds to magnetic fields
close to the upper bound and low powers, we find from Eqs.~(\ref{intens1})-(%
\ref{intens4}) 
\begin{equation}
I_{0}^{\prime}={\frac{\tilde P_{exc}^{2}}{N_{0}/\tau_{ph}^{\{I\}}}} \ll
I_{0}=\tilde P_{exc},\quad n_{0}\sim N_{0}\gg n_{0}^{\prime}  \label{upbound}
\end{equation}
This means that 
\begin{equation}
I_{0}^{\prime}\propto P_{exc}^{2}, \; \left(1-{\frac{B_{1}}{B}}%
\right)(B_{E}-B)^{5}  \label{upbound1}
\end{equation}
while 
\begin{equation}
I_{0}\propto P_{exc},\; \left(1-{\frac{B_{1}}{B}}\right).  \label{upbound2}
\end{equation}
Now it is clear from Eqs.~(\ref{intrel1}) and (\ref{intrel2}) that at $B$
near $B_{1}$ both intensities $I_{0}$ and $I_{0}^{\prime}$ increase linearly
with $B$. As $B$ increases further, $I_{0}$ continues to increase in
accordance with (\ref{upbound2}) but not so sharply as near $B_{1}$. 
\begin{figure}[htb]
\epsfxsize=11cm
\epsfysize=13cm
\mbox{\hskip 1.5cm}\epsfbox{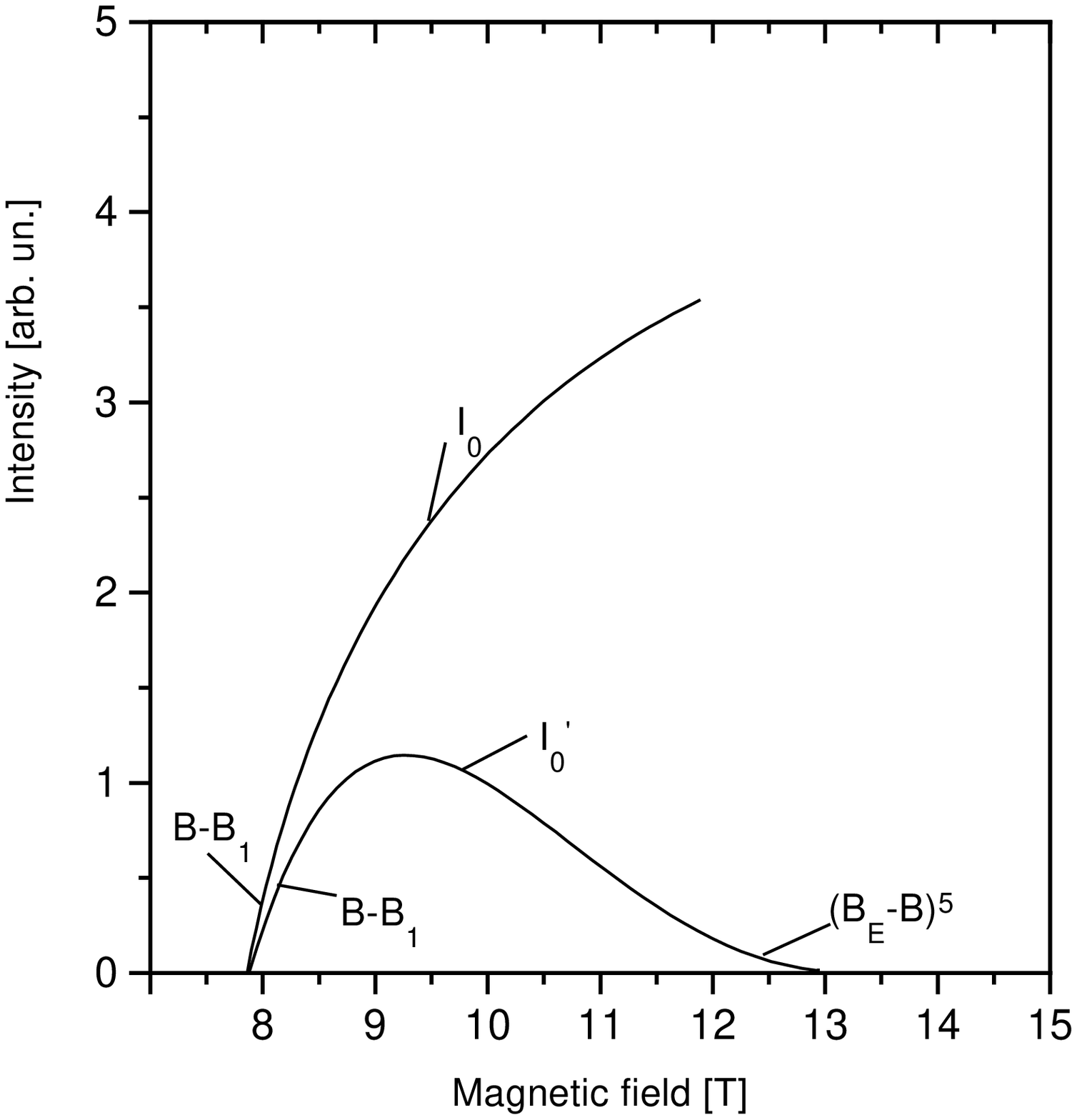}
\caption{Magnetic field dependence of the luminescence intensities $I_{0}$
and $I_{0}^{\prime}$ for recombination from the Landau levels $L_{0}$ and $%
L_{0}^{\prime}$, respectively.}
\label{intdep}
\end{figure}
However, $I_{0}^{\prime}$ shows another behavior given by Eq.~(\ref
{upbound1}) and decreases with $B$ as $(B_{E}-B)^{5}$ when $B$ is near $B_{E}
$ (Fig.~\ref{intdep}). At low powers $I_{0}^{\prime}$ depends quadratically
on $P_{exc}$ (Eq.~(\ref{upbound1})) while at high powers for $I_{0}^{\prime}$
(Eq.~(\ref{intrel2})) and in the whole range of the power variation for $%
I_{0}$ (Eqs.~(\ref{intrel2}) and (\ref{upbound2})) this dependence is linear.

In conclusion, our calculation provides an understanding of the main
features of the experiments reported in \cite{potemski1} and \cite{potemski2}.
In the light of our investigations it would be interesting to have
detailed experimental information on the intensities $I_{0}$ near $B_{1}$
and $I_{0}^{\prime}$ near $B_{E}$ for which also our theory provides
definite information.

\chapter{Magneto-transport in a non-planar 2DEG}
\section{Introduction}

The last decade has seen the many of different and creative new 
environments under which quantum nanosystems with the 2DEG are
investigated. Particularly, the 2DEG exposed to a homogeneous 
magnetic field has proved to be an extremely rich subject for 
investigations in theory and experiment \cite{landwehr92}.
In these nanosystems in addition to the lateral confinement, the carrier 
motion becomes quantized also in the plane normal to the magnetic 
field. The quantization length is varied with magnetic field which 
assures an easily way to obtain information on the properties of charge
carriers. Considerable efforts have been devoted to study transport 
properties in such effectively zero dimensional systems \cite{prange,%
beenakker}. 
Already, a set of remarkable phenomena such as the integer and 
the fractional quantum Hall effects \cite{chakrabarty}, the 
Aharonov-Bohm effect \cite{aharonov,washburn}, and the 
magneto-resistance 
oscillations of the 2DEG subjected to periodic electric field weak
modulations along one or two directions, also called {\it Weiss} 
oscillations \cite{weiss,gerhardts,winkler} (to mention just a few) has 
been uncovered.

In the last several years a more complex situation of the 2DEG in
a non-uniform magnetic field has attracted considerable interest 
\cite{peeters}.
Depending on the strength of the local magnetic field, the electron 
motion in the plane of the 2DEG can be tuned from regular to chaotic
\cite{roess1,roess2}. 
The motion of ballistic electrons in a periodic magnetic field is also 
believed to be closely related to the motion of {\it composite fermions}
in a density modulated 2DEG in the fractional QHE regime 
\cite{stomer,pffeifer}.
Such magnetic supersystems offer the possibility of producing 
magnetic confinement.
In a non-zero magnetic field region, the lowest energy state for an electron
is the lowest Landau level with energy $\h\o_B$ ($\o_B$ is the cyclotron 
frequency). In a zero-field region, the lowest energy state is zero. Therefore,
when electrons starts to move away from the zero-field region, the 
non-zero-field region acts as a barrier. Moreover this barrier differs 
from a 
usual potential barrier in two ways. First, the barrier is kinetic, {\it i.e.}
the electron gains kinetic energy as the electron overcomes the barrier.
Second, tunneling through the magnetic barrier is an inherently 
two-dimensional process so that the transmission probability depends
on the angle at which the electron hits the non-zero-field region. 
This last property has been exploited to predict a wave-vector selective 
filter for electrons \cite{peetmatvas}. Other interesting feature can
manifest magnetic "dots" coupled to the 2DEG. These dots could be used 
to produce magnetic fields that could confine electrons to a disk region. 
Such magnetic dots could be explored as memory elements in future 
electronics.

Theoretically the transport properties of a 2DEG subjected to a 
spatial dependent perpendicular magnetic field have been addressed
in several works.
The possibilities of the creation of periodic superstructures by a 
non-homogeneous magnetic field have been investigated in 
\cite{dubrovin,yoshnag,vilms}
Distinct theoretical predictions have been achieved for the limit of 
a weak one-dimensional magnetic modulation.
Transport of a 2DEG in a weakly modulated periodic magnetic field normal
to the electron sheet and its collective excitation spectra have been
studied in \cite{peetmatvas,peetvasil} and \cite{wu}. The single-particle
energy spectra of a 2DEG have been calculated in a non-homogeneous 
magnetic field for different step-like \cite{peetmat}, linearly varying with 
position \cite{muller} and other functional magnetic field profiles \cite{calvo}. 
The magnetic field dependence of the conductance of a ballistic QWr 
\cite{yoshnag} and of a 2DEG through an orifice \cite{avishai} has been 
studied. 
The properties of wave-vector dependent electron tunneling \cite{peetmatvas} 
and electron moving \cite{ibrahim} in step-like magnetic structures have been
investigated. Analysis of the weak localization and calculation of the Hall and 
diagonal resistivities of the 2DEG in an inhomogeneous magnetic field have been
presented in \cite{rammer,khaetskii,brey}. 
Quite recently it has been shown that the spatial distribution of electron
and current densities in a linearly varying magnetic field has very rich
structure related to the energy quantization \cite{hofstetter}.

Until recently experimental attempts to produce non-homogeneous 
magnetic fields on the micrometer or nanometer scale have failed.
Now, however experimental groups in Germany, Japan, and 
the UK have succeeded in coupling the 2DEG to the 
non-homogeneous magnetic field. High mobility 2DEGs are formed 
in standard GaAs/AlGaAs heterojunctions. The spatial modulation
of the magnetic field is made possible by depositing patterned gates
of superconducting or ferromagnetic materials on the surface of 
heterostructures.

A group at the Max-Planck-Institute in Stuttgart has used ferromagnetic
dysprosium (Dy) metal stripes \cite{ye1,ye2}, 
while a group at Tokyo University has used ferromagnetic nickel gates 
\cite{izawa} to modulate the 
magnetic field. The strength of these micromagnets can be increased via an
external magnetic field. A group at Nottingham University has used a 
superconducting stripes on the surface of the heterostructure 
\cite{carmona}.
Using this technique of patterned gates, it has become possible to realize 
experimentally magnetic dots \cite{mccord,ye1}.

However, such scheme of creating inhomogeneous magnetic
fields has two disadvantages: i) the variation in the magnetic 
field is very small in the 2DEG plane since gates are a few hundred 
angstr\"oms above the plane of the 2DEG and dies away rapidly a 
short distance below the gates, 
ii)  the patterned gate layers will also cause strain and electrostatic 
iii)  variations 
which are in general stronger than the effects of the varying magnetic
field which makes it hard to attribute effects unambiguously to the 
magnetic field.

Recently a research group from the Toshiba Cambridge Research 
Center Ltd. and Cavendish Laboratory have reported an alternative 
approach to produce spatially varying magnetic fields \cite{foden1}.
They have proposed to take advantage of the regrowth technology of 
$III-V$ semiconductors on patterned substrates and offer a more 
flexible and potentially fruitful solution of this problem. 
A remotely doped GaAs/Al$_x$Ga$_{1-x}$As heterojunction is grown 
over wafer previously patterned with series of facet. The electron gas is 
confined to a sheet at the GaAs/Al$_x$Ga$_{1-x}$As heteroface which
is no longer planar but follows the contour of the original wafer.
The use of {\it in situ} cleaning technique enables to regrow uniform
high quality 2DEGs over etched substrates \cite{mark1,mark2}.

\begin{figure}[htb]
\epsfxsize=13cm
\epsfysize=12cm
\mbox{\hskip 0.5cm}\epsfbox{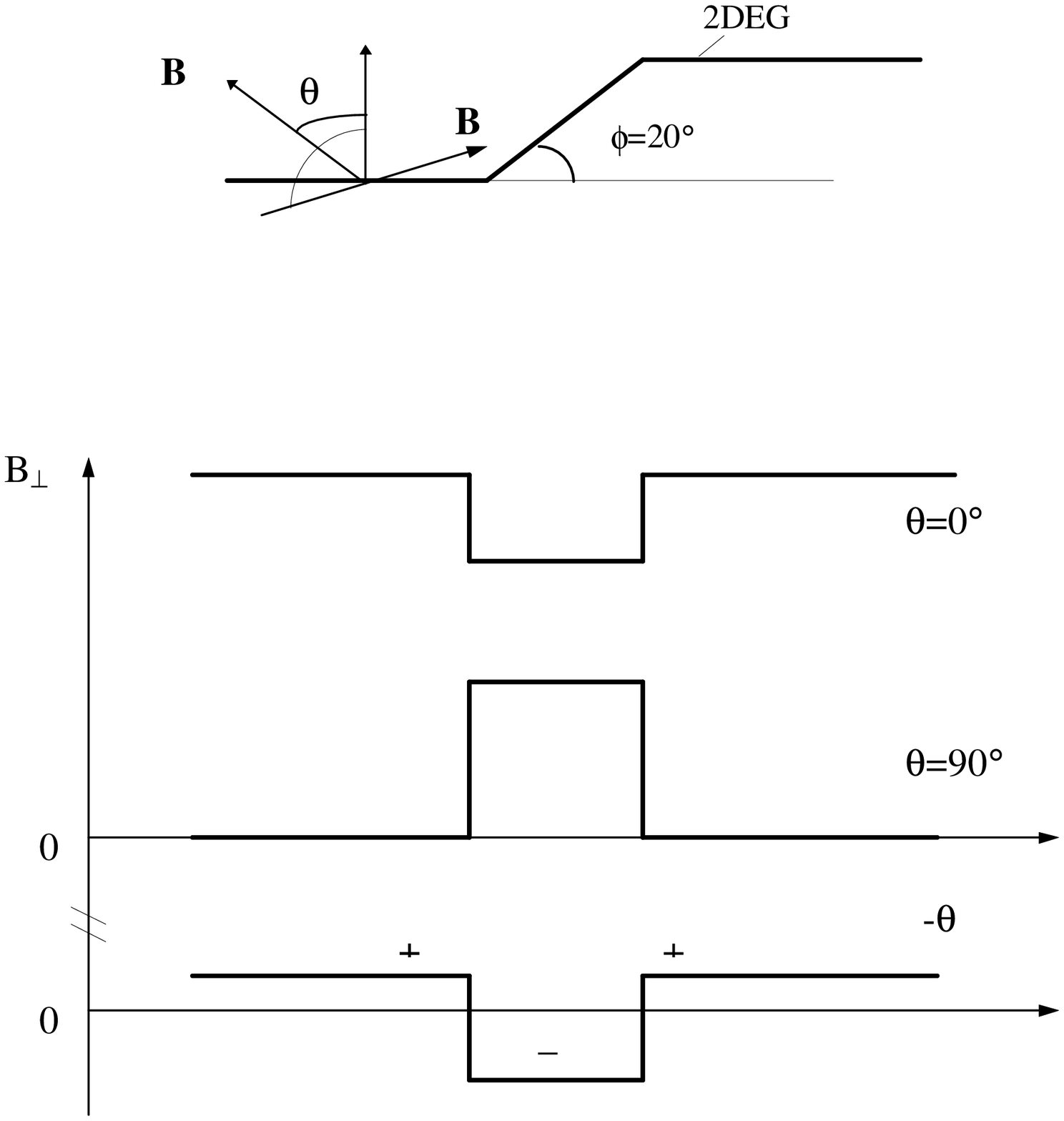}
\caption{Applying a uniform magnetic field produces a spatially 
non-homo\-ge\-neous field component normal to the 2DEG.
Different magnetic superstructures can be obtained, depending on the 
$\theta$ angle between the magnetic field and the substrate normal, 
magnetic barriers ($\theta=90^\circ$), magnetic wells ($\theta=0^\circ$), 
and novel situations when the normal component of the field changes its 
sign on the magnetic interface.}
\label{fg38}
\end{figure}

Application of a homogeneous magnetic field to this structure results
a spatially varying field component normal to the 2DEG. 
This has firstly been demonstrated for samples with a single planar facet
\cite{foden1,mark1,mark2,mark3}.
Rotating the plane of the sample allows us to find different non-homogeneous 
magnetic superstructures: magnetic barriers,  magnetic wells and 
completely novel situations where the normal component of the field 
changes its sign on the facet (see Fig.~\ref{fg38}).
Improved control over the topography of the electron gas will allow
magnetic-field variations in more than one dimension. In principle,
this technique offers the possibility of investigating the behavior of
electrons in a curved quasi-two-dimensional space, and the effects 
of varying that curvature. Quite recently a theoretical study of quantum
magnetic confinement and transport of electrons in spherical and 
hemispherical 2DEGs has been presented \cite{foden2}

Thus, using this {\it in situ} cleaning technique, we can now investigate 
the effect of varying {\it the topography} of an electron gas in addition 
to varying the {\it dimensionality}. 
This new technology will open up a new dimension for investigations in
nanophysics. 
Characterizing and understanding transport properties of 
magnetic superstructures with elementary cells containing a 2DEG
with well-defined topographical features are crucial both for the 
basic phenomena and device applications.

The aim of the present chapter is to investigate theoretically the 
magneto-transport of the non-planar 2DEG 
\cite{badaltrs1,badaltrs2,badaltrs3,badaltrs4,ibrahim}
to explain recent experimental results obtained in the Toshiba 
Cambridge Center and Cavendish laboratory
\cite{foden1,mark1,mark2,mark3,foden2}.
As an example, the electric field distribution
has been calculated in the presence of a magnetic tunnel barrier of 
$\mu$m width. The system satisfies the Poisson equation in which
line charges develop at the magnetic/non-magnetic-field interface.
We have found that most of the electrons are injected at the edges
of the magnetic barrier. The magneto-resistances across the facet as
well as in the planar regions of the 2DEG have been calculated which
provide an understanding of the main features of the magnetic field 
dependencies observed experimentally by M L Leadbeater {\it et al}
\cite{mark1}.
They have constructed a non-planar 2DEG which has been fabricated
by growth of a GaAs/AlGaAs heterojunction on a wafer pre-patterned
with facets at $20^\circ$ to the substrate. Applying a uniform magnetic
field, ${\bf B}$, produces a spatially non-uniform field component 
perpendicular to the 2DEG (see the $\theta=90^\circ$ case in 
Fig.~\ref{fg38}). 
With the field in the plane of the substrate an effective magnetic 
barrier has been created located at the facet. 
The resistance across such an etched facet has shown oscillations
which are periodic in $1/B$ and which are on top of a positive 
magneto-resistance background which increases quadratically with 
the magnetic field for small fields $B$ and linear in $B$ for large $B$.
In experiment the dimensions of the facet have been $40$ $\mu$m
wide and $3$ $\mu$m long, the voltage probes close to the facet are 
situated $10$ $\mu$m apart across the facet ({\it quod vide} 
Ref.~\cite{mark1} for more details).
The magneto-resistance has been measured using also the voltage 
probes on the planar regions of the 2DEG.
There is no perpendicular component of the magnetic field in these
regions, however, for probes directly adjacent to the facet, a strong 
magneto-resistance has been observed. Pairs of probes on opposite the
sides of the mesa and on the opposite sides of the facet have shown
the same symmetry. While there has been a pronounced asymmetry
in each of traces with reversal of the magnetic field direction.

Another motivation for our theory is the necessity of reexamining
the standard Hall analyses for very small device geometry 
\cite{leshouches,bauer}.
As device sizes have shrunk to the submicron scale,
it has become harder to make contact probes small relative to the
size of the sample to be characterized. Due to the very high current
densities involved, point-like contacts come to introduce their own
sources of error. Particularly, in the QHE geometry, singularities of 
the current density associated with the breakdown of the QHE 
\cite{ebert,komiyama}. Hall voltage distribution has been measured
in modulation-doped GaAs/AlGaAs heterojunctions while observing 
the QHE \cite{salerno} and has been calculated in an ideal 2DEG 
\cite{wieck,macdonald}
The geometric effect of contacts has been studied by many authors
\cite{wick,lippmann1,lippmann2,newsome,kawaji,rendell,altshul}
(see also the review article \cite{heremans}). 
First such a theory has been developed by Wick \cite{wick}, who 
assumes that the contacts have a much higher conductivity
than the sample being measured. Therefore the contacts can be modeled
as an equipotential at the sample boundary which allow to treat many 
geometries by the method of Schwartz-Christoffel mappings. These 
results obtained by Wick have been applied directly to the QHE regime
by Kawaji \cite{kawaji}. The same approach has been exploited by other
authors to analyze geometric contact effects in the QHE regimes 
\cite{rendell,altshul}. As we are aware, all of the exact treatments 
except the recent work \cite{jou}, treat a single isolated contact interface
which can be usefully applied to Hall bars with high aspect ratios while
for samples with lower aspect ratios this approach does not work.

Our theory presented in this work can be applied for samples with arbitrary 
aspect ratios and with any number of magnetic interfaces.

\section{Theoretical model}
\la{transport}

To explain, qualitatively, the main features of the experimental measurements
in Ref.~\cite{mark1}, namely the smooth background of the magnetic
field dependence of the resistance across the facet and the symmetries
of the resistance traces measured between various probes in the 
planar regions, we rely on a classical model since the width of the 
magnetic field barrier is much larger than the magnetic length.
Recall, that the present situation is different from that of discussed in 
Ref.~\cite{peetmatvas} where quantum tunneling through magnetic
nanostructures was treated.

Due to its topography, the non-planar 2DEG is embedded in ordinary
three-dimensional space so that the electron position is determined by
three Cartesian coordinates. In reality, however, we are dealing with a
two-dimensional problem. By a simple coordinate parameterization,
the problem could be reduced to the geometry with 2D-electrons 
located in a $(x,y)$-plane and the magnetic field normal component 
applied in $z$-direction.
We take the magnetic field profile $B(x)=0$ ($0<x<a$ and 
$b<x<L$) and $B(x)=B$ ($a<x<b$) which is 
appropriate to the experimental situation.
Assuming the electric field and the current density to be independent 
on the $z$-coordinate, from 2D classical electrostatics we obtain 
the spatial electric field distribution in the $(x,y)$-plane which 
determines the magneto-resistance which we compare with experiment.

In a steady state, the spatial distribution of an electric current density 
${\bf J}$ is independent of time, and satisfies the discontinuity equation 
\begin{equation}
\mbox{\bf div}\, {\bf J}(x,y)=0  \label{1}
\end{equation}
which means that total charge in any volume of the 2DEG remains constant.
The electric field {\bf E} in the 2DEG in which a steady current flows is
constant, and therefore from Maxwell's equations we have 
\begin{equation}
\mbox{\bf curl}\, {\bf E}(x,y)=0,
\label{222}
\end{equation}
{\it i.e.} it is a potential field. Eqs.~\r{1} and \r{222} should be
supplemented by a system equation relating {\bf J} and {\bf E}. In a 
zero magnetic field in the most of cases this relation is linear (Ohm's 
law). In a normal quantizing magnetic field charge carriers are moved 
due to the Coulomb and Lorenz forces so that the basic system equation
is given by 
\begin{equation}
{\bf J}(x,y)=\sigma \, {\bf E}(x,y)-\tan\zeta(x)\,{\bf J}(x,y) \times{\bf B}(x)
\label{333}
\end{equation}
where the conductivity $\sigma=ne\mu$ is determined by the sample 
mobility $\mu$ ($n$ is the surface concentration of electrons), the 
spatial dependent Hall angle $\zeta(x)$ is defined by 
\begin{equation}
\tan\,\zeta(x)=\sigma\, R_B\, B(x)
\label{444}
\end{equation}
$R_B$ is the Hall coefficient. The spatial dependent magnetic field can be
expressed by the Heaviside step function in the following way 
\begin{equation}
{\bf B}(x)=\hat{{\bf z}}B[\theta(x-a)-\theta(x-b)],\, \theta(x)=\cases{1,
\,\mbox{if\, $x>0$},\cr 0, \,\mbox{if\, $x<0$}}. 
\label{555}
\end{equation}
Using Eqs.~\r{1} and \r{222} it is easy to show that by virtue of 
Eqs.~\r{333}, \r{444} and \r{555}, the electric field satisfies the 
following equation at all points within the whole region of the 2DEG 
\begin{eqnarray}
\mbox{\bf div}\, {\bf E}(x,y)&=&\tan\delta\,\{\sin[2\zeta(x)]\,
E_y(x,y)-\cos[2\zeta(x)]  
\nonumber \\
&\times& E_x(x,y)\}\{\delta(x-a)-\delta(x-b)\}  \nonumber \\
&\equiv&4\pi\rho_{s}\{\delta(x-a)-\delta(x-b)\}. 
\label{666}
\end{eqnarray}
with $\delta(x)$ the Dirac-delta function.

At the boundary of the 2DEG, the normal component of the current 
density and the tangential component of the electric field must be 
continuous which follows from Eqs.~\r{1} and \r{222}, respectively.
\begin{figure}[htb]
\epsfxsize=14cm
\epsfysize=12cm
\mbox{\hskip 0cm}\epsfbox{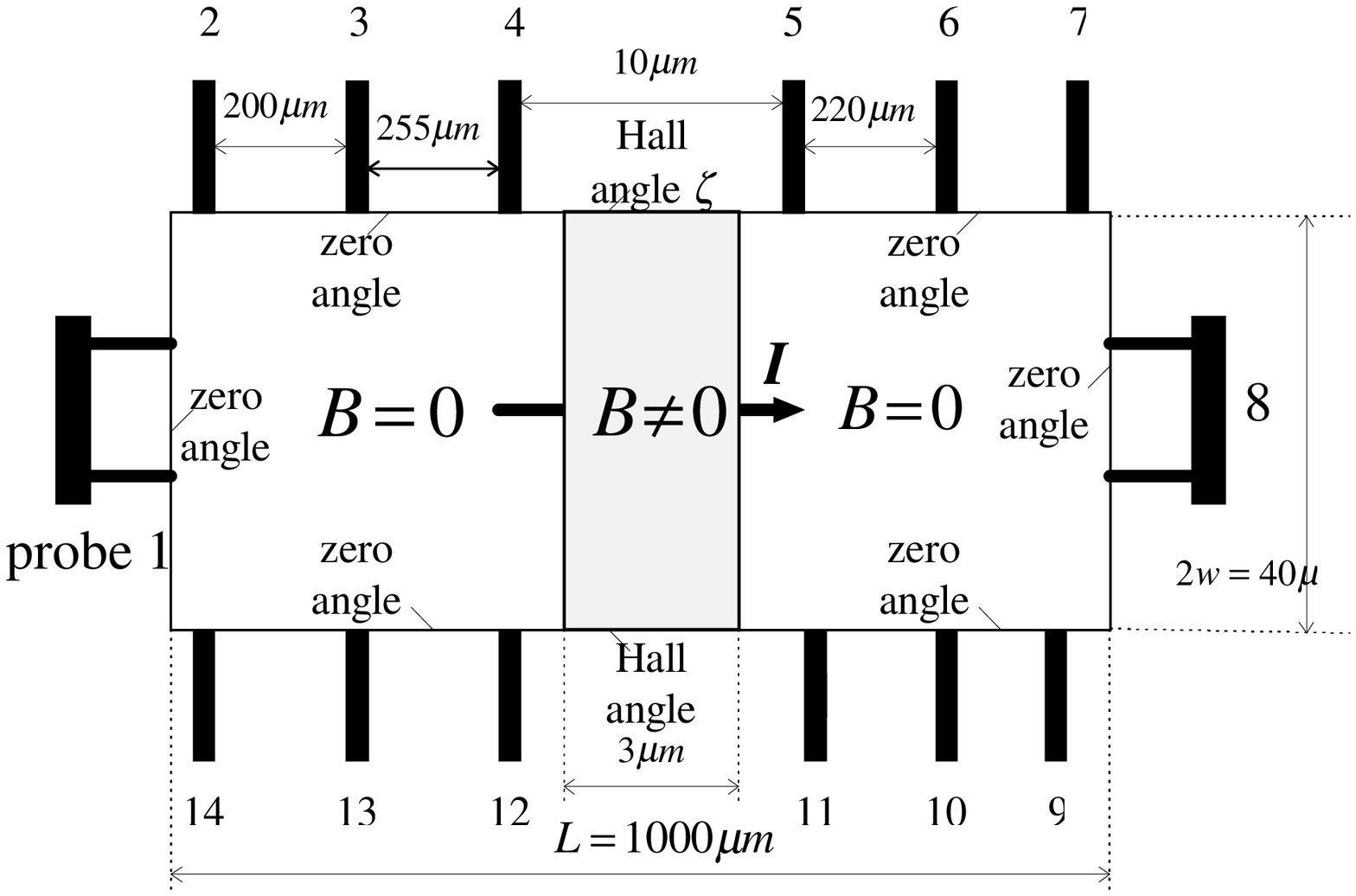}
\caption{A magnetic barrier is created in a non-planar 2DEG exposed to the
uniform magnetic field. Dimensions and voltage probe locations correspond to 
the experimental situation.
Boundary conditions are given in terms of angle which makes the electric 
field with respect to the sides and ends of the 2DEG. The current flows 
between voltage probes $1$ and $8$.}
\label{fg39}
\end{figure}
This means that there be no current flow out the sides ($x=-w$ and 
$x=w$) and no electric field parallel to the ends ($y=0$ and $y=L$)
of the 2DEG. For the magnetic field configuration given by Eq.~\r{555},
the system equation (\ref{222}) implies that the current and electric 
field make an angle $0$ and $\zeta$ in the regions $B=0$ and $B\neq0$,
respectively. Therefore the boundary condition ${\bf J}=0$ on the sides 
of the 2DEG can be replaced by the condition that the electric field 
makes an angle $0$ and $\zeta$ with respect to the sides,
respectively in the regions $B=0$ and $B\neq0$ (see Fig.~\ref{fg39}).
Now it is clear that the problem can be treated as a field problem, 
{\it i.e.} in order to solve the problem, it is necessary to obtain the 
solution of the set of Eqs.~\ref{333} and \ref{666} with boundary 
conditions shown on Fig.~\ref{fg39}. Unlike to the usual type 
electric potential problem, here boundary conditions do not involve 
specifications of the potential or its normal derivative but of the 
electric field angle with respect to the boundary of the 2DEG.

When the electric field represented as the gradient of a potential function 
it satisfies the 2D Laplace equation in the {\it separate} planar and 
non-planar regions of the 2DEG. However, this potential {\it does not} 
satisfy the Laplace equation in the whole 2DEG region.
It satisfies the Poisson equation with a non-trivial right side part 
which is determined by the field itself. Thus at the magnetic interfaces 
$y=a$ and $y=b$ there is an accumulation of linear charges with a charge 
density $\rho_{s}$.

Direct way to solve the such electric potential problem is to obtain the
solutions of three Laplace's equations separately in the both planar and
in the non-planar regions, introducing additional four unknown angles
which the electric field makes on both sides of the magnetic interfaces 
with respect to the normal of the interfaces. 
Then using the continuity of the normal current and the tangential
field components we have to match these solutions and eliminate these 
four angles. However, because of a singularity of the electric field at the
magnetic interfaces, inconveniences occur in such procedure (see also 
Ref.~\cite{jou}). To get round this difficulty, first we have solved 
directly Laplace's equation for whole region of the 2DEG, {\it i.e.} 
taking $\rho_s=0$ in Eq.~\r{666} with the same boundary conditions specified 
above on Fig.~\ref{fg39}.
Then, we have included the line charges at $y=a$ and $y=b$ which implies that 
the electric field exhibits jumps at these points and we have modified the 
solution of the Lalace equation such that it satisfies the Poisson equation 
corresponding to Eqs.~\r{222} and \r{666}.
To construct the solution of the 2D Laplace equation, the conformal mapping 
method \cite{fuchsshab} has been exploited. 
From the Cauchy-Riemann conditions it follows that the desired electric field 
related to the complex field via ${\bf E^\prime}=iE^{\prime}_x+E^\prime_y$ 
is solenoidal and irrotational, {\it i.e.} there is a complete correspondence 
between plane electrostatics field and regular functions. 
\begin{figure}[htb]
\epsfxsize=12cm
\epsfysize=12cm
\mbox{\hskip 0cm}\epsfbox{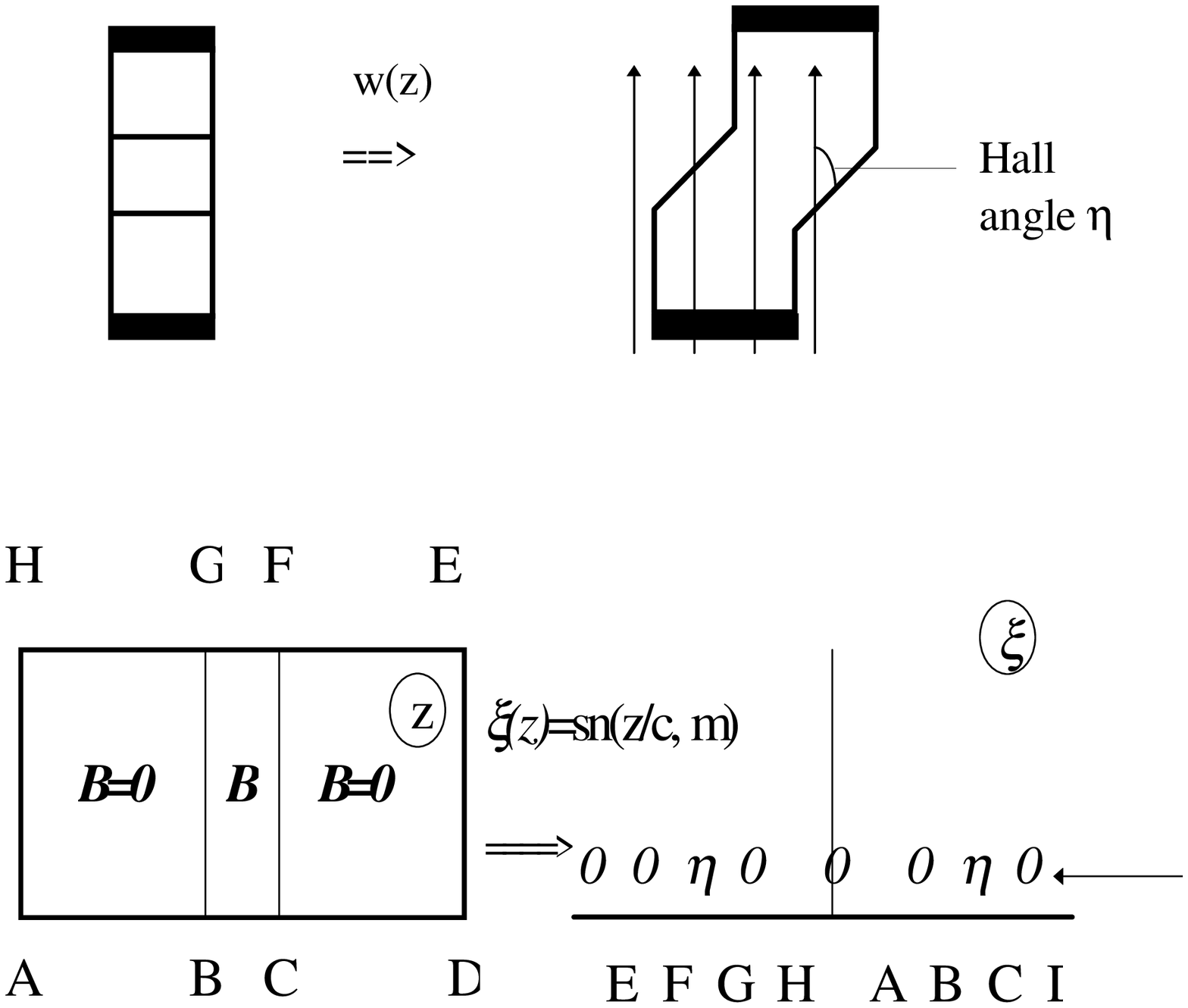}
\caption{ Conformal mapping $w(z)$ of the rectangular 
domain of the 2DEG onto the domain of a parallelogram form where boundary 
conditions are satisfied by a uniform electric field (the upper picture). 
Conformal mapping $\xi(z)$ of the rectangular domain of the 
2DEG onto the upper half plane of the Im$\xi>0$ such that the perimeter
of the rectangle goes into the real axis (the lower picture).}
\label{fg40}
\end{figure}
If the geometry of the 2DEG were that of including rectangular shown in 
Fig.~\ref{fg40} then the boundary conditions would be satisfied by a uniform 
electric field, ${\bf E}={\bf {\hat y}} E_0$. Therefore, if we let 
$-E'(z)=\lambda dw(z)/dz=exp[f(z)]$ ($\lambda$ is an arbitrary constant 
which should be determined by the potential difference between the ends of 
the 2DEG) then in order to map the rectangle into the such 
parallelogram in complex plane $w(z)$ shown in the upper picture in 
Fig.~\ref{fg40}, it is necessary to find an analytical function $f(z)$ 
whose imaginary part satisfies the boundary conditions:
\bea
&&{\bf Im} f(z)=0 \quad \mbox{on the ends of the 2DEG,}
\nonumber \\
&&{\bf Im} f(z)=0\; \mbox{or}\; \zeta \quad \mbox{on the sides of the 2DEG.}
\eea
Using the conformal mapping $\xi(z)$, we map the rectangular domain of the 
2DEG in the complex $z=x+iy$ plane onto the upper half plane of the 
Im$\xi>0$ such that the boundary of the rectangle goes into the real axis 
(the lower picture in Fig.~\ref{fg40}). 
This is accomplished by the Jacobi elliptic function, $\xi(z)=sn(z/c,m)$ 
where the constant $c=L/K^{\prime}(m)=w/K(m)$ and $m$ are determined by 
the sample 
aspect ratio, $K$ and $K^{\prime}$ are the complete elliptic integrals
of first kind \cite{abramovitz}, $2w$ is the width of the 2DEG. Thus the 
problem is reduced to the standard electrostatics problem of finding the 
potential for the upper half plane bordered by electrodes 
represented by the open intervals along real axis Im$\xi=0$: 
\begin{eqnarray}
&& \left(-\infty, -sn(1,\beta)\right);\, \left(-sn(1,\beta),-sn(1,\alpha)
\right);  \nonumber \\
&& \left(-sn(1,\alpha),sn(1,\alpha)\right);\,\left(sn(1,\alpha),
sn(1,\beta)\right);\, \left(sn(1,\beta),\infty\right)  \label{777}
\end{eqnarray}
where $\alpha=a/L$ and $\beta=b/L$, and
\begin{eqnarray}
sn(x,y)\equiv sn\left(K(m)x+iK^{\prime}(m)y,\,m\right).
\label{898}
\end{eqnarray}
Here the normalized length and width variables $y$ and $x$, respectively, are 
introduced so that $y\div(0,1),\,x\div(-1,1)$. 
These electrodes are at the respective potentials 
$0;\zeta,0,\zeta,0$ determined by the boundary conditions. Using the
Poisson's integral formula of the Dirichlet's problem for the upper half
plane, the electric field has been obtained which satisfies to the 2D Laplace 
equation
\begin{equation}
{\bf E^{\prime}}(x,y)= \left({\frac{[sn(x,y)+sn(1,\alpha)][sn(x,y)-sn(1,%
\beta)]}{[sn(x,y)-sn(1,\alpha)][sn(x,y)+sn(1,\beta)]}}\right)^{\delta/\pi}
\label{888}
\end{equation}
For the $a=0$ and $b=L$ limiting case, Eq.~\r{888} gives the electric field
distribution for a uniform magnetic field in terms of the Jacobi elliptic 
functions which is in agreement with that of obtained in Ref.~\cite{rendell} 
in the form of an expansion into hyperbolic functions. 
The field (\ref{888}) has singularities in the corners of the magnetic 
interfaces. The same behavior exhibits the electric field in the Hall effect
regime. In both cases an analytic function changes its phase from $0$ to 
$\zeta$ in one point which results a power-law singularity of this function 
at that point.

The above field $E^{\prime}$ does not conserve current and the normal 
component of $\vec{J}$ jumps at the magnetic interfaces. We can remedy
this as follows: in the planar regions the real electric field 
${\vec E}(x,y)$ is given by Eq.~\r{888} while in the non-planar region, 
due to the jump of the $y$-component of the field, we have 
\begin{equation}
E_{y}(x,y)={1\over  \cos^2\zeta}{\bf Re}\,{\vec{E}^{\prime}}(x,y)-
\tan\zeta \ {\bf Im}\,{\vec{E}^{\prime}(x,y)},\,
E_{x}(x,y)={\bf Im}\,{\vec{E}^{\prime}}(x,y).
\label{999}
\end{equation}
The above equation together with the auxiliary Eq.~\r{888} gives the 
solution of our problem and satisfies the Poisson equation 
corresponding to Eqs.~\r{222} and \r{666}. 
Now it is clear that in the whole region we have $J_{n}(\vec{E})=
\sigma E^{\prime}_{n}$, 
and since ${\bf div} \vec{E}^{\prime}=0$, the current is conserved at the
magnetic interfaces.

\section{Results and Discussion}
\la{results}

\begin{figure}[htb]
\epsfxsize=14cm
\epsfysize=13cm
\mbox{\hskip 0cm}\epsfbox{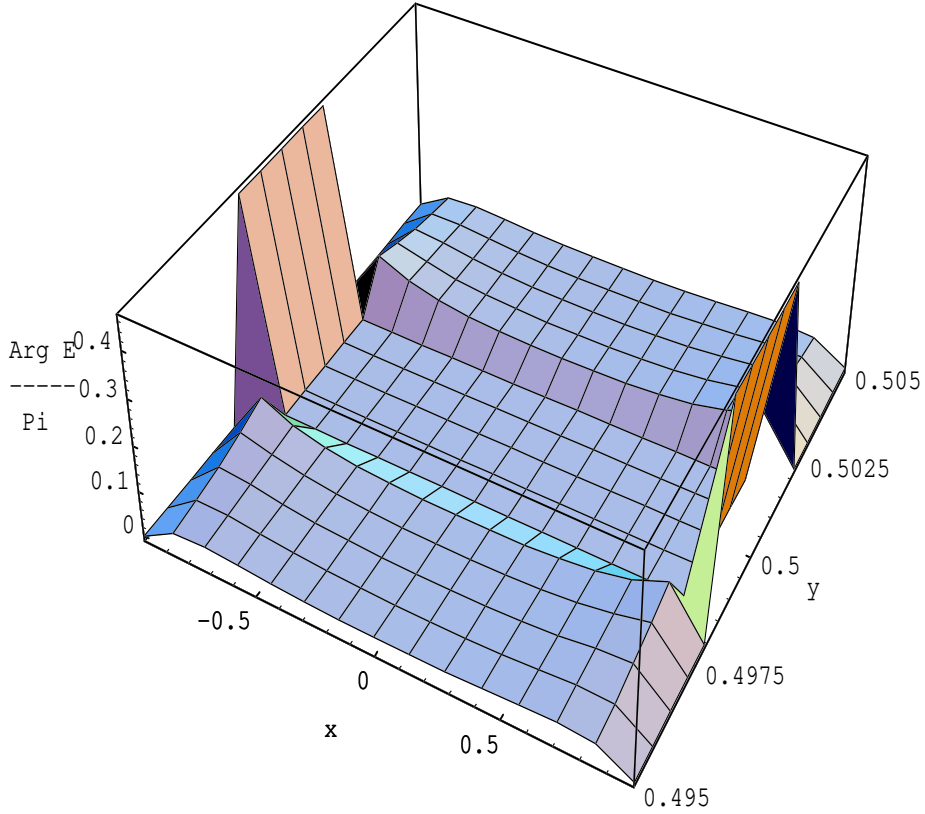}
\caption{Spatial distribution of the argument (in units of $\pi$) 
of $\vec{E}$ for the facet situated between the points $498.5$ and 
$501.5\;\mu$m. $L=1000\;\mu$m,  $2w=40\;\mu$m, $B=1$ T.} 
\label{fgarg} 
\end{figure}

\begin{figure}[htb]
\epsfxsize=14cm
\epsfysize=13cm
\mbox{\hskip 0cm}\epsfbox{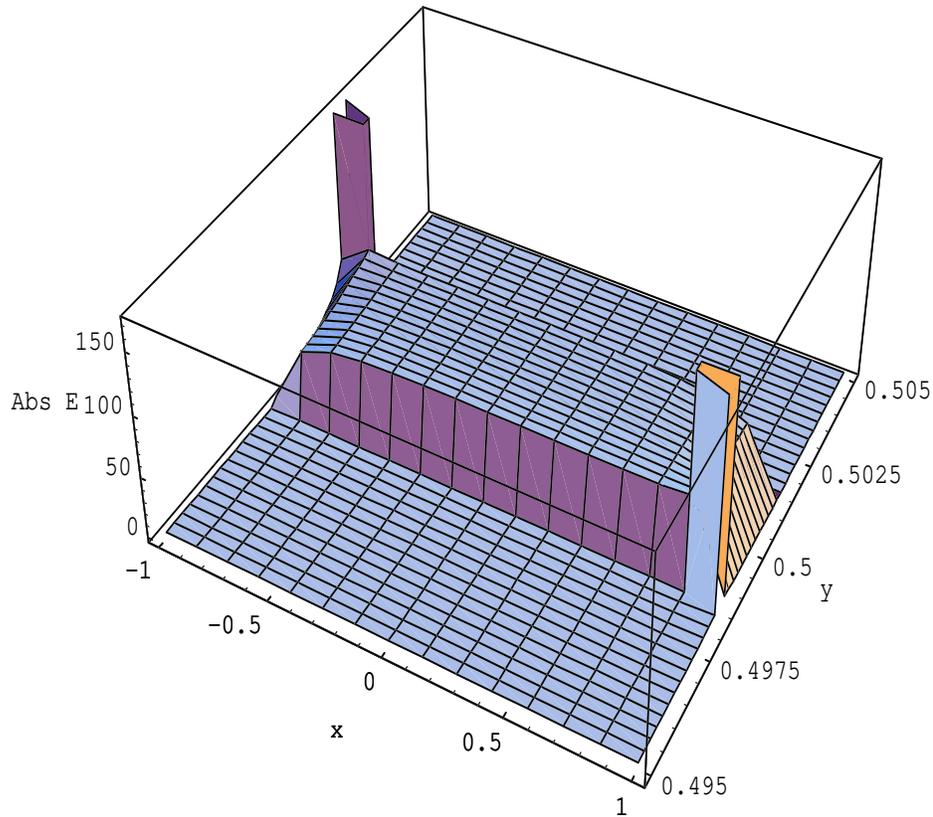}
\caption{Spatial distribution of the absolute value of $\vec{E}$ for 
the facet situated between the points $498.5$ and $501.5\;\mu$m. 
$L=1000\;\mu$m,  $2w=40\;\mu$m, $B=1$ T.} 
\label{fgabs} 
\end{figure}

The spatial distributions of the argument ${\bf Arg}{\vec E}(x,y)$ and the 
absolute value ${\bf Abs}{\vec E}(x,y)$ of the electric field 
${\vec E}(x,y)$ in the neighborhood of the facet are shown in 
Figs.~\ref{fgarg} and \ref{fgabs}. 
One can see that 
both components of the electric field are small outside  the facet region.
The field  exhibits power law singularity in the diagonally opposite 
corners of the facet while in the two other corners ${\vec E}(x,y)=0$. 
Near the sides of the planar and non-planar regions, ${\bf Arg}{\vec E}(x,y)$
is near to zero and $\zeta$, respectively. ${\bf Arg}{\vec E}(x,y)$ has 
local maximum along the $y$-axis near the edges of the magnetic barrier 
and sharply drops at the magnetic interface remaining near to zero in the 
whole non-planar region. The current flows between the ends of the 2DEG 
crossing the magnetic interfaces mainly in the singular points, {\it i.e.} it 
passes along the points for which ${\bf Arg}{\vec E}(x,y)$ is close to
zero. Electrons entering or exiting the small regions of the facet corners 
will have large velocities proportional to the electric field at these 
locations to account for current conservation with a large number of
electrons drifting with slow and uniform velocities in the middle of the 
facet were the electric field is smaller and uniform. Such an electric 
field distribution differs from the one in a uniform magnetic field.

\begin{figure}[htb]
\epsfxsize=14cm
\epsfysize=13cm
\mbox{\hskip 0cm}\epsfbox{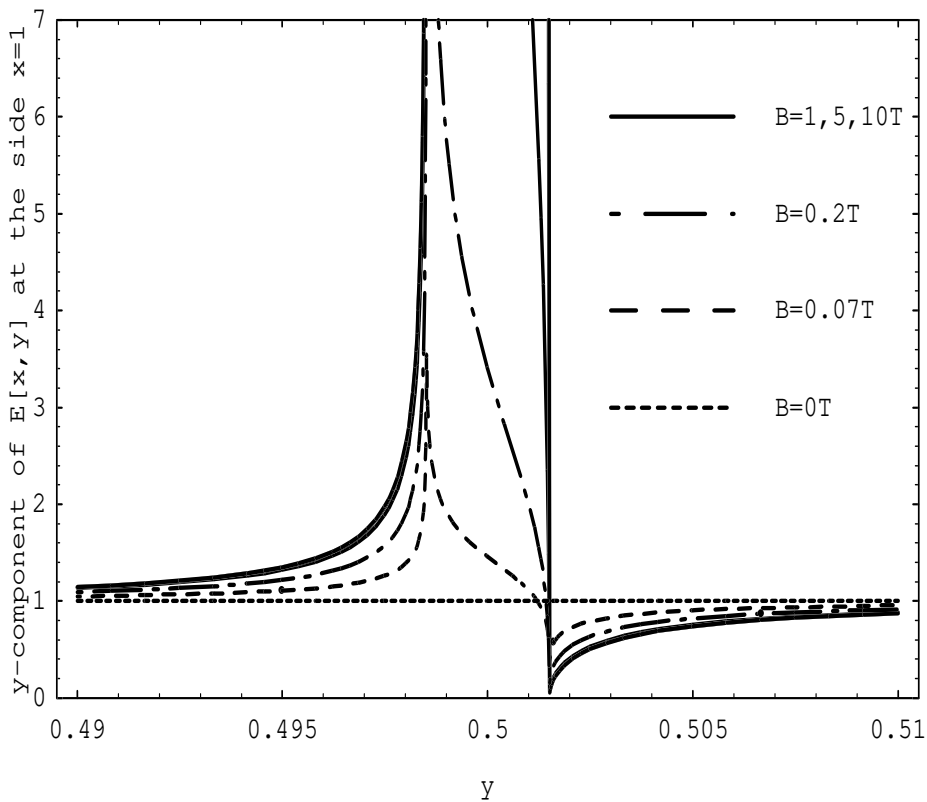}
\caption{The electric field profile at the edge of the 2DEG for different 
values of the magnetic field.} 
\label{fg43} 
\end{figure}

\begin{figure}[htb]
\epsfxsize=12cm
\epsfysize=12cm
\mbox{\hskip 0cm}\epsfbox{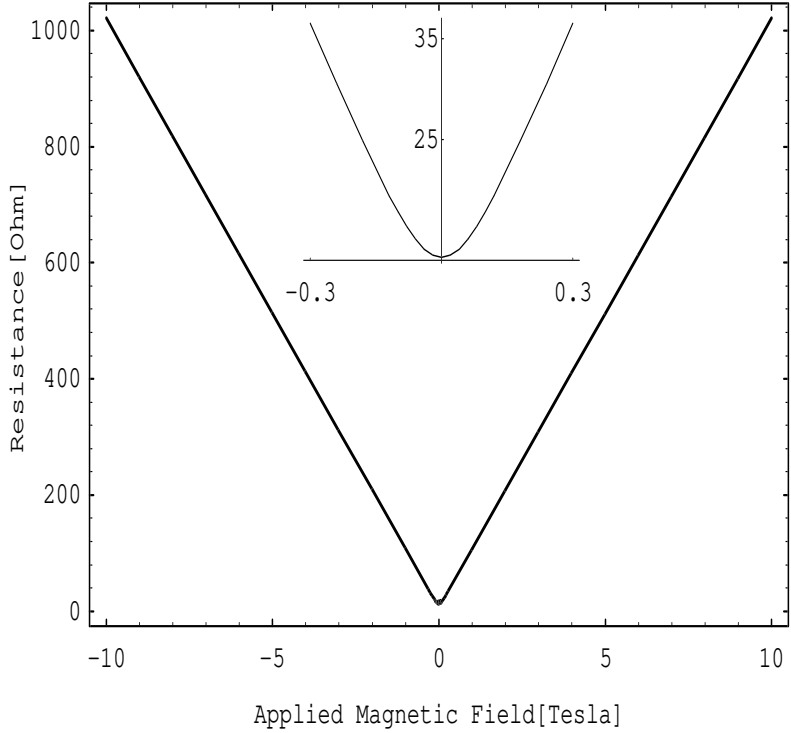}
\caption{Magneto-resistance across the facet corresponding to 
for $L(3-4)=255$ $\mu$m, $L(5-6)=220$ $\mu$m, and $L(2-3)=200$ $\mu$m.} 
\label{fg44} 
\end{figure}

The magneto-resistance is determined by the electric field edge profile
which is shown in Fig.~\ref{fg43} for different values of the magnetic field.
The magnetic field dependence of the magneto-resistance across the 
facet is shown in Fig.~\ref{fg44}. The classical origin of the 
positive background has been confirmed experimentally where it has 
been found that it persists at temperatures higher than $100$ K. Note 
that experimental configuration is effectively a two terminal measurement
where the measured resistance is determined both by the Hall resistance
and magneto-resistance. For small magnetic fields the Hall resistance is 
small and thus the resistance is determined by the magneto-resistance 
and consequently the resistance increases as $B^2$. For larger magnetic
fields a quasi-linear behavior of the resistance as a function of $B$ is 
found which is due to the fact that now it is Hall resistance which mainly
limits the current.
However, the resistance increases from $13.2$ to $1020.8$ $\Omega$ 
when $B: 0\rightarrow 10$ T, which is approximately four times less than 
observed experimentally. The reason for such a difference could be the low 
ratio of the facet ($3$ $\mu m$) to the mesa ($1000$ $\mu m$) length.
\begin{figure}[htb]
\epsfxsize=12cm
\epsfysize=12cm
\mbox{\hskip 0cm}\epsfbox{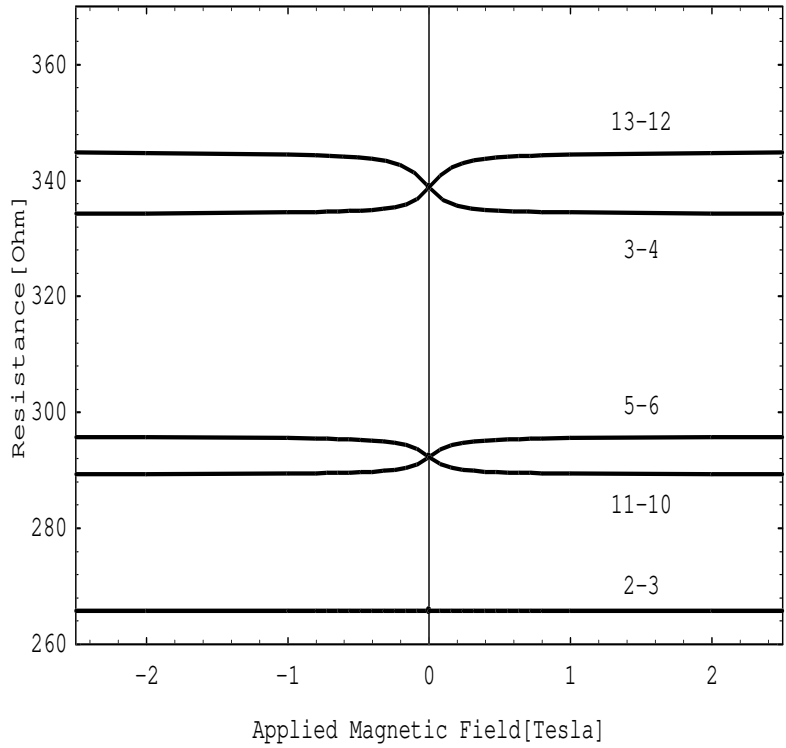}
\caption{Magneto-resistance in the planar regions corresponding to 
for $L(3-4)=255$ $\mu$m, $L(5-6)=220$ $\mu$m, and $L(2-3)=200$ $\mu$m.} 
\label{fg45} 
\end{figure} 
The figure Fig.~\ref{fg45} shows the magneto-resistance calculated in the 
planar regions. The top two curves are the resistance for probes above the 
facet on the left (12-13) and right (3-4) of the mesa and the center two 
curves show the pairs (10-11) and (5-6) below the facet (see 
Ref.~\cite{mark1}). 
In agreement with the experiment, pairs of probes on opposite sides of the 
mesa and on opposite sides 
of the facet show the same symmetry while there is a strong asymmetry in 
each of the traces with reversal of the direction of the  magnetic field. 
These results agree qualitatively with experiment but there are problems 
with the quantitative values of the resistance.
The main variation of the resistance takes place for small $B$ and the 
magnitude of the variation strongly decreases with distance from the facet. 
Therefore, in the scale of Fig.~\ref{fg45} one cannot see the resistance 
variation for probes (2-3) situated $300$ $\mu$m away from the facet. 

\chapter{ Summary}

Carrier interaction with phonons, photons, impurities, and electrons have 
been addressed in semiconductor nanoscale systems with carrier confinement 
in one and two dimensions subjected to quantizing magnetic fields and 
without it. Most importantly, our calculations allow to better understand
phonon signature in optical, thermalization, and transport experiments
that can be used to identify and characterize the basic phenomena of 
quantum confinement in these quantum nanostructures.
\begin{itemize}
\item{
In the frames of the polaron problem, the peculiarities of the magneto-polaron
spectrum near the longitudinal optical phonon emission threshold have been 
investigated in the 2DEG in the QHE geometry. In spite of weak electron-phonon
coupling, an infinite set of new complex quasiparticles, electron-phonon bound 
states, exists in the magneto-polaron spectrum which is coagulated to the 
threshold both above and below from it \cite{badalo2a,badalbs}.
}
\item{
The fine structure of cyclotron-phonon resonance due to the electron-phonon
bound states has been revealed \cite{badalcpr}.
Absorption of the electromagnetic field entirely governed by the bound states 
with the total angular momentum $\pm1$. According to the perturbation theory, 
photon absorption was to be expected at the phonon emission threshold. In reality,
the absorption spectrum consists of two groups of peaks which constitute an 
asymmetric "doublet" relative to the threshold.
}
\item{
A new method for calculating the probability of electron scattering from the 
deformation potential of acoustic phonons has been proposed \cite{badalinteff}. 
Such a probability summed over all phonon modes of the layered elastic medium
is expressed in terms of the elasticity theory Green function which contains all 
information about structure geometry.
}
\item{
Exploiting this method, the relaxation rates for a Fermi 2DEG located
in the vicinity of an interface between elastic semi-spaces have been
calculated \cite{badalsemispc}.
Analysis of limiting cases for an interface between solid and liquid 
semi-spaces, for a free and rigid surfaces has shown that there are situations 
when the phonon reflection from various interfaces alters the energy 
(or electron temperature) dependence of the relaxation times and 
leads to a strong reduction of the relation rates.
}
\item{
The interface effect is obtained to be highest in quantizing magnetic fields 
\cite{badalfree} since the 2DEG interacts with almost monochromatic cyclotron 
phonons in this case. 
The electron transition probability between discrete Landau levels of the 2DEG 
has an oscillating behavior of the magnetic field and of the distance from the 
2DEG to the interface.
}
\item{In quantizing magnetic fields, scattering from the piezoelectric potential of 
acoustic phonons is strongly suppressed with respect to the deformation 
interaction mechanism \cite{badalrspt}. 
}
\item{ Emission of ballistic acoustic phonons at electron transitions between 
fully discrete Landau levels in a 2DEG with account of the phonon reflection 
from a GaAs/AlGaAs type interface has been studied \cite{badalbal,badalsurf}. 
In accordance with the experimental results, the angular distribution of emitted 
phonons has a sharp expressed peak for small angles around the magnetic field.
Account for the interface effect affects essentially the intensity and the 
composition of the detected phonon field.
Under the deformation electron-phonon interaction on the sample reverse face,
the detector records both the {\it interference} field of the LA phonons and, 
which is most intriguing for experiment, the {\it conversion} field of the TA 
phonons \cite{badalbal}.
Emission of surface acoustic phonons is suppressed exponentially in a wide 
range of the magnetic field variation \cite{badalsurf}.
}
\item{
Polar optical PO-phonon assisted electron relaxation in the 2DEG in the
QHE geometry has been calculated \cite{badalpoda}. The surface optical SO 
phonon relaxation has been obtained to be at least by an order weaker than 
relaxation via bulk PO phonon emission. 
Account for the Landau level broadening  and for the PO phonon dispersion 
results to the finite relaxation rates associated with one-phonon emission.
The dispersion contribution gives rise to a sharp peak with the peak value
approximately $0.17$ fs$^{-1}$. The broadening contribution has a rather 
broad peak with relatively lower peak value. 
}
\item{
Two-phonon emission is a controlling relaxation mechanism above 
the phonon energy $\h\o_{PO}$ \cite{badalpoda}. 
Immediately above $\h\o_{PO}$, the PO+DA phonon relaxation rate 
increases as a $B^5$ achieving to the peak value exceeding $1$ ps $^{-1}$ 
at energy separations of the order of $\h s/a_{B}$ .
In the energy range $sa^{-1}_{B}\lesssim\Delta l\omega_{B}-\o_{PO}
\lesssim sd^{-1}$, the two-phonon peak decreases linearly in $B$. 
so that within the wide energy range of the order of $\h\o_B$, 
subnanosecond relaxation between Landau levels can be achieved.
}
\item{
We have calculated the inter edge state scattering length for an arbitrary
confining potential \cite{bad,maslov}. Phonon (deformation acoustic DA 
and piezoelectric PA interactions) and impurity scatterings are discussed 
and analytical expressions are derived. As follows from energy and momentum 
conservation, only phonons with frequencies above some threshold can 
participate in the transitions between edge states. As a result, phonon 
scattering is exponentially suppressed at low temperatures. The observed 
temperature dependence of the scattering length cannot be attributed to 
phonon scattering.
}
\item{Ballistic acoustic phonon emission (both for DA and PA interactions) 
by quantum edge states has been investigated \cite{badaledbal,badaledbal1}. 
At low temperatures the emitted acoustic phonon field is predominantly
concentrated within a narrow cone around the direction of the 
edge state propagation, while at high temperatures -- around 
the magnetic field normal to the electron plane.
At low temperatures the emission intensity decreases exponentially 
with decreasing filling of Fermi level.
The relative contributions of PA and DA interactions depend on 
the magnetic field, the shape of confining potential, and 
temperature.}
\item{
In recent magneto-luminescence experiments by Potemski {\it et al.} 
\cite{potemski1,potemski2} on one-side modulation doped GaAs/AlGaAs
quantum wells, an up-conversion has been observed and interpreted as 
being due to an Auger process. 
We have developed a theory of Auger up-conversion in quantum 
wells in quantizing magnetic fields to explain these experimental
results \cite{badalauger,badalrspt}. We have calculated the characteristic
times of electron-electron scattering processes between Landau levels of 
the lowest electric subband and of electron-acoustic phonon scattering 
between Landau levels of the two lowest electric subbands as well as the 
lifetime of a test hole with respect to both the Auger process and phonon 
emission. By analyzing rate equations for these processes as well as
for the pumping by interband excitation and the recombination of 
electrons with photo-induced holes, we have found the Auger process
time. As well the magnetic field and the excitation power 
dependencies of the two luminescence peaks have been obtained
which are consistent with the experimental findings. Thus, an 
understanding of the Auger up-conversion observed in the 
magneto-luminescence in quantum wells is provided.
}
\item{
Recently a research group from the Toshiba Cambridge Research 
Center Ltd. and Cavendish Laboratory have proposed a new technique
to produce non-homogeneous magnetic fields 
\cite{foden1,mark1}.
A remotely doped GaAs/Al$_x$Ga$_{1-x}$As heterojunction is grown 
over wafer previously patterned with series of facet. 
The 2DEG is no longer planar but follows the contour of the original wafer.
Application of a homogeneous magnetic field to this structure results
a spatially varying field component normal to the 2DEG. 
We have investigated theoretically the magneto-transport of the 
non-planar 2DEG \cite{badaltrs1,badaltrs2,badaltrs3,badaltrs4,ibrahim}.
The magneto-resistance across the facet as well as in the planar regions 
of the 2DEG has been calculated which explain the main features of the 
magnetic field dependence observed experimentally by M L Leadbeater 
{\it et al} \cite{mark1,mark2}.
}
\end{itemize}

\prefacesection{Acknowledgements}  
I thank Prof. Y B Levinson, U R\"ossler, and F. Peeters for the 
common work on the problems presented in the dissertation. These were 
the important stages in my research experience.\\
I am grateful to Prof. V M Harouthounian for his support during my work
in the semiconductor division of YSU.\\
I am thankful to all my colleagues from Radiophysics and Solid state 
physics departments of YSU, from the theoretical group of YrPhI, 
from Institute of Microelectronics in Chernogolovka, from Regensburg
and Antwerp Universities.
I acknowledge especially to Prof. G W Bryant, L J Challis, A V Chaplik, 
E M Kazaryan, J Keller, A A Kirakosyan, D L Maslov, A H Melikyan, H R Minasyan 
S G Petrosyan, \'E I Rashba, K F Renk, A G Sedrakyan, and A Shik, for
helpful discussions, suggestions, and comments.

\appendix
\addcontentsline{toc}{chapter}{Appendix}
\chapter{}
\label{appa} 

In this appendix we collect in a table some physical and material parameters 
for GaAs/Al$_x$Ga$_{1-x}$As heterojunction based quantum nanostructures.
\begin{table}[h]
\caption{Some physical and material parameters used in the dissertation.}
\label{tb4}
\begin{center}
\begin{tabular}{|c|cc|cc|cc|}
\hline\hline
{Physical and material parameters} 
& \multicolumn{6}{|c|}{GaAs} \\ 
& \multicolumn{5}{c}{} &  \\ \hline\hline
{Electron effective mass, $m_c$} 
& \multicolumn{6}{|c|}{$m_c=0.066m_e$
\footnote{$m_e$ is the free electron mass}} \\ \hline
{Crystal mass density, $\rho[\mbox{kg m}^{-3}]$} 
& \multicolumn{6}{|c|}{$\rho=5.31\cdot 10^{-3}$} \\ \hline
{Long. ac. LA ph. vel. av. in angles, $s[\mbox{m s}^{-1}]$} 
& \multicolumn{6}{|c|}{$s=5.14\cdot 10^3$} \\ \hline
{Trans. ac. TA ph. vel. av. in angles, $c[\mbox{m s}^{-1}]$} 
& \multicolumn{6}{|c|}{$c=3.13\cdot 10^3$} \\ \hline
{Cyclotron energy, $\h\o_B[\mbox{meV}]$} 
& \multicolumn{6}{|c|}{$1.75 B[\mbox{T}]$} \\ \hline
{Magnetic length, $a_B[\mbox{nm}]$} 
& \multicolumn{6}{|c|}{$25.66 B^{-1/2}[\mbox{T}]$} \\ \hline
{Landau level degeneracy, $(2\pi a^2_B)^{-1}[\mbox{m}^{-2}]$} 
& \multicolumn{6}{|c|}{$2.417\cdot 10^{14} B[\mbox{T}]$} \\ \hline
{Energy scale, $\h s/a_B[\mbox{meV}]$} 
& \multicolumn{6}{|c|}{$0.13 B^{1/2}[\mbox{T}]$} \\ \hline
{Energy scale, $\h s/d[\mbox{meV}]$} 
& \multicolumn{6}{|c|}{$3.37/ d[\mbox{nm}]$} \\ \hline
{Subband spacing, ${\h^2\over2m_c}{\pi^2\over d^2}[\mbox{meV}]$} 
& \multicolumn{6}{|c|}{$3.76{m_e\over m_c}{10\over d[\mbox{nm}]}^2$} 
\\ \hline
& \multicolumn{6}{|c|}{Al$_x$Ga$_{1-x}$As} \\ 
& \multicolumn{2}{|c}{$x=0$} 
& \multicolumn{2}{c}{$x=0.3$} 
& \multicolumn{2}{c|}{$x=1$} \\ \hline
{High frequency dielectric constant, $\kappa_{\infty}$} 
& \multicolumn{2}{|c|}{10.9} 
& \multicolumn{2}{|c|}{$12.0$} 
& \multicolumn{2}{|c|}{$8.5$} \\ \hline
{Long. opt. LO ph. energy, $\h\o_{LO}[meV]$} 
& \multicolumn{2}{|c|}{$36.62$} 
& \multicolumn{2}{|c|}{$49.80$} 
& \multicolumn{2}{|c|}{$50.00$} 
\\ \hline 
{Trans. opt. TO ph. energy, $\h\o_{TO}[meV]$} 
& \multicolumn{2}{|c|}{$33.30$} 
& \multicolumn{2}{|c|}{$45.10$} 
& \multicolumn{2}{|c|}{$44.90$} \\ \hline
& \multicolumn{2}{|c|}{} 
& \multicolumn{2}{|c|}{34.57} 
& \multicolumn{2}{|c|}{34.82}\\ 
{Surf. opt. SO ph. energy, $\h\o_{SO}[meV]$} 
& \multicolumn{2}{|c|}{} 
& \multicolumn{2}{|c|}{$47.87$} 
& \multicolumn{2}{|c|}{$47.49$} \\ \hline
& \multicolumn{2}{|c}{$\alpha_{PO}$}
& \multicolumn{4}{c|}{$\alpha_{SO}/\alpha_{PO}$} \\
{Fr\"olich coupling constant} 
& \multicolumn{2}{|c}{$0.07$}
& \multicolumn{2}{c}{$1.96$} 
& \multicolumn{2}{c|}{$1.95$} \\
& \multicolumn{2}{|c}{}
& \multicolumn{2}{c}{$0.044$} 
& \multicolumn{2}{c|}{$0.035$} \\
\hline\hline
\end{tabular}
\end{center}
\end{table}

\chapter{}
\label{appndphi}
The integrals $\Phi_m$ in Eqs.~\r{phihat} are given explicitly by the
following formulas.

For the solid state-liquid contact
\be
{\hat{\Phi}_m}(\xi_S,\nu,\mu,\delta,r)=
{\hat{\Phi}^{B_1}_m}(\nu,\mu,\delta)+
{\hat{\Phi}^{B_2}_m}(\nu,\mu,\delta)+
{\hat{\Phi}^L_m}(\nu,\mu,\delta)
+{\hat{\Phi}^S_m}(\xi_S,\nu,\mu,\delta,r)
\la{sllq}
\ee
where
\bea
{\hat{\Phi}^{B_1}_m}&=&{2^{2+\,m}\over\pi}\,\int_0^1dt\, {t^{2\,m}}\,
{{{{\left( {{\delta \,{{\nu }^4}}\over {{\sqrt{{{\mu }^2} - {t^2}}}}} + 
4\,{t^2}\,{\sqrt{{{\nu }^2} - {t^2}}} \right) }}}\over 
{{{ {{\left( {{\nu }^2}-2\,{t^2} \right) }^2} +  {{\delta \,{{\nu }^4}\,
{\sqrt{1 - {t^2}}}}\over {{\sqrt{{{\mu }^2} - {t^2}}}}}+ 4\,{t^2}\,
{\sqrt{1 - {t^2}}}\, {\sqrt{{{\nu }^2} - {t^2}}}}}}},
\la{sllq1}
\\
{\hat{\Phi}^{B_2}_m}&=&{2^{2 + m}\over\pi}\,\int_1^\nu dt\,t^{2\,m}\,
{{\left({{\delta \,{{\nu }^4}}\over{{\sqrt{{{\mu }^2} - 
{t^2}}}}} + 4\,{t^2}\,{\sqrt{{{\nu }^2} - {t^2}}}\right)
\,\left({{\nu }^2} - 2\,{t^2} \right)^2}
\over
{\left({\delta\,\nu^4\sqrt{t^2-1}\over\sqrt{\mu^2-t^2}}+ 
4\,t^2\,\sqrt{t^2-1}\sqrt{\nu^2 -t^2}\right)^2 +
\left( {{\nu }^2} - 2\,{t^2} \right)^4}},
\la{sllq2}
\\
{\hat{\Phi}^{L}_m} &=& {2^{2 + m}\over\pi}\,\int_\nu^\mu dt\,t^{2\,m}\,
{{{{\delta \,{{\nu }^4}}\over{{\sqrt{{{\mu }^2} - 
{t^2}}}}}\,\left({{\nu }^2} - 2\,{t^2} \right)^2}
\over
{\left(4\,t^2\,\sqrt{t^2-1}\sqrt{t^2-\nu^2}-
\left({{\nu }^2} - 2\,{t^2} \right)^2 \right)^2 +
\delta^2\,\nu^8{t^2-1\over \mu^2-t^2}}}
\la{sllq3}
\\
{\hat{\Phi}^{S}_m}&=&
{2^{2 + m}}\;{\left({{{{\it \xi_S}}\over {\nu }}}\right)^{2\,m}}\;
{{\left( {\it q_S} - {{\delta \,{{{\it \xi_S}}^4}}\over {4\,r}}\right) }\over 
{{{{{\left( {\it p_S} - {\it q_S} \right) }^2}}\over 
{{\it p_S}\,{\it q_S}}} + 2\,\left( {\it p_S} - {\it q_S} \right) \,
{\it q_S} + {{\delta \,\left( {{{\it p_S}}^2} - {r^2} \right) \,
{{{\it \xi_S}}^4}}\over {4\,{\it p_S}\,{r^3}}}}}.
\la{sllq4}
\eea

For the free crystal surface
\be
{\hat{\Phi}_m}(\xi_R,\nu )={\hat{\Phi}^{B_1}_m}(\nu )+
{\hat{\Phi}^{B_2}_m}(\nu )+{\hat{\Phi}^R_m}(\xi_R,\nu)
\la{free}
\ee
where
\bea
{\hat{\Phi}^{B_1}_m}&=&{{2^{2 + m}}\over \pi}\int_0^1 dt\,{t^{2\,m}}
{\,4\,{t^{2}}\,{\sqrt{{{\nu }^2} - {t^2}}}\over 
{ {{\left( {{\nu }^2}-2\,{t^2} \right) }^2} + 
4\,{t^2}\,{\sqrt{1 - {t^2}}}\,{\sqrt{{{\nu }^2} - {t^2}}} }},
\la{free1}
\\
{\hat{\Phi}^{B_2}_m}&=&{2^{2 + m}\over\pi}\,\int_1^\nu dt\,{t^{2\,m}}
{{{4\,t^{2}}\,{{\left( {{\nu }^2} - 2\,{t^2} \right) }^2}\,
{\sqrt{{{\nu }^2} - {t^2}}}}\over { {{\left( {{\nu }^2} - 
2\,{t^2} \right) }^4} + 16\,{t^4}\,\left( {{\nu }^2} - {t^2} \right) \,
\left({t^2} - 1 \right)}},
\la{free2}
\\
{\hat{\Phi}^R_m}&=&
{2^{2 + m}}\;{\left(\xi_R \over \nu \right)^{2\,m}}\; 
{p_R\,{q_R}^2 \over 2\,p_R\,\left( p_R - q_R\right) \,{{q_R}^2} + 
\left( p_R - q_R \right)^2}.
\la{free3}
\eea

For the rigid boundary
\be
{\hat{\Phi}_m}(\nu )={\hat{\Phi}^{B_1}_m}(\nu )+{\hat{\Phi}^{B_2}_m}(\nu ) 
\la{rgd}
\ee
where
\bea
{\hat{\Phi}^{B_1}_m}&=&{{2^{2 + m}}\over \pi}
\int_0^1 dt\,{t^{2\,m}}\,{{\sqrt{{{\nu }^2} - {t^2}}}\over 
{ {{t^2} } + {\sqrt{1 - {t^2}}}\,{\sqrt{{{\nu }^2} - {t^2}}} }},
\la{rgd1}
\\
{\hat{\Phi}^{B_2}_m}&=&{2^{2 + m}\over\pi}\,\int_1^\nu dt\,{t^{2\,m}}
{{{t^{2}}\,{{\sqrt{{{\nu }^2} - {t^2}}}}\over 
{ {t^4}+\left( {{\nu }^2} - {t^2} \right) \,\left({t^2} - 1 \right) }}}.
\la{rgd2}
\eea

In formulas \r{sllq}-\r{rgd2} following notations are introduced
\bea
&&\delta={{\rho '}\over {\rho }},\: \mu={s\over {s'}},\:\xi_S={{{\it c_R}}\over c},\\
&&q={\sqrt{1 - {{\xi }^2}}},\: p={\sqrt{1 - {{{c^2}\,{{\xi }^2}}\over {{s^2}}}}},\:
r={\sqrt{1 - {{{c^2}\,{{\xi }^2}}\over {{{s'}^2}}}}}.
\la{notations}
\eea
In the case of the solid state-liquid contact, when the densities of the 
contacted semi-spaces differ strongly, we can calculate $\hat{\Phi}_m $ 
for the extreme value of the parameter $\nu=1$
\bea
&&\hat{\Phi}_0=1+2\delta(1-2\mu^2+3\mu^4/2),\:\hat{\Phi}^S_0=0 ,
\\
&&\hat{\Phi}_0=1+2\delta\mu^2(1-3\mu^2+5\mu^4/2),\:\hat{\Phi}^S_1=0 .
\la{asymptot}
\eea
Notice that in this case, relaxation is determined by the bulk and "leaky"
waves. The contribution of the "leaky" waves is proportional to $\delta\ll1$.

For a free crystal surface taking $\nu=0.59$ \cite{gantlev} for GaAs, we 
obtain from the dispersion equation of the Rayleigh waves that 
$\xi_R\approx 0.917$. Then numerical evaluation of \r{free}-\r{free3} gives:
\be
{\hat{\Phi}^{B_1}_0}\approx0.625,\: {\hat{\Phi}^{B_2}_0}\approx 0.297, \:{\hat{\Phi}^{R}_0}\approx 1.706,\:
{\hat{\Phi}_0}\approx 2.628,
\la{freenum0}
\ee
and
\be
{\hat{\Phi}^{B_1}_0}\approx 0.873,\: {\hat{\Phi}^{B_2}_0}\approx 0.954, \:
{\hat{\Phi}^{R}_0}\approx 0.999,\:
{\hat{\Phi}_0}\approx 2.826.
\la{freenum1}
\ee
Thus, we see that in the case of energy relaxation, the largest contribution 
comes from the surface Rayleigh waves while in the case of momentum 
relaxation, contributions of different phonon modes differ slightly.

Similar calculations for the rigid boundary give
\be
{\hat{\Phi}^{B_1}_0}\approx1.265,\: {\hat{\Phi}^{B_2}_0}\approx 0.452, 
\:{\hat{\Phi}_0}\approx 1.717,
\la{rgdnum0}
\ee
and
\be
{\hat{\Phi}^{B_1}_1}\approx 0.852,\: {\hat{\Phi}^{B_2}_1}\approx 1.407, 
\:
{\hat{\Phi}_1}\approx 2.259.
\la{rgdnum1}
\ee
In this case the interface effect is stronger for momentum relaxation.

\bibliographystyle{unsrt}
\bibliography{thes}

\newpage

\section*{List of publications underlying the dissertation} 
\addcontentsline{toc}{chapter}{List of publications underlying the dissertation}
\begin{enumerate}
\item{S~M Badalyan and A~H Melikyan.
 Threshold anomalies in disintegration of an optic phonon into two
  acoustic phonons in one dimensional anharmonic crystal.
 {\em Sov. Phys. Low Temp.}, {\bf 13}:670, 1987.}
\item{S~M Badalyan and Y~B Levinson.
 Bound states of electron and optical phonon in a quantum well.
 {\em Sov. Phys. JETP}, {\bf 67}:641--645, 1988.}
\item{S~M Badalyan and Y~B Levinson.
 Cyclotron-phonon resonance in a two dimensional electron gas.
 {\em Sov. Phys. Semicond.}, {\bf 22}:1278--1281, 1988.}
\item{S~M Badalyan and Y~B Levinson.
 Effect of an interface on the scattering of a two dimensional
  electron gas from acoustic phonons.
 {\em Sov. Phys. Solid State}, {\bf 30}:1592--1597, 1988.}
\item{S~M Badalian and Y~B Levinson. Interface effect on the interaction 
of a two-dimensional electron gas with acoustic phonons in a quantizing 
magnetic field. Proceedings of 14th All-Union Conference on Theory of 
Semiconductors. Donetsk, 1989.}
\item{S~M Badalyan.
{\em Electron-Phonon Interaction in Quasi Two Dimensional Electron
 Systems}. PhD thesis, Yerevan State University, Yerevan, 1989.}
\item{S~M Badalyan.
 Scattering of electrons in a two dimensional Fermi gas by acoustic
phonons near an interface between elastic half-spaces.
 {\em Sov. Phys. Semicond.}, {\bf 3}:1087--1091, 1989.}
\item{S~M Badalian and Y~B Levinson.
 Free surface effect on the interaction of a two dimensional electron
  gas with acoustic phonons in a quantizing magnetic field.
 {\em Phys. Lett. {\bf A}}, {\bf 140}:62--66, 1989.}
\item{S~M Badalian and Y~B Levinson.
 Ballistic acoustic phonon emission by a two dimensional electron gas
  in a quantizing magnetic field with account of the phonon reflection from a
  {G}a{A}s/{A}l{G}a{A}s interface.
 {\em Phys. Lett. {\bf A}}, {\bf 155}:200--206, 1991.}
\item{S~M Badalian, Y~B Levinson, and D~L Maslov.
 Scattering of electron edge states in a magnetic field by impurities
  and phonons. {\em Sov. Phys. JETP LETT.}, 53:595--599, 1991.}
\item{D~L Maslov, Y~B Levinson, and S~M Badalian.
 Interedge relaxation in a magnetic field.
 {\em Phys. Rev. {\bf B}},{\bf 46}:7002--7010, 1992.}
\item{S~M Badalian and Y~B Levinson.
 Suppression of the emission of surface acoustic phonons from a two
  dimensional electron gas a quantizing magnetic field.
 {\em Phys. Lett. {\bf A}}, {\bf 170}:229--231, 1992.}
\item{S~M Badalian, U~R\"ossler, and M~Potemski.
 Theory of Auger up-conver\-sion in quantum wells.
 page 1433, Regensburg, Germany, 1993. 13th General Conference of the
  Condensed Matter Division, European Physical Society.}
\item{S~M Badalian, U~R\"ossler, and M~Potemski.
 Theory of Auger up-conver\-sion in quantum wells in a quantizing
  magnetic field. {\em J. Physics Cond. Mat.}, {\bf 5}:6719--6728, 1993.}
\item{S~M Badalian.
 Ballistic acoustic phonon emission by quantum edge states.
 Abstract 1022, Madrid, 1994. 14th General Conference of the
  Condensed Matter Division, European Physical Society.}
\item{S~M Badalian.
 Emission of ballistic acoustic phonons by quantum edge states.
 {\em J. Physics Cond. Mat.}, {\bf 7}:3929--3936, 1995.}
\item{S~M Badalian.
 Electron relaxation in the quantum {H}all effect geometry: One- and
  two-phonon processes.
 {\em Phys. Rev. {\bf B}}, {\bf 52}:14 781--14 788, 1995.}
\item{S~M Badalian, I~S Ibrahim, and F~M Peeters.
 Theory of the magneto-transport of electrons in a non-planar two
  dimensional electron gas.
 Abstract TuP 75, W\"urzburg, Germany, 1996. 12th International
  Conference on the Application of High Magnetic Fields in Semiconductor
  Physics.}
\item{S~M Badalian, I~S Ibrahim, and F~M Peeters.
 Magneto-transport of electrons in a non-homogeneous magnetic field.
 Abstract TuP--52, Li\'ege, Belgium, 1996. 9th International
  Conference on the "Superlattices, Microstructures and Microdevices, July
  14-19, 1996, (ICSMM-9).}
\item{I~S Ibrahim, S~M Badalian, and F~M Peeters.
 Magneto-transport of electrons through a magnetic barrier.
 Eindhoven, Holland, 1996. Semiconductor days Meeting in Eindhoven.}
\item{I~S Ibrahim, V~A Schweigert, S~M Badalian, and F~M Peeters.
 Magneto-transport of electrons in a non-homogeneous magnetic field.
 {\em Superlattices and Microstructures}, {\bf 22}:203-207, 1997.
}
\item{S~M Badalian, I~S Ibrahim, and F~M Peeters.
 Theory of the magneto-transport of electrons in a non-planar two
  dimensional electron gas.
 In G~Landwehr and W~Ossau, editors, {\em High Magnetic Fields in the
  Physics of Semiconductors II}, pages 327--330. World Scientific, Singapore,
  1997.}
\end{enumerate}

\newpage
\section*{Biographical Sketch}
\addcontentsline{toc}{chapter}{Biographical Sketch}

\begin{tabular}{lp{3.4in}}
{\bf Present employment}&\\
From 1995 & Leading researcher, super\-vi\-sor of a scientific theme, 
Yerevan State University, email:badalyan@lx2.yerphi.am\\[4pt]
{\bf Research in other}\\
{\bf institutions} &\\[4pt]
1997, 91-93& Institute for Theoretical Physics, University of Regensburg, 
Germany\\[4pt] 
1996& Department of Condensed Matter Theory, University of Antwerp, 
Belgium\\[4pt]
1996, 95& International Center for Theoretical physics, Italy\\[4pt]
1987-1991& Institute of Microelectronics Technology and 
High Purity Materials, Chernogolovka, Russia\\[4pt]
{\bf Doctorate}& \\
1989& Candidate of Physics and Mathematics (Ph.D)\\ [4pt] 
Subject of examination:&Electron-phonon interaction in 2D-electron systems \\[4pt]
Scientific advisor:& Prof. Y. B. Levinson, Institute of Microelectronics,
Chernogolovka,  Russia\\[4pt]
Official opponents:& Prof. A. V. Chaplik, 
Institute of Semiconductor Physics, Novosibirsk, Russia, 
Prof. A. A. Kirakosyan, Yerevan State University, Armenia\\[4pt]
Leading organization & Institute of Solid State Physics, Chernogolovka, 
Russia\\[4pt]
{\bf Awards}& \\
1997&Civilian research and development foundation, principle 
co-investigator, grant No. AP1-375.\\[4pt]
1994&International science foundation, grant No. RYU000.\\[4pt]
1994& "Young scientists 93" Honour of "Armenia" foundation for 
academic research.\\[4pt]
{\bf Postgraduate studies}& \\
1985-1988& Ph.D student, Institute for Physical Researches, Armenia\\[4pt]
{\bf Graduate studies}&\\
1977-1982&Department of Physics (Theoretical physics), 
Yerevan State University (Honour degree)\\[4pt]
{\bf Ten-year school}&\\
1967-1977&The secondary school No.118 (Honour degree)\\[4pt]
{\bf Languages} &Armenian, Russian, English, German\\[4pt]

\end{tabular}

\end{document}